\documentclass[12pt]{iopart}
\usepackage{cite}
\usepackage{graphicx}
\usepackage{lipsum} 
\usepackage{acronym,xspace}
\usepackage{macros}

\usepackage{color}

\usepackage{lineno}

\usepackage{setspace}
\renewcommand{\arraystretch}{1} 

\usepackage{array, multirow}   
\usepackage{booktabs}
\usepackage{tabularx}
\usepackage{rotating}
\usepackage{lscape} 
\usepackage{geometry}
\usepackage{color, colortbl}
\definecolor{Gray}{gray}{0.9}
\usepackage{makecell}
\newcommand{\mycc}{\cellcolor{Gray}}

\usepackage{hyperref}
\hypersetup{
    colorlinks=true,
    linkcolor=blue,
    filecolor=magenta,      
    urlcolor=cyan,
}

\DeclareFontFamily{OT1}{pzc}{}
\DeclareFontShape{OT1}{pzc}{m}{it}{<-> s * [1.10] pzcmi7t}{}
\DeclareMathAlphabet{\mathpzc}{OT1}{pzc}{m}{it}

\usepackage{setspace}
\renewcommand{\arraystretch}{0.6} 

\usepackage{iopams}  

\begin{document}

\title[Glitch Subtraction]{Reduction of transient noise artifacts in gravitational-wave data using deep learning}

\author{Kentaro Mogushi}

\address{Institute of Multi-messenger Astrophysics and Cosmology, Missouri University of Science and Technology, Physics Building, 1315 N.\ Pine St., Rolla, MO 65409, USA}

\vspace{10pt}
\begin{indented}
\item[]\today
\end{indented}

\begin{abstract}
Excess transient noise artifacts, or ``glitches'' impact the data quality of ground-based \ac{GW} detectors and impair the detection of signals produced by astrophysical sources. Mitigation of glitches is crucial for improving \ac{GW} signal detectability. However, glitches are the product of short-lived linear and non-linear couplings among the interrelated detector-control systems that include optic alignment systems and mitigation of seismic disturbances, generally making it difficult to model noise couplings. Hence, typically time periods containing glitches are vetoed to mitigate the effect of glitches in \ac{GW} searches at the cost of reduction of the analyzable data. To increase the available data period and improve the detectability for both model and unmodeled \ac{GW} signals, we present a new machine learning-based method which uses on-site sensors/system-controls monitoring the instrumental and environmental conditions to model noise couplings and subtract glitches in \ac{GW} detector data. We find that our method reduces 20-70\% of the excess power in the data due to glitches. By injecting software simulated signals into the data and recovering them with one of the unmodeled \ac{GW} detection pipelines, we address the robustness of the glitch-reduction technique to efficiently remove glitches with no unintended effects on \ac{GW} signals.
\end{abstract}

\section{Introduction} \label{intro}


The first detection of a \ac{GW} signal from a \ac{BBH} merger \cite{Abbott:2016GW150914} on September 14th, 2019 was the breakthrough of \ac{GW} astronomy. Since then, about 50 high confident detections \cite{GWTC-1, Abbott:2020niy, LIGOScientific:2020stg, PhysRevLett.125.101102, Abbott_2020, Abbott:2020khf} have been made by the network of \ac{GW} detectors currently consisting of the two LIGO detectors in the United States \cite{advLigo2015}, the Virgo detector in Italy \cite{advVirgo2015}, the GEO-HF detector in Germany \cite{Luck:2010rt}, the KAGRA detector in Japan \cite{Akutsu:2020his}, and eventually the LIGO-India in India \cite{LIGO-India}. One of the dominating factors for the breakthough in 2015 was the improvement of the detector sensitivity by a factor of $\sim 10$ at the detector's most sensitive frequency compared to the initial LIGO \cite{advLigo2015}. Reducing transient and periodic noise sources in instrumental and environmental origin may enhance the identification of \ac{GW} signals by increasing the significance of candidate events of astrophysical signals and avoiding classifying them as sub-threshold triggers \cite{GWTC-1, Riles:2012yw}. 

The first detection of a \ac{GW} signal from a \ac{BBH} merger \cite{Abbott:2016GW150914} on September 14th, 2019 was the breakthrough of \ac{GW} astronomy. One of the dominating factors for the breakthrough in 2015 was the improvement of the detector sensitivity by a factor of $\sim 10$ at the detector's most sensitive frequency compared to the initial LIGO \cite{advLigo2015}. Reducing transient and periodic noise sources in instrumental and environmental origin may enhance the identification of \ac{GW} signals by increasing the significance of candidate events of astrophysical signals and avoiding classifying them as sub-threshold triggers \cite{GWTC-1, Riles:2012yw}.


The physical couplings due to the detector design are noise sources such as the fluctuating amplitude of laser light in the arm cavities, fluctuations of the photon arrival time at the output-port photodiode, thermal fluctuations of mirror coatings, and optic suspension \cite{TheLIGOScientific:2014jea}. Besides, during observation runs, environmental and instrumental noise sources including wind or ground motions as well as optic-controlling systems limit the detector sensitivity \cite{TheLIGOScientific:2016zmo}. Reducing the effect of these noise sources on the detector's output, \textit{strain} channel, is crucial to improve the detectability of any \ac{GW} signals and better understand physics in the universe. 

The \ac{aLIGO} detector use approximately twenty hundred thousand auxiliary \textit{channels}, or sensors/control-subsystems that monitor different aspects of environmental and instrumental conditions inside and around the detectors in parallel to the strain channel in the time domain. These auxiliary channels can be potential witnesses of couplings of noise sources and can be used to subtract the noise in the strain channel. 

Long-lasting noise sources with a duration longer than $\sim 4$ seconds have linear and/or non-linear non-stationary coupling mechanisms (\cite{Ormiston:2020ele} set the analysis window to be 8 seconds for the long-lasting noise). Among others, the main technique to subtract linearly coupled long-lasting noise sources is to calculate coherence between the witness channels and the strain channel. For example, the main source in $\sim50-1000$ Hz frequency band coupled linearly with the jitter of the pre-stabilization laser beam in angle and size was subtracted \cite{Driggers:2018gii, Kwee:12, Robert:2016gii, Davis:2018yrz}. For non-linearly and non-stationary coupled noise sources in environmental and instrumental origin, techniques with machine learning algorithms show successful subtractions and improve the detector sensitivity \cite{Tiwari:2015ofa, Meadors:2013lja, Mukund:2020lby, Vajente:2019ycy, Ormiston:2020ele}. 


Other than long-lasting noise, transient noise artifacts, or \textit{glitches} significantly impair the quality of \ac{GW} detector data and reduce confidence in the significance of candidate \ac{GW} events because glitches may resemble astrophysical signals. Glitch removal is crucial for \ac{GW} signal searches. However, glitches are the product of \textbf{short-lived} linear and non-linear couplings among interrelated detector subsystems, generally making it difficult to find their coupling mechanisms. To mitigate the effect of glitches on \ac{GW} searches, LIGO-Virgo collaboration vetoes time periods where glitches are present \cite{Davis:2021ecd, Abbott:2020niy} or reduces the significance of \ac{GW} candidates based on the probability of the glitch presence \cite{Godwin:2020weu}. Software engines such as \ac{UPV} \cite{Isogai:2010zz(UPV)}, \ac{hVETO} \cite{Smith:2011an(hVETO)}, Pointy Poisson \cite{Essick:2020cyv}, PyChChoo \cite{mogushi2021application}, and \textsc{iDQ} \cite{Biswas:2013wfa, Essick:2020qpo} find the statistical correlation of glitches between the detector's output and numerous auxiliary channels to identify the presence of glitches while avoiding the mitigation of astrophysical signals. 

The above mitigation techniques have disadvantages: reduction of the analyzable data which might contain \ac{GW} signals or remaining the effect of the glitch contamination on the data. Therefore, it is desirable to subtract glitches for overcoming the above disadvantages. The functional forms to model the coupling mechanisms of glitches can require detailed knowledge of instrumental subsystems and a large number of parameters \cite{Was:2020ziy}. Also, there exist situations where functional forms can not be obtained because of unknown physical mechanisms about glitches. In addition to subtraction techniques for long-lasting noise sources noted above, other machine learning techniques have shown promising applications to \ac{GW} astronomy \cite{Cuoco:2020ogp}. For instance, \cite{Mukund:2016thr} present a method using various techniques including the wavelet decomposition to classify transient excess-power waveforms injected into the simulated Gaussian data and real data during the LIGO's sixth science run. \cite{Soni:2021cjy} show updated results of a glitch-classification software \textsc{Gravity Spy} \cite{Zevin:2016qwy} to the \ac{O3} data with the help of citizen scientists. \cite{Biswas:2013wfa} show the comparison of various machine-learning algorithms to predict the presence of glitches based on auxiliary channels \cite{Biswas:2013wfa}. \cite{Mogushi:2021cpw} introduce a method to estimate the data containing a \ac{CBC} signal lost due to the presence of an overlapping glitch. These successes motivate us to develop a new machine learning-based method to subtract glitches using auxiliary channels with no dependency from any astrophysical-signal waveforms and no precise prior knowledge of all the system configurations, allowing the method to be easily adaptive to changes in the detector settings.

In this paper, we present a machine learning-based algorithm that subtracts glitches in the detector's output using auxiliary witness channels. Using two classes of glitches that are adversely affecting unmodeled \ac{GW} detection searches, we characterize the performance of our subtraction technique. Adding simulated \ac{GW} signals to the data before subtraction, we validate the algorithm not to manipulate or introduce biases to the resultant estimates of the astrophysical parameters.

\section{Glitch subtraction pipeline} \label{glitch_subtraction_pipeline}
We introduce the analysis pipeline applied to the ground-based \ac{GW} data for glitch subtraction. The pipeline processes the spectrograms magnitude of time series recorded in the strain channel, $h(t)$, and a set of witness auxiliary channels for a class of glitches, $a_i(t)$. Those witness auxiliary channels do not causally follow from the detector's output where astrophysical signals are present. Choosing such channels allows the pipeline to only subtract glitches while preserving astrophysical signals. The algorithm uses a 2-dimensional \ac{CNN} which uses a user-chosen set of witness channels as the input and then outputs the predicted spectrogram magnitude of glitches in $h(t)$. The output of the \ac{CNN} is then converted to time series using the \ac{FGL} transformation \cite{1164317,6701851} and conditioned before being subtracted from $h(t)$.                   
\subsection{FORMALISM AND LOSS FUNCTION} \label{formalism_and_loss_function}
The time series $h(t)$ in the strain channel can be formulated as
\begin{equation} \label{eq:strain_equal_glitch_noise}
    s(t) = h(t) + g(t) + n(t) \,,
\end{equation}
where $h(t)$ is the astrophysical signal that may be present in the data, $g(t)$ is a user-targeted glitch waveform that is coupled with witness channels, and $n(t)$ represents the sum of untargeted glitch waveforms and the noise that are not wanted to be subtracted. We design the pipeline to produce an estimate of $g(t)$ from a set of witness channels $a_i(t)$ and subtract it from $s(t)$.

Because the amplitude of a glitch waveform varies rapidly and differently in a given frequency bin over its duration, it is more efficient to build the frequency-dependent features before the data being fed into the neural network to learn the glitch-couplings. We create the \ac{STFT} from the time series. The \ac{STFT} divides the time series into small segments and calculates a discrete Fourier transform of each of divided segments. The \ac{STFT} comprises the real and imaginary parts, which can be inverted to the corresponding time series. However, we find that the neural network more efficiently learns the wanted output with the \ac{mSTFT} than using both real and imaginary parts together because the \ac{mSTFT} has a simpler pattern than the real and imaginary parts individually have. The limitation of using the \ac{mSTFT} is that it is not invertible. To compensate for this limitation, we use the \ac{FGL} transformation \cite{1164317, 6701851} to invert the \ac{mSTFT} to the corresponding time series by estimating the phase evolution. Because the phase is not fully recovered in this transformation, the phase of the estimated glitch waveform might slightly shift from the phase of the true waveform and the estimated waveform might have the opposite sign of the amplitude. Using the least square fitting method, we allow the amplitude of the estimated glitch waveform to change up to a factor of $\pm 3$ and allow the phase to shift up to $\pm 0.02$ seconds. We find that correcting the amplitude of the estimated waveform produces a better subtraction while preserving potential astrophysical signals. Also, changing the magnitude of the estimated waveform helps us avoid unwanted data manipulation because the amplitude correction factor tends to be zero when the estimated waveform mismatches the true glitch waveform.

Because of the complexity in estimating $g(t)$ directly, we design the neural network to uses the witness channels and produce an estimate of the \ac{mSTFT} $G(t, f)$ of a glitch waveform, where $f$ denotes the frequency. The neural network can be represented as a function $\mathcal{F}(A_i(t, f); \vec{\theta})$ which maps the magnitude \acp{mSTFT} $A_i(f,t)$ of the witness channels to $G(t, f)$ given a set of parameters $\vec{\theta}$. The parameters are obtained by minimizing a loss function $J$ which denotes the difference between the real glitch \ac{mSTFT} and the predicted counterpart. The operation in the network can be formulated as
\begin{equation} \label{eq:obtain_parameters}
    \vec{\theta} = {\rm argmin}_{\vec{\theta'}}J\left[G(t, f), \mathcal{F}(A_i(t, f); \vec{\theta'} )\right]\,.
\end{equation}

In the analysis, we choose the loss function to be the \ac{MSE} across each pixel of \ac{mSTFT} as 
\begin{equation}\label{eq:loss_func}
    J = \frac{1}{N}\sum_{k=1}^{N}(G[k] - \hat{G}[k;\vec{\theta}])^2 \,,
\end{equation}
where $k$ denotes each pixel, $N$ is the total number of pixels, and $\hat{G}[k;\vec{\theta}]:=\mathcal{F}(A_i(t,f), \vec{\theta})[k]$ is an estimate of \ac{mSTFT} obtained with the neural network. 

In practice, the dimension of \ac{mSTFT} depends on the duration of $g(t)$, its sampling rate, and a user-chosen frequency resolution for \ac{mSTFT}. Using an appropriate combination of the neural internal window function called \textit{kernel} and the \ac{mSTFT} dimension, the neural network can efficiently minimize the loss function. 

\subsection{DATA PRE-PROCESSING} \label{data_preprocessing}
The target class of glitches and its witness channels are chosen to create the data set for the neural network. We select a target glitch class that is labeled by a machine-learning classification tool called \textsc{Gravity Spy} \cite{Zevin:2016qwy} which classifies glitches based on their \ac{TFR} morphology using the Q-transform \cite{ROBINET2020100620}. \textsc{Gravity Spy} might misclassify some glitches by assigning inappropriate names. Also, some glitches are not loud enough to be worth to be subtracted. Because of the above two reasons, we select a set of glitches in the class with a relatively high \ac{SNR} threshold (e.g., 10) and classification confidence level (e.g., 0.9) for the glitches. 

To find the witness channels for these glitches, we use the software called \tool which allows us to analyze safe auxiliary channels in the coincident windows for the glitches and discover witness channels statistically. Because of the computational efficiency for training the neural network and the achievement of an efficient prediction made by the network, using only witness channels without non-witness channels is sufficient. We use the auxiliary channels having the probability that the glitch set is louder than the quiet set greater than 0.9 as the witness channels. Because different noise couplings might produce a similar \ac{TFR} morphology, not all of the glitches in the class have a strong correlation with excess power recorded in the witness channels. Therefore, we select the subset of glitches in a class with their top-ranked witness channel has the probability of an excess-power measure belonging to the glitch set, above 0.9 \cite{mogushi2021application} to make sure that the subset of glitches has a strong correlation with the witness channel. Around the time of glitches, we use a set of time series of the strain and the witness channels with a duration of 36 seconds. Glitches are excess power transients that are distinctively different from long-term varying noise sources. To filter out the long-term varying noise and magnify the characteristic of glitches, we whiten the time series by calculating the convolution between the time series and the time-domain {\it finite-impulse-response} filter created with the median value of \acp{ASD}, where each \ac{ASD} is the square root of \ac{PSD} obtained by calculating the ratio of the square of the \ac{FFT} amplitude with the Hanning window \cite{essenwanger1986elements} of the divided time series with a duration of 2 seconds, to a given frequency-bin width (see details in \cite{duncan_macleod_2020_4301851}). The choice of 36-second duration for the time series is motivated such that the median value of \ac{ASD} appropriately captures the long-term varying noise characteristic.

To have the same dimension for the \acp{mSTFT} of the glitch waveform and $a_i(t)$ and save the computational cost, we re-sample the whitened time series with the same sampling rate. For the glitch class used in the results in Sec. \ref{pipeline_performace_on_ligo_data}, we choose the sampling rate to be the lowest sampling rate of witness channels. This choice of sample rate has the Nyquist frequency $\sim 2$ times higher than the highest frequency of the glitch so that all the characteristics of glitches are captured. To subtracting a glitch waveform from $s(t)$, this choice also makes the resolution of the predicted glitch waveform small enough to apply a small time shift ($\sim 0.02$ seconds) for the phase correction.   

To extract a glitch waveform $g(t)$, we consider two distinct time-frequency regions. Glitches are expected to be present in one of the regions and not present in the other region. For example, for one of the glitch classes called {\it Scattered light} glitches in Sec.\ref{pipeline_performace_on_ligo_data}, we consider the two frequency regions above or below the highest frequency of the glitch (e.g., 100 Hz). We assume that the upper-frequency region represents the \acp{STFT} of $n(t)$ in Eq.\ (\ref{eq:strain_equal_glitch_noise}) and the lower-frequency region represents the \ac{STFT} containing glitches (see \ref{apx:comp_sclight_quiet} for the verification of this assumption). We keep pixels in both real and imaginary parts in the lower frequency region of the \ac{STFT} with their magnitude values above 99 percentile (see the study regarding this threshold choice in \ref{apx:pixel_percentile_threhold}) of the \ac{mSTFT} value in the upper-frequency region, otherwise, set the pixel values to be zero. Subsequently, we set the pixel values in the upper region to be zero as well. After extracting the \ac{STFT} of a glitch waveform, we invert the \ac{STFT} to the time series to obtain the extracted glitch waveform. For {\it Scattered light} glitches, we find that the median value of the overlap $O$ (defined in Eq.\ (\ref{eq:overlap})) between the predicted \acp{mSTFT} from the network and the \acp{mSTFT} of the extracted glitch waveform with the cutoff frequency of 100 Hz is 1\% greater than the overlap with the cutoff frequency of 200 Hz. As \ref{apx:comp_sclight_quiet} shows that the \ac{mSTFT} does not contain excess power above the Gaussian fluctuations in the frequency region above 100 Hz in the data of {\it Scattered light} glitches, choosing the cutoff frequency of 200 Hz lets the data keep a few pixels of Gaussian fluctuations above 200 Hz. Therefore, the overlap with the cutoff frequency of 100 Hz is larger than that with the cutoff frequency of 200 Hz. We choose a different choice in splitting the time-frequency region in \acp{STFT} for the other class of glitches (see details in Sec. \ref{pipeline_performace_on_ligo_data}). To determine choices of splitting the time-frequency region in \acp{STFT}, one can use the method shown in \ref{apx:comp_sclight_quiet} and/or use the peak times and peak frequencies of \textsc{Omicron} triggers \cite{ROBINET2020100620} to find if glitches are isolated in the time domain (see details in \ref{apx:peak_time}).

To help the network learn the excess-power couplings more efficiently, we then divide each time series into smaller overlapping segments. Each training sample comprises segments from multiple witness channels. Larger overlap durations increase the data-set size so that the network can have more learnable resources and hence predict the output efficiently at the cost of computational time and memory.

Some of the segments might contain no glitches or only glitches that are not coupled with a set of chosen witness channels. Removing such segments from the data set helps the network learn the couplings more robustly. Using \ac{CQT} in \textsc{nnAudio} \cite{9174990}, we only select segments with the peak pixel in the \ac{CQT} of the strain channel being loud enough and being near the peak pixels of at least one of the witness channels within a coincidence window (e.g., less than 1 second). During the above process, we further select the subset of segments with the peak frequency within an expected frequency range. As a criterion of the loudness of the pixel, we consider the peak pixel to be loud enough compared to be the Gaussian fluctuations when the peak pixel value is louder than the 90 percentile of the pixel values by a factor of 3.        

After selecting the subset of segments, we finally create the \acp{mSTFT} of training sample segments and the corresponding \ac{mSTFT} of a glitch waveform. To let the network learn efficiently, we normalize the \acp{mSTFT} of the strain and witness channels with their mean $\mu$ and standard deviation $\sigma$, then use these normalized \acp{mSTFT} as the input and the true output in Eq.\  (\ref{eq:obtain_parameters}) to build the network model parameters.  

\subsection{NEURAL NETWORK ARCHITECTURE} \label{neural_network_architecture}
As mentioned in the earlier section, the pipeline uses a 2-dimensional \ac{CNN} that uses the witness channels and predicts the \ac{mSTFT} of a glitch waveform in the detector's output. The input for the network is a multi-dimensional image with the width (height, depth) corresponding to time-bins (frequency-bins, channels). The \ac{CNN} typically consists of a series of convolutional layers, where each layer uses discrete window functions, or kernel with trainable weight. After taking the input image, the layer slides its kernel through the input image and then computes the dot products between the kernels and the portion of the image inside of the kernel. Typically, the kernel's dimension is smaller than that of the image so that the \ac{CNN} learns local features in the image, making the network suitable to process locally outstanding characteristics of excess-power transients in the image. The output of each layer is passed to a non-linear activation function and becomes the input for the subsequent layer. Because the output of each layer is down-sampled, the subsequent layer represents the input image with a fewer number of features. This sequence of layers is known as an \textit{encoder} that can extract the glitch-coupled excess-power characteristics and suppress the slowly-varying noise in the image \cite{bank2021autoencoders}. After the convolutional layer, the network consists of transposed convolutional layers known as a \textit{decoder}  \cite{RUMELHART1988399, pmlr-v27-baldi12a, 10.1007/978-3-642-21735-7_7, JMLR:v11:vincent10a}. Each transposed convolution layer inserts pixels of zero between the pixels in the input image and then slides the kernel to computes the dot products of kernels and the modified input image within the kernel. As a consequence, transposed convolutional layers recast the encoded image to have a higher number of pixels by up-sampling. Comprising the encoding and decoding layers (so-called \textit{autoencoder}) \cite{RUMELHART1988399, pmlr-v27-baldi12a, 10.1007/978-3-642-21735-7_7, JMLR:v11:vincent10a}, the output of the network will have the same dimension as the glitch image with extracted glitch-coupled features.  

The above considerations motivate us to use the fully connected convolutional autoencoder in our network that provides an estimate of the glitch image from the witness channels' counterpart. In addition to the autoencoder, we employ a convolutional layer before the encoder to normalize the input images, and use a convolution layer after the decoder to make the output-image dimension to be the same as that of the glitch image. More specifically, the input images of the witness channels are first passed to the input convolutional layer and then normalized with \textit{Batch Normalization} \cite{ioffe2015batch}. To make the dimension of the input and output images for the network, the input layer uses a stride of 1 and an appropriate zero padding arrangement. In the encoder, the width and height of the image are reduced by a factor of 2 while the depth (or the number of channels) is increased by a factor of 2. Instead of using pooling layers (e.g., Max Pooling \cite{Yamaguchi1990ANN}), each layer makes use of a stride of 2 with an appropriate zero padding arrangement. The output of the encoding layer is passed to the decoding layer. In the decoding layer, the width and height of the image are increased by a factor of 2 while the depth is reduced by a factor of 2 by using a stride of 2 with an appropriate zero padding arrangement. The output of the decoding layer is fed into the output convolutional layer, in which an appropriate kernel with a stride of 1 is chosen to make the final output image has the same dimension as that of the glitch image. Except for the output layer, the output of each layer is passed to an activation function before being fed into the subsequent layer. 

We adopt the symmetrical structure for the autoencoder similar to \cite{Ormiston:2020ele} because each convolutional layer is commonly known to learn a different level of characteristics of the input images. An earlier (later) layer in the encoder tends to learn a lower (higher) level of characteristics. Hence, the first (last) layer in the encoder extracts a lower (higher) level of characteristics that are then recast by the last (first) layer in the decoder. Likewise, the intermediate levels of characteristics are also extracted and recast by a pair of layers in the encoder and decoder.          

In our analysis, the network comprises four convolutional layers for both the encoder and decoder. We choose different sizes ($\sim 10$) of a kernel for different classes of glitches. For the activation function, we the \textit{ReLU} \cite{Hahnloser2000,Nair2010RectifiedLU} in the input and the autoencoder layers. We do not use any activation function after the output layer. Each encoding layer increases the number of channels from the value in the input image to 8, 16, 32, and 64, respectively. Each decoding layer decreases the number of channels in inverse order.

\subsection{TRAINING AND VALIDATION} \label{training_inference}
The analysis of the network can be divided to be two parts; training and validation. During training, the data set are divided into smaller chunks of data, or so-called \textit{mini-batches} to reduce the computational memory. Data in each mini-batch are fed into the network and the loss function in Eq.\ (\ref{eq:loss_func}) is computed by averaging over each mini-batch. The network parameters $\vec{\theta}$ are updated according to the gradient of the loss function with respect to the parameters. For calculating the gradient, we use one of the first-order stochastic gradient descent methods, \textit{ADAM} \cite{adam}. We iteratively update $\vec{\theta}$ by repeating the above calculations over a number of cycles, or \textit{epochs}. After each epoch, the loss value and the coefficient of determination are calculated using the validation set to prevent over-fitting; the network parameters are tuned with the training set so that the network tends to represent the glitch couplings contained in the training set instead of representing the couplings in a broader data set such as a validation set which is not used to turn the network parameters. To avoid over-fitting, the validation set is chosen not to be overlapping with any data in the training set. We stop the iteration if the network shows over-fitting according to the coefficient of determination of the validation set.        

\subsection{OUTPUT DATA POST-PROCESSING} \label{output-data_postprocessing}
The output of the network is conditioned before subtracting glitches in the detector's output data. Because the glitch \ac{mSTFT} is normalized with the mean $\mu$ and the standard deviation $\sigma$ of the pixel values across the training set before being fed into the network, we invert the normalization by multiplying the predicted \ac{mSTFT} by $\sigma$ and add $\mu$. While the true \ac{mSTFT} has no negative values by definition of \ac{mSTFT}, the predicted \ac{mSTFT} could have negative values due to the imperfection of the network prediction. We find that the negative values of the predicted \ac{mSTFT} are typically anti-correlated with the values of the true \ac{mSTFT}. 
The median values of the overlap $O$ in \ref{eq:overlap} between negative pixels of the predicted \acp{mSTFT} and the corresponding pixels in the true \acp{mSTFT} are approximately $- 0.1$ and $- 0.35$ for {\it Scattered light} and {\it Extremely loud} glitches, respectively. Taking absolute values of the predicted \acp{mSTFT} makes these negative pixels have a positive effect on the overlap of all pixels though the effect is small because the average of the absolute values of the negative pixels is only $\sim 5$\% and $\sim9$\% of the average value of the positive pixels for {\it Scattered light} and {\it Extremely loud} glitches, respectively. Using the \ac{FGL} transformation \cite{1164317, 6701851}, we estimate the glitch waveform from the absolute value of \ac{mSTFT} predicted by the network. 

Because the estimated waveform has a slightly larger or smaller amplitude compared to the extracted glitch waveform and the phase of the estimated waveform is slightly shifted due to the network-prediction imperfection and the phase-estimation uncertainty in the \ac{FGL} transformation, we let the estimated waveform change only its amplitude by a factor up to $\pm 3$ and change the phase up to $\pm 0.04$ seconds to subtract glitches efficiently and avoid introducing unintended effects on astrophysical signals potentially present. To determine values of the amplitude and phase correction factors, we use the least square fitting method within the time periods where the glitch is present. To determine the portion of the glitch presence, we (1) calculate the absolute values of the estimated glitch waveform, (2) then smooth the curve with the convolution with the rectangular function, and (3) finally determine the time window where the absolute values are above a threshold. 

Because the glitch waveforms (including both the estimate and extracted) are fluctuated instead of that their values are smoothly increased towards the peak of amplitude, taking the time window where the absolute values are above a threshold without smoothing the curves makes the time windows to be divided and to cover only sub-portions of glitches (corresponding to high thresholds) or the time window to cover wide portions including the region with no glitches (corresponding to low thresholds). Therefore, we employ the convolution with a rectangular function as one of the smoothing methods. We typically set this threshold to be the $\sim 90$ percentile of the absolutes values of a set of the estimated glitch waveforms (see details for each glitch class in \ref{apx:time_window_subtract_glitches} and Sec. \ref{pipeline_performace_on_ligo_data}). To subtract glitches even more efficiently, we divide this time portion and apply the least square fitting against the strain data within the divided portions (see details for each glitch class in Sec. \ref{pipeline_performace_on_ligo_data}). Smaller lengths of the divided portions allow us to subtract glitches more efficiently but less robustly preserve the waveform of astrophysical signals. To balance the above two factors, we divide the time portion finer around the center time of the glitch waveform because typically glitches have higher frequencies and larger amplitudes around the center time of the waveform. In the least square fitting, we find that the subtraction is better by separating the fitting with the bounds of the amplitude correction factor either in the range $(0, 3)$ or $(-3, 0)$ and then choose the better fitting result based on its coefficient of determination $R^2$.

\section{Pipeline performance on LIGO data} \label{pipeline_performace_on_ligo_data}

We apply our glitch subtraction pipeline to the data of the \ac{L1} detector from January 1st, 2020 to February 3rd, 2020. We choose two district classes of glitches comprising different types of noise couplings that are dominant events for creating \textit{background} \ac{cWB} \cite{Klimenko:2008fu,Klimenko:2015ypf} triggers with high-ranking statistics to study and quantify the performance of our pipeline.    

To quantify the network-prediction accuracy, we use the overlap of the true \ac{mSTFT} and the predicted \ac{mSTFT} of the glitch given as  
\begin{equation}\label{eq:overlap}
    O = \frac{\sum_{i=1}^{N} G[i]\hat{G}[i]}{\sqrt{\sum_{j=1}^{N} G^2[j]\sum_{k=1}^{N} \hat{G}^2[k]}}\,,
\end{equation}
where $G[i]$ and $\hat{G}[i]$ are the true and estimated \acp{mSTFT} of the glitch, respectively, and $(i,j,k)$ run over pixel indices, and $N$ is the total number of pixels. The overlap indicates the direct measure of the network prediction accuracy, ranging from 0 (mismatched) to 1 (perfect matched). 

\ac{SNR} is the dominant factor for the detection of \ac{GW} signal searches. Lowering \ac{SNR} of noise artifacts improves the detectability of \ac{GW} signals. . We quantify the performance of the glitch reduction with our pipeline by calculating the \ac{FNR} after the subtraction as
\begin{equation}\label{eq:fraction_snr_reduction}
    {\rm \ac{FNR}} = \frac{{\rm SNR}_b - {\rm SNR}_a}{{\rm SNR}_b} \,,
\end{equation}
where ${\rm SNR}_b$ and ${\rm SNR}_a$ are the matched-filter \acp{SNR} \cite{Usman:2015kfa} obtained using the extracted glitch waveform as a template and the data before and after the subtraction, respectively (see how to obtain the extracted glitch waveform in Sec \ref{data_preprocessing}). Values of \ac{FNR} close to 1 indicate efficient glitch reductions while negative values of \ac{FNR} imply the increase of the glitch energy in the data.  

\subsection{SCATTERED LIGHT GLITCHES} \label{scattered_light_result}

We first apply our subtraction technique to a class of glitches called \textit{Scattered light} glitches. Typically, winds and/or earthquakes shake the detector and move the test mass mirrors in the detector arms. The displacements of the mirrors in the longitudinal or rotational directions differ the relative arm lengths to produce arch-like glitches in the \ac{mSTFT} in $10-90$ Hz, with a duration of $\sim 2$ seconds \cite{Davis:2021ecd, Soni:2021cjy,Soni:2020rbu}. 

To identify the witness channels for this glitch class, we use \tool \cite{mogushi2021application} on the list of \textit{Scattered light} glitches with \ac{SNR} above 10 between January 1st, 2020 and February 3rd, 2020 in the \ac{L1} detector from the \textsc{Gravity Spy} catalog. We use two witness channels with high confidence: L1:ASC-CSOFT\_P\_OUT\_DQ and L1:ASC-X\_TR\_B\_NSUM\_OUT\_DQ which monitor the common length of the two arms and the transmitted light from the mirror at the end of the $x$-arm, respectively. 

\begin{figure}[!ht]
    \begin{minipage}{0.5\hsize}
  \begin{center}
   \includegraphics[width=1\textwidth]{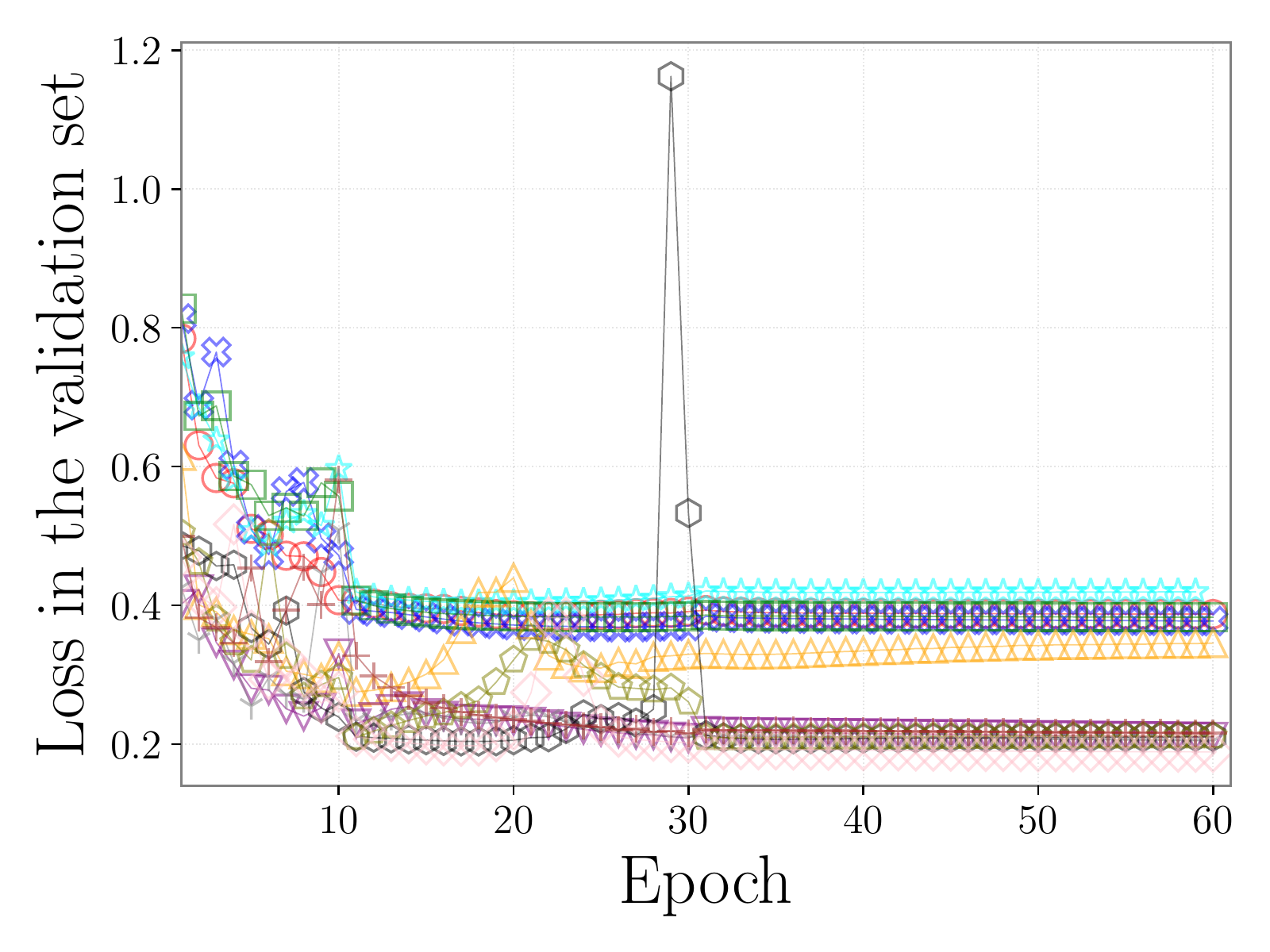}
  \end{center}
    \end{minipage}
    \begin{minipage}{0.5\hsize}
  \begin{center}
   \includegraphics[width=1\textwidth]{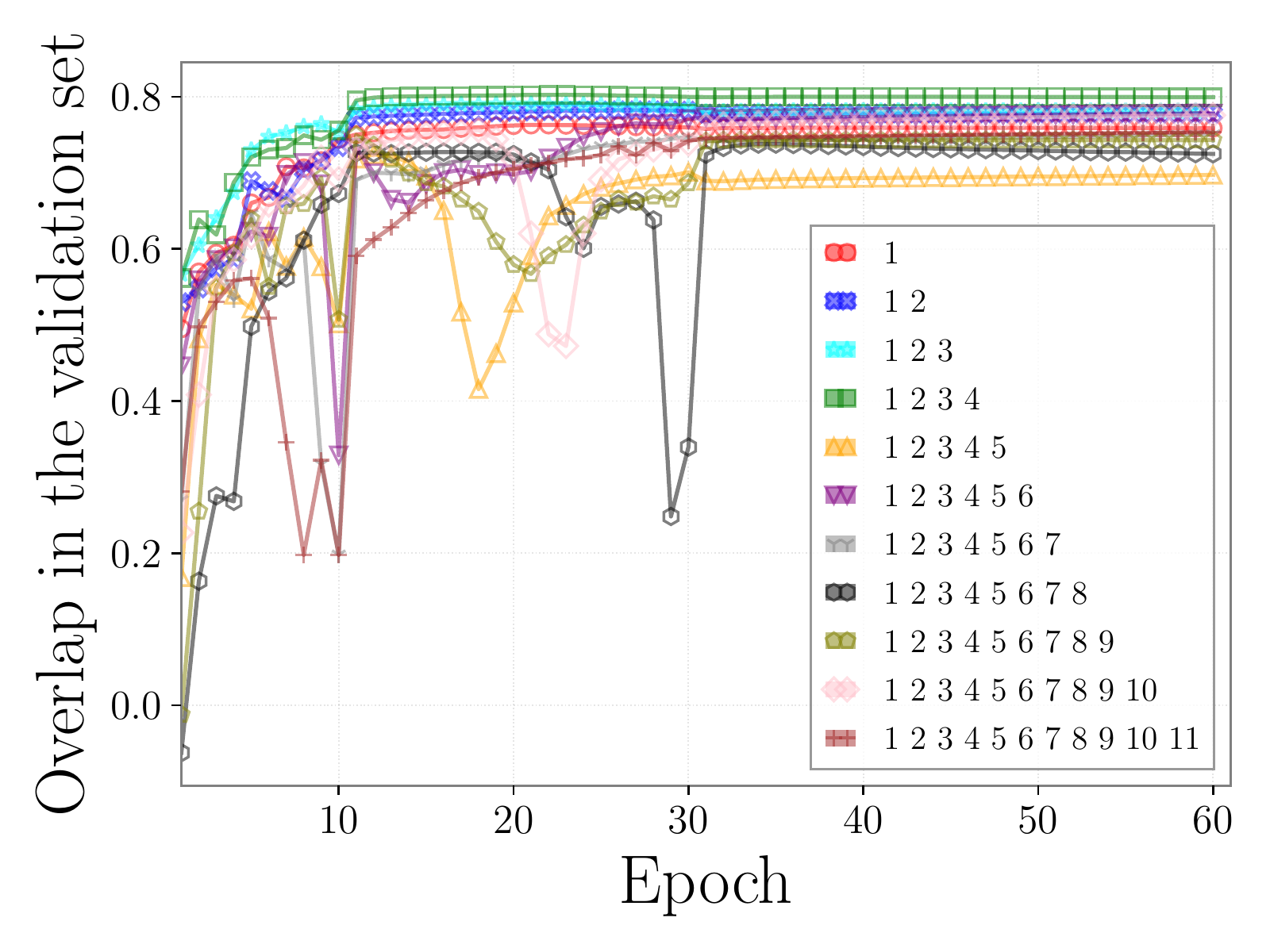}
  \end{center}
    \end{minipage}
    \begin{minipage}{0.5\hsize}
  \begin{center}
   \includegraphics[width=1\textwidth]{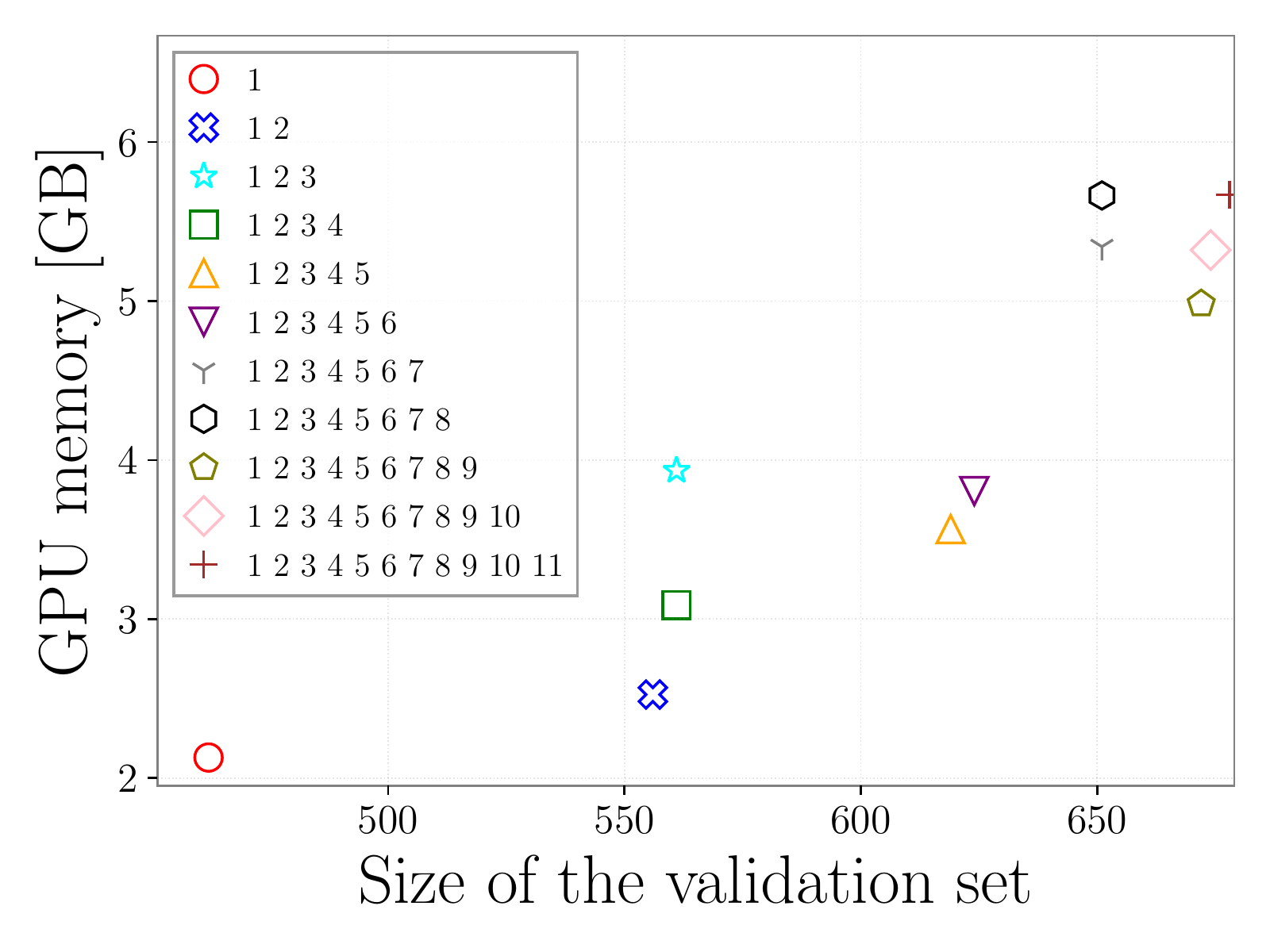}
  \end{center}
    \end{minipage}
    \begin{minipage}{0.5\hsize}
  \begin{center}
   \includegraphics[width=1\textwidth]{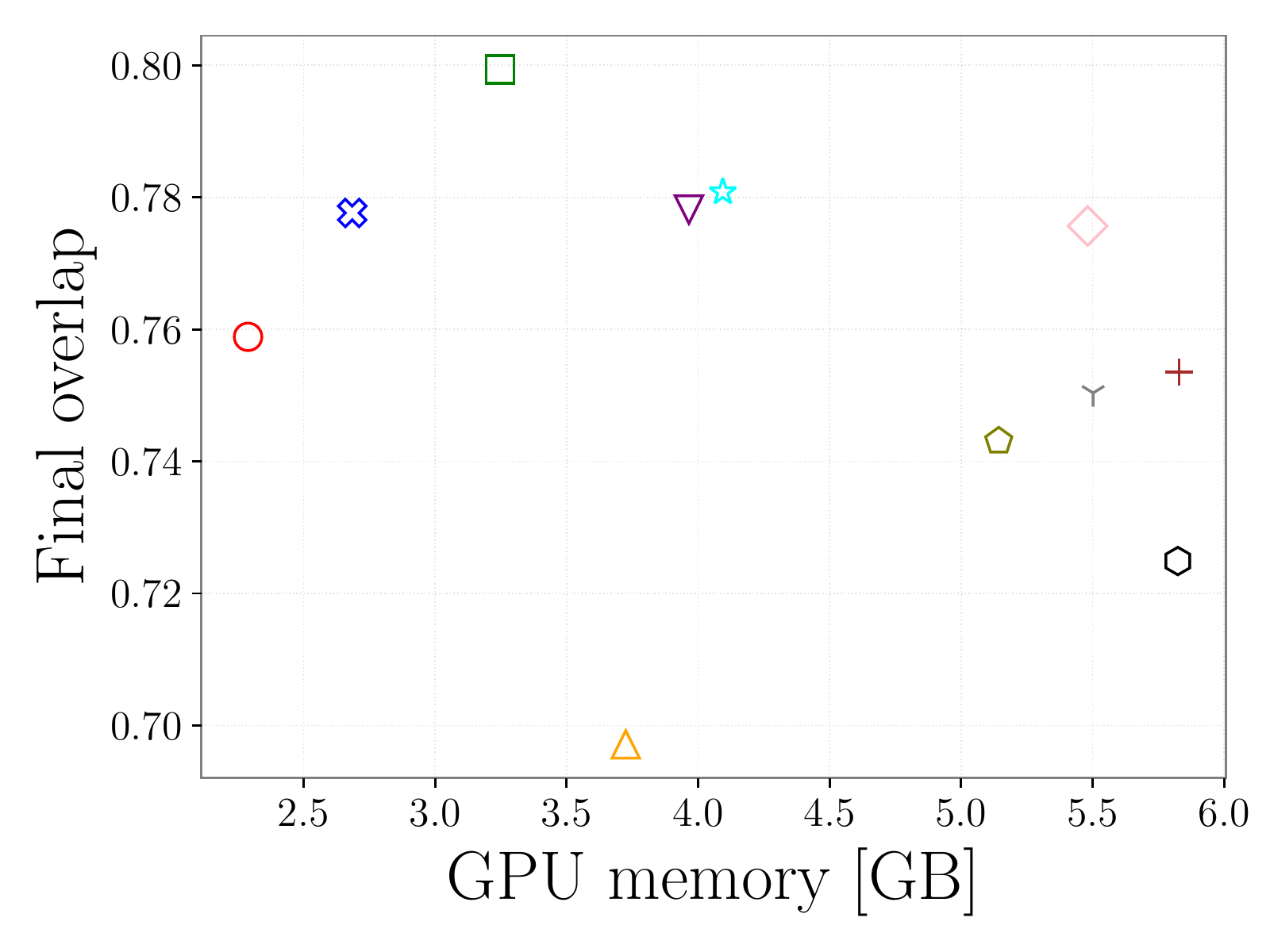}
  \end{center}
    \end{minipage}
\caption{Losses defined in Eq.\ (\ref{eq:loss_func}) (top-left) and overlaps (top-right) over epochs in the validation sets using channel sets containing $i^{\rm th}$ ranked channels for {\it Scattered light} glitches. The bottom panels show sizes of the validation sets, overlaps at the end of the training, and used GPU memories.}
\label{fig:sclight_varation_channel}
\end{figure}

To check that the top two witness channels are sufficient, we train the network with various sets of channels with the confidence $p_g$ of being witnesses of this glitch class from $p_g = 0.93$ up to $p_g = 0.71$ \cite{mogushi2021application}. We consider 11 different channel sets, where $i^{\rm th}$ set contains up to $i^{\rm th}$ ranked channels. We train the network with a learning rate of $10^{-3}$ ($10^{-4}$/$10^{-5}$) for the 1-10 (11-30/31-60) epochs, where learning rates determine the gradient to update the network parameters and smaller learning rates correspond to smaller gradients. We terminate the training process if a value of $R^2$ in the validation set plateaus, i.e., a value of $R^2$ in the current epoch does not differ from the value in the previous epoch greater than $\pm 0.001$\%. Figure \ref{fig:sclight_varation_channel} shows losses in Eq.\ (\ref{eq:loss_func}) and overlaps in the validation sets, the validation sample size, and the GPU memories used to train the network for various channel sets. Overlaps at the termination for all channel sets range in $\sim0.7\sim0.8$, where the GPU memory for the $11^{\rm th}$ channel set is greater than the memory for the $1^{\rm st}$ channel set by a factor of 2.5. Using a higher number of witness channels with high confidence provides a larger amount of glitch-coupling information to the network and let the performance better whereas adding low confident witness channels only provides non-glitch-coupling information to the network and does not improve the performance because the network seems not to use those low confident channels. Furthermore, using low confident channels might add data samples that have chance coincident excess power between these channels and the strain channels and/or subdominant glitch-coupling information (see details for the coincident selection in Sec. \ref{data_preprocessing}), where the size of the validation set for the $11^{\rm th}$ channel set is greater than the size for the $1^{\rm st}$ channel set by a factor of $~1.5$. Therefore, the termination overlap tends to decrease with the use of redundant GPU memories by adding channels above the $5^{\rm th}$ rank as shown in the top-right panel in Fig.\ \ref{fig:sclight_varation_channel}. Using the first two ranked channels is sufficient because the termination overlap for the $2^{\rm nd}$ channel set is only less than 3\% smaller than the largest value of the termination overlap (obtained by the $4^{\rm th}$ channel set) and saves 18\% GPU memory. In the following, we use the top-tow ranked channels for {\it Scattered light} glitches.

We pre-process the data as described in the previous section with the time series of the strain and the witness channels being re-sampled to a sampling rate of 512 Hz. During the pre-processing, we consider the frequency range above 100 Hz to be the background-noise region and use the 99 percentile pixel-value to extract the glitch waveform below 100 Hz. We also apply a high-pass filter to the strain channel at 10 Hz. We set the sample dimensions of the (training/validation/testing) sets to be (9131/2233/678) with the segment overlaps of (93.7/93.7/75)\%, where the segment overlap is the fraction of the time-window overlap between segments created by sliding a time window to divide a larger segment into smaller segments (so-called {\it data augmentation} \cite{data_aug1, Perez2017TheEO, Lemley2017SmartAL}). We set that there is no overlapping time between the three sets and the testing set is later than the other sets. We create the \ac{mSTFT} with a duration of 8 seconds, a frequency range up to 256 Hz, and (time/frequency) resolutions of 0.0625 seconds and 2 Hz, respectively. We use a square kernel with a size of $(8,8)$ in the autoencoder in the network. During the post-processing, we consider the region where the glitches are present to be the time when the absolute value of the estimated glitch waveform is above the 75 percentile of the corresponding values across the testing set. We choose the divided window length for the least square fitting to be as small as 0.1 seconds and expand the window length from the center of the glitch by a factor of 1.1. 

\begin{figure}[ht!]
    \centering
    \includegraphics[width=0.5\linewidth]{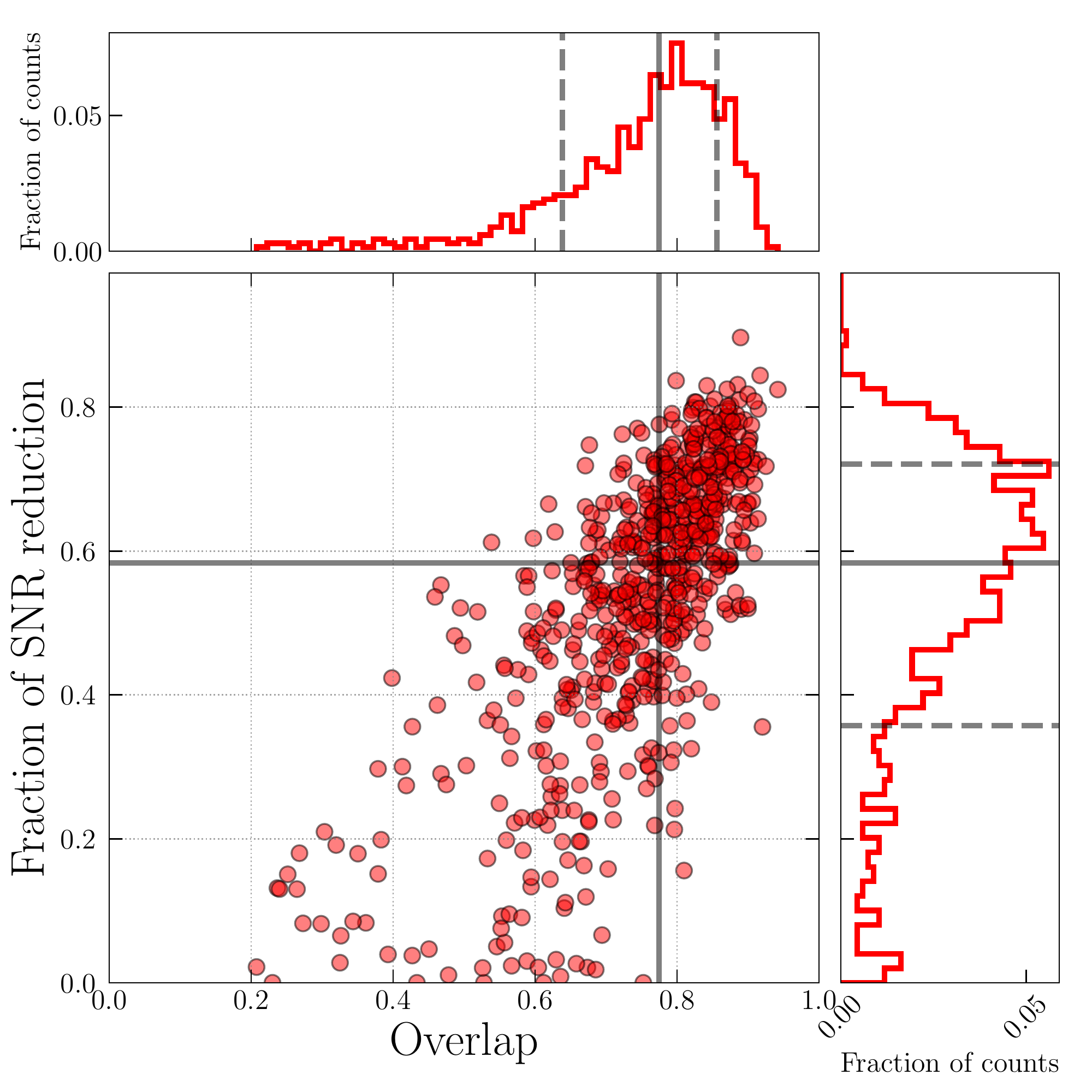}
    \caption[Distribution of the overlap of the extracted and estimated \acp{mSTFT} of \textit{Scattered light} glitches in the testing set.]%
    {Distribution of the overlap of the extracted and estimated \acp{mSTFT} of \textit{Scattered light} glitches in the testing set. The black solid and dashed lines denote the median and 1-$\sigma$ percentiles.}
    \label{fig:overlap_vs_snr_sclight} 
\end{figure}

Figure \ref{fig:overlap_vs_snr_sclight} shows the distribution of the overlap of the \ac{mSTFT} and the \ac{FNR} of the testing set of \textit{Scattered light} glitches. More similar \acp{mSTFT} correspond to more similar waveforms after the \ac{FGL} transformation, resulting in more efficient reductions of glitch \acp{SNR}. We find the overlap ranging from $\sim$0.6-0.9 and the \ac{FNR} ranging $\sim 0.4-0.7$ in 1-$\sigma$ percentiles and no negative \ac{FNR}, indicating no additional glitch energy added to the strain data. 

\begin{figure}[ht!]
    \begin{minipage}{0.5\hsize}
  \begin{center}
   \includegraphics[width=1\textwidth]{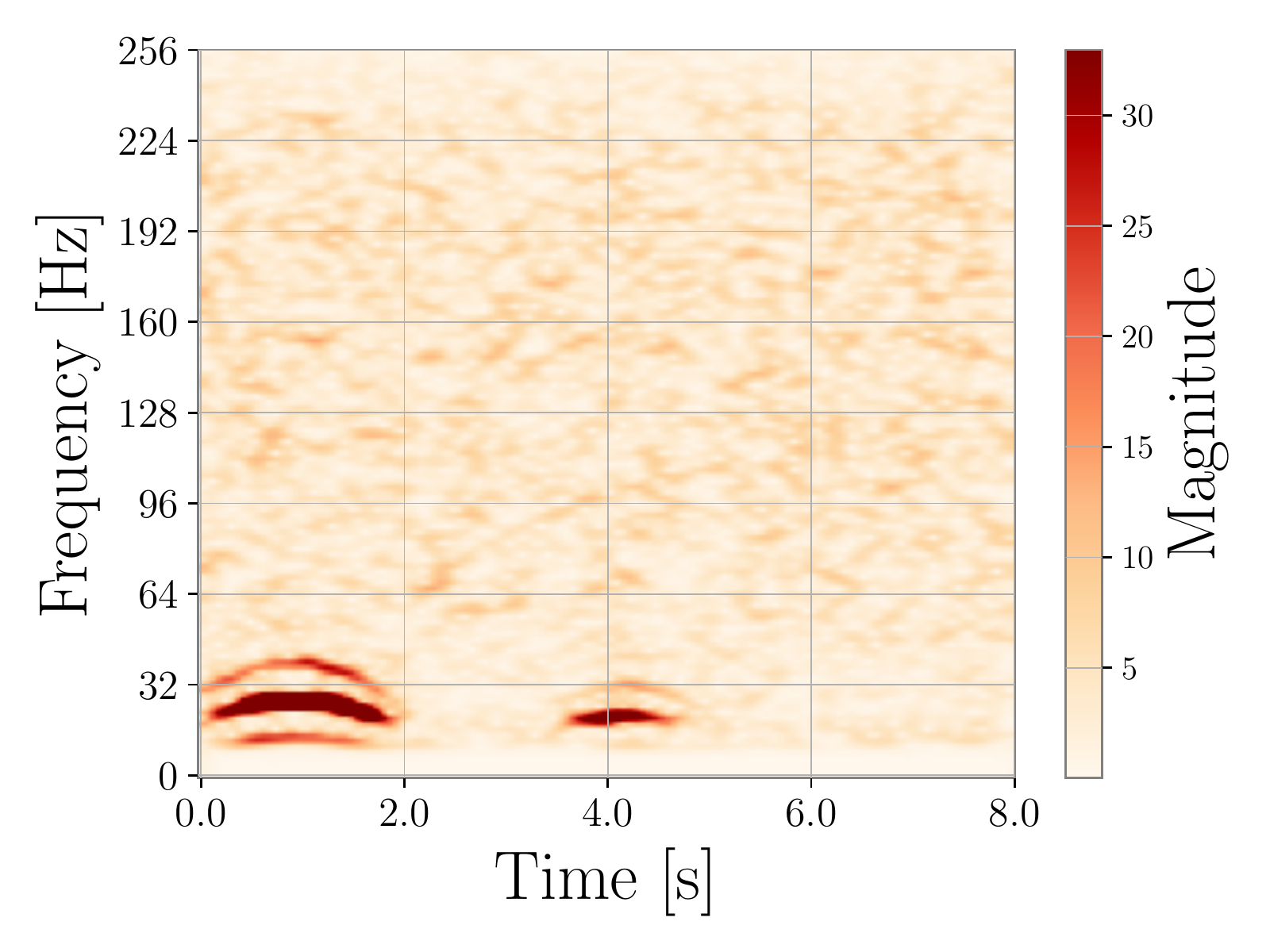}
  \end{center}
    \end{minipage}
    \begin{minipage}{0.5\hsize}
  \begin{center}
   \includegraphics[width=1\textwidth]{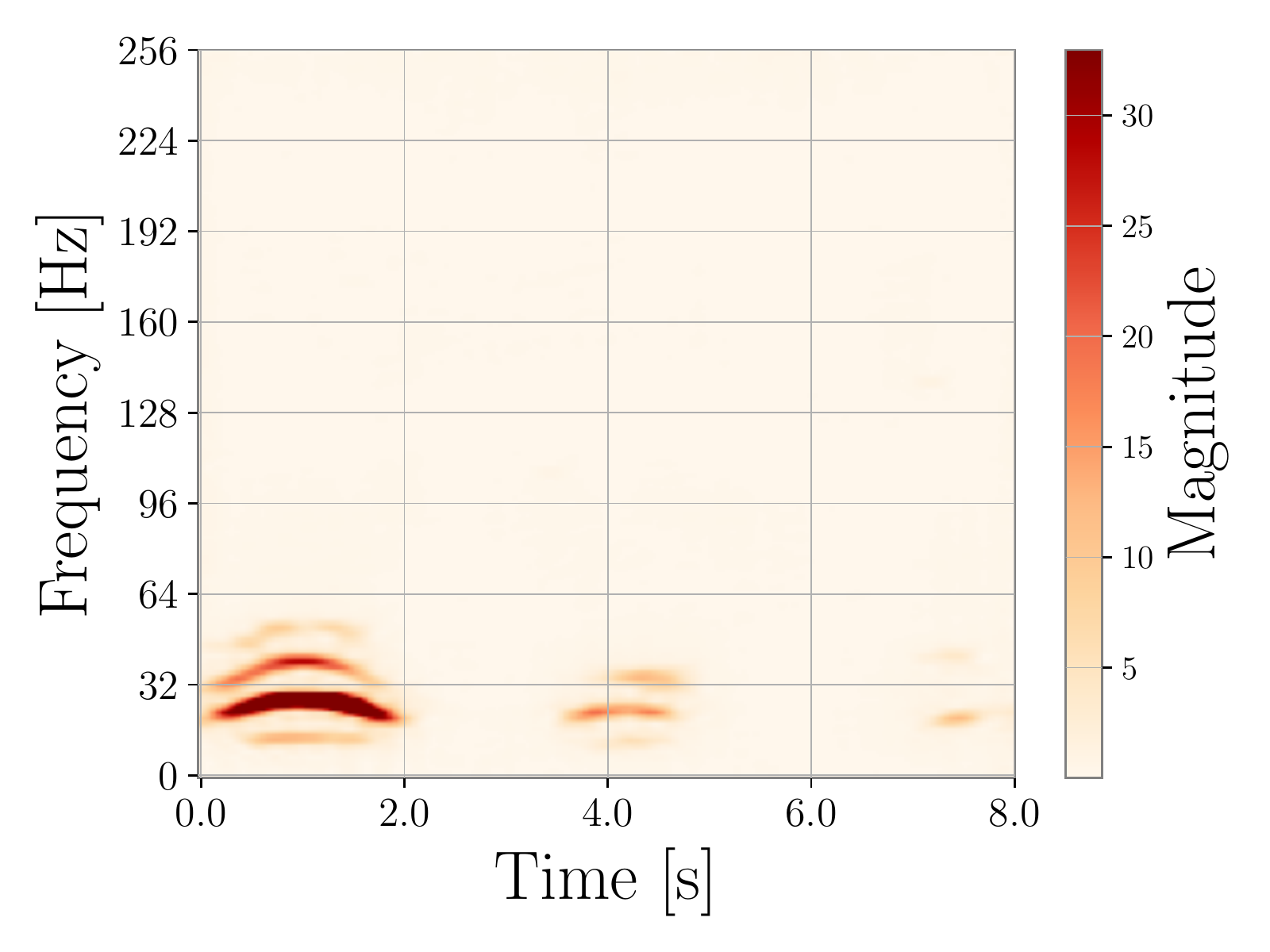}
  \end{center}
    \end{minipage}
    \begin{minipage}{0.5\hsize}
  \begin{center}
   \includegraphics[width=1\textwidth]{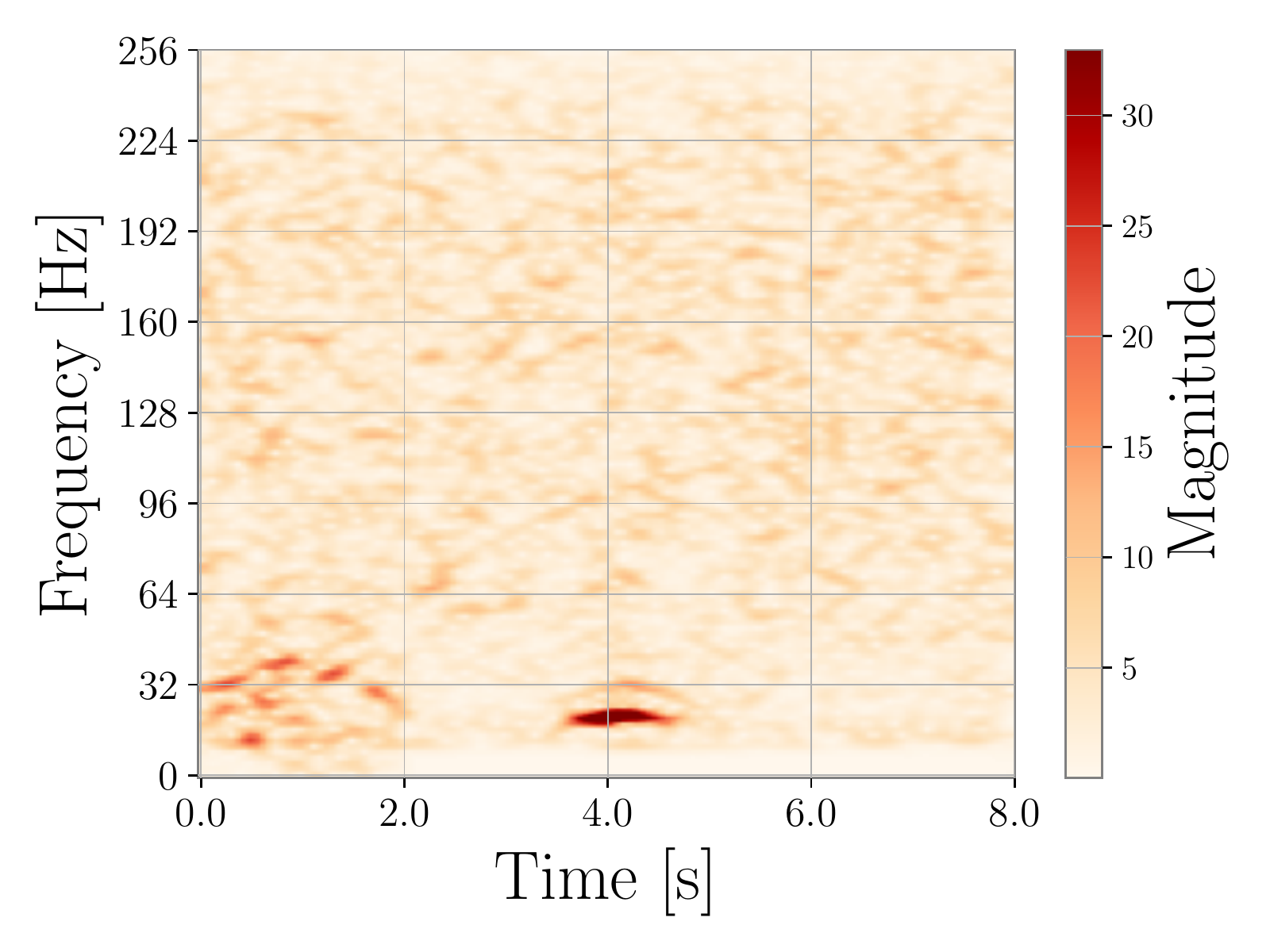}
  \end{center}
    \end{minipage}
    \begin{minipage}{0.5\hsize}
  \begin{center}
   \includegraphics[width=1\textwidth]{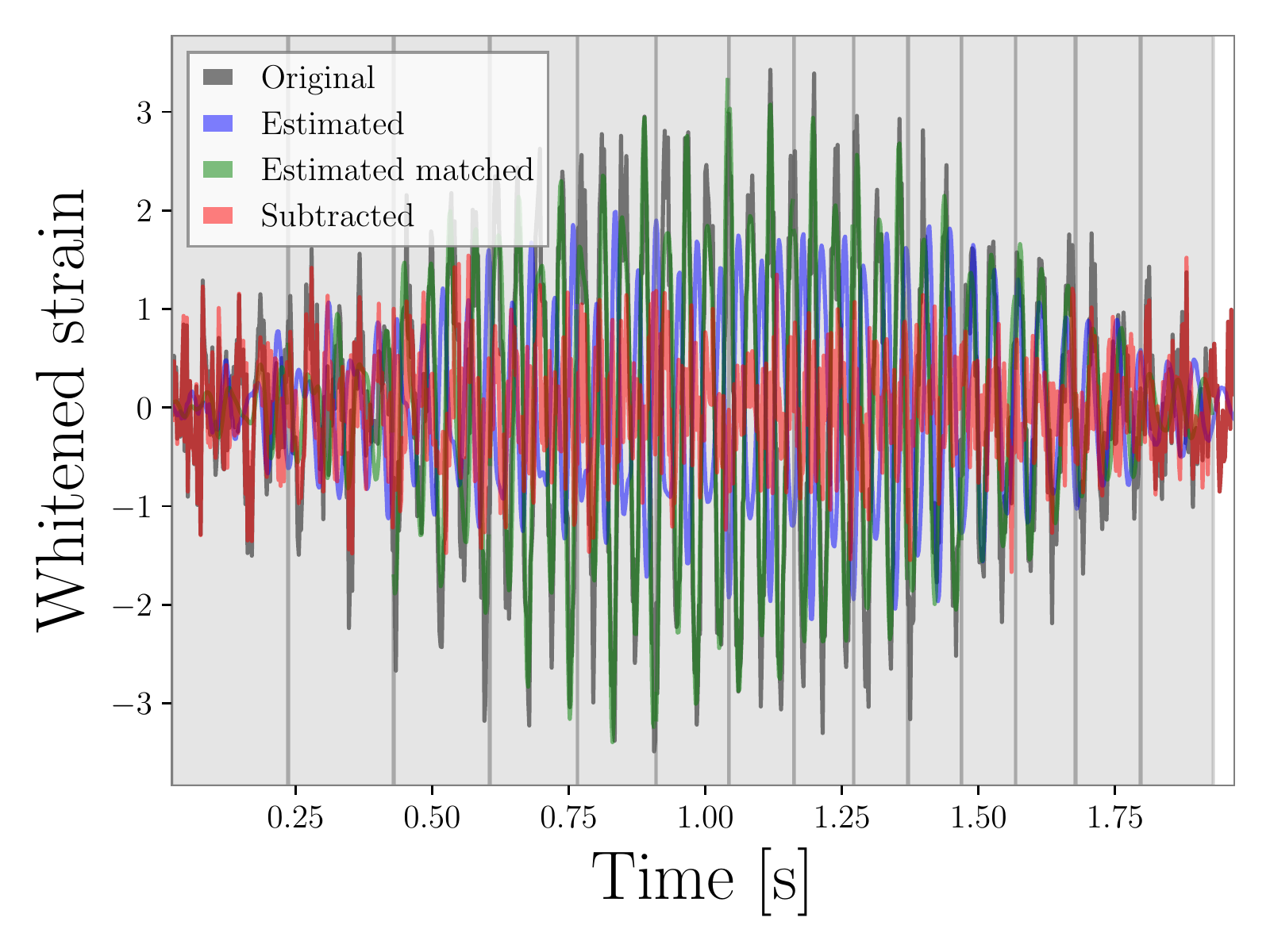}
  \end{center}
    \end{minipage}
\caption[The \acp{mSTFT} and time series of the optimal case in the testing set of {\it Scattered light} glitches.]%
{The \ac{mSTFT} of the strain data (top-left), the network estimated \ac{mSTFT} (top-right), and the \ac{mSTFT} of the strain data after glitch subtraction (top-left) in the optimal testing sample of \textit{Scattered light} glitches. In the top-left panel, the gray (blue/green/red) curves denote the original (estimated/estimated-matched/subtracted) whitened time series, where the estimated-matched time series is created after the amplitude and phase corrections with the least square fitting within divided segments shown as the gray bands. The overlap between \ac{mSTFT} of the extracted glitch waveform and the estimated \ac{mSTFT} is $O=0.92$ and ${\rm \ac{FNR}}=0.84$.}
\label{fig:sclight_optimal}
\end{figure}

\begin{figure}[ht!]
    \begin{minipage}{0.5\hsize}
  \begin{center}
   \includegraphics[width=1\textwidth]{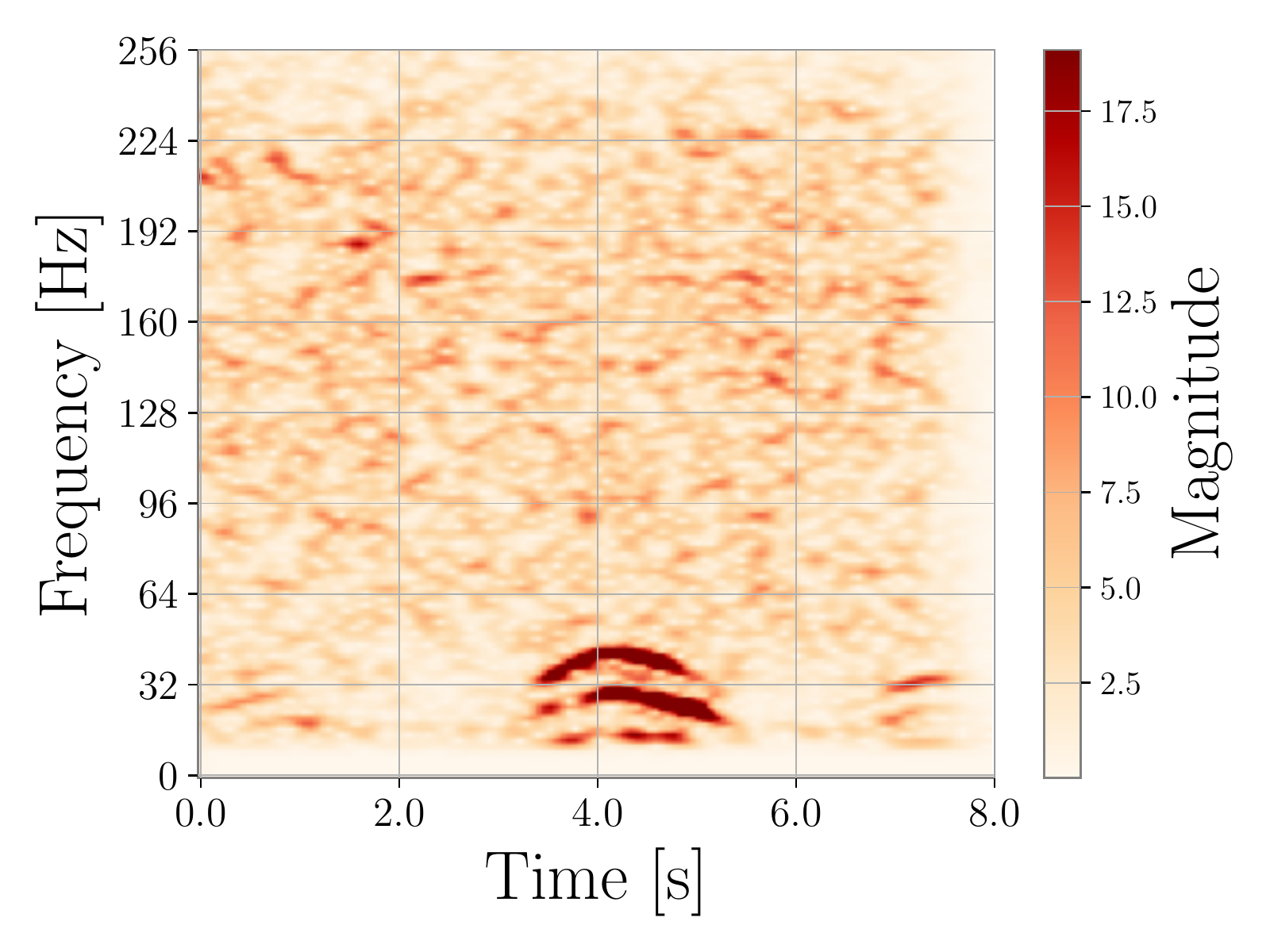}
  \end{center}
    \end{minipage}
    \begin{minipage}{0.5\hsize}
  \begin{center}
   \includegraphics[width=1\textwidth]{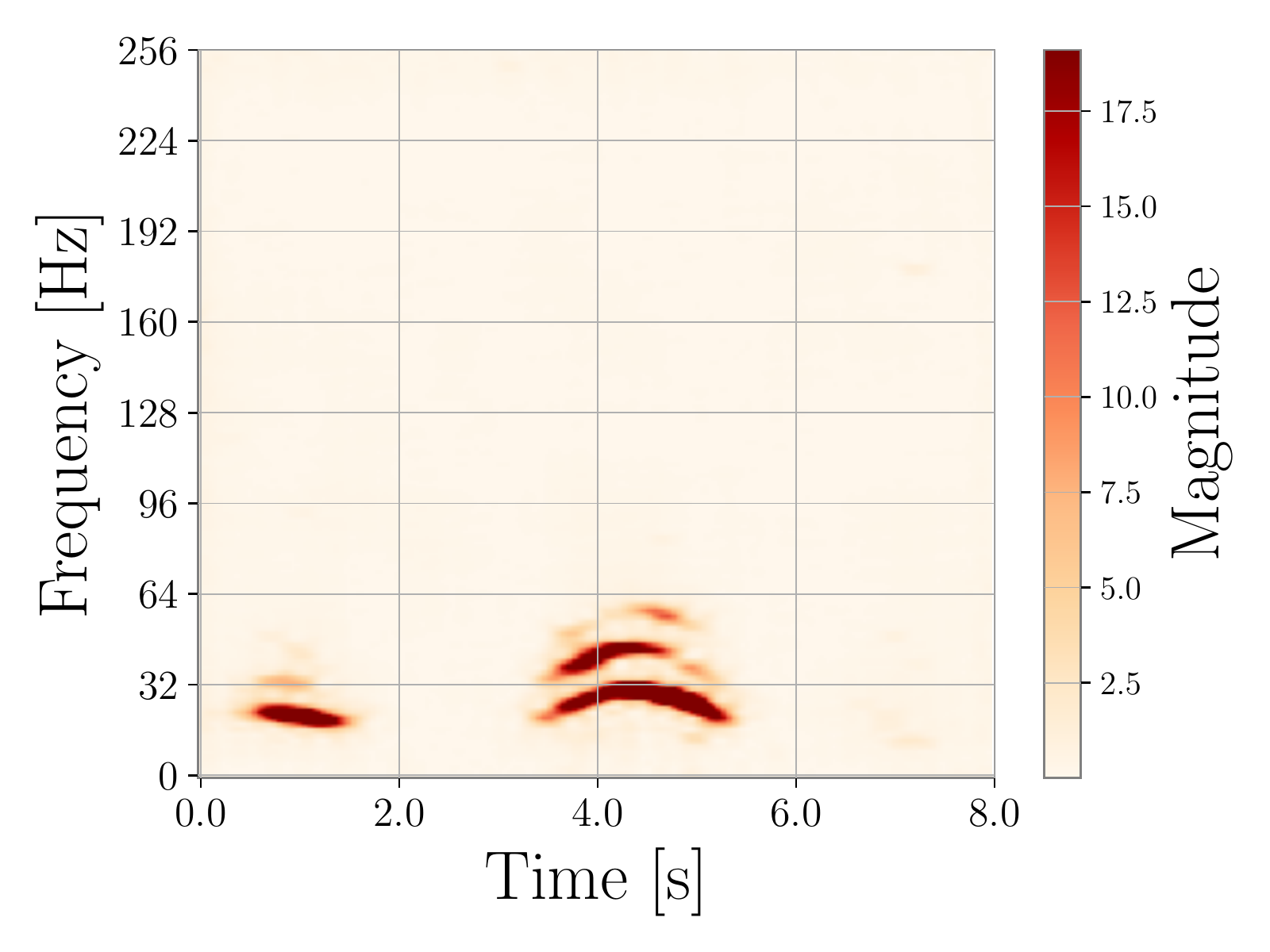}
  \end{center}
    \end{minipage}
    \begin{minipage}{0.5\hsize}
  \begin{center}
   \includegraphics[width=1\textwidth]{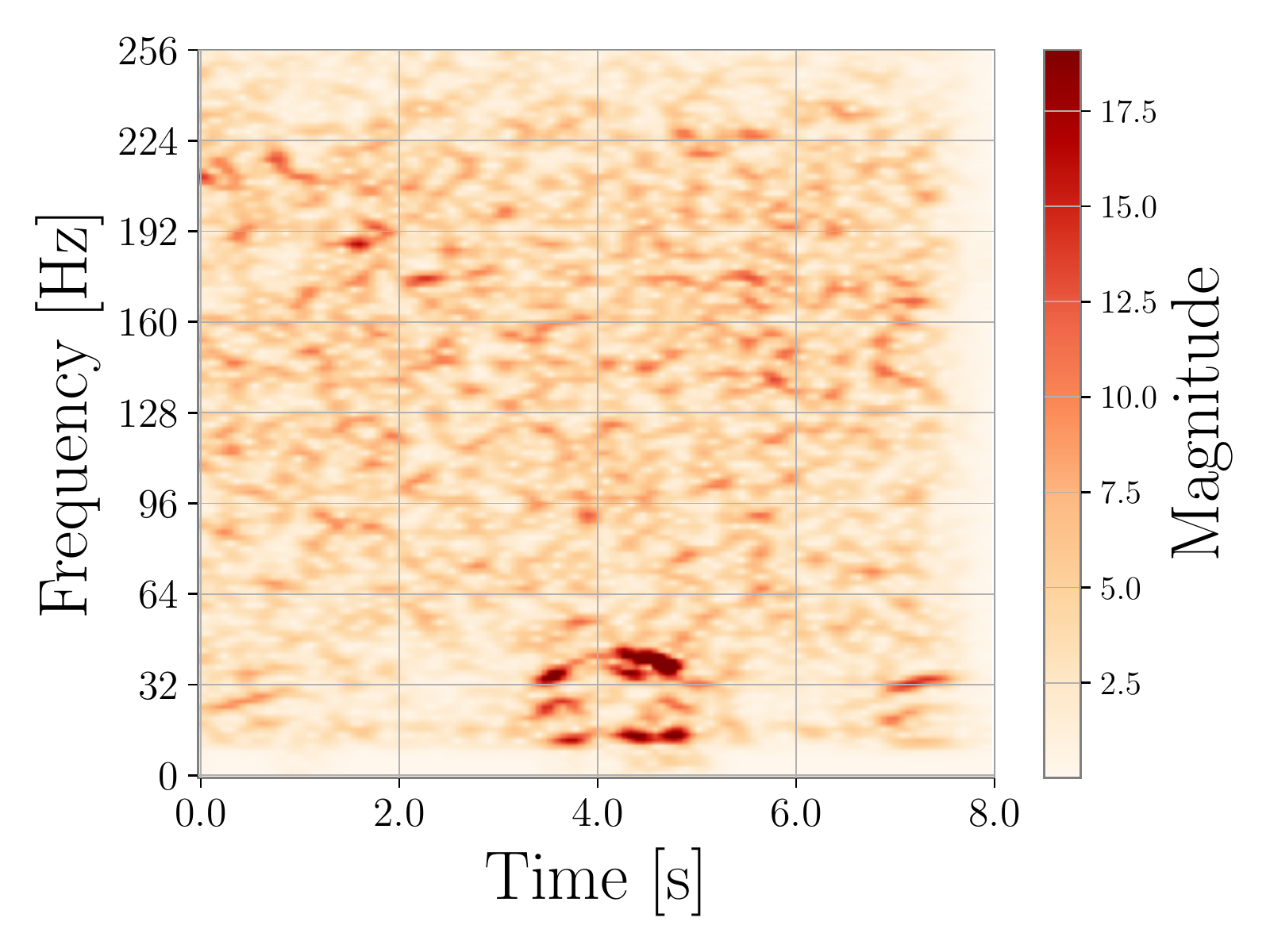}
  \end{center}
    \end{minipage}
    \begin{minipage}{0.5\hsize}
  \begin{center}
   \includegraphics[width=1\textwidth]{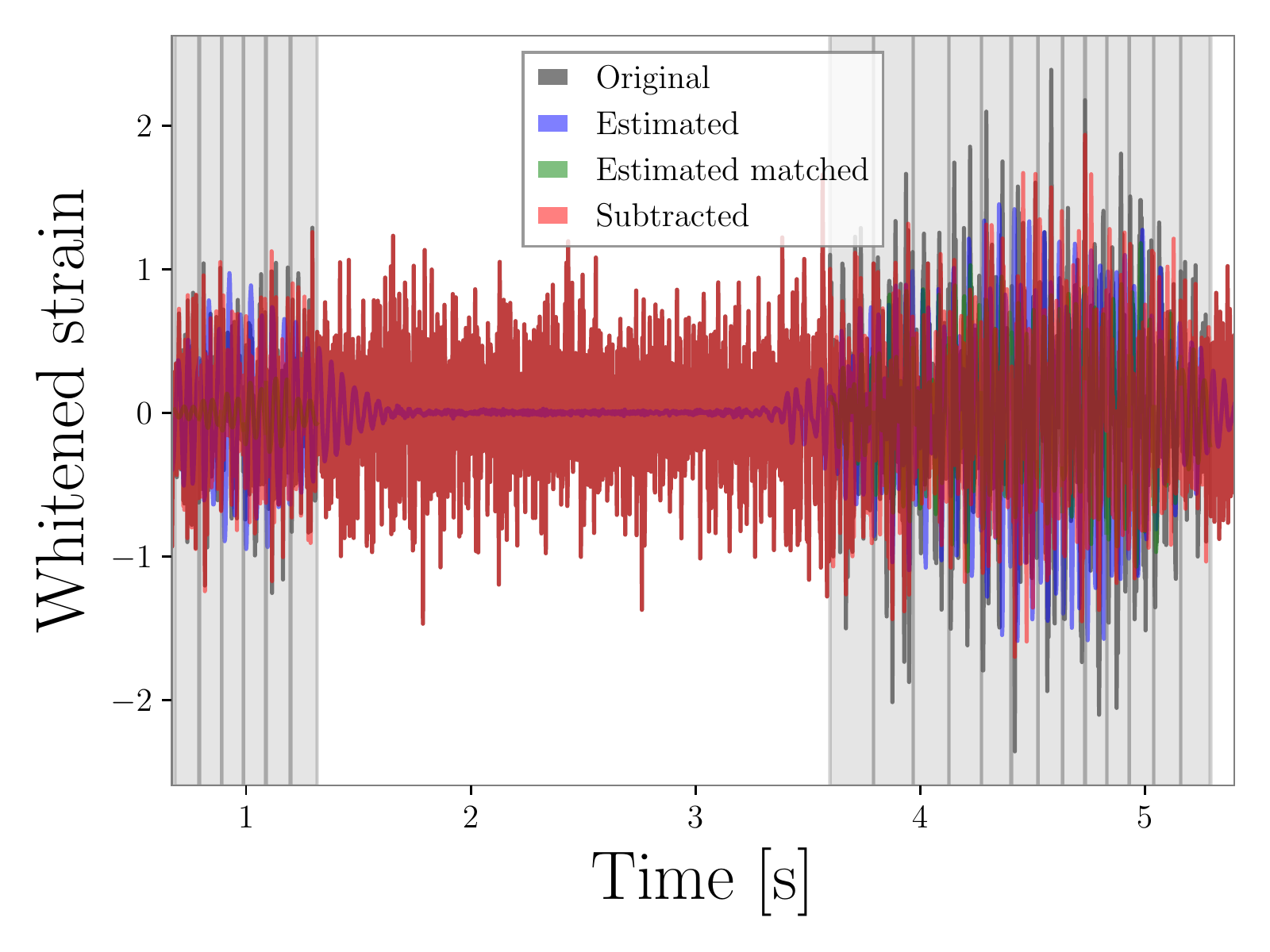}
  \end{center}
    \end{minipage}
\caption[The \acp{mSTFT} and time series of the median case in the testing set of {\it Scattered light} glitches.]%
{The \ac{mSTFT} of a down-sampled strain data at 512 Hz with a high-pass filter at 10 Hz (top-left), the network estimated \ac{mSTFT} (top-right), and the \ac{mSTFT} of the strain data after glitch subtraction (top-left) in the median testing sample of \textit{Scattered light} glitches. In the top-left panel, the gray (blue/green/red) curves denote the original (estimated/estimated-matched/subtracted) whitened time series, where the estimated-matched time series is created after the amplitude and phase corrections with the least square fitting within divided segments shown as the gray bands. The overlap between \ac{mSTFT} of the extracted glitch waveform and the estimated \ac{mSTFT} is $O=0.65$ and ${\rm \ac{FNR}}=0.58$.}
\label{fig:sclight_median}
\end{figure}

\begin{figure}[ht!]
    \begin{minipage}{0.5\hsize}
  \begin{center}
   \includegraphics[width=1\textwidth]{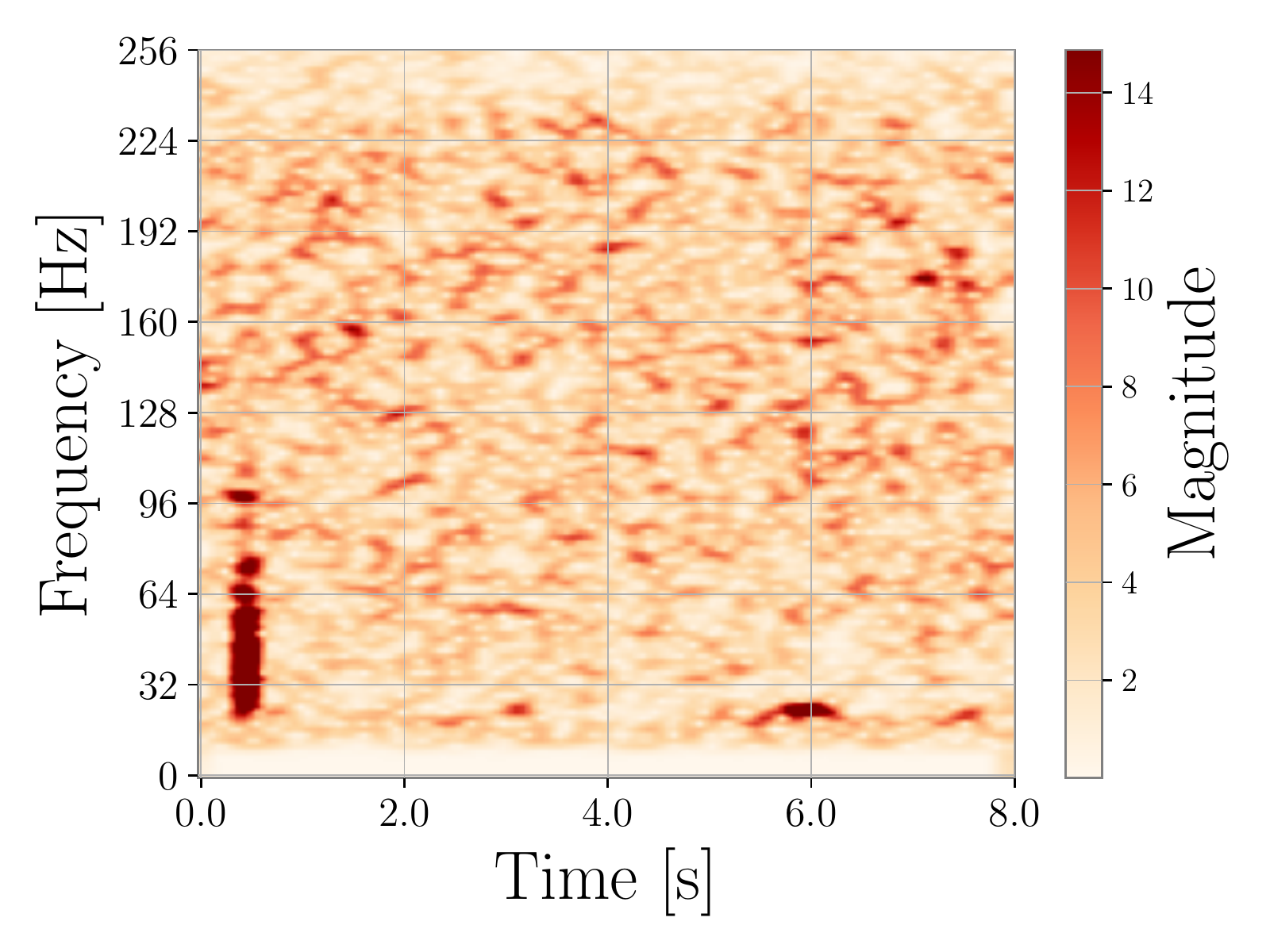}
  \end{center}
    \end{minipage}
    \begin{minipage}{0.5\hsize}
  \begin{center}
   \includegraphics[width=1\textwidth]{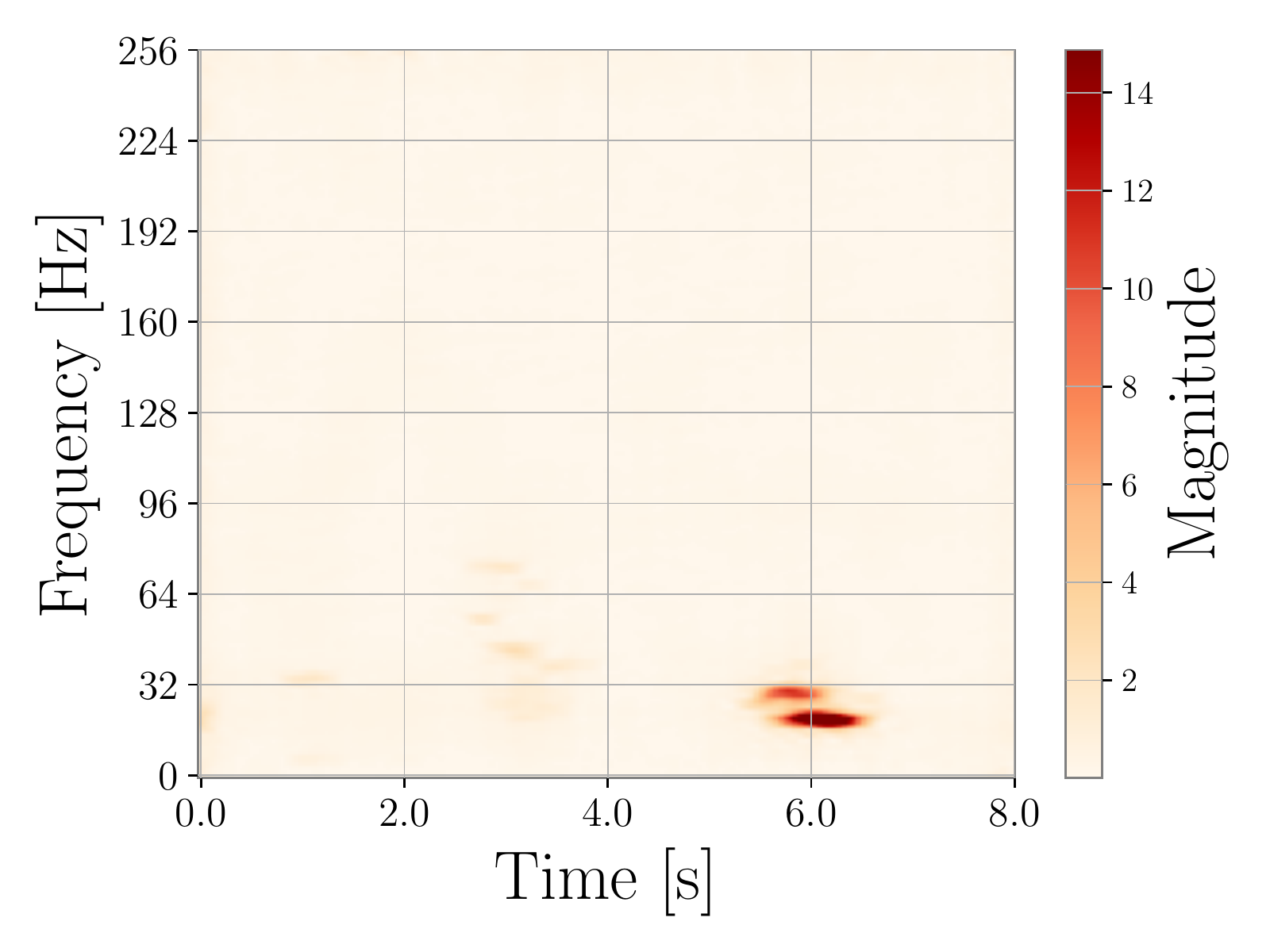}
  \end{center}
    \end{minipage}
    \begin{minipage}{0.5\hsize}
  \begin{center}
   \includegraphics[width=1\textwidth]{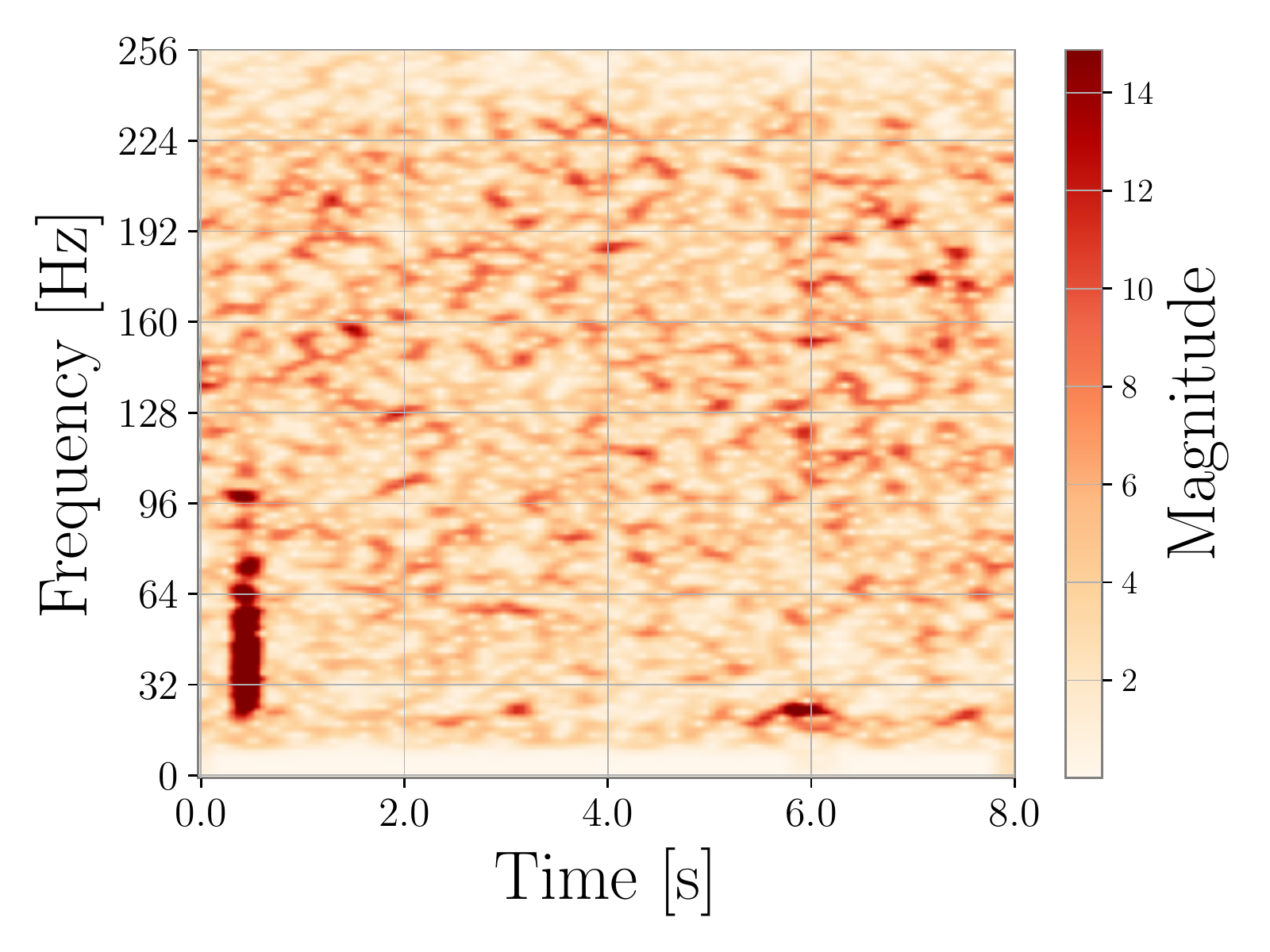}
  \end{center}
    \end{minipage}
    \begin{minipage}{0.5\hsize}
  \begin{center}
   \includegraphics[width=1\textwidth]{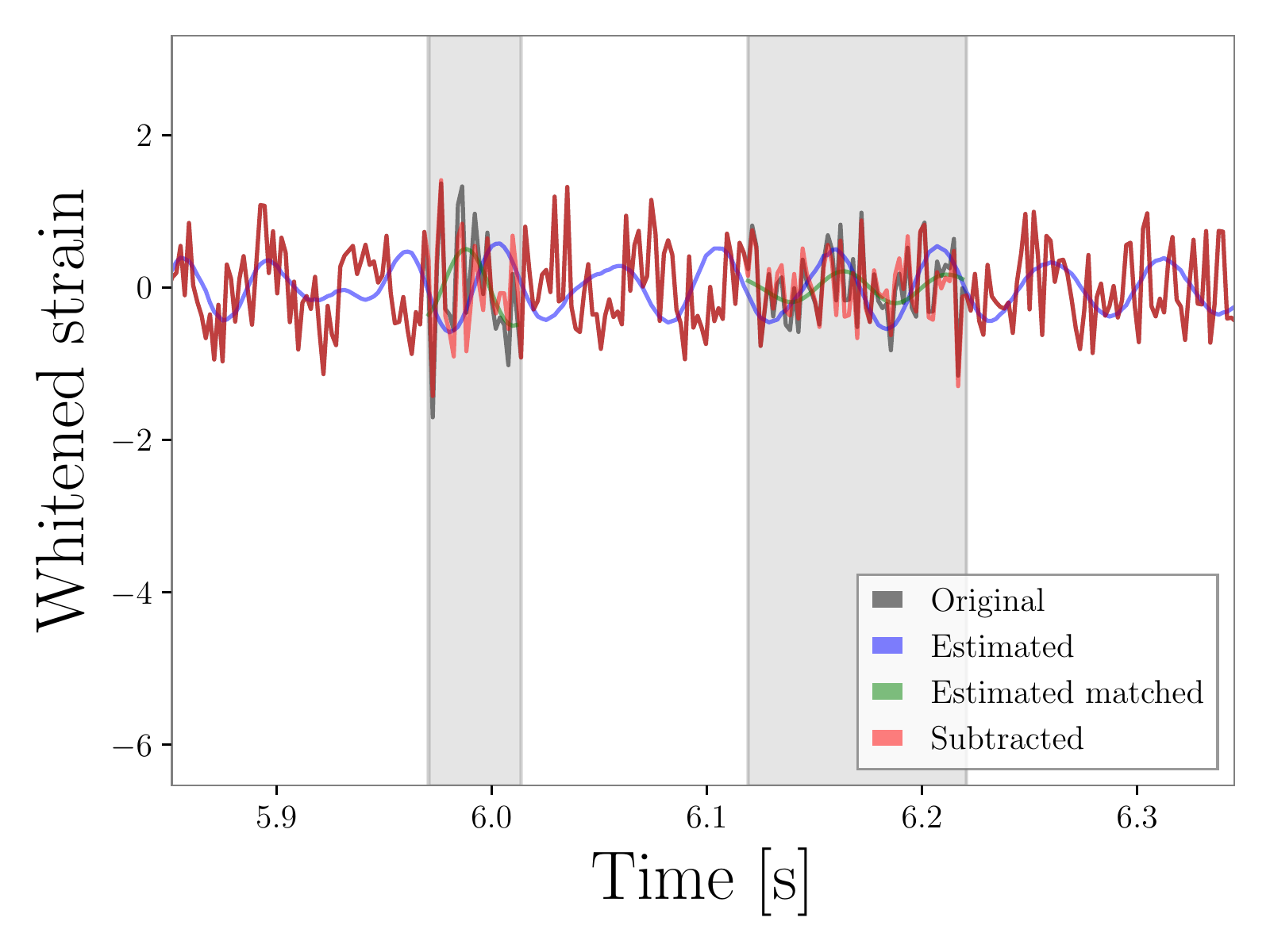}
  \end{center}
    \end{minipage}
\caption[The \acp{mSTFT} and time series of the least case in the testing set of {\it Scattered light} glitches.]%
{The \ac{mSTFT} of a down-sampled strain data at 512 Hz with a high-pass filter at 10 Hz (top-left), the network estimated \ac{mSTFT} (top-right), and the \ac{mSTFT} of the strain data after glitch subtraction (top-left) in the least testing sample of \textit{Scattered light} glitches. In the top-left panel, the gray (blue/green/red) curves denote the original (estimated/estimated-matched/subtracted) whitened time series, where the estimated-matched time series is created after the amplitude and phase corrections with the least square fitting within divided segments shown as the gray bands. The overlap between \ac{mSTFT} of the extracted glitch waveform and the estimated \ac{mSTFT} is $O=0.21$ and ${\rm \ac{FNR}}=0.02$.}
\label{fig:sclight_least}
\end{figure}

Figures \ref{fig:sclight_optimal} (\ref{fig:sclight_median}/\ref{fig:sclight_least}) shows the \ac{mSTFT} of a strain data down-sampled at 512 Hz and high-passed at 10 Hz in the testing set, the corresponding estimated \ac{mSTFT}, and the time series used to subtract the glitches of the optimal (median/least) case, where the overlap between the \ac{mSTFT} of the extracted glitch waveform and the estimated \ac{mSTFT} is $O=0.92$ (0.65/0.21) and ${\rm \ac{FNR}} = 0.84$ (0.58/0.02). In the least case, a short-lived arch-like glitch at $\sim 6$ seconds in the top-left panel in Fig.\ \ref{fig:sclight_least} is estimated by the network. However, only the fractions of this glitch are subtracted due to our selection criterion about the glitch presence mentioned above. In this case, the small value of ${\rm \ac{FNR}}=0.02$ is also due to the presence of a non-{\rm Scattered light} glitch at $\sim 0.5$ seconds because this glitch contributes to ${\rm SNR}_b$ dominantly. We note that we build the network for a particular class of glitches so that the glitch at $\sim 0.5$ seconds in the least case is consistent with the performance of our built network.          


\subsection{EXTREMELY LOUD GLITCHES} \label{extremely_loud_results}
Unlike \textit{Scattered light} glitches where the waveforms in the strain can be analytically modeled with monitored mirror motions and the suspension systems \cite{Was:2020ziy}, many other glitches are so far not modeled because of an incomplete understanding of their physical non-linear noise-coupling mechanisms. The non-linear activation function used in the network allows us to model non-linear noise couplings and subtract glitches. We apply our method to a class of \textit{Extremely loud} glitches with \ac{SNR} above 7.5 between January 1st, 2020 and February 3rd, 2020 in the \ac{L1} detector from the \textsc{Gravity Spy} catalog. We use 4 witness channels with high confidence: L1:LSC-POP\_A\_LF\_OUT\_DQ, L1:LSC-REFL\_A\_LF\_OUT\_DQ, L1:ASC-X\_TR\_A\_NSUM\_OUT\_DQ, and L1:ASC-Y\_TR\_B\_NSUM\_OUT\_DQ, identified by \tool \cite{mogushi2021application}. This glitch class is expected to be produced by the laser intensity dips and have extremely high excess power in 10- 4096 Hz, lasting $\sim 0.2$ seconds. 

\begin{figure}[!ht]
    \begin{minipage}{0.5\hsize}
  \begin{center}
   \includegraphics[width=1\textwidth]{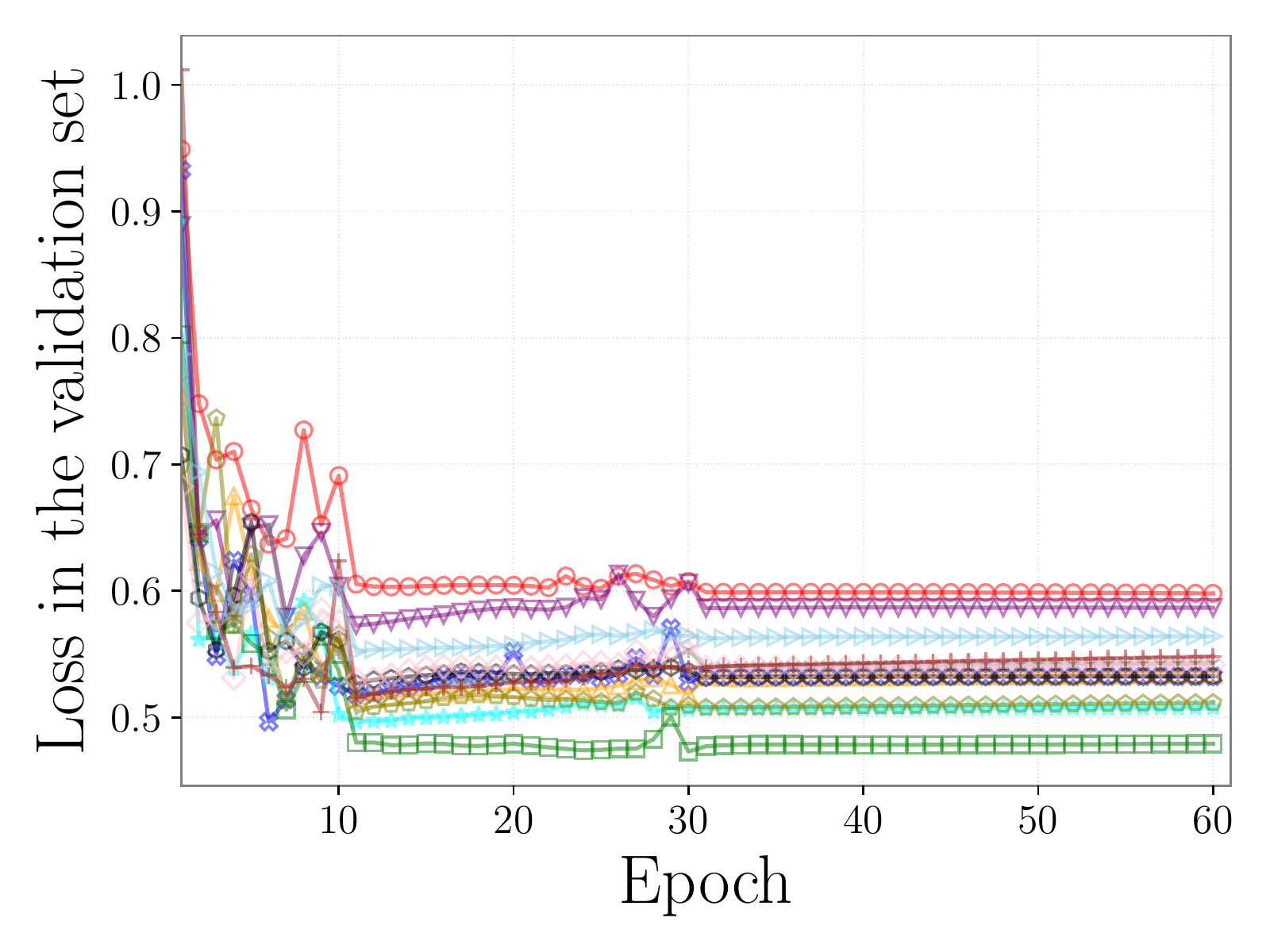}
  \end{center}
    \end{minipage}
    \begin{minipage}{0.5\hsize}
  \begin{center}
   \includegraphics[width=1\textwidth]{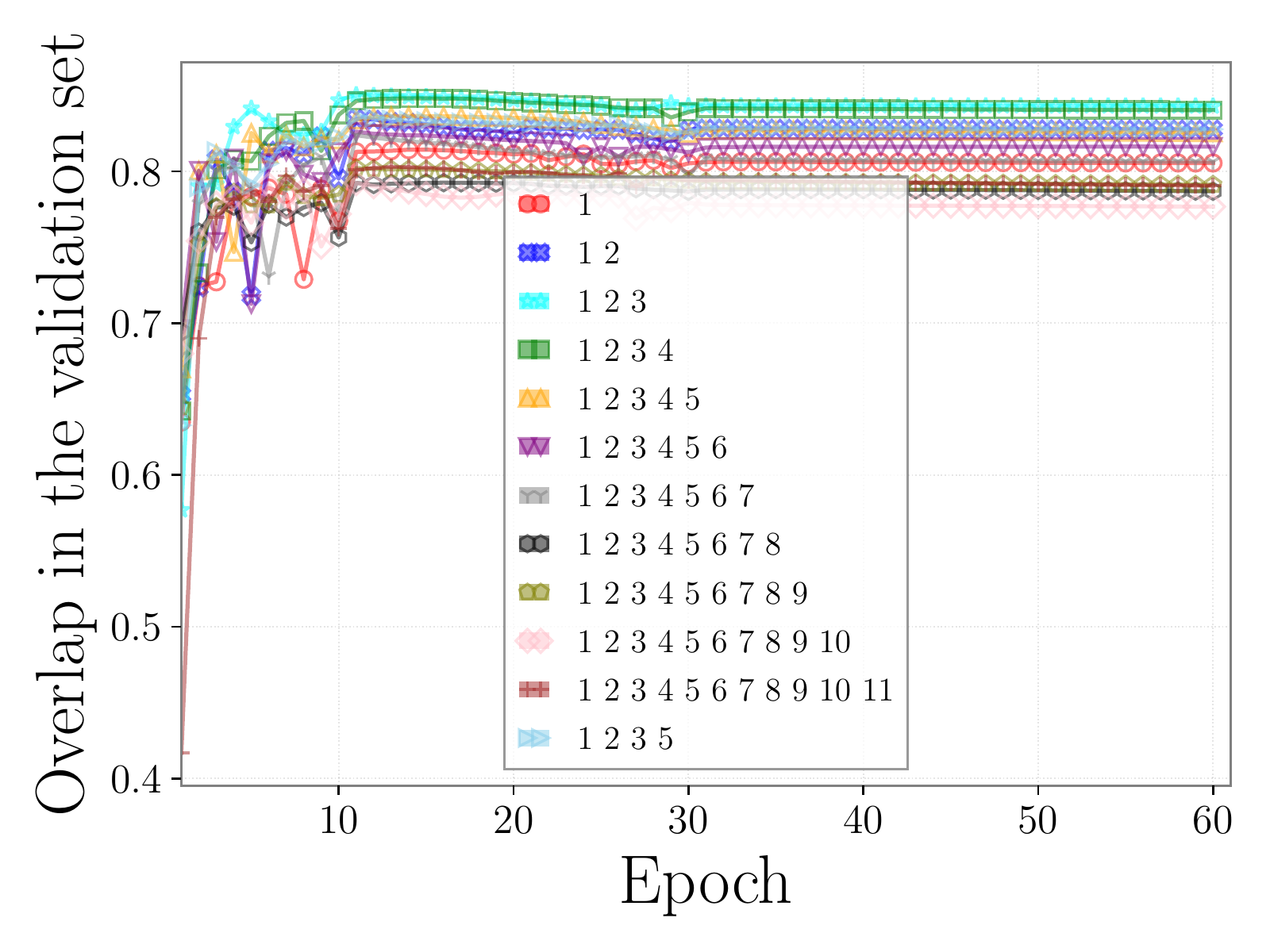}
  \end{center}
    \end{minipage}
    \begin{minipage}{0.5\hsize}
  \begin{center}
   \includegraphics[width=1\textwidth]{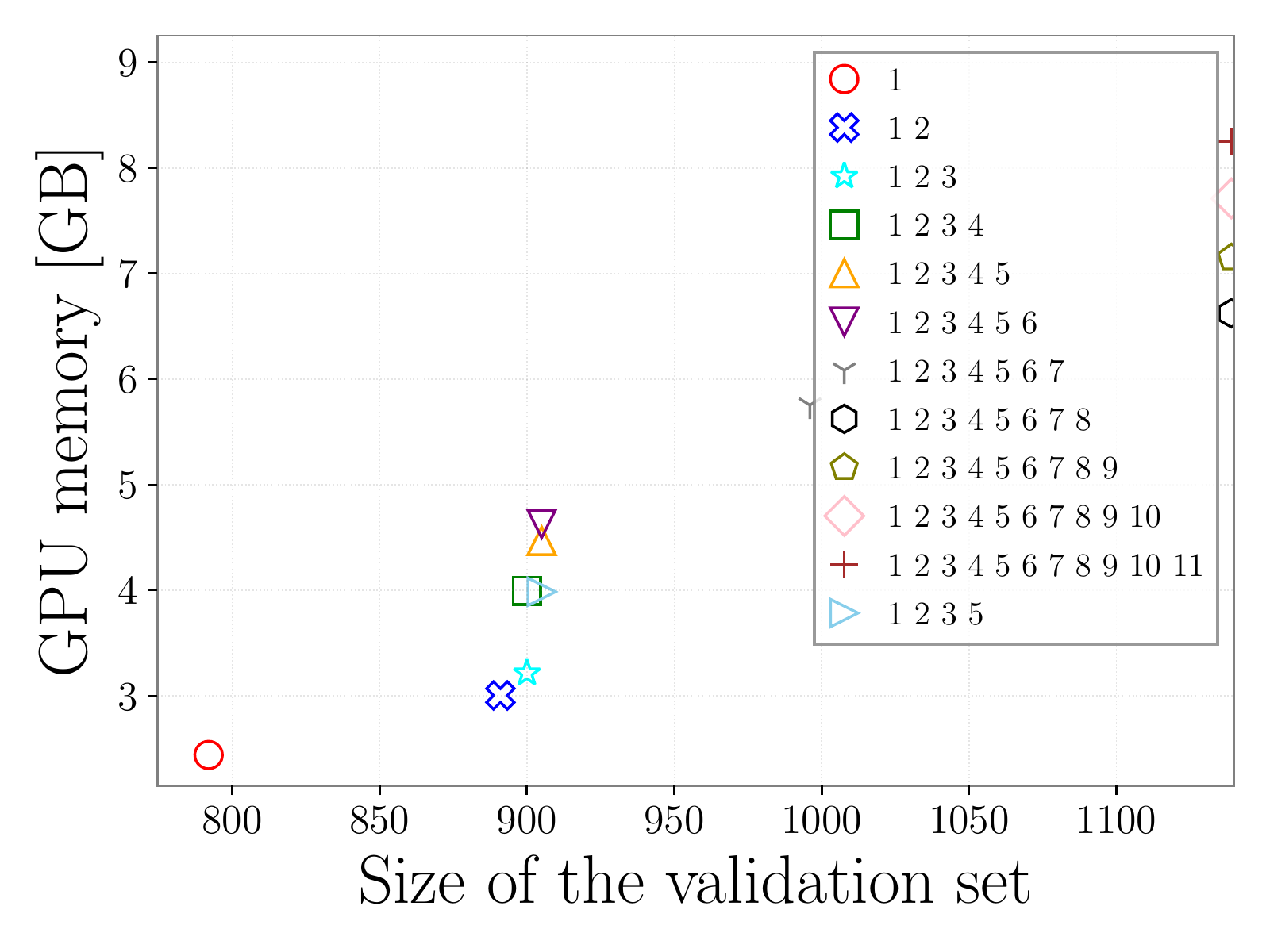}
  \end{center}
    \end{minipage}
    \begin{minipage}{0.5\hsize}
  \begin{center}
   \includegraphics[width=1\textwidth]{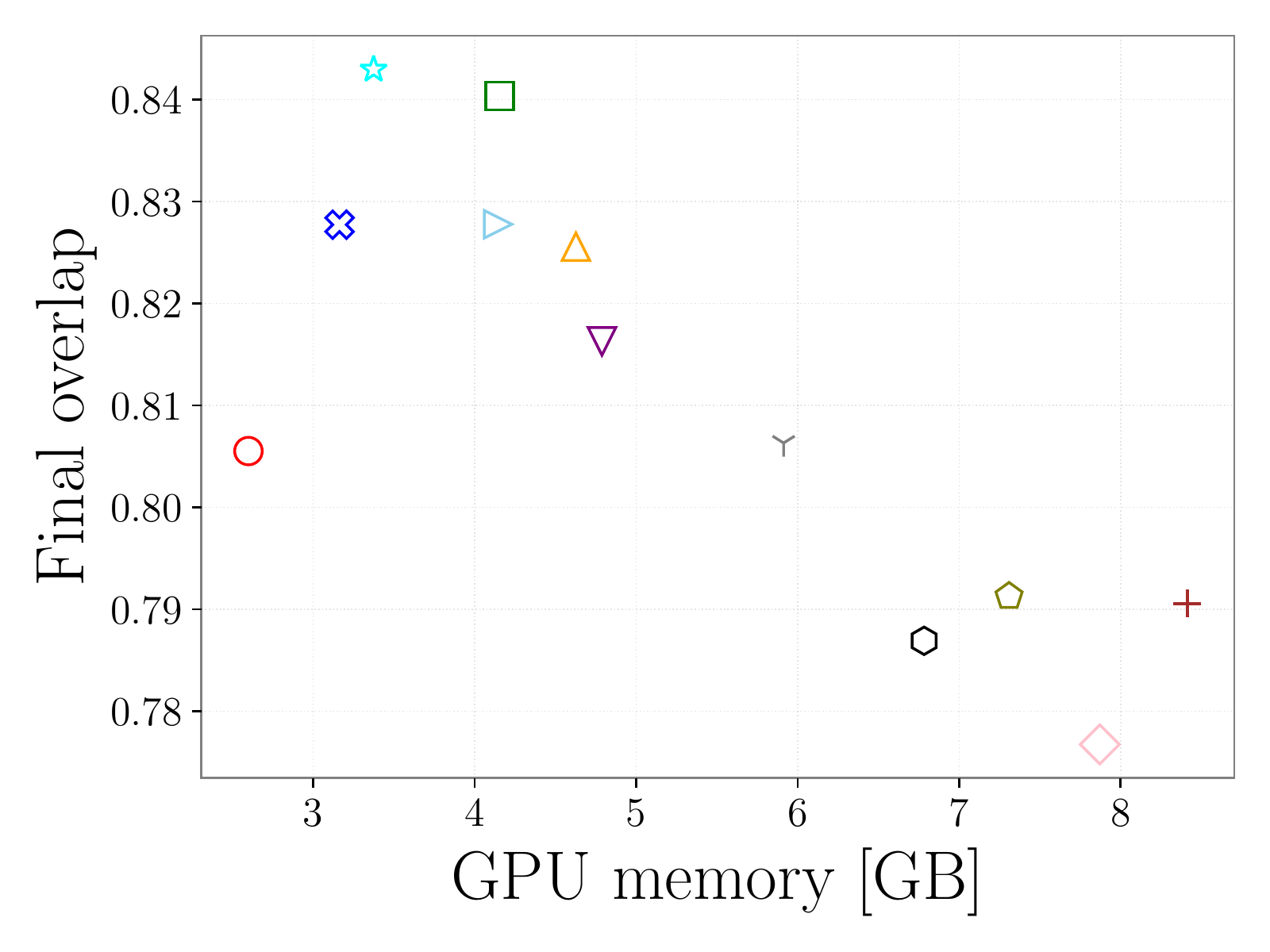}
  \end{center}
    \end{minipage}
\caption[Variation the network performance due to different sets of channels for {\it Extremely loud} glitches.]%
    {Losses defined in Eq.\ (\ref{eq:loss_func}) (top-left) and overlaps (top-right) over epochs in the validation sets using channel sets containing $i^{\rm th}$ ranked channels for {\it Extremely loud} glitches. The bottom panels show sizes of the validation sets, overlaps at the end of the training, and used GPU memories.}
\label{fig:exloud_varation_channel}
\end{figure}

To check that the choice of witness channels noted above is sufficient, we train the network with various sets containing channels with $p_g = 0.99$ up to $p_g = 0.55$. We consider 11 different channel sets, where $i^{\rm th}$ set contains up to $i^{\rm th}$ ranked channels. Also, we consider the $12^{\rm th}$ channel set containing the 1-3$^{\rm th}$ ranked channels and $5^{\rm th}$ ranked channels because the $3^{\rm rd}$ ranked channel (L1:ASC-Y\_TR\_B\_NSUM\_OUT\_DQ) with $p_g=0.98$ and the $4^{\rm th}$ ranked channel (L1:ASC-Y\_TR\_A\_NSUM\_OUT\_DQ) with $p_g=0.97$ both record the transmitted light in yaw-direction in the alignment length control sub-system and have almost the same glitch-coupling information. The $5^{\rm th}$ ranked channel (L1:LSC-POP\_A\_LF\_OUT\_DQ) records the transmitted light in low frequencies from the power recycling cavity. With the same procedure in Sec. \ref{scattered_light_result}, we compare losses, overlaps, GPU memories, and size of the validation sets for all channel sets as shown in Fig.\ \ref{fig:exloud_varation_channel}. The termination overlaps range from 0.77 obtained with the $10^{\rm th}$ channel set to 0.84 obtained with the $3^{\rm rd}$ channel set. The overlap decreases by adding channels 4-11$^{\rm th}$ ranked channels. In particular, 6-11$^{\rm}$ channels have values of $p_g<0.57$, indicating no evidence of being witnesses so that the network obtain no significant glitch-coupling information from these low confidence channels with the use of redundant GPU memories up to 8.4 GB. In the following, we choose the $12^{\rm th}$ channel set containing witness channels noted in the previous paragraph because its termination overlap is only less than 2\% compared to the largest value obtained with the $3^{\rm rd}$ channel set, and 0.5\% increase of the data set.        

During the data pre-processing, we re-sample the time series of the strain and the witness channels to a sampling rate of 2048 Hz and apply a high-pass filter at 10 Hz. We consider the time range outside of the 5-second window around the glitch time from the \textsc{Gravity Spy} catalog to be the background-noise region and use the 99 percentile pixel-value to extract the glitch waveform within the 5-second window because these glitches are isolated and not repeating, unlike \textit{Scattered light} glitches. We set the sample dimensions of the (training/validation/testing) sets to be (3879/940/1233) with the segment overlaps of (96.8/96.8/87.5)\%, where there is no overlapping time between the three sets and the testing set is later than the other sets. We create the \ac{mSTFT} with a duration of 2 seconds, a frequency range up to 1024 Hz, and (time/frequency) resolutions of 0.0156 seconds and 8 Hz, respectively. We use a rectangular kernel with a size of $(13,4)$ in the autoencoder in the network. During the post-processing, we consider the region of the glitch presence to be the time when the absolute value of the estimated glitch waveform is above the 90 percentile of the corresponding values across the testing set. We choose the divided window length for the least square fitting to be as small as 0.02 seconds and expand the window length from the center of the glitch by a factor of 1.1. 

\begin{figure}[ht!]
    \centering
    \includegraphics[width=0.5\linewidth]{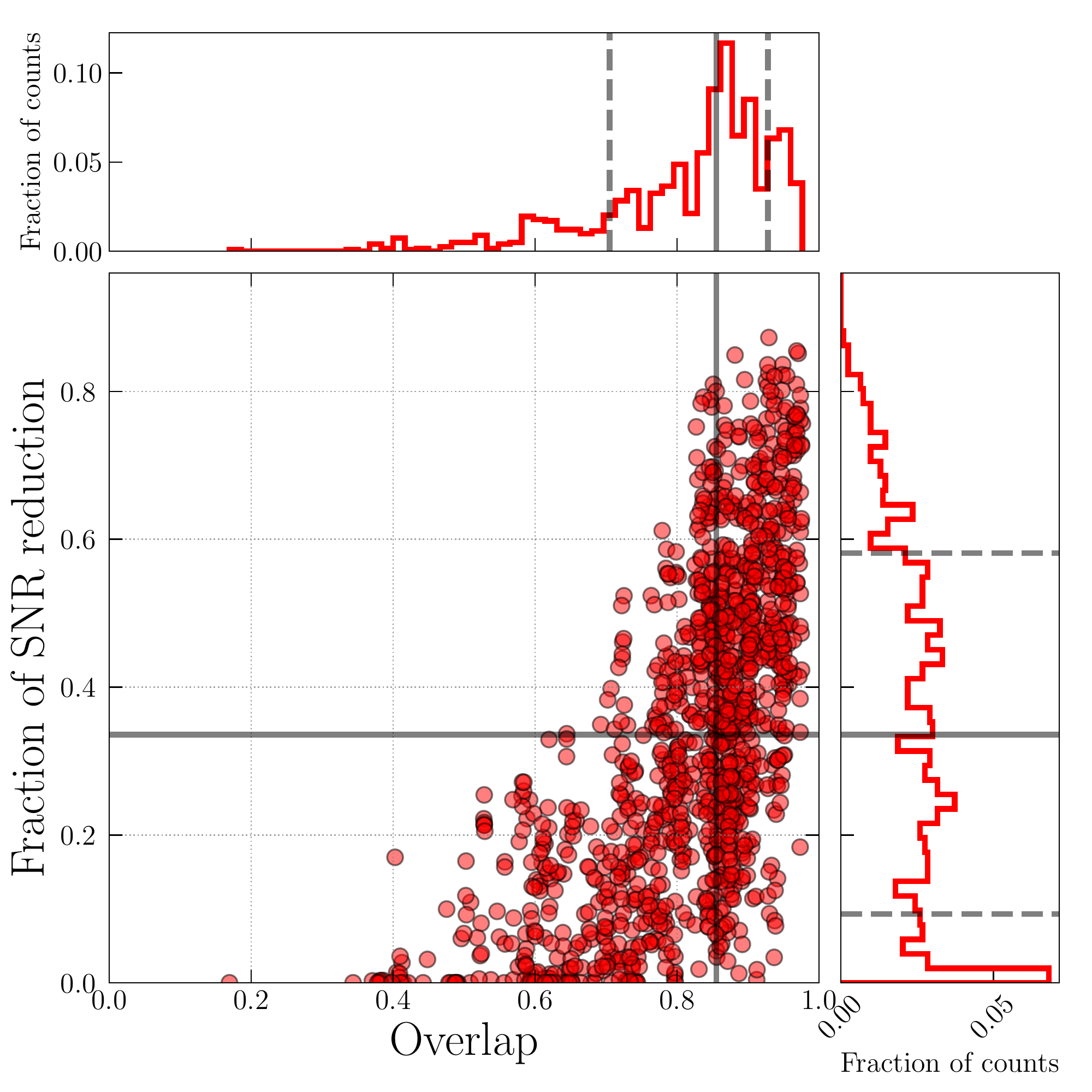}
    \caption[Distribution of the overlap of the extracted and estimated \acp{mSTFT} of \textit{Extremely loud} glitches in the testing set.]%
    {Distribution of the overlap of the extracted and estimated \acp{mSTFT} of \textit{Extremely loud} glitches in the testing set. The black solid and dashed lines denote the median and 1-$\sigma$ percentiles.}
    \label{fig:overlap_vs_snr_exloud} 
\end{figure}

Figure \ref{fig:overlap_vs_snr_exloud} shows the distribution of the overlap of the \ac{mSTFT} and the \ac{FNR} of the testing set of \textit{Extremely loud} glitches. We find the overlap ranging from $\sim$0.7-0.9 and the \ac{FNR} ranging $\sim 0.1-0.6$ with 1-$\sigma$ percentiles and no negative \ac{FNR} indicating no additional glitch added to the strain data.

\begin{figure}[ht!]
    \begin{minipage}{0.5\hsize}
  \begin{center}
   \includegraphics[width=1\textwidth]{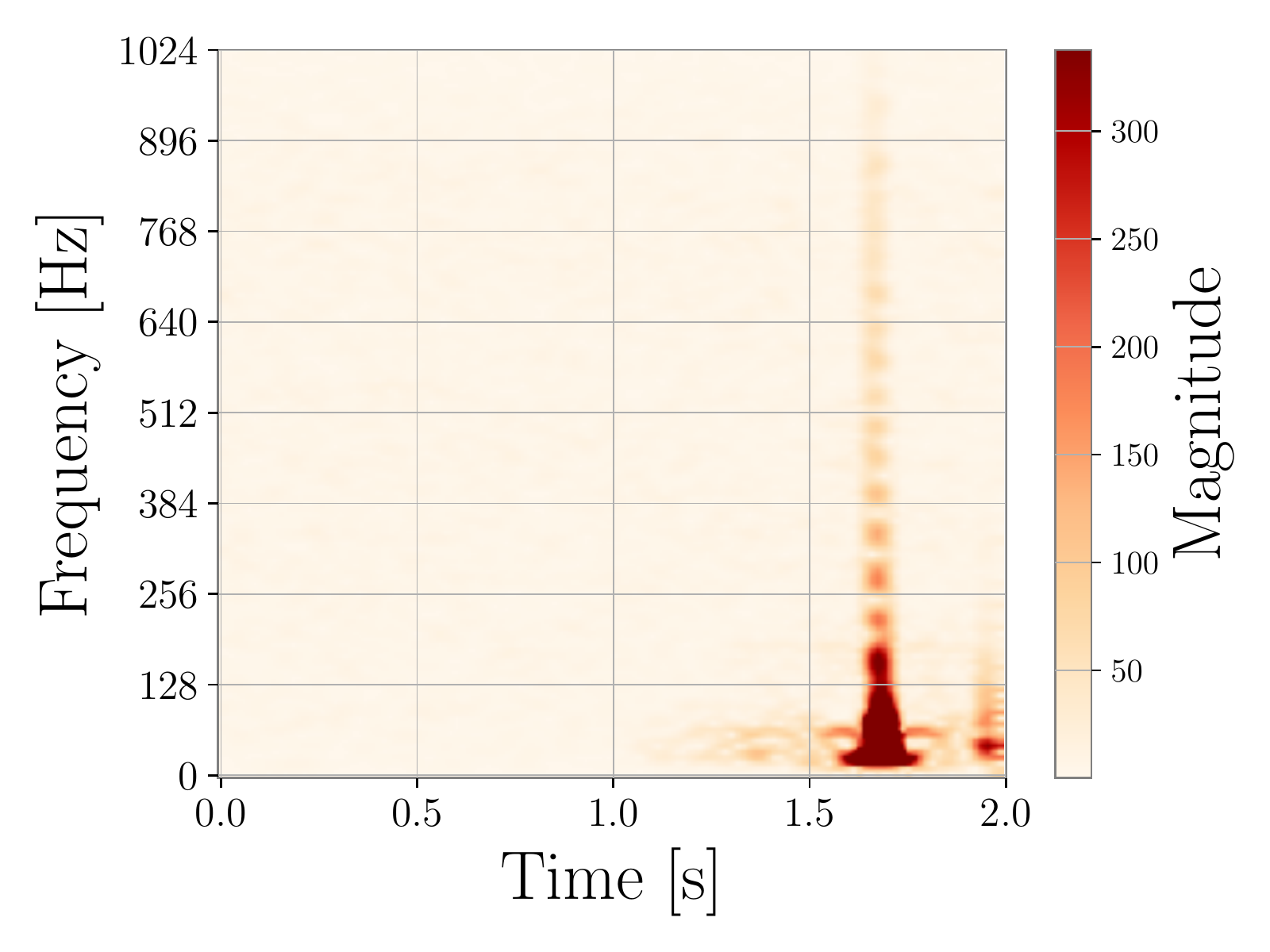}
  \end{center}
    \end{minipage}
    \begin{minipage}{0.5\hsize}
  \begin{center}
   \includegraphics[width=1\textwidth]{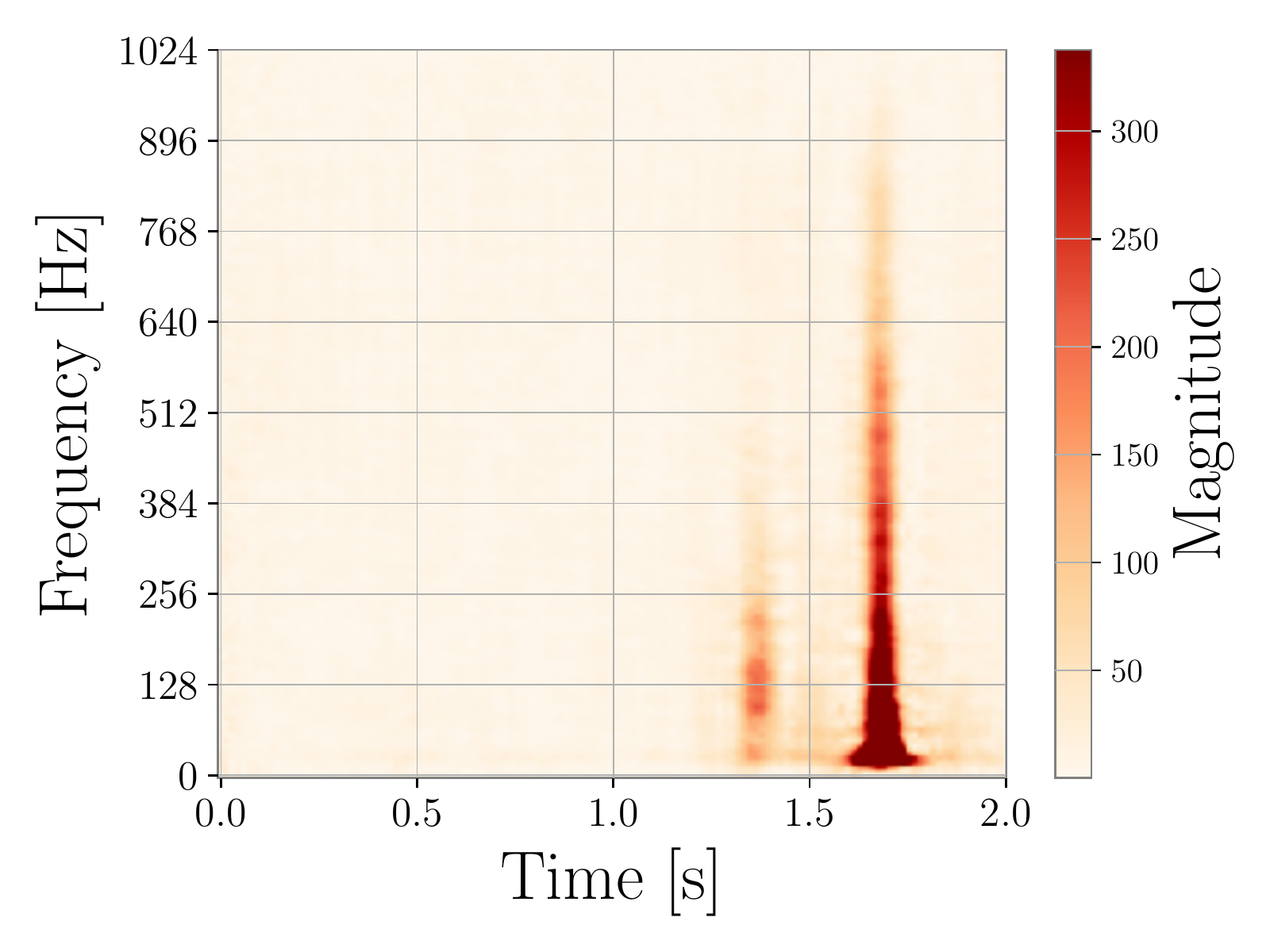}
  \end{center}
    \end{minipage}
    \begin{minipage}{0.5\hsize}
  \begin{center}
   \includegraphics[width=1\textwidth]{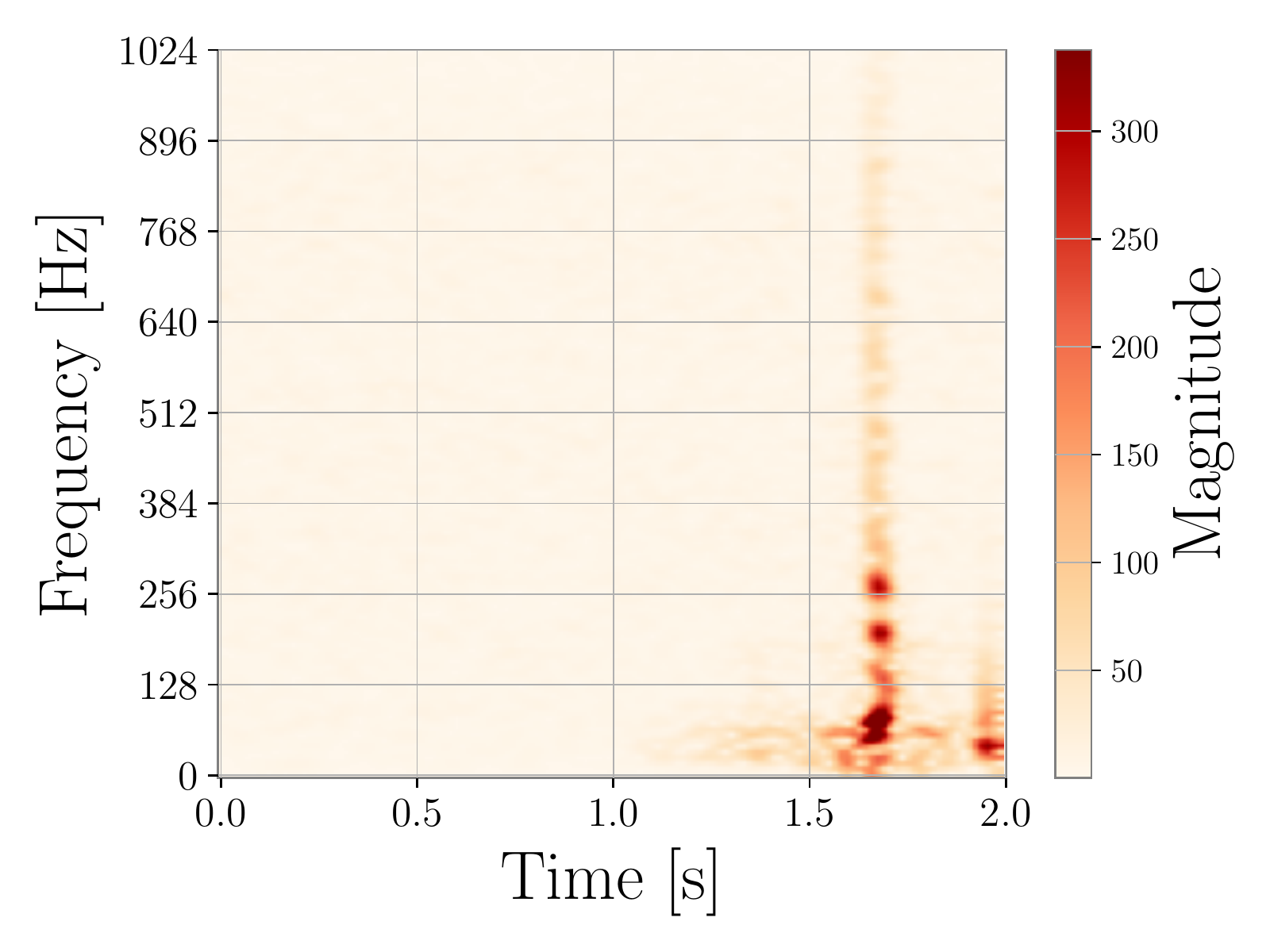}
  \end{center}
    \end{minipage}
    \begin{minipage}{0.5\hsize}
  \begin{center}
   \includegraphics[width=1\textwidth]{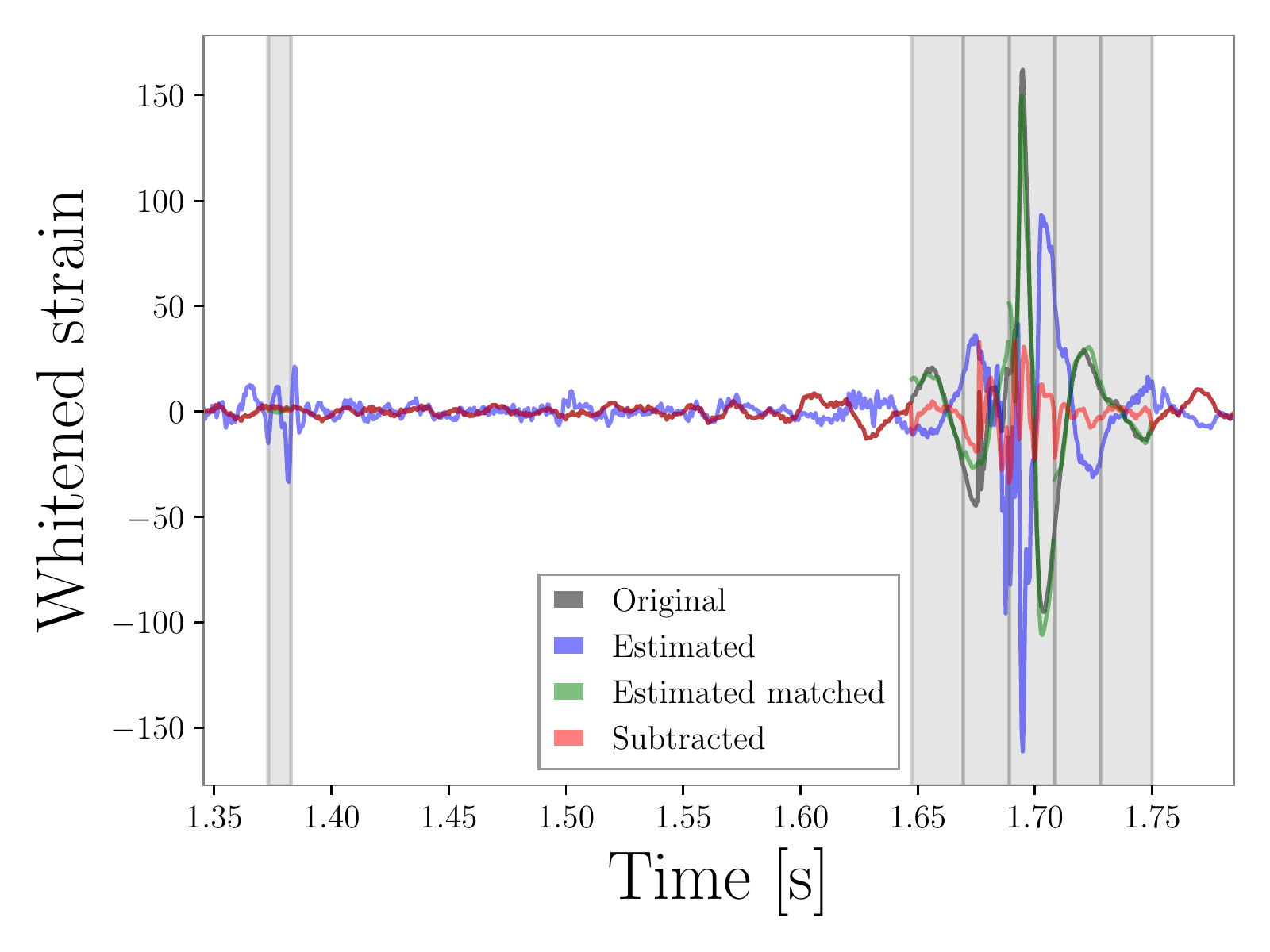}
  \end{center}
    \end{minipage}
\caption[The \acp{mSTFT} and time series of the optimal case in the testing set of {\it Extremely loud} glitches.]%
    {The \ac{mSTFT} of a down-sampled strain data at 2048 Hz with a high-pass filter at 10 Hz (top-left), the network estimated \ac{mSTFT} (top-right), and the \ac{mSTFT} of the strain data after glitch subtraction (top-left) in the optimal testing sample of \textit{Extremely loud} glitches. In the top-left panel, the gray (blue/green/red) curves denote the original (estimated/estimated-matched/subtracted) whitened time series, where the estimated-matched time series is created after the amplitude and phase corrections with the least square fitting within divided segments shown as the gray bands. The overlap between \ac{mSTFT} of the extracted glitch waveform and the estimated \ac{mSTFT} is $O=0.93$ and ${\rm \ac{FNR}}=0.87$.}%
    \label{fig:exloud_optimal}
\end{figure}

\begin{figure}[ht!]
    \begin{minipage}{0.5\hsize}
  \begin{center}
   \includegraphics[width=1\textwidth]{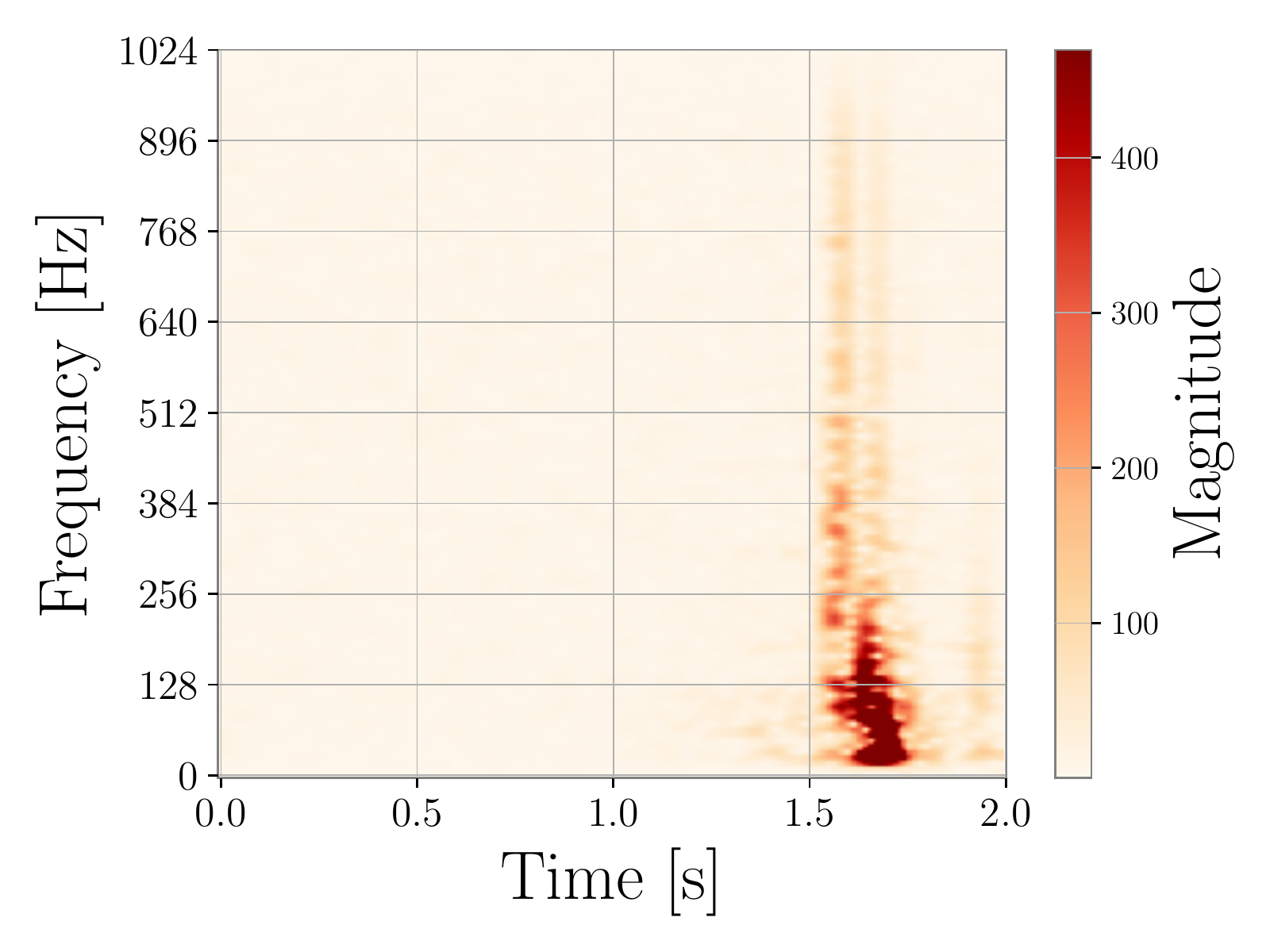}
  \end{center}
    \end{minipage}
    \begin{minipage}{0.5\hsize}
  \begin{center}
   \includegraphics[width=1\textwidth]{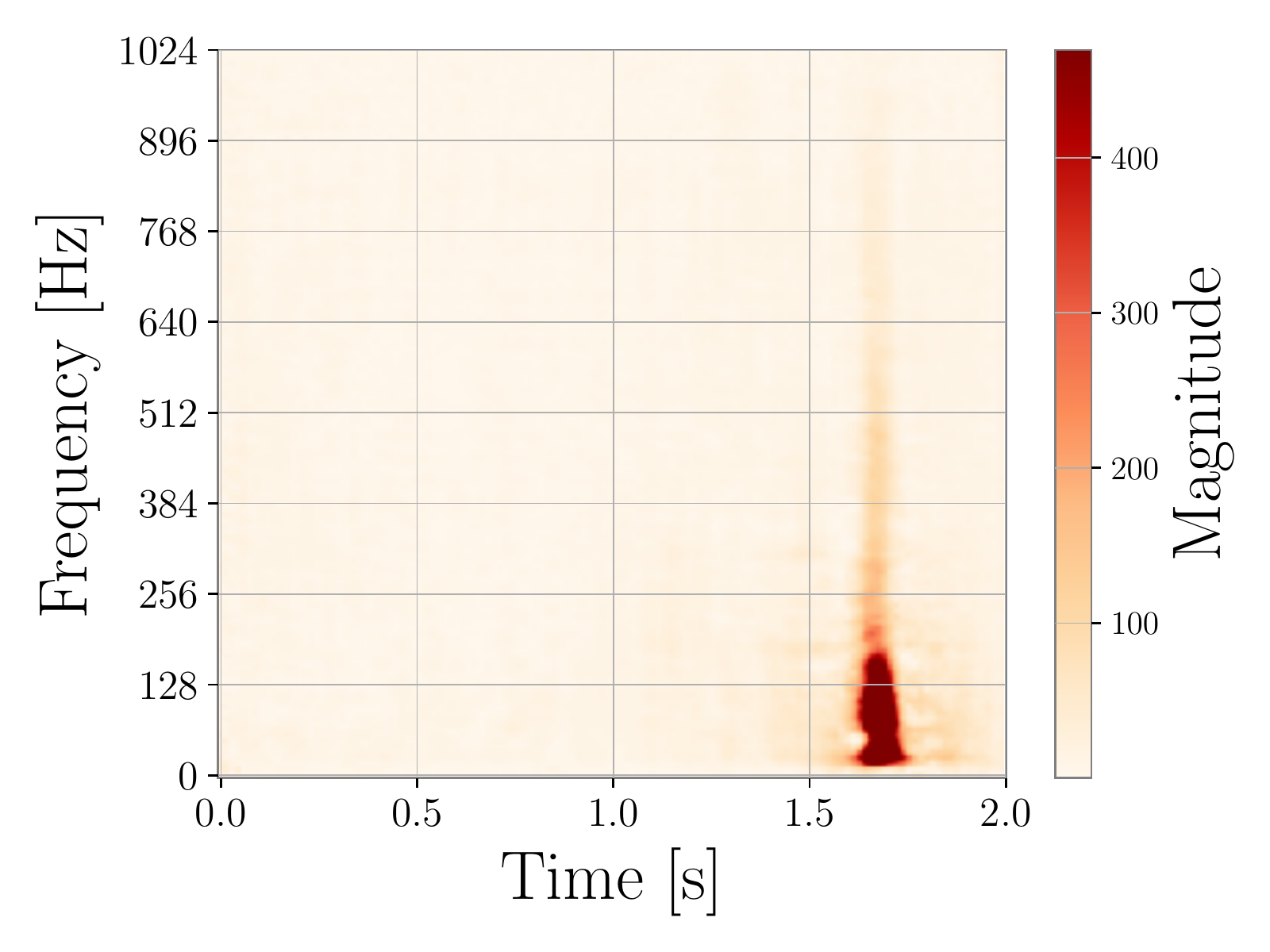}
  \end{center}
    \end{minipage}
    \begin{minipage}{0.5\hsize}
  \begin{center}
   \includegraphics[width=1\textwidth]{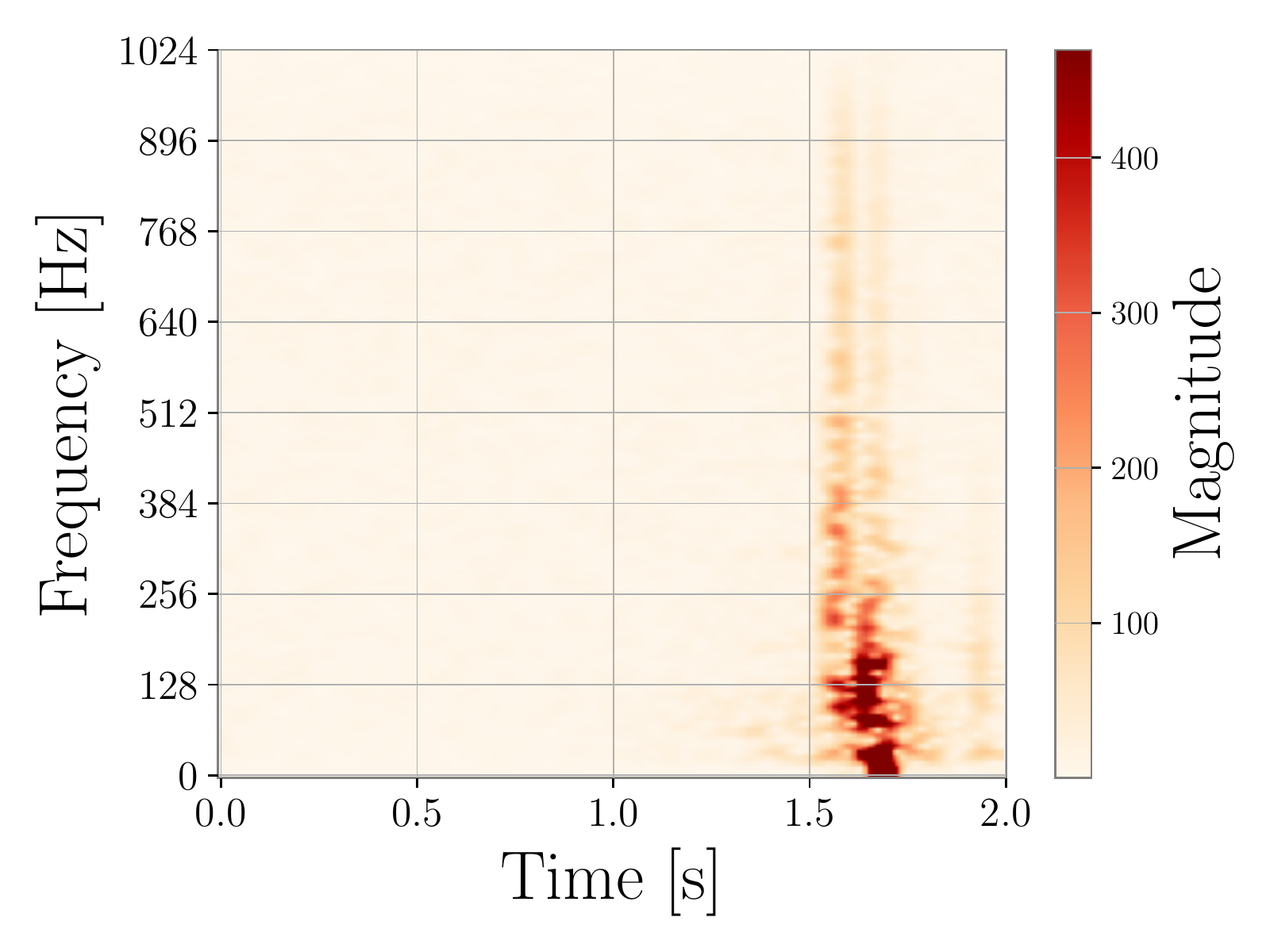}
  \end{center}
    \end{minipage}
    \begin{minipage}{0.5\hsize}
  \begin{center}
   \includegraphics[width=1\textwidth]{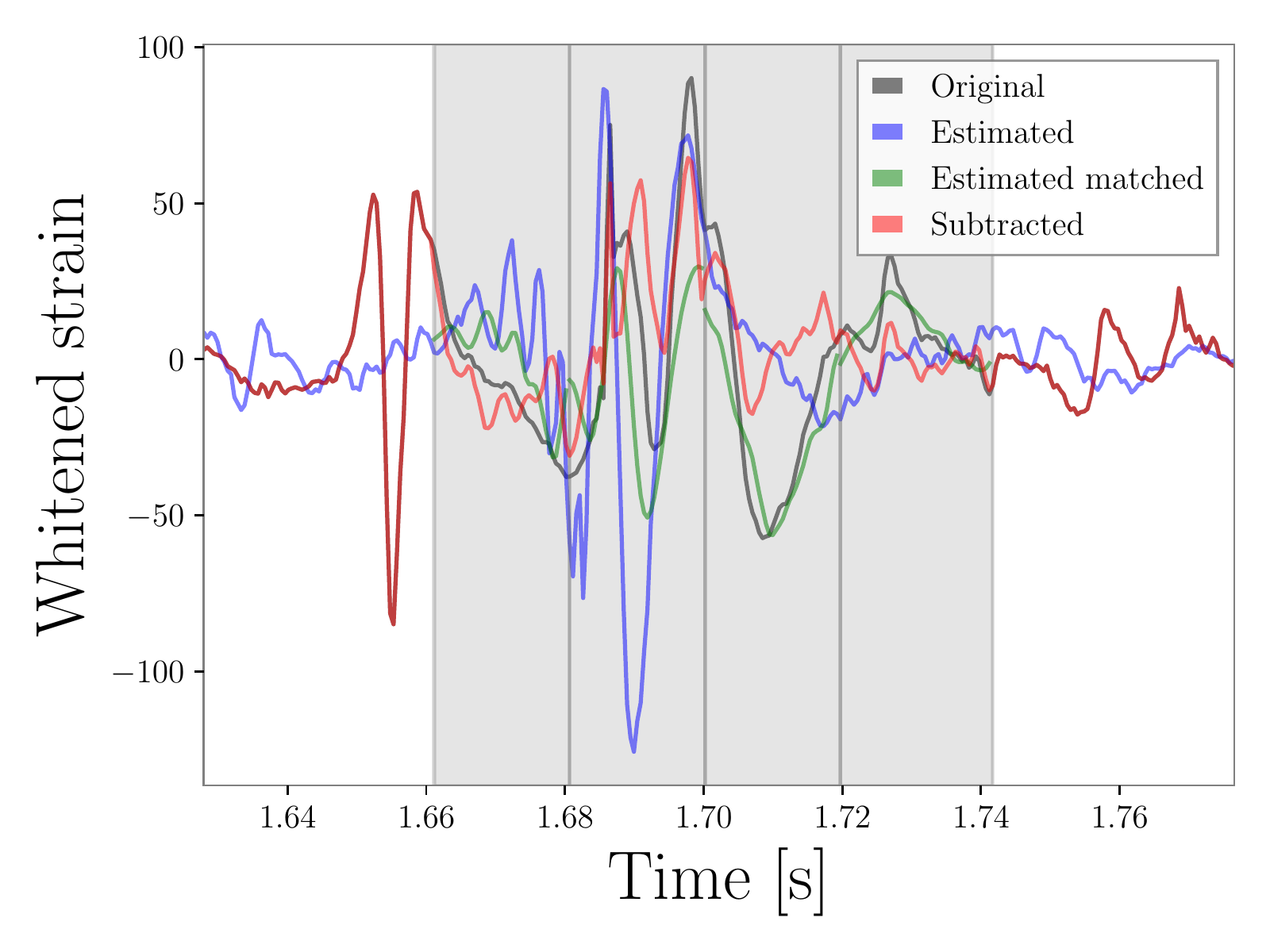}
  \end{center}
    \end{minipage}
\caption[The \acp{mSTFT} and time series of the median case in the testing set of {\it Extremely loud} glitches.]%
    {The \ac{mSTFT} of a down-sampled strain data at 2048 Hz with a high-pass filter at 10 Hz (top-left), the network estimated \ac{mSTFT} (top-right), and the \ac{mSTFT} of the strain data after glitch subtraction (top-left) in the median testing sample of \textit{Extremely loud} glitches. In the top-left panel, the gray (blue/green/red) curves denote the original (estimated/estimated-matched/subtracted) whitened time series, where the estimated-matched time series is created after the amplitude and phase corrections with the least square fitting within divided segments shown as the gray bands. The overlap between \ac{mSTFT} of the extracted glitch waveform and the estimated \ac{mSTFT} is $O=0.86$ and ${\rm \ac{FNR}}=0.33$.}%
    \label{fig:exloud_median}
\end{figure}

\begin{figure}[ht!]
    \begin{minipage}{0.5\hsize}
  \begin{center}
   \includegraphics[width=1\textwidth]{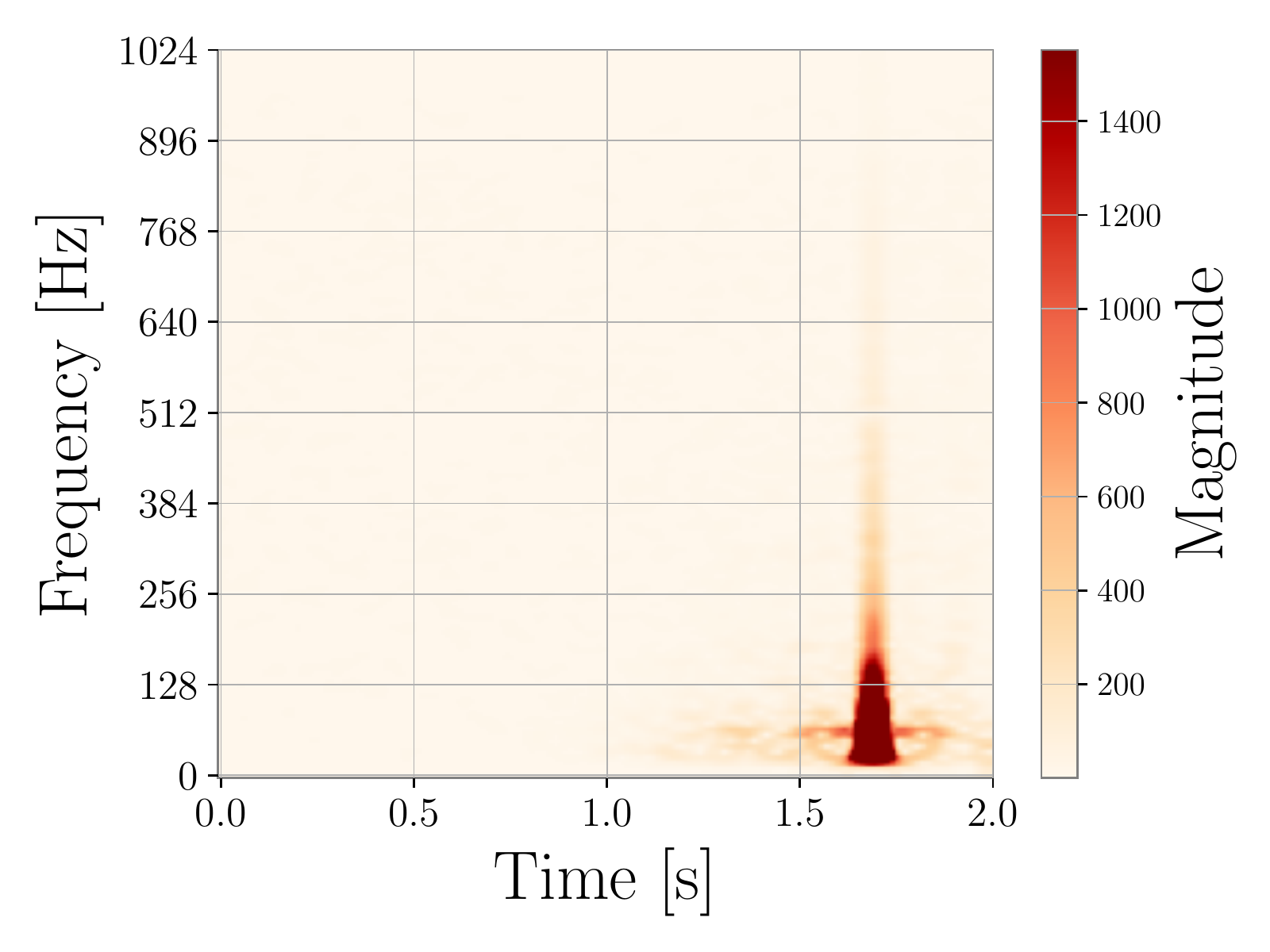}
  \end{center}
    \end{minipage}
    \begin{minipage}{0.5\hsize}
  \begin{center}
   \includegraphics[width=1\textwidth]{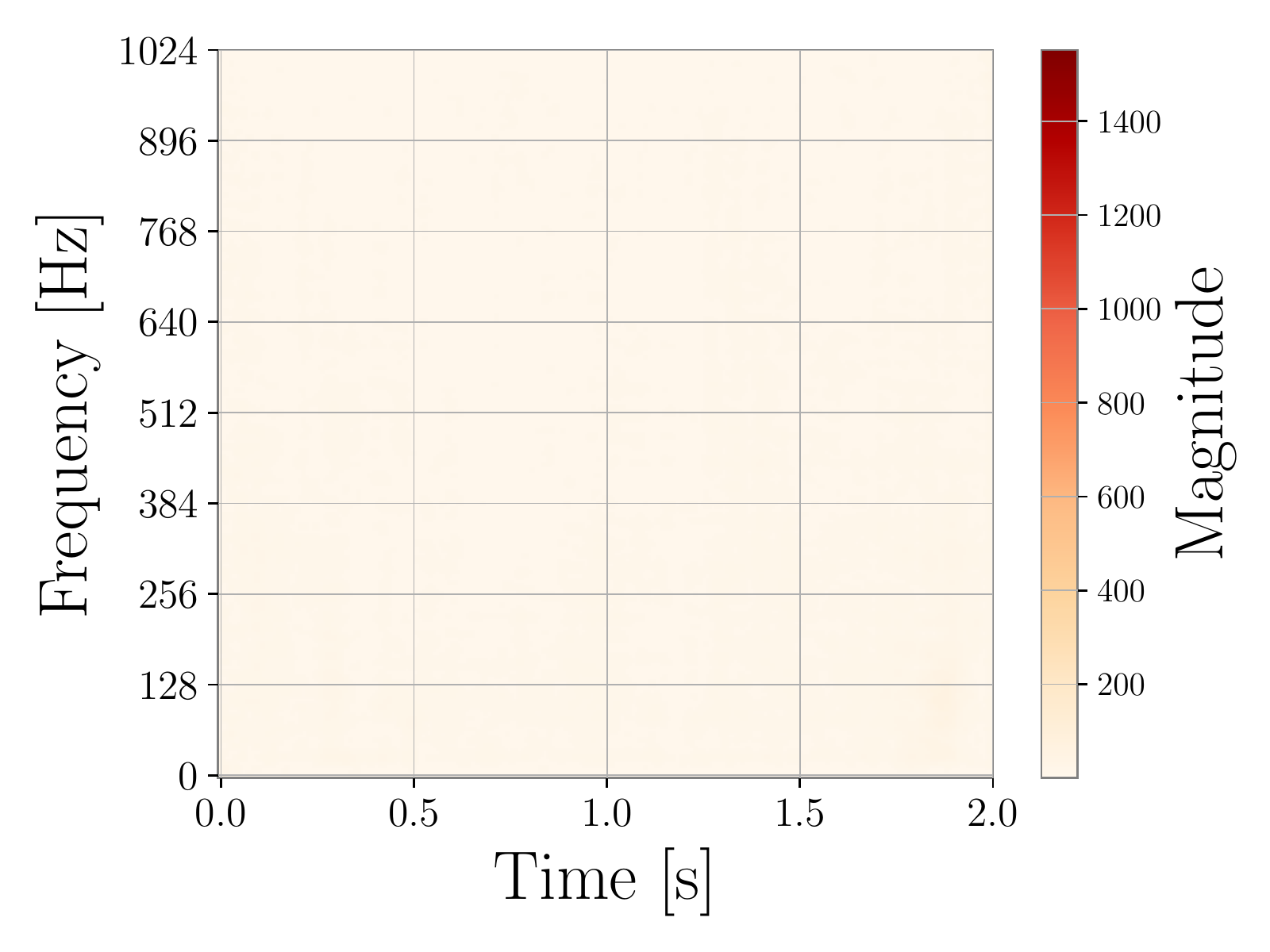}
  \end{center}
    \end{minipage}
    \begin{minipage}{0.5\hsize}
  \begin{center}
   \includegraphics[width=1\textwidth]{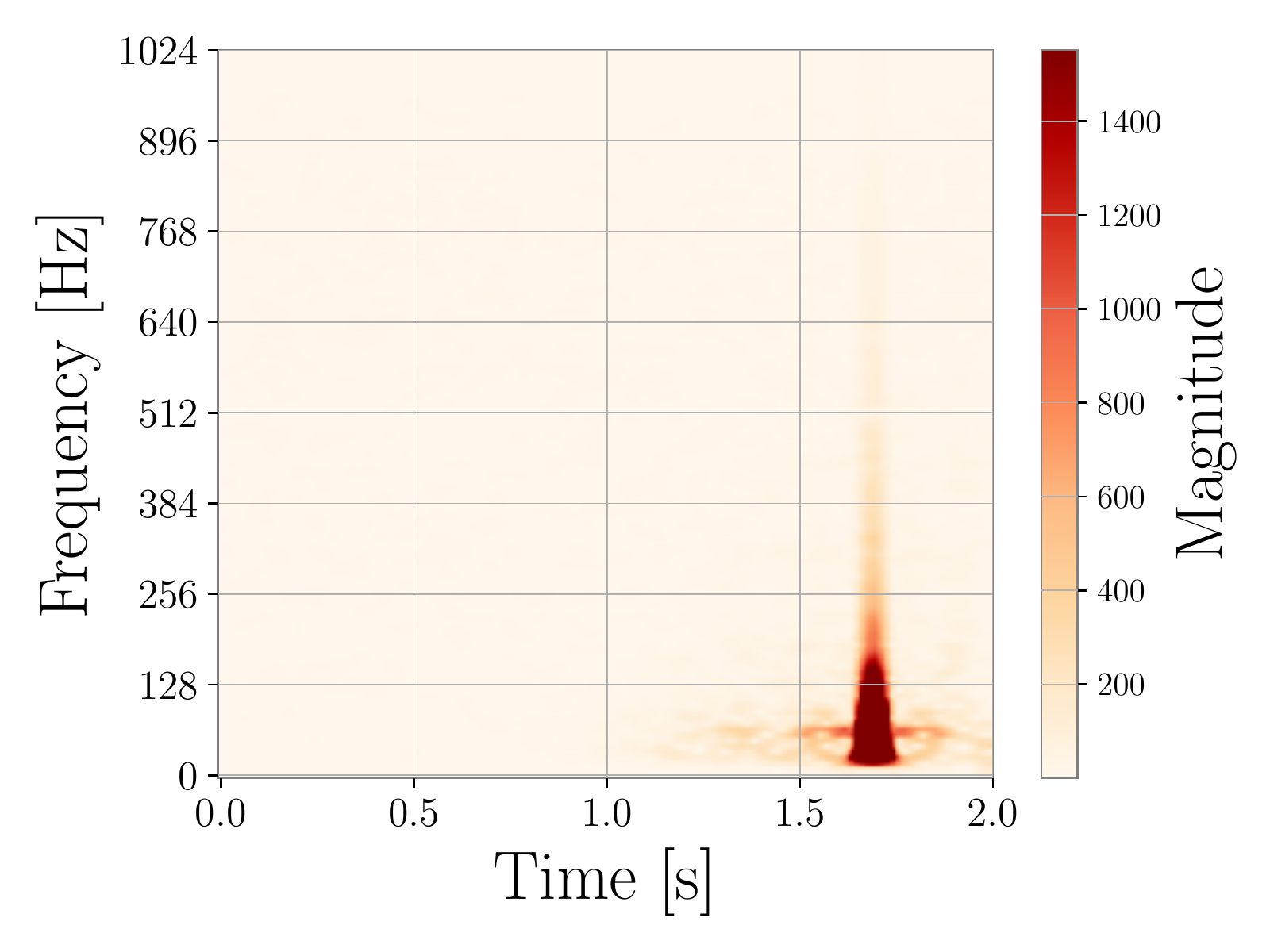}
  \end{center}
    \end{minipage}
    \begin{minipage}{0.5\hsize}
  \begin{center}
   \includegraphics[width=1\textwidth]{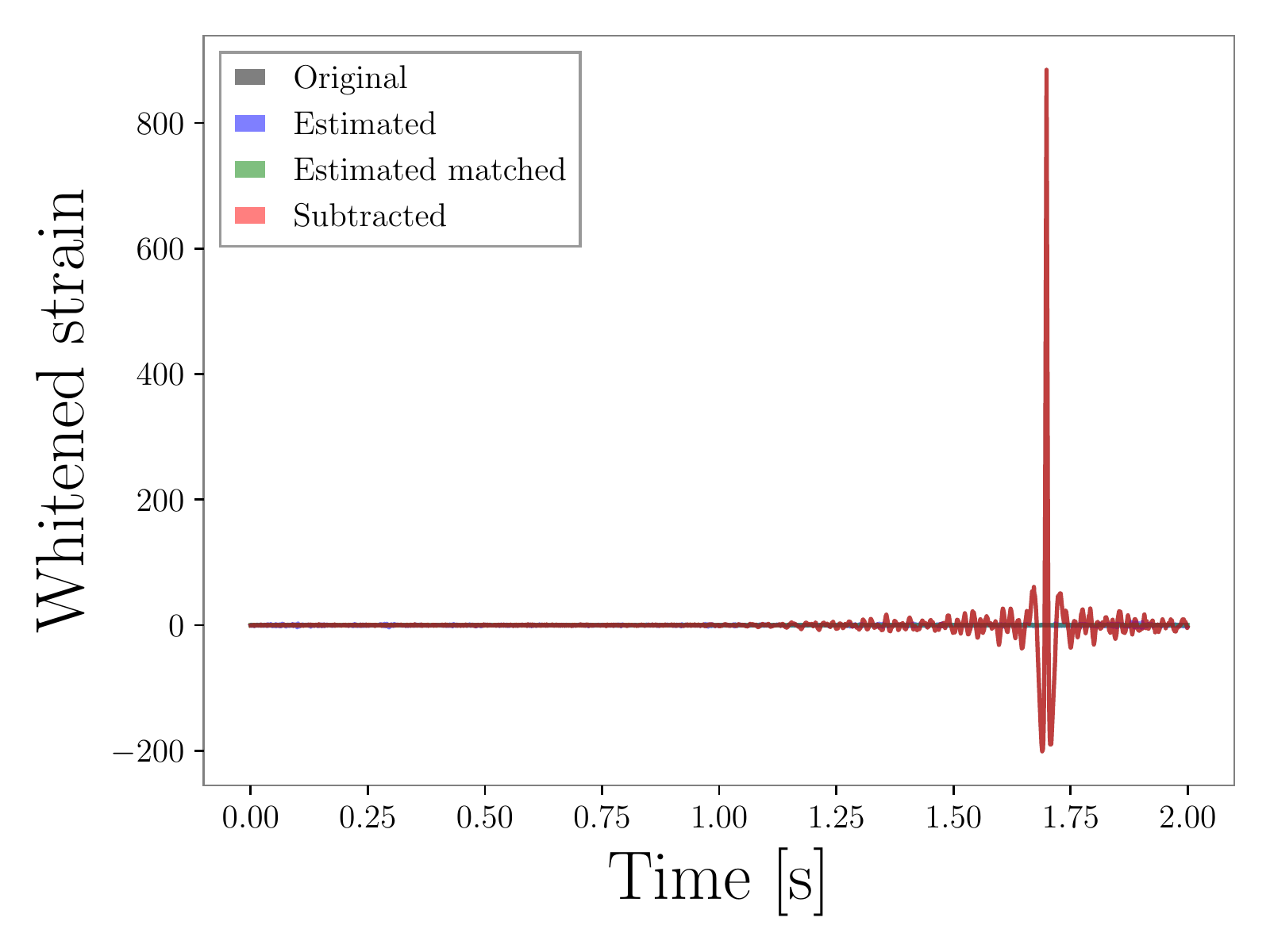}
  \end{center}
    \end{minipage}
\caption[The \acp{mSTFT} and time series of the least case in the testing set of {\it Extremely loud} glitches.]%
    {The \ac{mSTFT} of a down-sampled strain data at 2048 Hz with a high-pass filter at 10 Hz (top-left), the network estimated \ac{mSTFT} (top-right), and the \ac{mSTFT} of the strain data after glitch subtraction (top-left) in the least testing sample of \textit{Extremely loud} glitches. In the top-left panel, the gray (blue/green/red) curves denote the original (estimated/estimated-matched/subtracted) whitened time series, where the estimated-matched time series is created after the amplitude and phase corrections with the least square fitting within divided segments shown as the gray bands. The overlap between \ac{mSTFT} of the extracted glitch waveform and the estimated \ac{mSTFT} is $O=0.17$ and ${\rm \ac{FNR}}=0$.}%
    \label{fig:exloud_least}
\end{figure}

Figures \ref{fig:exloud_optimal} (\ref{fig:exloud_median}/\ref{fig:exloud_least}) shows the \ac{mSTFT} of a strain data down-sampled at 2048 Hz and high-passed at 10 Hz in the testing set, the corresponding estimated \ac{mSTFT}, and the time series used to subtract the glitches of the optimal (median/least) case, where the overlap between the \ac{mSTFT} of the extracted glitch waveform and the estimated \ac{mSTFT} is $O=0.93$ (0.86/0.17) and ${\rm \ac{FNR}} = 0.84$ (0.33/0). In the least case, our chosen four witness channels seem not to witness no excess power coincident with the glitch so that the network estimates no glitches.


Our method subtracts \textit{Scattered light} glitches more efficiently than \textit{Extremely loud} glitches because the network finds it difficult to model short-lived ($\sim0.2$ seconds) non-linear couplings for \textit{Extremely loud} glitches. In values of the overlap binned from 0.5 to 1.0 with a bin width of 0.1, the averaged value of \ac{FNR} for \textit{Scattered light} glitches are greater than the corresponding values for \textit{Extremely loud} glitches by a factor ranging from 1.3 for the bin $\mathcal{O}=0.9-1.0$ to 3.8 for the bin $\mathcal{O}=0.6-0.7$.

\subsection{INJECTION RECOVERY WITH COHERENT WAVEBURST} \label{injection_cwb}

Subtracting glitches results in a new strain data which is expected to contain smaller energy due to the presence of glitches, leading to better detectability of astrophysical signals. One way of examining the robustness of our glitch-subtraction method is to add software-simulated signals with known astrophysical parameters into the strain data before subtraction and use \ac{GW} detection pipelines to recover the injected signals. In this process, we can assess whether the glitch subtraction technique reduces only the targeted glitches without manipulating the measured astrophysical signals. We use our glitch-subtraction method after injecting a signal in coincidence with a glitch.  

The presence of glitches adversely affects the unmodeled \ac{GW} signal searches that do not rely on known waveforms in priori. In \ac{O3a}, the percentages of the single-detector observing time removed by the data-quality vetoes for the unmodeled searches are greater than the percentages for modeled searches by a factor of $\sim2.7$ and $\sim7.9$ for the \ac{H1}-detector and \ac{L1}-detector, respectively \cite{Abbott:2020niy}. Therefore, it is more beneficial to apply  glitch-subtraction techniques for unmodeled searches. We use \ac{cWB} to recover injections before and after subtraction and compare these recovered signals as well as the recovered signal injected in a simulated colored Gaussian noise with the \ac{PSD} of the \ac{L1} data when the glitch subtraction is applied.

To account for the performance of our glitch subtraction operated on only the \ac{L1} data, we create a simulated colored Gaussian noise for the \ac{H1} data with the same sensitivity as the \ac{L1} data and inject signals both in the \ac{L1} and \ac{H1} detectors where the signal coincides with a glitch in the \ac{L1} data. The ranking statistic $\rho$ of \ac{cWB} accounts for the correlation of a signal injected on the two-detector data so that higher-ranking statistics for a given injection imply that the recovered signal in the \ac{L1} data is more similar to the signal in the \ac{H1} data, indicating successful glitch subtraction and a better detectability.

\subsubsection{Gaussian-modulated Sinusoid Injections} \label{injection_cwb_sin_gaussian}

Following studies of unmodeled \ac{GW} signal searches \cite{Was:2012zq, Abbott:2009kk, AdrianMartinez:2012tf, Abadie:2012bz, Briggs:2012ce, Abbott:2008zzb}, we inject a circularly polarized Gaussian-modulated sinusoid signal: 
\begin{equation}\label{eq:sin_gau_waveform}
    \left[
         \begin{array}{c}
            h_+(t)           \\
            h_\times (t) \\
        \end{array}
    \right]
    =
    \left[
        \begin{array}{c}
            \cos\{2\pi f_0 (t-t_0)\} \\
            \sin\{2\pi f_0 (t-t_0)\} \\
        \end{array}
    \right]
    \exp\left\{-\frac{2\pi f_0 (t-t_0)}{2Q^2} \right\}\,,
\end{equation}
where $f_c$ is the central frequency, $t_0$ is the center time, $d$ is the distance to the source, $A$ is an arbitrary amplitude scaling factor, $Q$ determines the length of the signal. 

Motivated by studies \cite{Was:2012zq, Abbott:2009kk, AdrianMartinez:2012tf, Abadie:2012bz, Briggs:2012ce, Abbott:2008zzb} that use $f_c=150$ Hz, we also  choose $f_c=150$ Hz. To validate the glitch subtraction technique successfully subtract glitches in the presence of signals with duration compatible glitches, we choose $Q=30$, where the injected signal lasts $\sim1.5$ seconds which is compatible with the duration of {\it Scattered light} glitches. In addition to the above motivation, we consider signals similar to the first detection of \ac{IMBBH} \cite{PhysRevLett.125.101102}, where the detected signal has a peak frequency of $\sim 50$ Hz and $\sim 3$ wave cycles. The improvement in the \ac{IMBBH} detection by subtracting glitches might be useful to understand the mechanism of astrophysical populations \cite{Abbott:2020mjq}. Therefore, we choose $f_c=50$ Hz and $Q=5$ for the second choice.     

Using the two representative signal waveforms with different sets of parameters: $f_c=50$ Hz, $Q=5$ and $f_c=150$ Hz, $Q=30$, and choosing the injected \acp{SNR} to be uniformly sampled from a set of \acp{SNR}, the source direction to be isotropically sampled in the sky, the injected time to be uniformly sampled in a given time window, we examine the pipeline performance on the testing samples with an optimal set of values of ${\rm \ac{FNR}=0.84},O=0.92$ (shown in Fig.\ \ref{fig:sclight_optimal}) and a median value of ${\rm \ac{FNR}=0.58},O=0.65$ (shown in Fig.\ \ref{fig:sclight_median}) for \textit{Scattered light} glitches as well as an optimal set of values of ${\rm \ac{FNR}=0.87},O=0.93$ (shown in Fig.\ \ref{fig:exloud_optimal}) and a median value of ${\rm \ac{FNR}=0.33},O=0.86$ (shown in Fig.\ \ref{fig:exloud_median}) for \textit{Extremely loud} glitches. To assess the effect of the injection time on the pipeline performance, we consider two different injection-time windows: the subtracted portion in the testing-sample data or the full length of the testing-sample data. Because we apply the glitch subtraction in the partial data with excess power detected from the estimated glitch waveform and keep the original data for the rest of the data portion, we inject signals in the subtracted portion to study if our technique can subtract glitches that are overlapping with signals. We also consider the full length of the testing-sample data as the injection-time window because subtracting glitches may affect detections of signals near to glitches but not overlapping with them. The injection times are uniformly sampled in the full window of 0.4-7.6 and 0.1-1.8 seconds, and the partial window of 0.1-1.8 (3.5-5.4) and 1.65-1.75 (1.65-1.75) seconds for the optimal (median) case of the {\it Scattered light} and {\it Extremely loud} glitches, respectively, with a time step of 5\% of the window length. We use sets of injected \acp{SNR} of $\{2, 4, 8, 16, 30, 50, 100\}$ and $\{2, 4, 8, 16, 30, 50, 100, 200, 300 ,400 ,500\}$ {\it Scattered-light} and {\it Extremely-loud} glitch sets, respectively, where larger injected \acp{SNR} are chosen for {\it Extremely-loud} glitch set to assess the \ac{cWB} detection performance for injections overlapping with glitches with high excess power. We inject 500 (250) waveforms with either high or low $f_c$ in the full (partial) injection-time window for each testing sample in each glitch class such that we have 16 injection-test sets.    

\begin{figure}[!ht]
    \begin{minipage}{1\hsize}
  \begin{center}
   \includegraphics[width=1\textwidth]{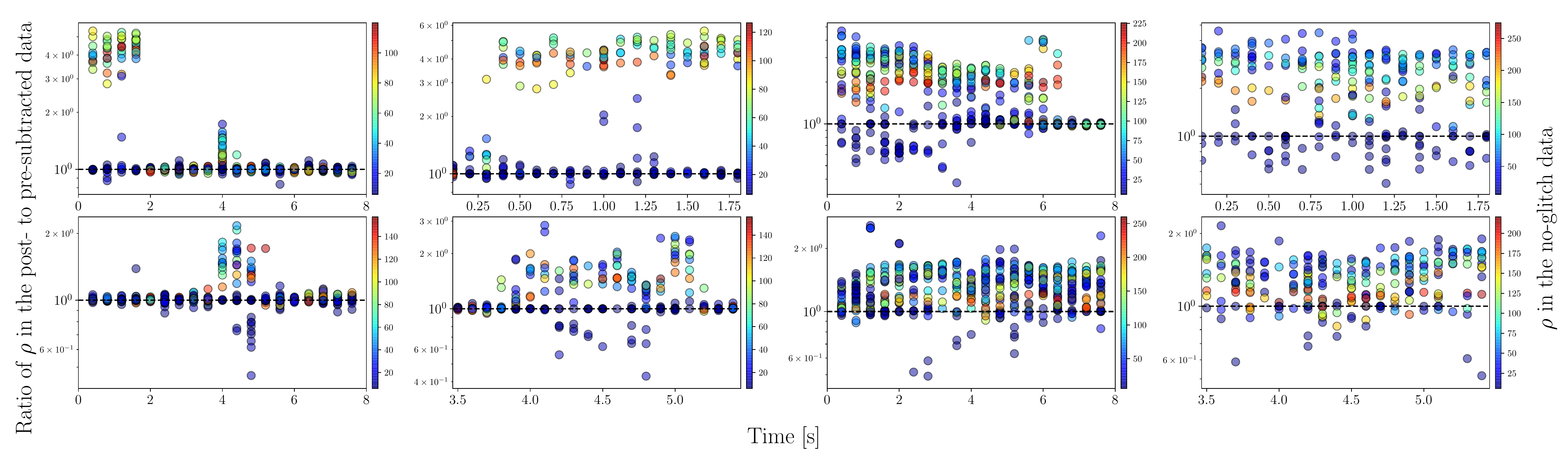}
  \end{center}
    \end{minipage}
    \begin{minipage}{1\hsize}
  \begin{center}
   \includegraphics[width=1\textwidth]{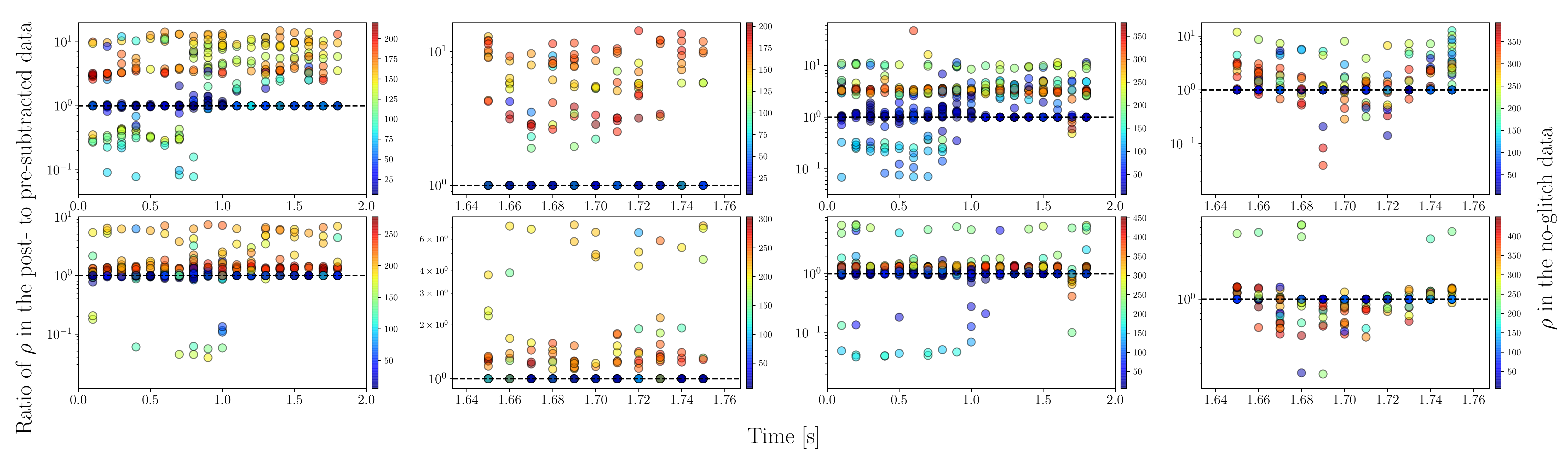}
  \end{center}
    \end{minipage}
\caption[Enhancements of $\rho$ after glitch subtraction for Gaussian-modulated
sinusoidal injections.]%
    {Enhancements of $\rho$ after glitch subtraction as a function of the injection time for high-frequency (1-2$^{\rm th}$ columns) and low-frequency (3-4$^{\rm th}$ columns) Gaussian-modulated sinusoidal waveforms injected in the full (1,3$^{\rm th}$ columns) and partial (2,4$^{\rm th}$ columns) windows of the optimal (1,3$^{\rm th}$ rows) and median (2,4$^{\rm th}$ rows) testing samples of {\it Scattered light} (1-2$^{\rm th}$ rows) and {\it Extremely loud} (2-4$^{\rm th}$ rows) glitches, respectively.}
\label{fig:singau_rho_enhance}
\end{figure}

Figure \ref{fig:singau_rho_enhance} shows the enhancement of $\rho$ after glitch subtraction for Gaussian-modulated sinusoidal injections. With the typical setting in \ac{cWB}, only a signal with a ranking statistic greater than 6 is reported. We set the statistic for those missed signals to be 6 to quantify the enhancement due to subtraction. The percentages of injections with values of $\rho$ after glitch subtraction greater or equal to the corresponding values before glitch subtraction ranges from 67\% (obtained from the set with high-frequency signals injected in the full window of the optimal testing samples of {\it Scattered light}) to 100\% (obtained from the set with high-frequency signals injected in the partial window of the optimal testing sample of {\it Extremely loud} glitches). Similarly, values the enhancement $\left<\rho_{\rm a}/\rho_{\rm b}\right>$, where $\rho_{\rm a}$ and $\rho_{\rm b}$ are $\rho$
obtained from the data after and before the glitch subtraction, respectively, averaged over injections range from 1.2 (obtained from the set with high-frequency signals injected in the full window of the optimal testing samples of {\it Scattered light}) to 3.5 (obtained from the set with high-frequency signals injected in the partial window of the optimal testing sample of {\it Extremely loud} glitches). 

Removing glitches with their characteristic frequencies close to that of signals typically improves values of $\rho$ effectively because \ac{cWB} reconstructs signals more effectively. Because \textit{Scattered light} glitches have the largest power at a frequency of $\sim30$ Hz, the $\rho$ enhancements for the low-frequency injection sets are larger than the enhancement for the high-frequency injection sets by a factor of up to $\sim1.3$. Similarly, {\it Extremely loud} glitches have a peak frequency of $\sim 110$ Hz on average so that values of the enhancement for the high-frequency injection sets are greater than values for the low-frequency injection sets by a factor of up to $\sim1.8$.   

Higher values \ac{FNR} indicate larger reductions of excess power due to glitches. Hence, values of $\rho$ obtained from the optimal testing samples are larger than values obtained from the median samples. The $\rho$ enhancements in optimal-sample sets are greater than the median-sample sets by a factor $\sim1.15\sim1.23$ ($\sim1.5\sim1.7$) and $\sim1.8\sim2.2$ ($\sim1.7\sim2.3$) for the full (partial) injected window for {\it Scattered light} and {\it Extremely loud} glitches. Because {\it Extremely loud} glitches typically have extremely loud \ac{SNR} $\sim 1500$ while {\it Scattered glitch} glitches have \ac{SNR} $\sim 17$, subtracting {\it Extremely loud} glitches improves $\rho$ more than subtracting {\it Scattered light} glitches. Also, the partial injection-window sets, where signals are overlapping with glitches tend to correspond to larger $\rho$ enhancements than the full injection-window sets. Values of $\rho$ are typically improved after glitch subtraction for the majority of injections near to glitches but not overlapping with them because the incoherent energy between detectors is reduced and the \ac{cWB} obtains higher correlations of signals between detectors. Table. \ref{table:percentage_enhance} shows values of the $\rho$ enhancement and percentages of injection with non-reduced $\rho$ after glitch subtraction for all sets.

\begin{figure}[!ht]
    \begin{minipage}{0.5\hsize}
  \begin{center}
   \includegraphics[width=1\textwidth]{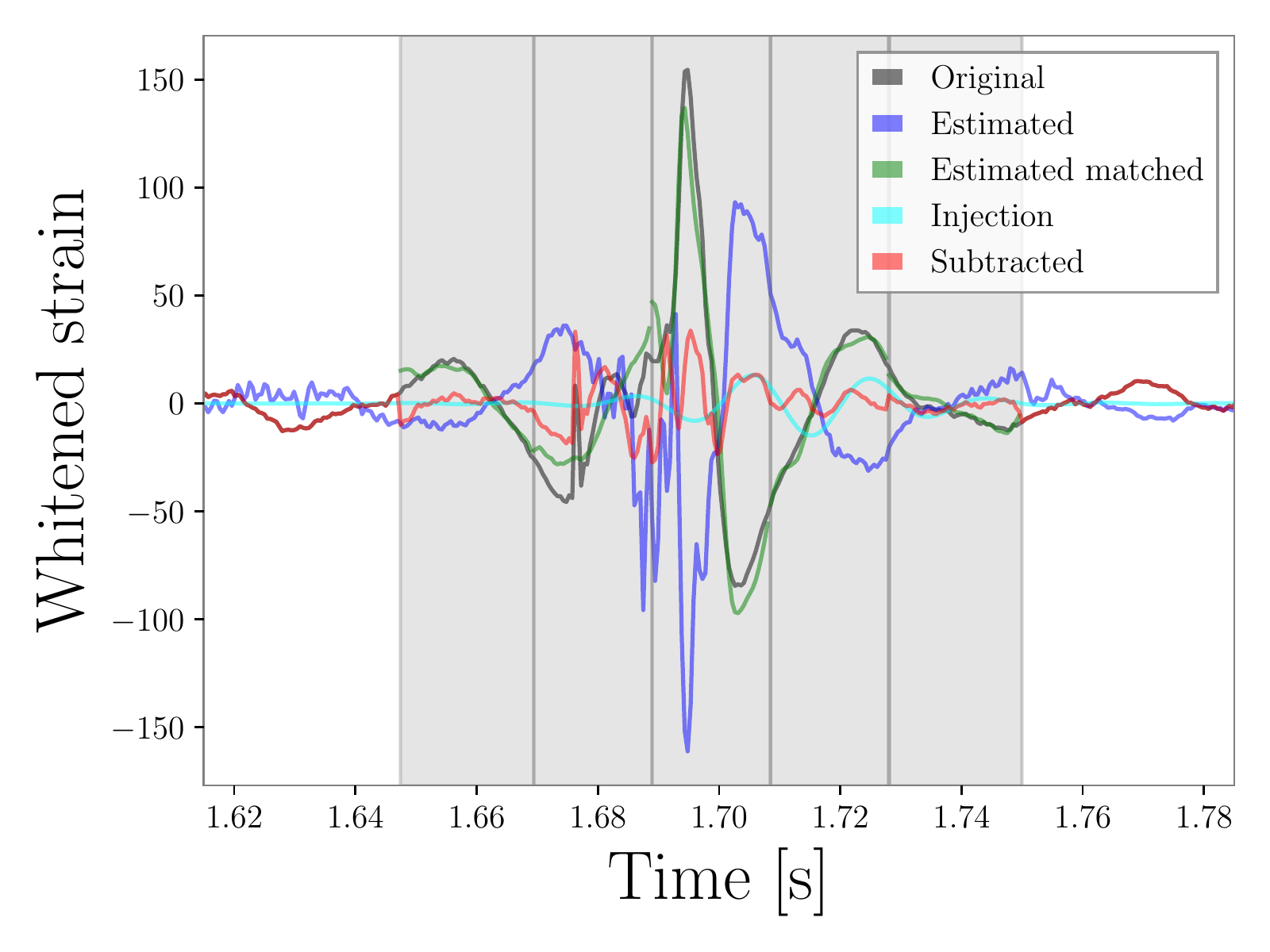}
  \end{center}
    \end{minipage}
    \begin{minipage}{0.5\hsize}
  \begin{center}
   \includegraphics[width=1\textwidth]{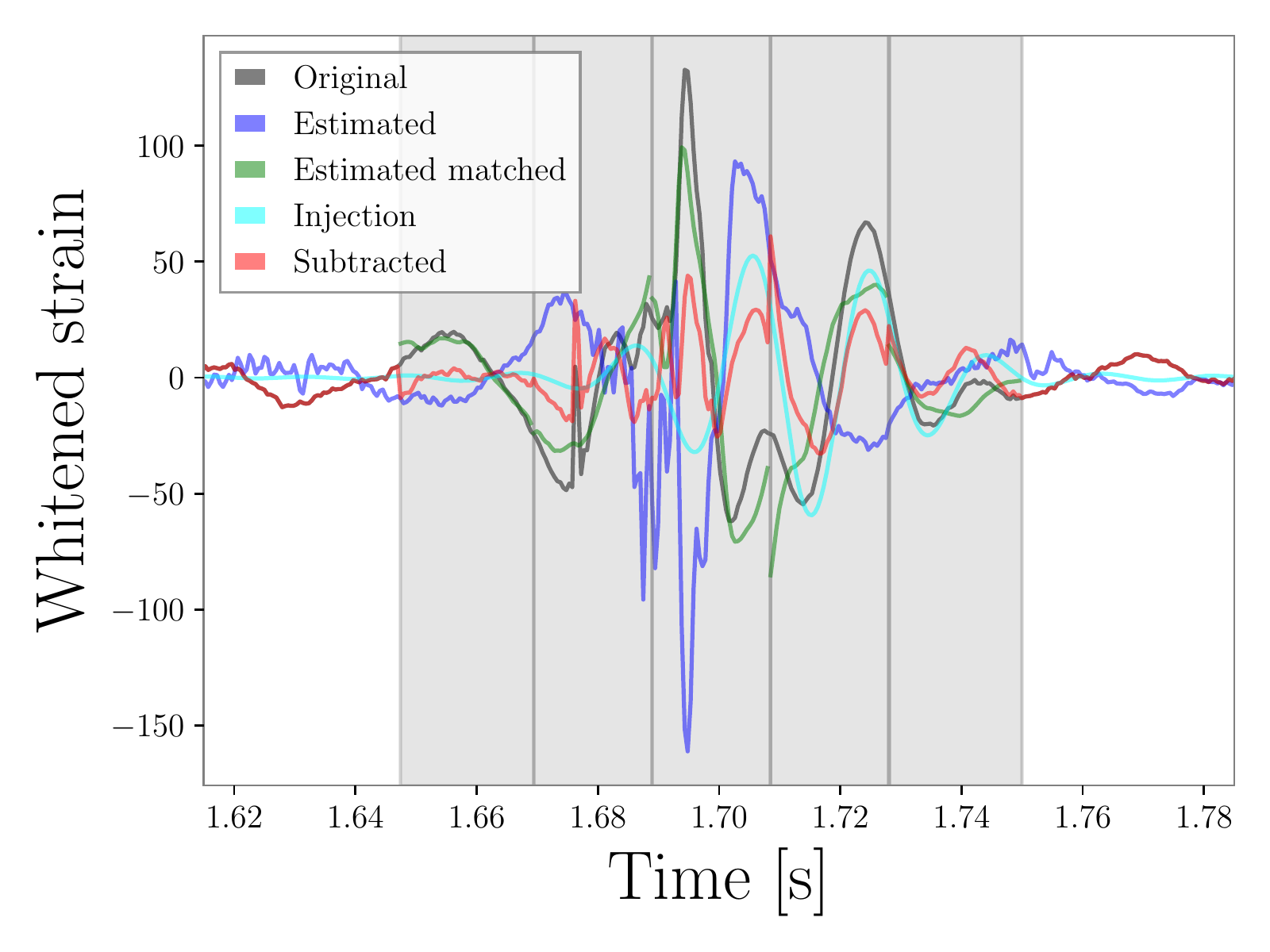}
  \end{center}
    \end{minipage}
\caption[Failure glitch subtraction when the amplitude of an injection is significantly large.]%
{Successful glitch subtraction for a low-frequency ($f_c=50$ Hz and $Q=5$) Gaussian modulated sinusoidal injection with \ac{SNR} of 50 (left) and failure subtraction for the same injection with \ac{SNR}=200 (right) in the optimal testing sample of {\it Extremely loud} glitches.}
\label{fig:failure1}
\end{figure}

\begin{figure}[!ht]
    \begin{minipage}{0.5\hsize}
  \begin{center}
   \includegraphics[width=1\textwidth]{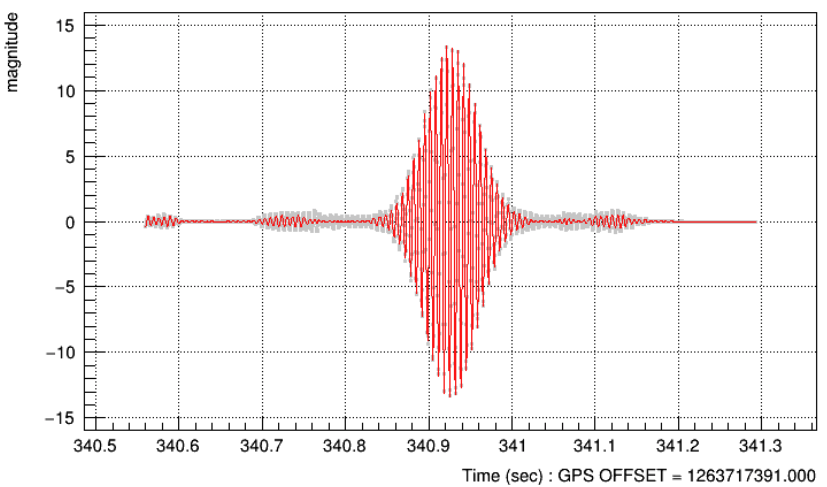}
  \end{center}
    \end{minipage}
    \begin{minipage}{0.5\hsize}
  \begin{center}
   \includegraphics[width=1\textwidth]{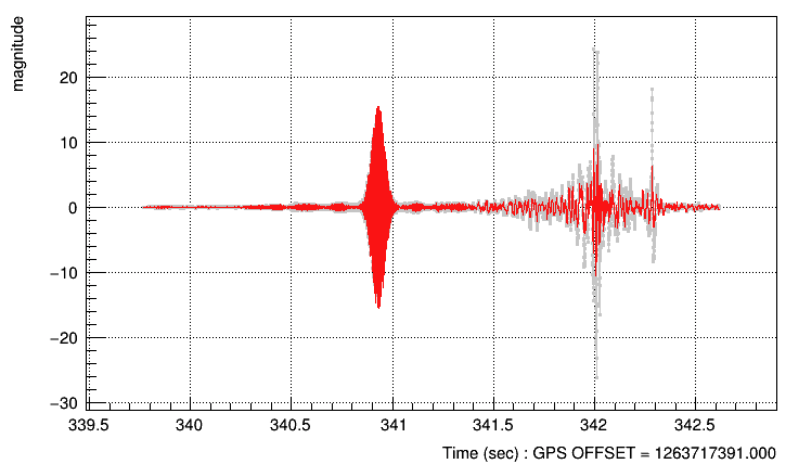}
  \end{center}
    \end{minipage}
\caption[Unsuccessful \ac{cWB} reconstruction for an injection near by an remaining glitch after subtraction.]%
{Successful \ac{cWB} reconstruction for a high-frequency ($f_c=150$ Hz and $Q=30$) Gaussian modulated sinusoidal injection before glitch subtraction (left) and unsuccessful \ac{cWB} reconstruction after glitch subtraction due to the nearby remaining glitch (right) in the optimal testing sample of {\it Extremely loud} glitches. The \ac{cWB} correlation coefficients are 0.99 and 0.65 for the reconstructed signals before and after glitch subtraction, respectively. The ratio of $\rho$ after and before subtraction is $\rho_{\rm a}/\rho_{\rm b} = 0.36$. Note that the time scales in the left and right panels are different due to the \ac{cWB} automated reconstruction process, where the glitch is outside of the reconstruction time window before glitch subtraction.}
\label{fig:failure2}
\end{figure}

Injections with reduced $\rho$ after glitch subtraction are mainly due to 1) the least square fitting process operated between estimated glitch waveforms and the data, or 2) the \ac{cWB} reconstruct process. The first reason is typically observed when the amplitude of signals is significantly large so that the least-square fitting method dominantly reduces the difference between a signal and an estimated glitch waveform in these cases. Hence, the signal energy is reduced. for example, these cases are observed when signals with high values of $\rho$ obtained from the no-glitch data are injected at the center of glitches. Figure \ref{fig:failure1} shows an example failure case due to this reason: when an injected signal has larger or comparable to the amplitude of the overlapping glitch, the least square fitting method dominantly minimizes the difference between the estimated glitch waveform and the injection. The second reason is observed when the amplitude of the remaining glitches after subtraction is comparable to the amplitude of the nearby non-overlapping injected signals so that \ac{cWB} reconstructs the sum of the remaining glitch and the true signal as a signal and the correlation of signals between detectors becomes smaller. The second reason can be seen in 0-1 seconds in the panels in the 3$^{rd}$-1,3$^{\rm th}$ columns in Fig.\ \ref{fig:singau_rho_enhance}, where the original data without subtraction is used (see Fig.\ \ref{fig:exloud_median} and \ref{fig:exloud_optimal} for the subtracted portions). Figure \ref{fig:failure2} shows an example of unsuccessful \ac{cWB} reconstruction for an injection nearby the remaining glitch after subtraction. In this case, the original data with injections is used around the injection time because no excess power is detected at the time of injections from the estimated glitch waveform. However, \ac{cWB} reconstructs the injection differently before and after glitch subtraction. Before glitch subtraction, the \ac{cWB} reconstruction process does not use the time portion containing the glitch because the amplitude of the glitch is not compatible with the signal amplitude. After glitch subtraction, the \ac{cWB} used the data portion containing the remaining glitch whose amplitude is compatible with the amplitude of the injection. As a result, the \ac{cWB} network correlation coefficient \cite{Klimenko:2008fu,Klimenko:2015ypf} is reduced to 0.65 from 0.99 after glitch subtraction, leading to $\rho_{\rm a}/\rho_{\rm b} = 0.36$.


More accurate signal reconstructions or higher values of the network correlation coefficient produce better estimates of the source direction. To assess the accuracy of the \ac{cWB} source-direction estimates, we calculate the overlap of sky maps obtained with the no-glitch data and sky maps obtained with pre-subtracted data, where the sky map is provided as the probability distribution over pixelized solid angles and the sky-map overlap is calculated by taking the inner product between two sky maps similar to Eq.\ (\ref{eq:overlap}). Also, we calculate the overlap of sky maps obtained with the no-glitch data and sky maps obtained with the post-subtracted data. For injection missed by \ac{cWB} with no reported sky maps, we set sky maps to be uniform probability distributions over solid angles according to the maximum entropy principle \cite{ratnaparkhi-1996-maximum, reynar-ratnaparkhi-1997-maximum} for the least amount of knowledge about the source direction. We count percentages $P_{\rm sky}$ of injections with the sky-map overlap of the no-glitch-post-subtraction data greater or equal to the sky-map overlap of the no-glitch-pre-subtracted data. Values of $P_{\rm sky}$ greater 50\% imply that estimates of the source direction become more accurate after glitch subtraction and $P_{\rm sky}~50$\% indicates that source-direction estimates are compatible before and after glitch subtraction. 

\begin{figure}[!ht]
    \begin{minipage}{1\hsize}
  \begin{center}
   \includegraphics[width=1\textwidth]{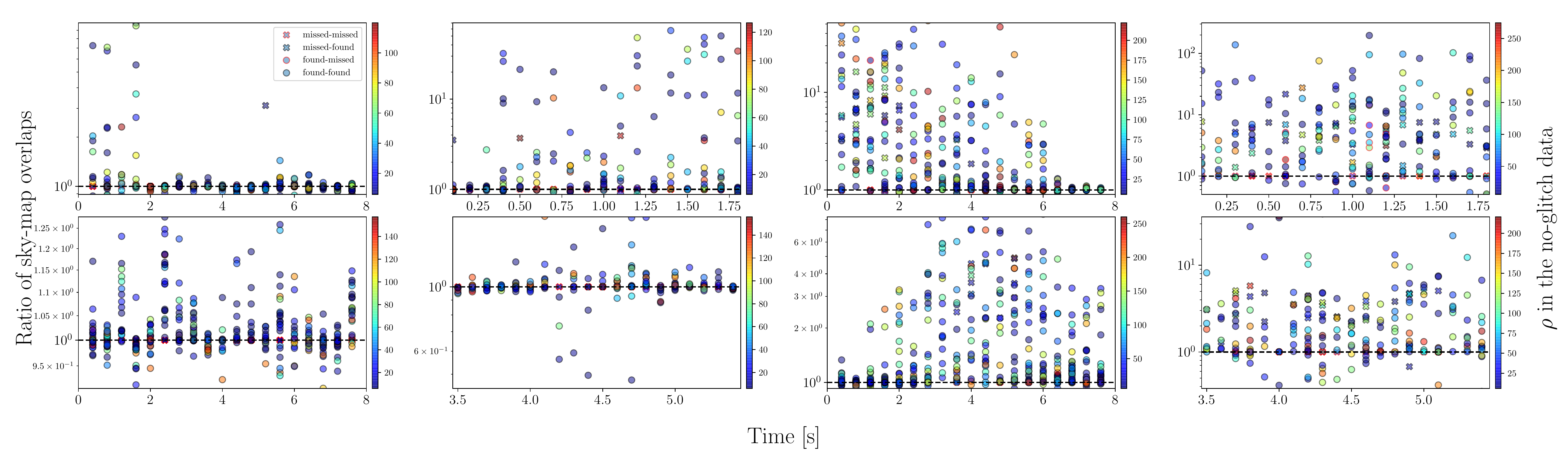}
  \end{center}
    \end{minipage}
    \begin{minipage}{1\hsize}
  \begin{center}
   \includegraphics[width=1\textwidth]{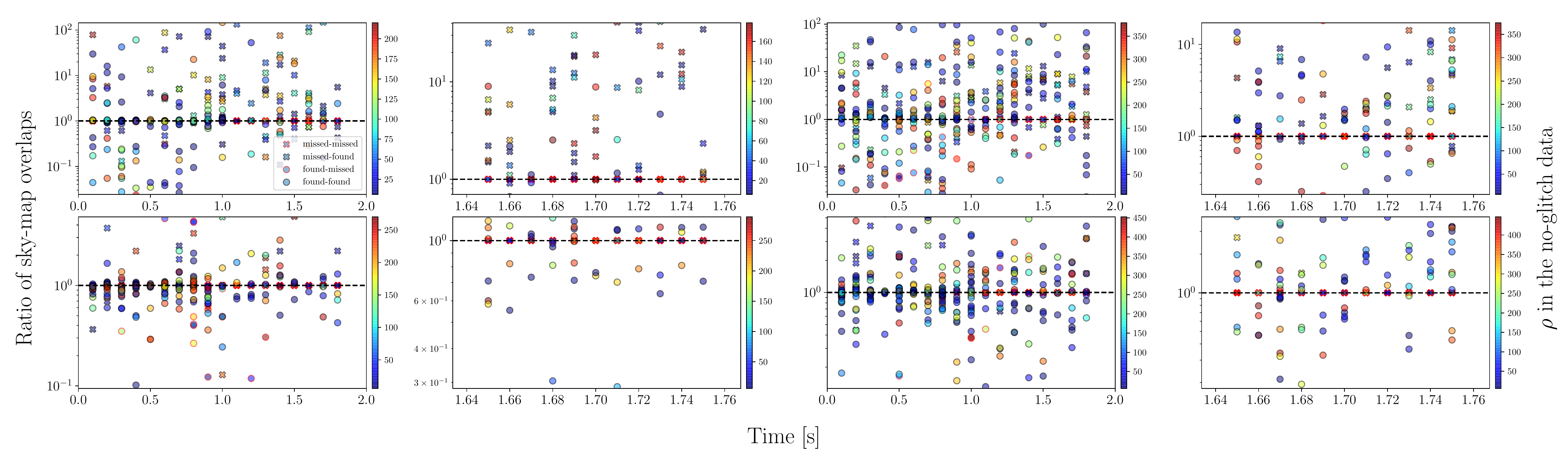}
  \end{center}
    \end{minipage}
\caption[Ratio of sky-map overlaps for Gaussian-modulated sinisoidal injections.]%
    {Ratio of sky-map overlaps between the no-glitch data and the post-subtracted data to sky-map overlaps with the former and the post-subtracted data, as a function of the injection time for high-frequency (1-2$^{\rm th}$ columns) and low-frequency (3-4$^{\rm th}$ columns) Gaussian-modulated sinusoidal waveforms injected in the full (1,3$^{\rm th}$ columns) and partial (2,4$^{\rm th}$ columns) windows of the optimal (1,3$^{\rm th}$ rows) and median (2,4$^{\rm th}$ rows) testing samples of {\it Scattered light} (1-2$^{\rm th}$ rows) and {\it Extremely loud} (2-4$^{\rm th}$ rows) glitches, respectively.}
\label{fig:singau_sky_enhance}
\end{figure}

Figure \ref{fig:singau_sky_enhance} shows ratios of sky-map overlaps between the no-glitch data and the post-subtracted data to sky-map overlaps with the former and the post-subtracted data for Gaussian modulated sinusoidal injections. Values of $P_{\rm sky}$ range from 60\% (obtained with the set with high-frequency injections in the full window of the optimal testing sample of {\it Scattered light} glitches) to 94\% (obtained with the set with high-frequency injections in the partial window of the optimal testing sample of {\it Extremely loud} glitches). Because better signal reconstructions correspond to more accurate source-direction estimates, Values of $P_{\rm sky}$ with optimal-testing-sample sets are greater than values with median-testing-sample sets by a factor of $\sim0.86\sim1.2$ ($\sim1.01\sim1.1$) for {\it Scattered light} and {\it Extremely loud} glitches. The exceptional sets with high-frequency injections in the full window for {\it Scattered light} glitches have $P_{\rm sky}=60$\% for the optimal-testing-sample set and $P_{\rm sky}=69$\% for the median-testing-sample set, respectively. However, they are compatible. 90\% of injections in the above two exceptional sets have the ratio of the sky-map overlap of the no-glitch-post-subtracted data to the sky-map overlap of the no-glitch-pre-subtracted data in 0.93-2.3 (0.94-1.2) for the optimal (median)-testing sample set because the central frequency $f_c=150$ Hz is distinctively different from the peak frequency ($\sim 30$ Hz) of {\it Scattered light} glitches. We find that the maximum value of the ratio of the sky-map overlaps to be 150 and 4.5 for the above optimal and median-testing-sample sets, respectively. Values of $P_{\rm sky}$ obtained with the low (high)-frequency sets are greater than values obtained with high (low)-frequency sets for {\it Scattered light} ({\it Extremely loud}) glitches by a factor of $\sim1.08\sim1.35$ ($\sim1.01\sim1.1$) because removing glitches with their characteristic frequencies compatible with central frequencies of injections improve the \ac{cWB} reconstructions more effectively. Table. \ref{table:percentage_enhance} shows percentages of injection with the non-reduced ratio of sky-map overlaps after glitch subtraction for all sets.  

To assess the accuracy of the \ac{cWB} estimated central frequency $\hat{f_c}$ across all injections, we calculate the normalized residual:
\begin{equation}\label{eq:resi_freq}
    \Delta{f}_c = \frac{\hat{f_c} - {f_c}}{f_c} \,,    
\end{equation}
where $f_c$ is the injected central frequency. To quantify the similarity between two distributions, we calculate the two-sided \ac{KS} statistic $S_{\rm nb}$ \cite{kolmogorov_1951} between values of $\Delta f_c$ obtained with the no-glitch data and the data before glitch subtraction as well as the \ac{KS} statistic $S_{\rm na}$ between values of $\Delta f_c$ obtained with the former and the data after glitch subtraction. \ac{KS} statistics are bounded between 0 and 1 and smaller values indicate two distributions are more similar. Values of the ratio $R^{\rm nb}_{\rm na}: = S_{\rm nb}/S_{\rm na}$ greater 1 imply that the \ac{cWB} estimated values of $\hat{f_c}$ in the post-subtracted data are more similar to corresponding values in the no-glitch data than the pre-subtracted data while smaller values indicate the \ac{cWB} estimates in the post-subtracted data are less accurate than the pre-subtracted data. $R^{\rm nb}_{\rm na}\sim1$ implies that the glitch-subtraction technique does not produce unintended effects on the data for the estimates of the central frequency. 

\begin{figure}[!ht]
    \begin{minipage}{1\hsize}
  \begin{center}
   \includegraphics[width=1\textwidth]{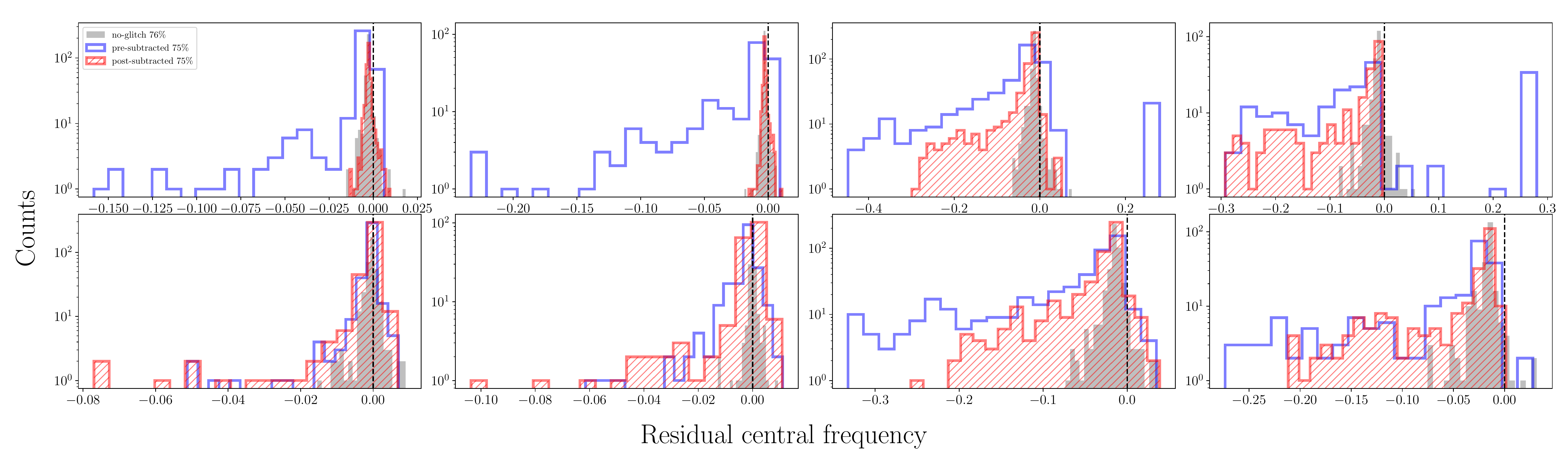}
  \end{center}
    \end{minipage}
    \begin{minipage}{1\hsize}
  \begin{center}
   \includegraphics[width=1\textwidth]{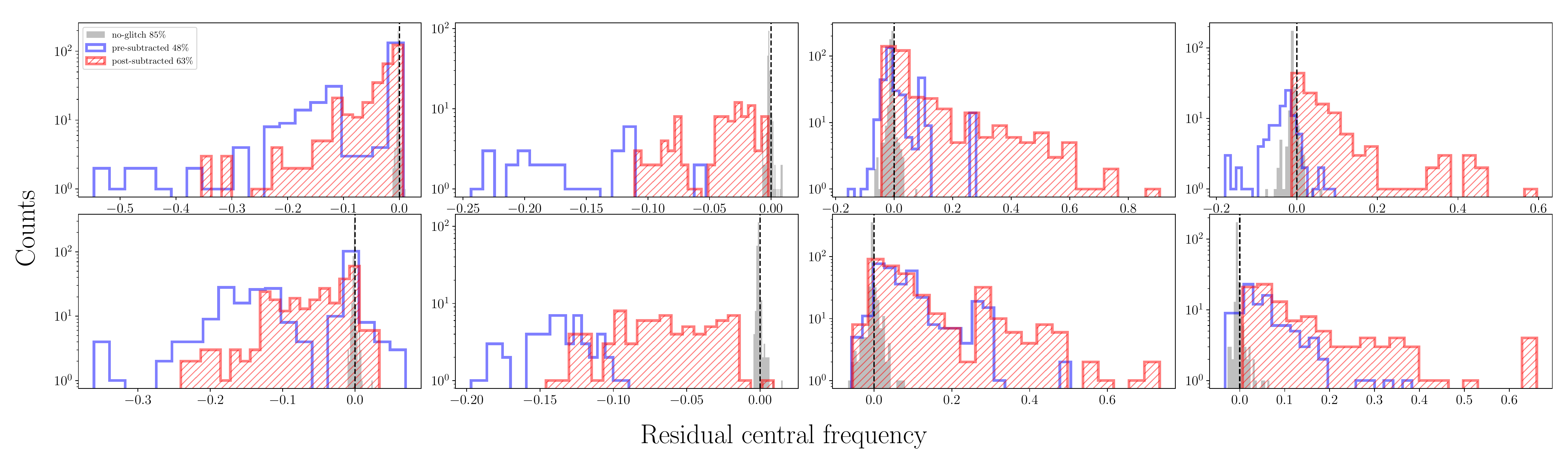}
  \end{center}
    \end{minipage}
\caption[Distributions of the residual central frequency for Gaussian-modulated sinusoidal injections.]%
    {Distributions of the residual central frequency $\Delta f_c$ for high-frequency (1-2$^{\rm th}$ columns) and low-frequency (3-4$^{\rm th}$ columns) Gaussian-modulated sinusoidal waveforms injected in the full (1,3$^{\rm th}$ columns) and partial (2,4$^{\rm th}$ columns) windows of the optimal (1,3$^{\rm th}$ rows) and median (2,4$^{\rm th}$ rows) testing samples of {\it Scattered light} (1-2$^{\rm th}$ rows) and {\it Extremely loud} (2-4$^{\rm th}$ rows) glitches, respectively.}
\label{fig:singau_freq_dist}
\end{figure}

Figure \ref{fig:singau_freq_dist} shows distributions of $\Delta f_c$ obtained with the no-glitch, the pre- and post-subtracted data. Values of $R^{\rm nb}_{\rm na}$ range from 0.41 (obtained from the set with low-frequency injections in the full window of the optimal testing sample of {\it Extremely loud} glitches) to 4.47 (obtained from the set with high-frequency injections in the partial window of the optimal testing sample of {\it Scattered light} glitches). When injections are overlapping with the remaining {\it Extremely} loud glitch after subtraction or the \ac{cWB} reconstructs the sum of the injection and near non-overlapping glitches as a signal, the estimated central frequency $\hat{f}_c$ deviates from the injected value $f_c$. For example, the set with the lowest $R^{\rm nb}_{\rm na}=0.41$ has 9.6\% of injections have values of $\Delta f_c=0.28$-0.91 from the injected value $f_c=50$ Hz (corresponding to $\hat{f}_c=64$-95 Hz) for the post-subtracted data and no injection above $\hat{f_c}=64$ Hz for the pre-subtracted data. For the set with the highest value $R^{\rm nb}_{\rm na}=4.47$, distribution of $\Delta f_c$ in the post-subtracted data differ from the distribution in the pre-subtracted data and compatible to the distribution in the no-glitch data (see the 1$^{\rm st}$ row-2$^{\rm nd}$ column in Fig.\ \ref{fig:singau_freq_dist}). Values of $R^{\rm nb}_{\rm na}$ for the low (high)-frequency injection sets are greater than values for the high (low)-frequency injection sets by a factor of $\sim1.04\sim1.68$ ($\sim0.95\sim2.4$) because subtracting glitches with their characteristic frequency compatible with signal frequency improves the \ac{cWB} reconstruction accuracy. For high-frequency injection sets, values of $R^{\rm nb}_{\rm na}$ obtained with optimal-testing-sample sets are greater than values obtained with corresponding median-testing-sample set by a factor $\sim1.04-2.8$ across the two glitch classes. For low-frequency injection sets, values of $R^{\rm nb}_{\rm na}$ obtained with median-testing-sample sets are greater than values obtained with optimal-testing-sample sets by a factor of $\sim1.05\sim2.43$ because of the contribution of high-frequency nearby remaining glitches to the \ac{cWB} signal reconstruction, mentioned above for the set with the lowest value $R^{\rm nb}_{\rm na}=0.41$. Table \ref{table:ks_ratio} shows percentages of found injections and values of $R^{\rm nb}_{\rm na}$. 

\subsubsection{Binary Black Hole Injections} \label{injection_cwb_bbh}

In addition to tests with Gaussian-modulated sinusoid signals in Sec. \ref{injection_cwb_sin_gaussian}, we also assess the performance of the \ac{cWB}-signal recovery by injecting non-spinning \textsc{IMRphenomD} \ac{BBH} merger waveforms \cite{IMRPhenomD}. Following the choice of injection parameters used in \cite{Ormiston:2020ele}, we choose the component masses to be uniformly distributed in [26, 64] \(M_\odot\) with a constraint of the primary-mass $m_1$ to the secondary-mass $m_2$ ratio in $ [0.125, 1]$, the source direction and binary orientation to be isotropically distributed, and the coalescence phase and the polarization angle to be uniformly distributed in $[0, 2\pi]$ and $[0, \pi]$, respectively. We choose the injected \ac{SNR} sampled from a set of \acp{SNR} used in Sec. \ref{injection_cwb_sin_gaussian} and the injection time sampled in the full length of the testing sample data. We have 500 \ac{BBH} injections for each set so that we have 4 \ac{BBH} sets. 

\begin{figure}[!ht]
    \begin{minipage}{0.5\hsize}
  \begin{center}
   \includegraphics[width=1\textwidth]{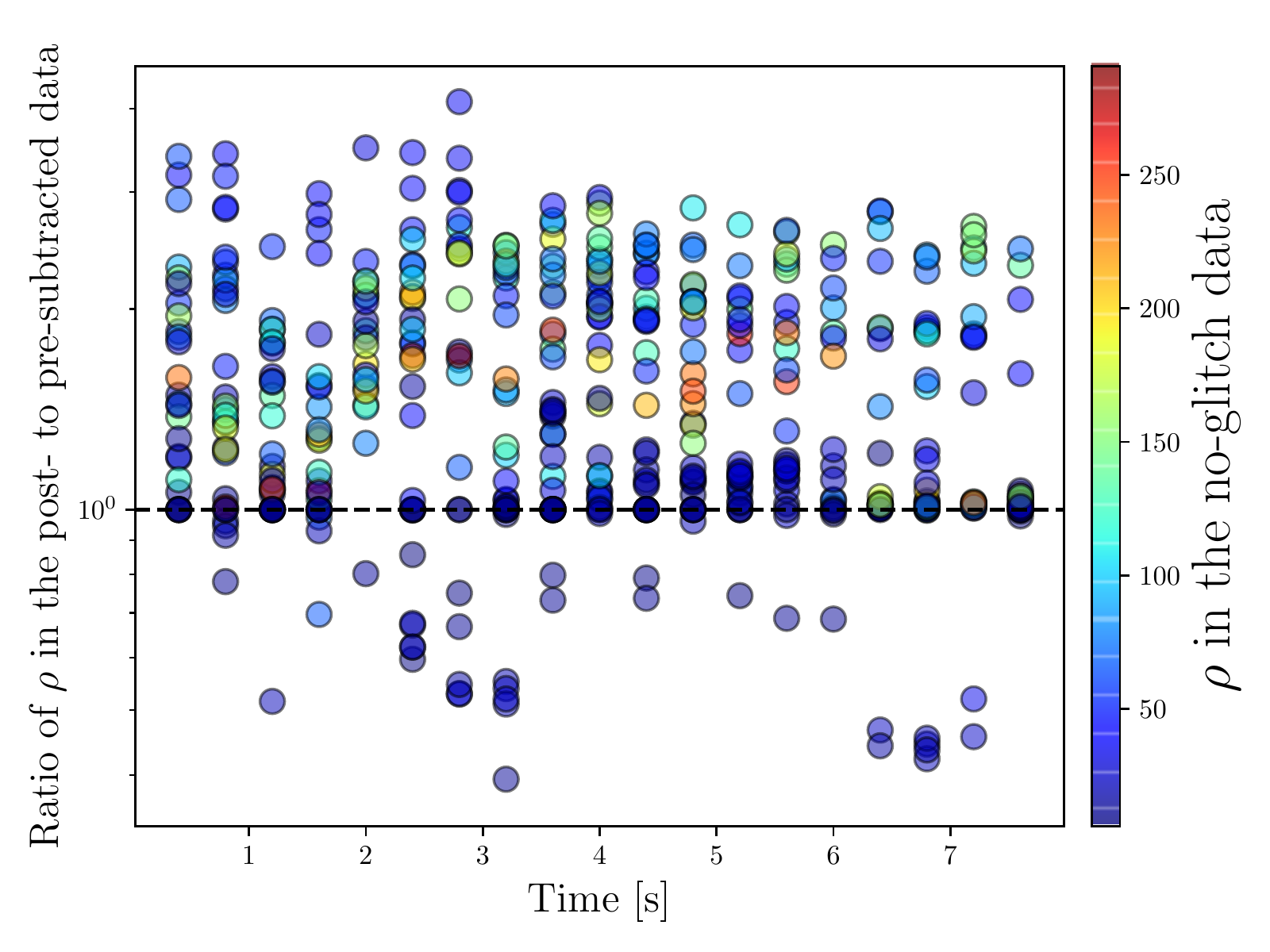}
  \end{center}
    \end{minipage}
    \begin{minipage}{0.5\hsize}
  \begin{center}
   \includegraphics[width=1\textwidth]{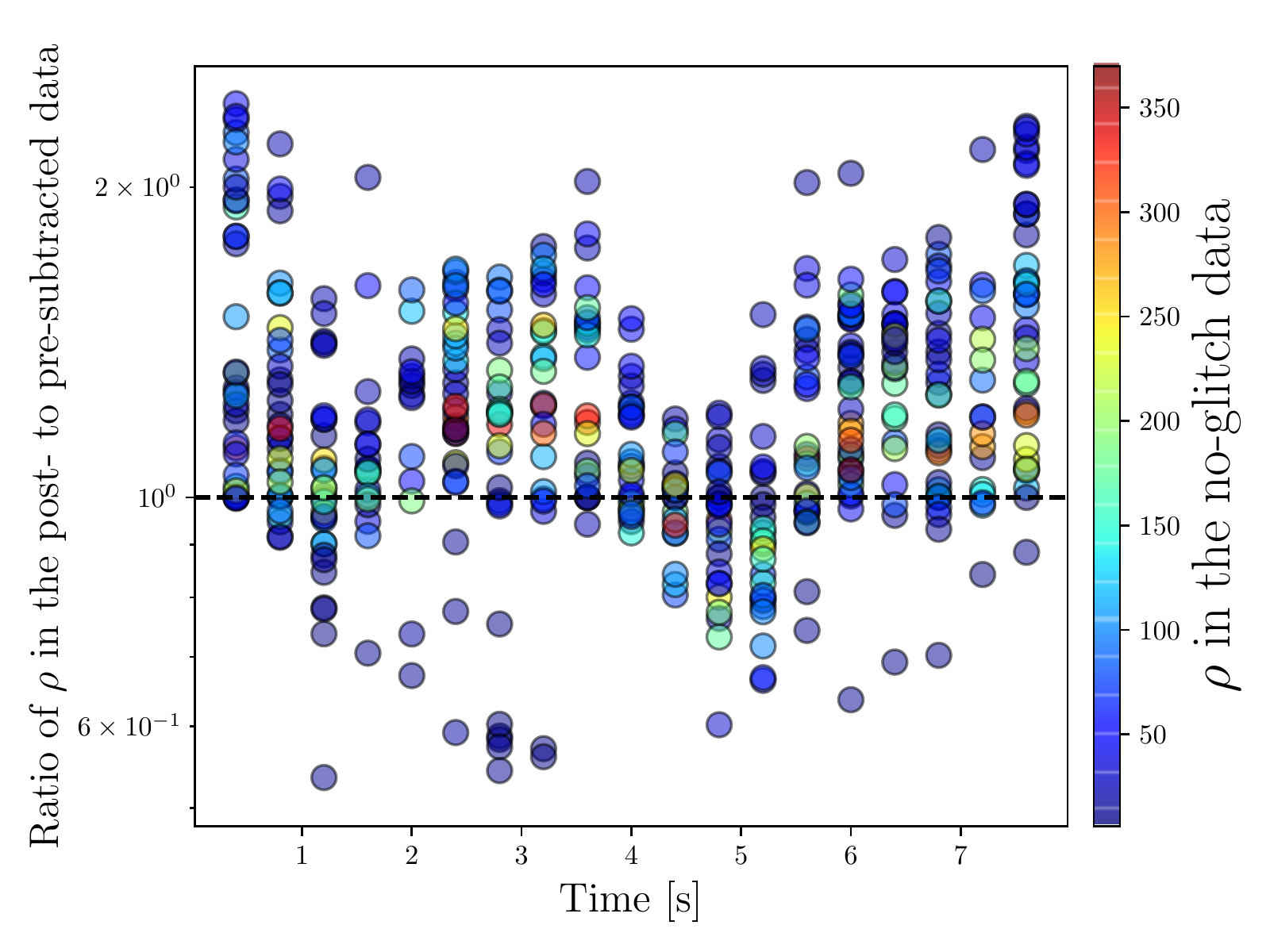}
  \end{center}
    \end{minipage}
    \begin{minipage}{0.5\hsize}
  \begin{center}
   \includegraphics[width=1\textwidth]{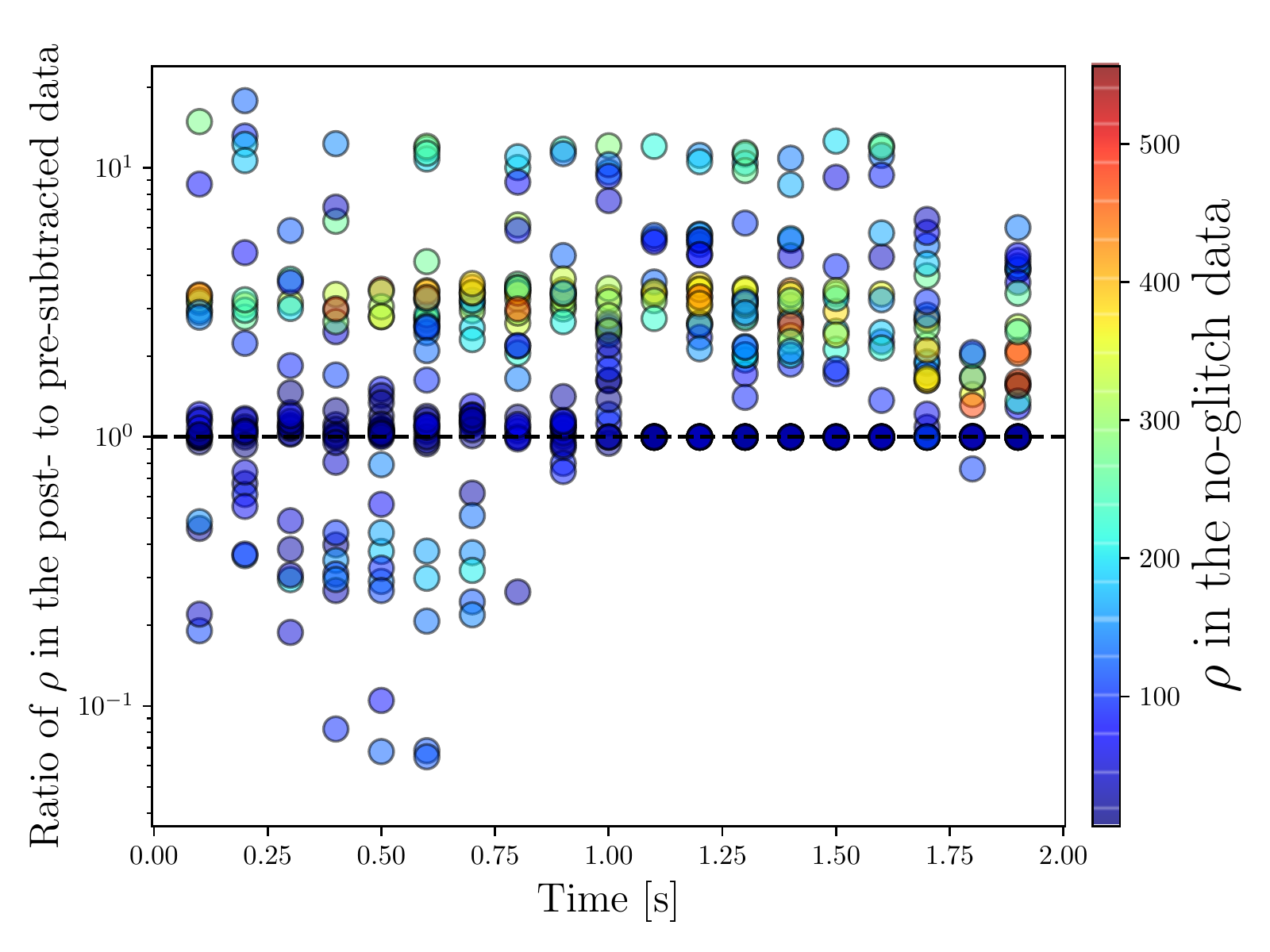}
  \end{center}
    \end{minipage}
    \begin{minipage}{0.5\hsize}
  \begin{center}
   \includegraphics[width=1\textwidth]{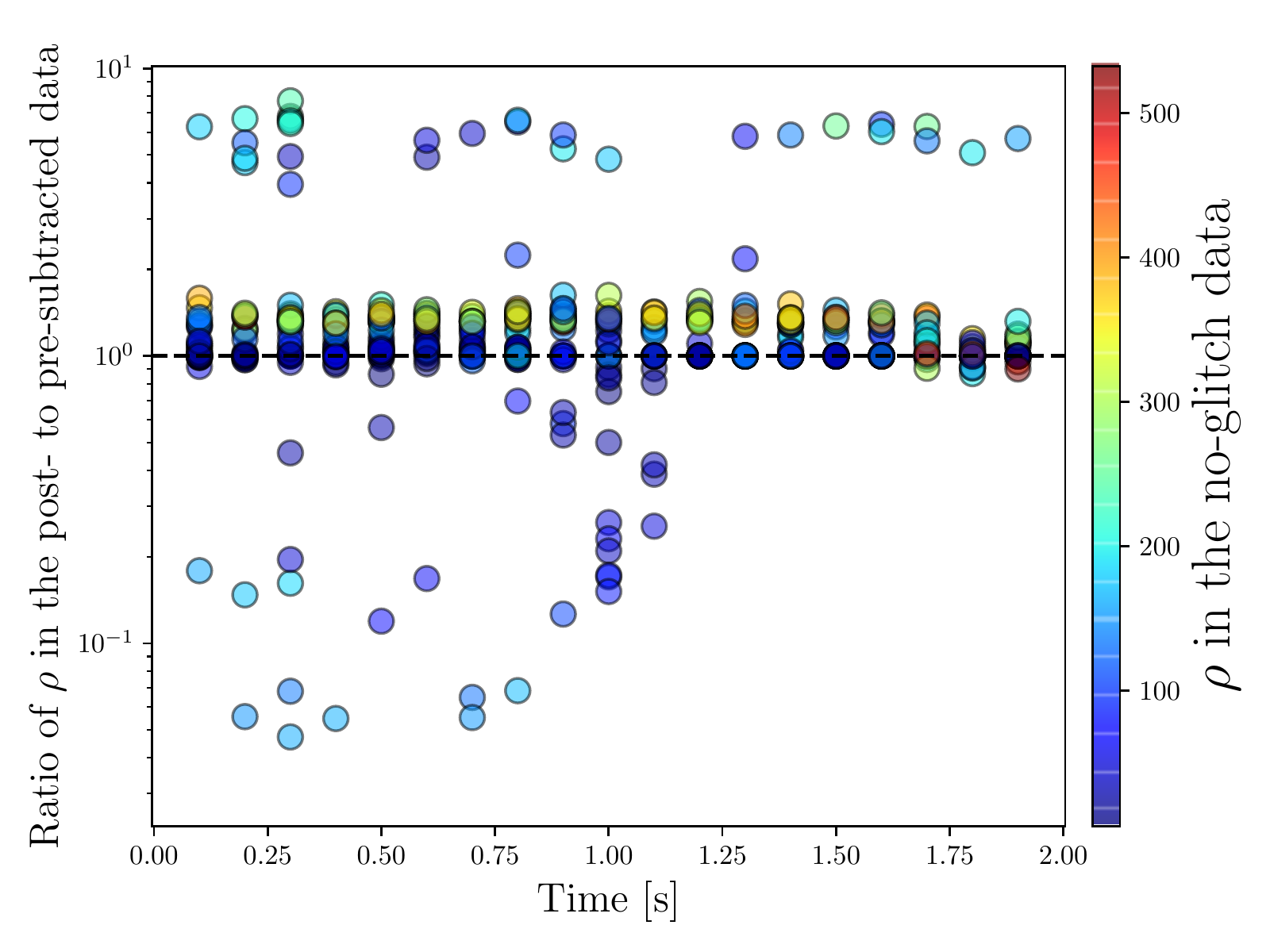}
  \end{center}
    \end{minipage}
\caption[Enhancement of $\rho$ after glitch subtraction for \ac{BBH} injections.]%
    {Enhancement of $\rho$ after glitch subtraction as a function of the injection time for \ac{BBH} waveforms injected in the optimal (left) and median testing (right) samples of {\it Scattered light} (top) and {\it Extremely loud} (bottom) glitches, respectively.}
\label{fig:cbc_rho_enhance}
\end{figure}

Figure \ref{fig:cbc_rho_enhance} shows the $\rho$ enhancement for \ac{BBH} injections. Percentages $P_\rho$ of injections with non-reduced $\rho$ after glitch subtraction range from 76\% (obtained with the median-testing-sample of {\it Scattered light} glitches) to 91\% (obtained with the optimal-testing-sample of {\it Scattered light} glitches). Values of the enhancement $\left<\rho_{\rm a}/\rho_{\rm b}\right>$ averaged over injections range from 1.2 (obtained with the median testing sample of {\it Scattered light} glitches) to 2.7 (obtained with the optimal testing sample of {\it Extremely loud} glitches). Subtracting significant energy due to glitches improves the \ac{cWB} reconstruction so that values of the enhancement in {\it Extremely-loud} sets are greater than values in {\it Scattered-light} sets by a factor of 1.25 and 1.8 for the optimal and median testing sample, respectively. In {\it Scattered-light} ({\it Extremely-loud}) sets, the value of the enhancement for the optimal test set is greater than the value of the median testing set by a factor of 1.25 (1.8). Table. \ref{table:percentage_enhance} shows values the $\rho$ enhancement and $P_\rho$ for \ac{BBH} injection sets.

\begin{figure}[!ht]
    \begin{minipage}{0.5\hsize}
  \begin{center}
   \includegraphics[width=1\textwidth]{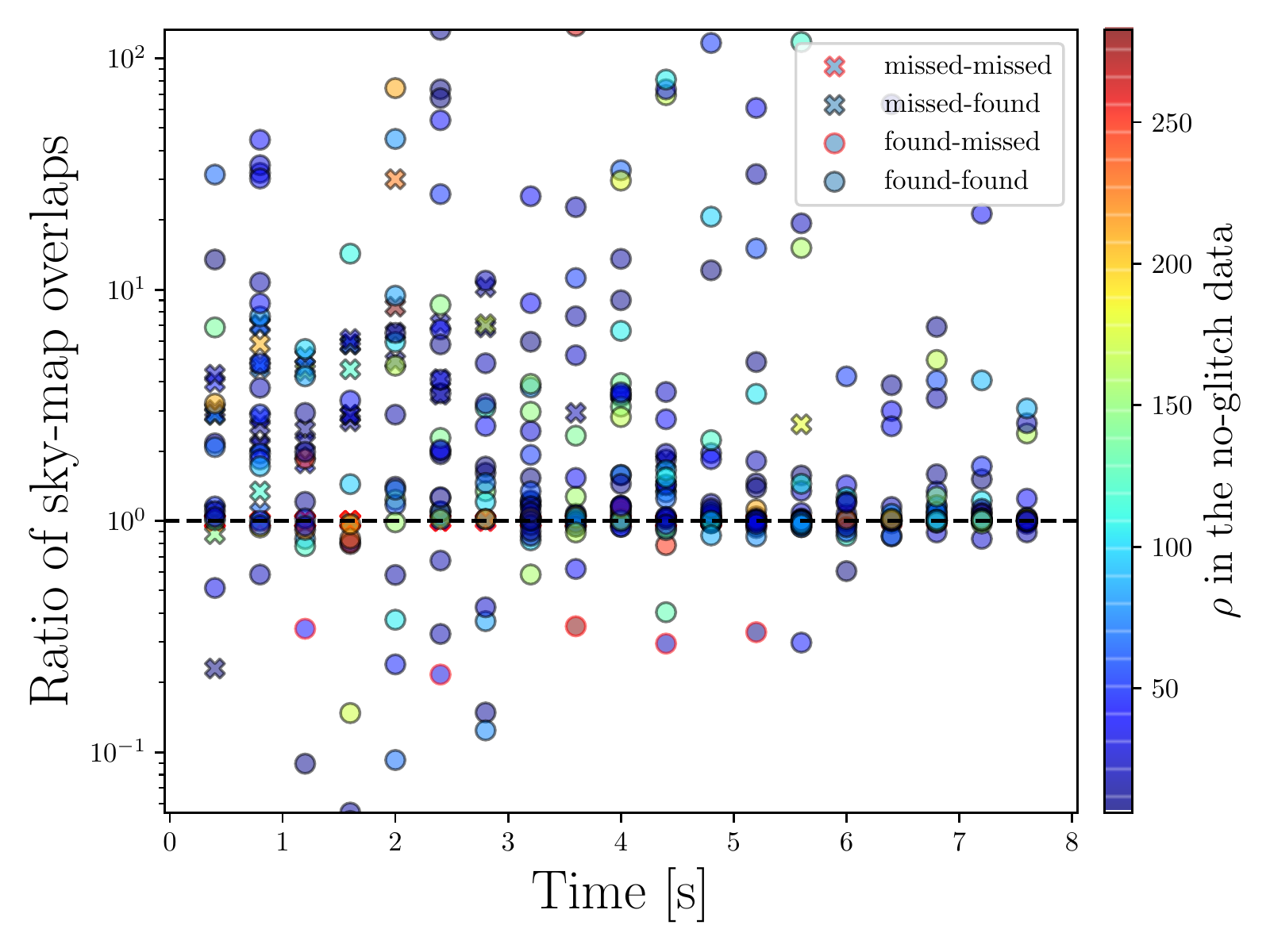}
  \end{center}
    \end{minipage}
    \begin{minipage}{0.5\hsize}
  \begin{center}
   \includegraphics[width=1\textwidth]{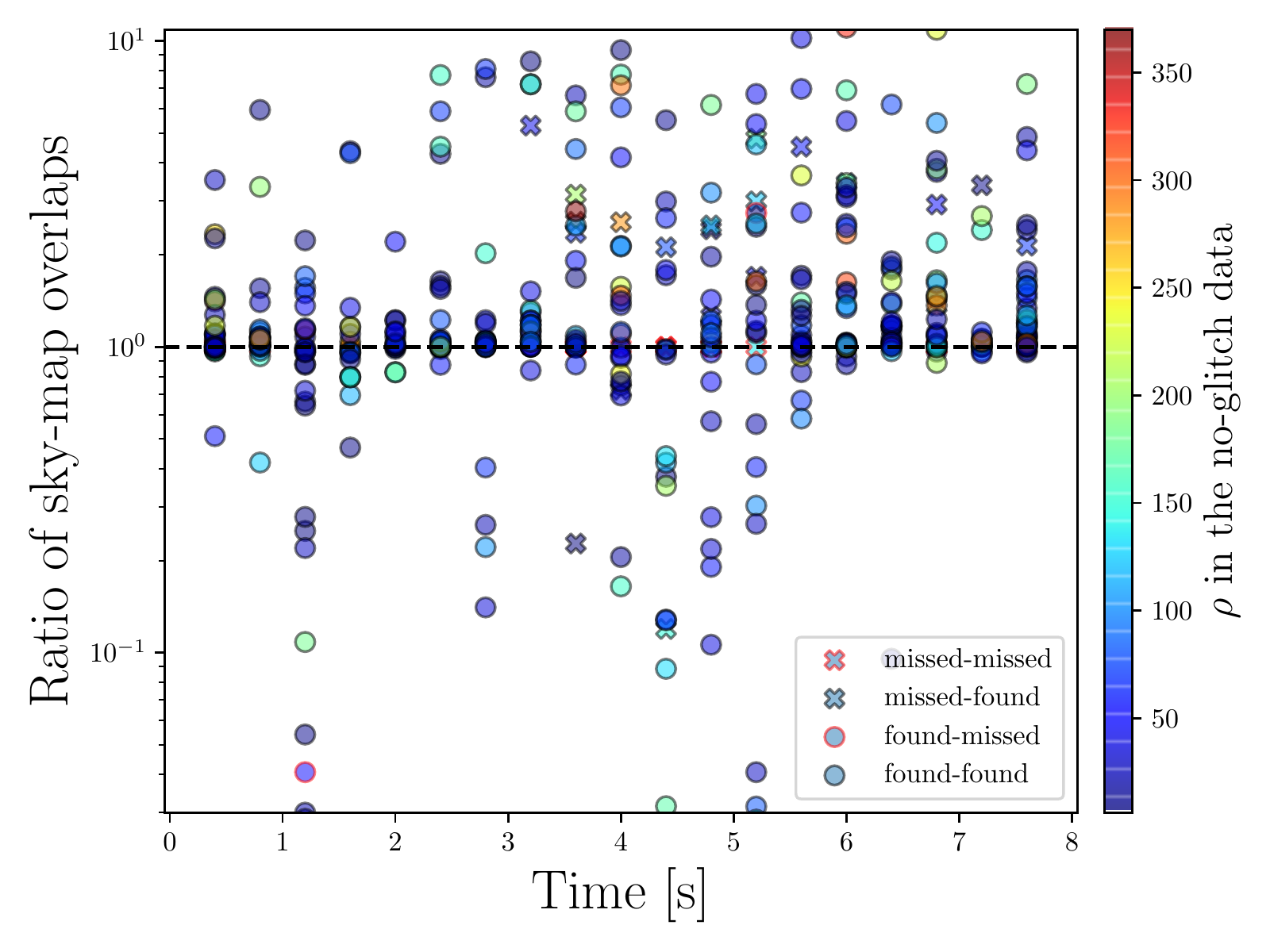}
  \end{center}
    \end{minipage}
    \begin{minipage}{0.5\hsize}
  \begin{center}
   \includegraphics[width=1\textwidth]{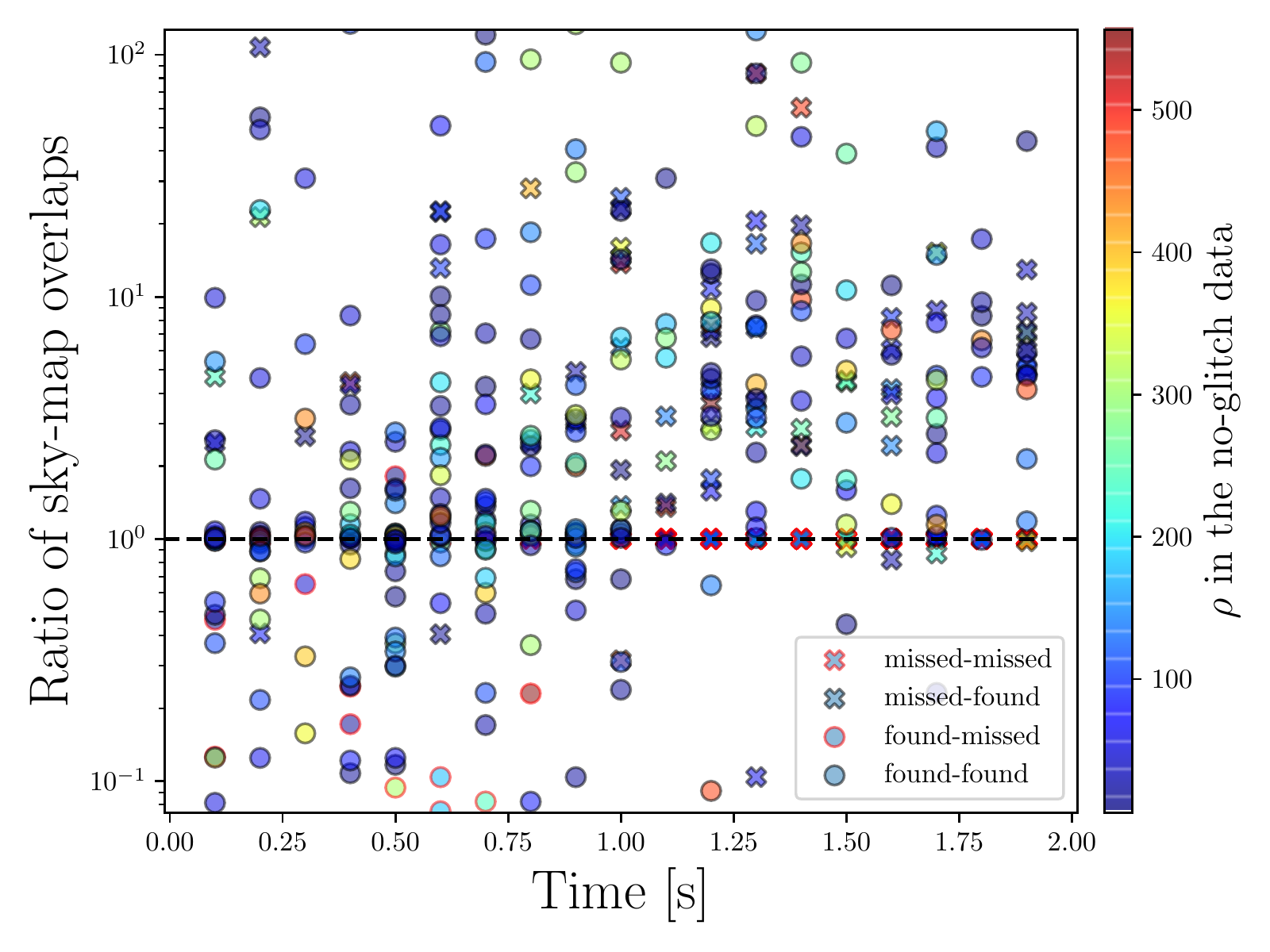}
  \end{center}
    \end{minipage}
    \begin{minipage}{0.5\hsize}
  \begin{center}
   \includegraphics[width=1\textwidth]{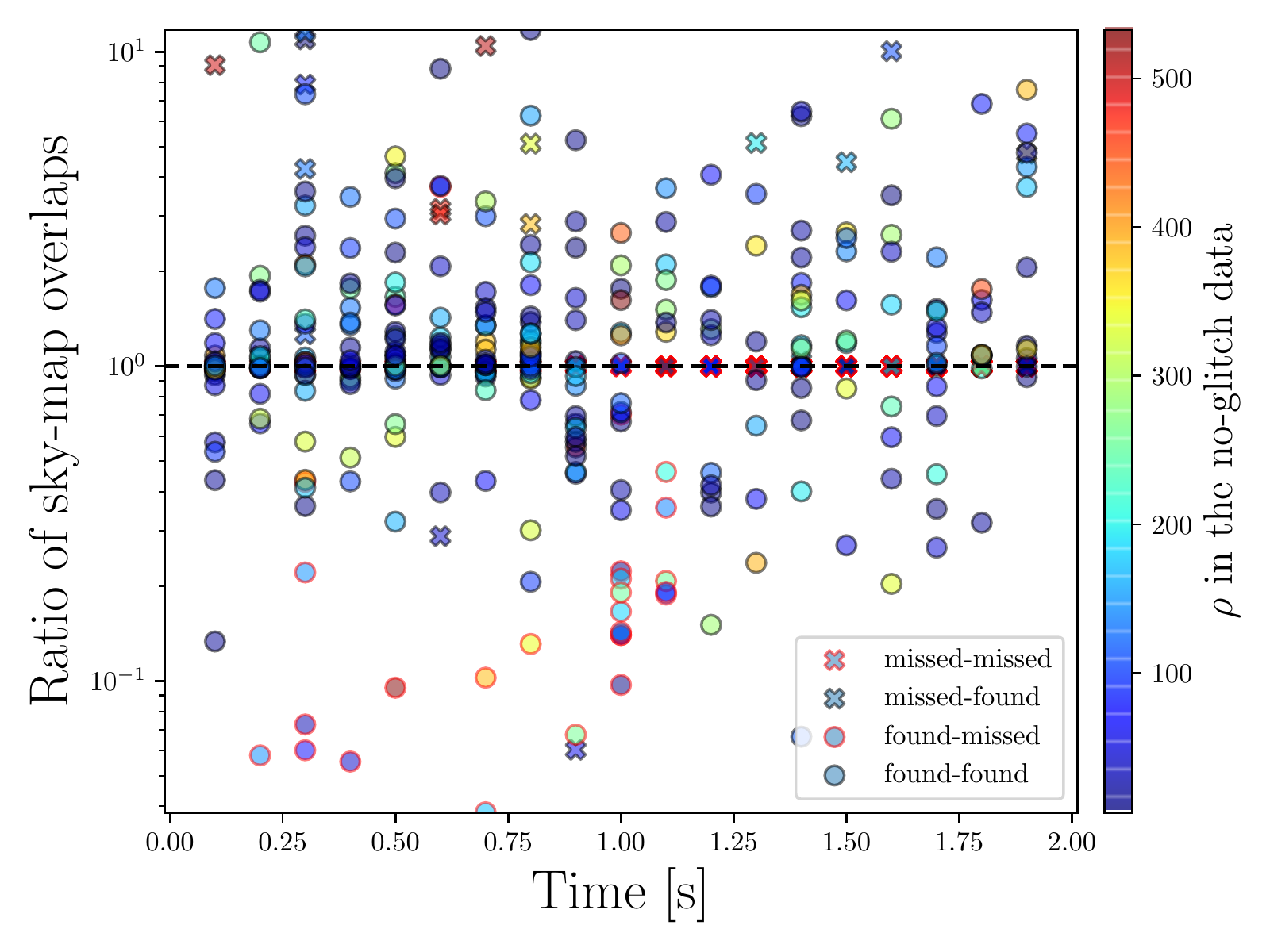}
  \end{center}
    \end{minipage}
\caption[Enhancement of sky-map overlaps after glitch subtraction for \ac{BBH} injections.]%
    {Ratio of sky-map overlap obtained with the no-glitch data and the pre-subtracted data to sky-map overlap obtained with the former and the post-subtracted data as a function of the injection time for \ac{BBH} waveforms injected in the optimal (left) and median testing (right) samples of {\it Scattered light} (top) and {\it Extremely loud} glitches, respectively. The first and second texts in labels in the legend denote that injections in the pre-subtracted and post-subtracted data are found or missed, respectively.}
\label{fig:cbc_skymap_enhance}
\end{figure}

Similar to $P_{\rm sky}$ for Gaussian modulated sinusoidal injections, we compute $P_{\rm sky}$ for \ac{BBH} injection set. Figure \ref{fig:cbc_skymap_enhance} shows ratios of sky-map overlaps between the no-glitch data and the post-subtracted data to sky-map overlaps with the former and the post-subtracted data for \ac{BBH} injections. Values of $P_{\rm sky}$ range from 73\% (obtained with the median testing sample of {\it Scattered light} glitches) to 80\% (obtained with the optimal testing sample of {\it Extremely loud} glitches). Values of $P_{\rm sky}$ in the optimal testing set are greater than values in the median testing sample set for {\it Scattered light} and {\it Extremely-loud} glitches by a factor of 1.06 and 1.14, respectively. Table. \ref{table:percentage_enhance} shows values of $P_{\rm sky}$ for \ac{BBH} injection sets.

To assess the accuracy of the \ac{cWB} estimated chirp mass across all injections, we calculate the normalized residual:
\begin{equation}\label{eq:resi_chirp}
    \Delta \mathcal{M} = \frac{\hat{\mathcal{M}} - \mathcal{M}}{\mathcal{M}} \,,    
\end{equation}
where $\hat{\mathcal{M}}$ and $\mathcal{M}=(m_1 m_2)^{3/5}/(m_1 + m_2)^{1/5}$ are the \ac{cWB} estimated and injected chirp mass, respectively. Similar to the procedure in the previous section, we calculate the two-sided \ac{KS} statistic $S_{\rm nb}$ \cite{kolmogorov_1951} between $\Delta \mathcal{M}$ obtained with the no-glitch data and the pre-subtracted data as well as the \ac{KS} statistic $S_{\rm na}$ obtained with the former and the post-subtracted data. We calculate values of the ratio $R^{\rm nb}_{\rm na}: = S_{\rm nb}/S_{\rm na}$.  

\begin{figure}[!ht]
    \begin{minipage}{0.5\hsize}
  \begin{center}
   \includegraphics[width=1\textwidth]{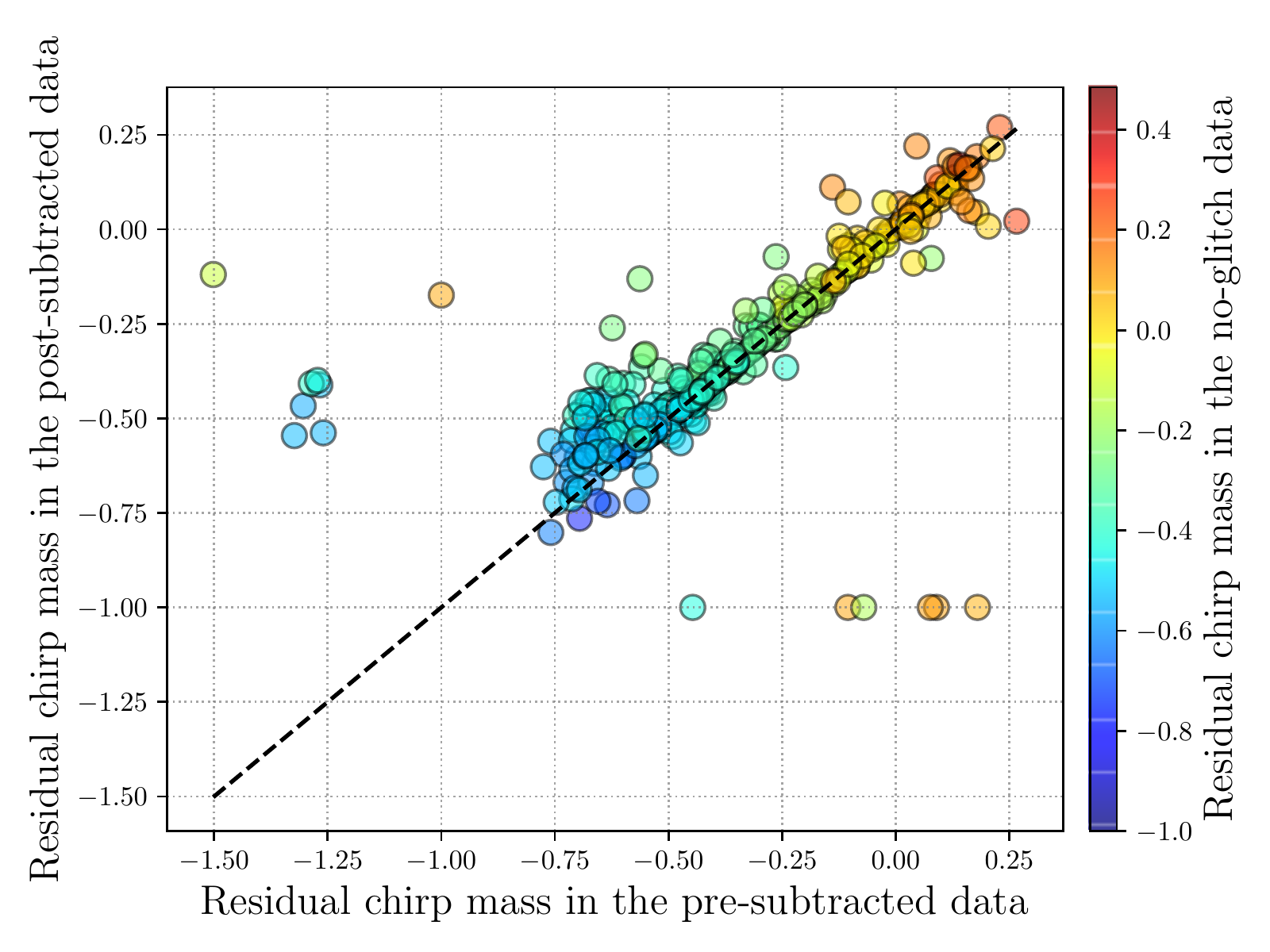}
  \end{center}
    \end{minipage}
    \begin{minipage}{0.5\hsize}
  \begin{center}
   \includegraphics[width=1\textwidth]{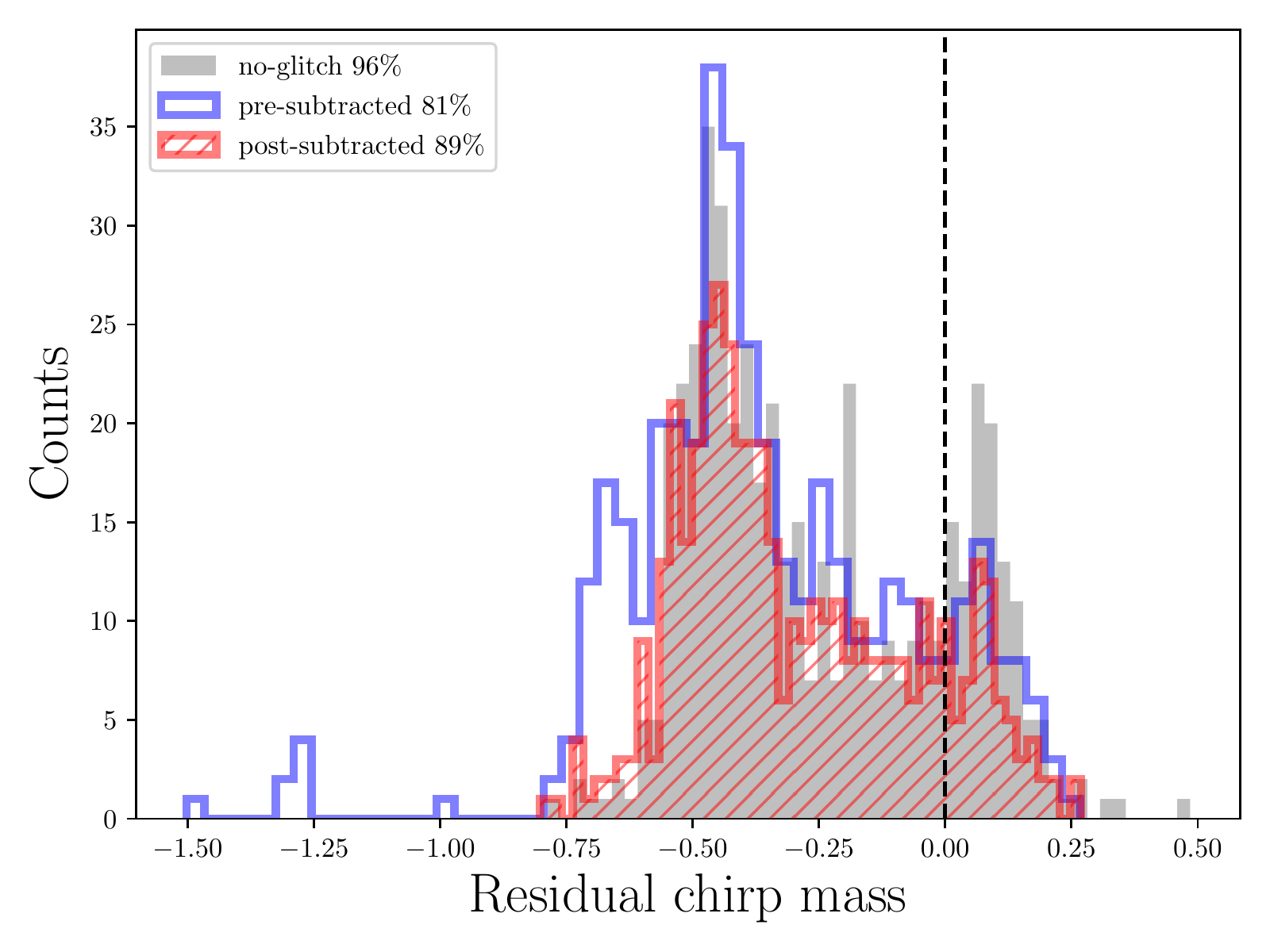}
  \end{center}
    \end{minipage}
    \begin{minipage}{0.5\hsize}
  \begin{center}
   \includegraphics[width=1\textwidth]{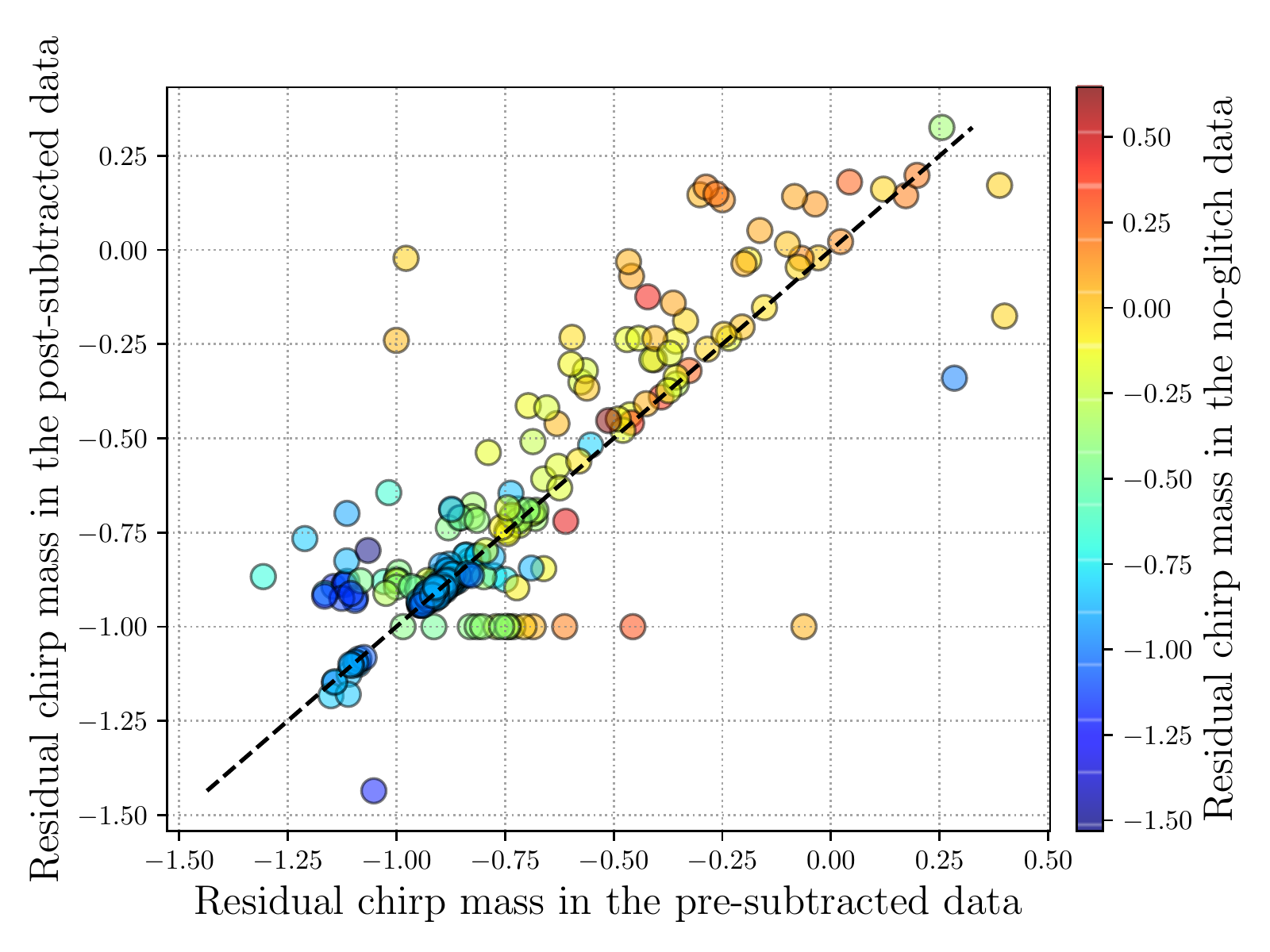}
  \end{center}
    \end{minipage}
    \begin{minipage}{0.5\hsize}
  \begin{center}
   \includegraphics[width=1\textwidth]{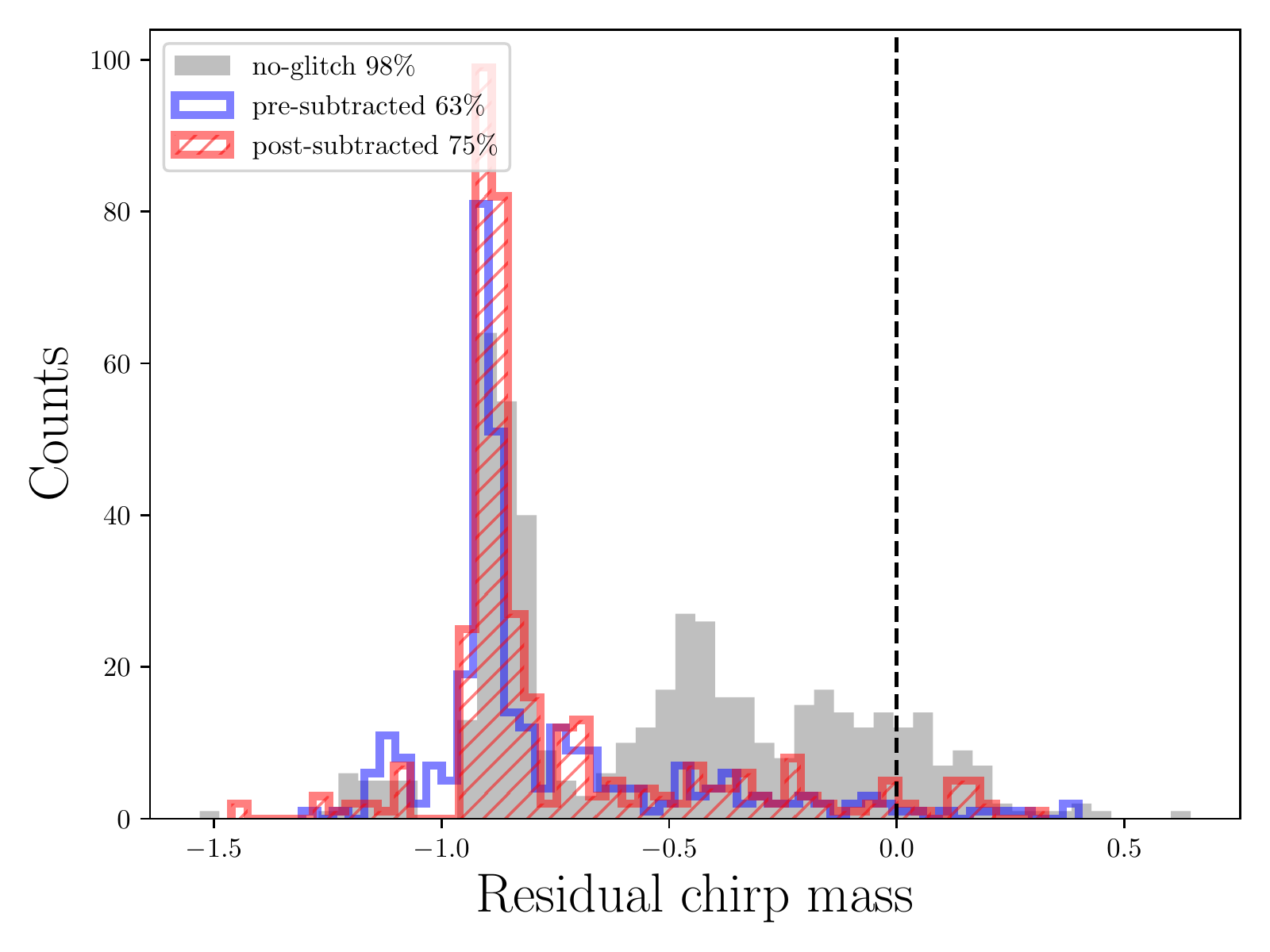}
  \end{center}
    \end{minipage}
\caption[Residual of chirp mass estimated by \ac{cWB}.]%
{Residual of chirp mass estimated by \ac{cWB} in the optimal testing sample of {\it Scattered light} (top) and {\it Extremely loud} (bottom) glitches. The dashed lines in the left panels denote the same value of residual chirp mass before and after glitch subtraction. The dashed lines in the right panels denote the injected values.}
\label{fig:residual_mchirp}
\end{figure}

\begin{figure}[ht!]
    \centering
    \includegraphics[width=0.8\linewidth]{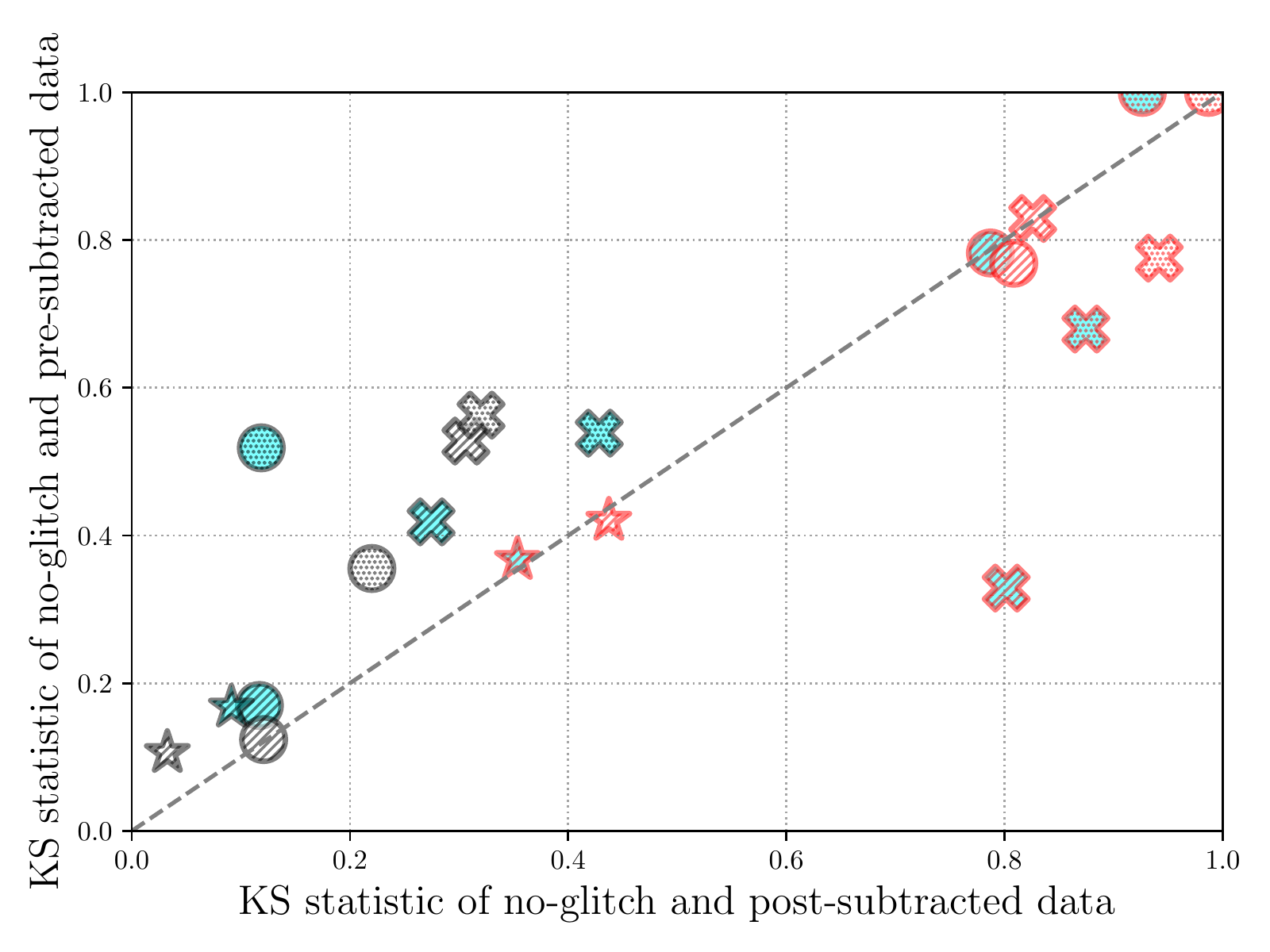}
    \caption[\ac{KS} statistic obtained with the no-glitch data and the data before glitch subtraction VS. \ac{KS} statistic obtained with the former and the data after glitch subtraction.]%
    {\ac{KS} statistic obtained with the no-glitch data and the data before glitch subtraction VS. \ac{KS} statistic obtained with the former and the data after glitch subtraction for residual peak frequency or chirp mass estimated by \ac{cWB} in the cases of high-frequency (circle marker), low-frequency (cross marker) Gaussian-modulated sinusoid waveforms and \ac{BBH} waveforms (star marker) injected in the full (slashed hatch) and partial window (dotted hatch) in the optimal (cyan face color) and median (white face color) testing sample of {\it Scattered light} (black edge color) and {\it Extremely loud} (red edge color) glitches, respectively.}
    \label{fig:residual_intri} 
\end{figure}

Figure \ref{fig:residual_mchirp} shows distributions of $\Delta \mathcal{M}$ obtained with the no-glitch, the pre- and post-subtracted data in the optimal testing sample sets. The distributions in the optimal testing sample sets are comparable with distributions in the median testing sample sets. As shown in Table \ref{table:ks_ratio}, values of $R^{\rm nb}_{\rm na}$ range from 0.96 (obtained from the median testing sample of {\it Extremely loud} glitches) to 3.26 (obtained from the median testing sample of {\it Scattered light} glitches) as shown in Table \ref{table:ks_ratio}. Values of $R^{\rm nb}_{\rm na}$ are close to or greater than 1, indicating the glitch subtraction technique produces no unintended effect on \ac{cWB} estimates for the chirp mass or improves the estimates. As shown in Fig.\ \ref{fig:residual_intri}, $S_{\rm na}$ and $S_{\rm nb}$ are compatible for \ac{BBH} sets so that the distribution of $\Delta \mathcal{M}$ in the post and pre-subtracted data are similar. Values of $S_{\rm na}$ and $S_{\rm nb}$ for \ac{BBH} sets are typically smaller than values for Gaussian modulated sinusoidal sets because \ac{BBH} waveforms distinctively differ from glitch waveforms and \ac{cWB} reconstructs \ac{BBH} injections more effectively than the Gaussian modulated sinusoidal injections.

\begin{table}[!ht]
\newcolumntype{g}{>{\columncolor{Gray}}c}
\renewcommand*{\arraystretch}{1.1}
\setlength{\tabcolsep}{2pt}
\caption[Enhancements of the \ac{cWB} ranking statistic and percentages of injections with non-reduced ranking statistics and sky-map overlaps after glitch subtraction.]%
{Enhancements of the \ac{cWB} ranking statistic $\left< \rho_{\rm a}/\rho_{\rm b} \right>$, i.e., ratios of the \ac{cWB} ranking statistic $\rho_{\rm a}$ after and $\rho_{\rm b}$ before glitch subtraction averaged over injections, as well as percentages $P_\rho$ of injections with non-reduced ranking statistics and percentages $P_{\rm sky}$ of non-reduced sky-map overlaps after glitch subtraction. The sky-map overlaps are calculated between sky maps obtained with the simulated colored Gaussian data and either of the data before or after glitch subtraction. Values of $\rho$ are set to be 6 for injections missed. The sky-localization estimate is set to be uniformly distributed in solid angles.}
\begin{center}
\begin{tabular}{c c c || c c c c c c c} \toprule
\multirow{2}{*}{Glitch class} & \multirow{2}{*}{\makecell{Testing \\sample}} & \multirow{2}{*}{Injection} & \multicolumn{3}{ c }{Full window} & \multicolumn{3}{ c }{Partial window} \\ 
\cmidrule(rl){4-6} \cmidrule(rl){7-9}
 & & & $\left<\rho_{\rm a}/\rho_{\rm b} \right>$ & $P_\rho$ & $P_{\rm sky}$ & $\left<\rho_{\rm a}/\rho_{\rm b} \right>$ & $P_\rho$ & $P_{\rm sky}$  \\ 
\midrule\midrule
\multirow{6}{*}{\makecell{{\it Scattered} \\ {\it light}}} & \multirow{3}{*}{Optimal} & High frequency & 1.2 & 67\% & 60\% & 2.0 & 84\% & 79\% \\
& & \mycc Low frequency & \mycc 1.6 & \mycc 69\% & \mycc 81\% & \mycc 2.0 & \mycc 89\% & \mycc 86\% \\
& & \ac{BBH} & 1.5 & 91\% & 78\% & -- & -- & -- \\
\cmidrule(rl){2-2} \cmidrule(rl){3-3} \cmidrule(rl){4-4} \cmidrule(rl){5-5} \cmidrule(rl){6-6} \cmidrule(rl){7-7} \cmidrule(rl){8-8} \cmidrule(rl){9-9}
&  \multirow{3}{*}{Median} & \mycc High frequency & \mycc 1.03 &\mycc 75\% & \mycc 69\% & \mycc 1.2 & \mycc 86\% & \mycc 65\% \\
&  & Low frequency & 1.3 & 90\% & 81\% & 1.3 & 88\% & 84\% \\
&  & \mycc \ac{BBH} & \mycc 1.2 &\mycc 76\% & \mycc 73\% & \mycc -- & \mycc -- & \mycc -- \\
\cmidrule(rl){1-1} \cmidrule(rl){2-2} \cmidrule(rl){3-3} \cmidrule(rl){4-4} \cmidrule(rl){5-5} \cmidrule(rl){6-6} \cmidrule(rl){7-7} \cmidrule(rl){8-8} \cmidrule(rl){9-9}
\multirow{6}{*}{\makecell{{\it Extremely} \\ {\it loud}}} & \multirow{3}{*}{Optimal} & High frequency & 2.8 & 84\% & 74\% & 3.5 &  100\% & 94\% \\
& & \mycc Low frequency & \mycc 2.8 & \mycc 87\% & \mycc 70\% & \mycc 2.0 & \mycc 88\% & \mycc 85\% \\
& & \ac{BBH} & 2.7 & 88\% & 80\% & -- & -- & -- \\
\cmidrule(rl){2-2} \cmidrule(rl){3-3} \cmidrule(rl){4-4} \cmidrule(rl){5-5} \cmidrule(rl){6-6} \cmidrule(rl){7-7} \cmidrule(rl){8-8} \cmidrule(rl){9-9}
& \multirow{3}{*}{Median} & \mycc High frequency & \mycc 1.5 & \mycc 83\% & \mycc 67\% & \mycc 1.5 & \mycc 100\% & \mycc 86\% \\
&  & Low frequency & 1.3 & 86\% & 66\% & 1.2 & 77\% & 84\% \\
&  & \mycc \ac{BBH} & \mycc 1.5 & \mycc 85\% & \mycc 70\% & \mycc -- & \mycc -- & \mycc -- \\
\bottomrule
\end{tabular}
\label{table:percentage_enhance}
\end{center}
\end{table}
\begin{table}[!ht]
\newcolumntype{g}{>{\columncolor{Gray}}c}
\renewcommand*{\arraystretch}{1.1}
\setlength{\tabcolsep}{2pt}
\caption[Percentages of found injections and ratios of the two-sided \ac{KS} statistics for with the residual of peak frequency or chirp mass estimated by \ac{cWB}.]%
{Percentages ($P_{\rm n}/P_{\rm b}/P{\rm a}$) of found injections in the (no-glitch data/data before subtraction/data after subtraction), as well as ratios $R^{\rm nb}_{\rm na}:=S_{\rm nb}/S_{\rm na}$ of the \ac{KS} statistic $S_{\rm nb}$ obtained with the no-glitch data and the data before glitch subtraction to the statistic $S_{\rm na}$ obtained with the former and the data after glitch subtraction for residual peak frequency and chirp mass.}
\begin{center}
\begin{tabular}{c c c || c c c c  } \toprule
\multirow{2}{*}{Glitch class} & \multirow{2}{*}{\makecell{Testing \\sample}} & \multirow{2}{*}{Injection} & \multicolumn{2}{ c }{Full window} & \multicolumn{2}{ c }{Partial window} \\ 
\cmidrule(rl){4-5} \cmidrule(rl){6-7}
 & & & $(P_{\rm n}/P_{\rm b}/P_{\rm a})$ & $R^{\rm nb}_{\rm na}$ &  $(P_{\rm n}/P_{\rm b}/P_{\rm a})$ & $R^{\rm nb}_{\rm na}$  \\ 
\midrule\midrule
\multirow{6}{*}{\makecell{{\it Scattered} \\ {\it light}}} & \multirow{3}{*}{Optimal} & High frequency & (76/75/75)\% & 1.45 & (78/77/78)\% & 4.47  \\
& & \mycc Low frequency & \mycc (98/92/96)\% & \mycc 1.52  & \mycc (97/74/88)\% & \mycc 1.26  \\
& & \ac{BBH} & (96/81/89)\% & 1.84 &  -- & -- \\
\cmidrule(rl){2-2} \cmidrule(rl){3-3} \cmidrule(rl){4-4} \cmidrule(rl){5-5} \cmidrule(rl){6-6} \cmidrule(rl){7-7} 
&  \multirow{3}{*}{Median} & \mycc High frequency & \mycc (75/75/75)\% &\mycc 1.02  & \mycc (76/76/78)\% & \mycc 1.61 \\
&  & Low frequency & (98/95/98)\% & 1.72  & (99/81/93)\% & 1.76  \\
&  & \mycc \ac{BBH} & \mycc (98/93/96)\%  &\mycc 3.26   & \mycc -- & \mycc --  \\
\cmidrule(rl){1-1} \cmidrule(rl){2-2} \cmidrule(rl){3-3} \cmidrule(rl){4-4} \cmidrule(rl){5-5} \cmidrule(rl){6-6} \cmidrule(rl){7-7} 
\multirow{6}{*}{\makecell{{\it Extremely} \\ {\it loud}}} & \multirow{3}{*}{Optimal} & High frequency & (85/48/63)\% & 0.99  & (88/12/38)\% & 1.08   \\
& & \mycc Low frequency & \mycc (99/66/78)\% & \mycc 0.41  & \mycc (99/39/49)\% & \mycc 0.78  \\
& & \ac{BBH} & (98/63/75)\% & 1.04   & -- & --  \\
\cmidrule(rl){2-2} \cmidrule(rl){3-3} \cmidrule(rl){4-4} \cmidrule(rl){5-5} \cmidrule(rl){6-6} \cmidrule(rl){7-7} 
& \multirow{3}{*}{Median} & \mycc High frequency & \mycc (88/52/55)\% & \mycc 0.95   & \mycc (88/24/31)\% & \mycc 1.01   \\
&  & Low frequency & (99/68/68)\% & 1.00 & (99/40/42)\% & 0.82  \\
&  & \mycc \ac{BBH} & \mycc (97/66/67)\% & \mycc 0.96  & \mycc -- & \mycc --\\  
\bottomrule
\end{tabular}
\label{table:ks_ratio}
\end{center}
\end{table}

\subsubsection{False Alarm Rate} \label{false_alarm_rate}

The confidence of a \ac{GW} signal candidate is quantified by the \ac{FAR}, or the rate of terrestrial noise events with their ranking statistics (e.g., $\rho$ in \ac{cWB}) equal or higher than the ranking statistic of an astrophysical candidate event. Lower values of \ac{FAR} indicate that \ac{GW} signal candidates are astrophysical in their origin with higher confidence. Similarly, higher values of the inverse \ac{FAR} (iFAR)\acused{iFAR} correspond to higher confidence. The glitch subtraction technique might reduce values $\rho$ of the noise events and increase values of $\rho$ for \ac{GW} signal candidates when they near or overlap with glitches.   

To assess the effect of the glitch subtraction technique on the \ac{FAR} of injections used in the previous section, we use isolated samples in the testing sets with sample sizes of 237 and 156 for {\it Scattered light} and {\it Extremely loud} glitches from January 7$^{\rm th}$, 2020 3:00 UTC to February 2$^{\rm rd}$, 2020 03:55 UTC and January 15$^{\rm th}$, 2020 17:44 UTC to February 3$^{\rm rd}$, 2020 23:55 UTC, respectively. Because the testing sets used in the previous section have segment overlaps to account for statistical errors with larger sample sizes, we use the isolated testing samples from the sets to avoid double subtraction. We use the data containing the above testing samples during the detectors are observing, corresponding to the 20.4- and 20.5-day data from the \ac{L1} and \ac{H1} detectors, respectively. The percentages of the total duration of the isolated testing samples are 0.07\% and 0.026\% of 20.4 days for {\it Scattered light} and {\it Extremely loud} glitches, respectively.   

Using the \ac{L1} data before glitch subtraction with the original \ac{H1} data and applying time shifts to the \ac{L1} data, we get the {\it background} trigger set, where time shits are applied to get triggers representing the noise events coincident between detectors by chance and enlarge the analysis time. Similarly, we also use the \ac{L1} data after glitch subtraction with the original \ac{H1} data to get another background trigger set. With time shift applied to the \ac{L1} data, we obtain 21.2-year equivalent background triggers both before and after glitch-subtracted data. Both trigger sets have the maximum values of $\rho=  53.8$ and the lowest \ac{FAR} of $1.5\times10^{-9}$ Hz (corresponding to \ac{iFAR} of 21.2 years). 

\begin{figure}[ht!]
    \centering
    \includegraphics[width=0.7\linewidth]{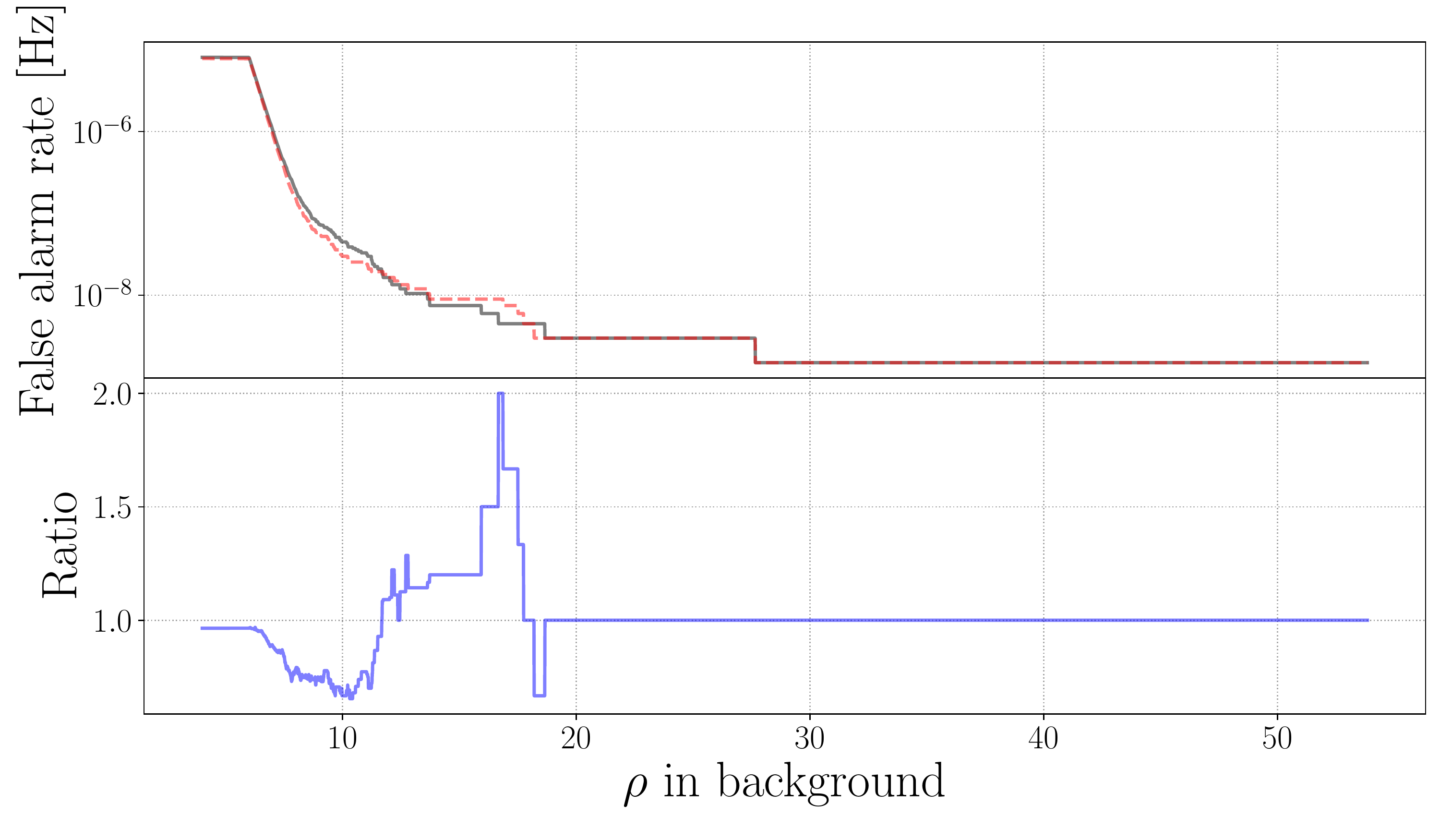}
    \caption[False alarm rates before and after glitch subtraction.]%
    {False alarm rates before (gray-solid) and after (red-dashed) glitch subtraction (top) and the ratio of the latter to the former.}
    \label{fig:bk_far_comparison} 
\end{figure}

\begin{figure}[!ht]
    \centering
    \includegraphics[width=0.6\linewidth]{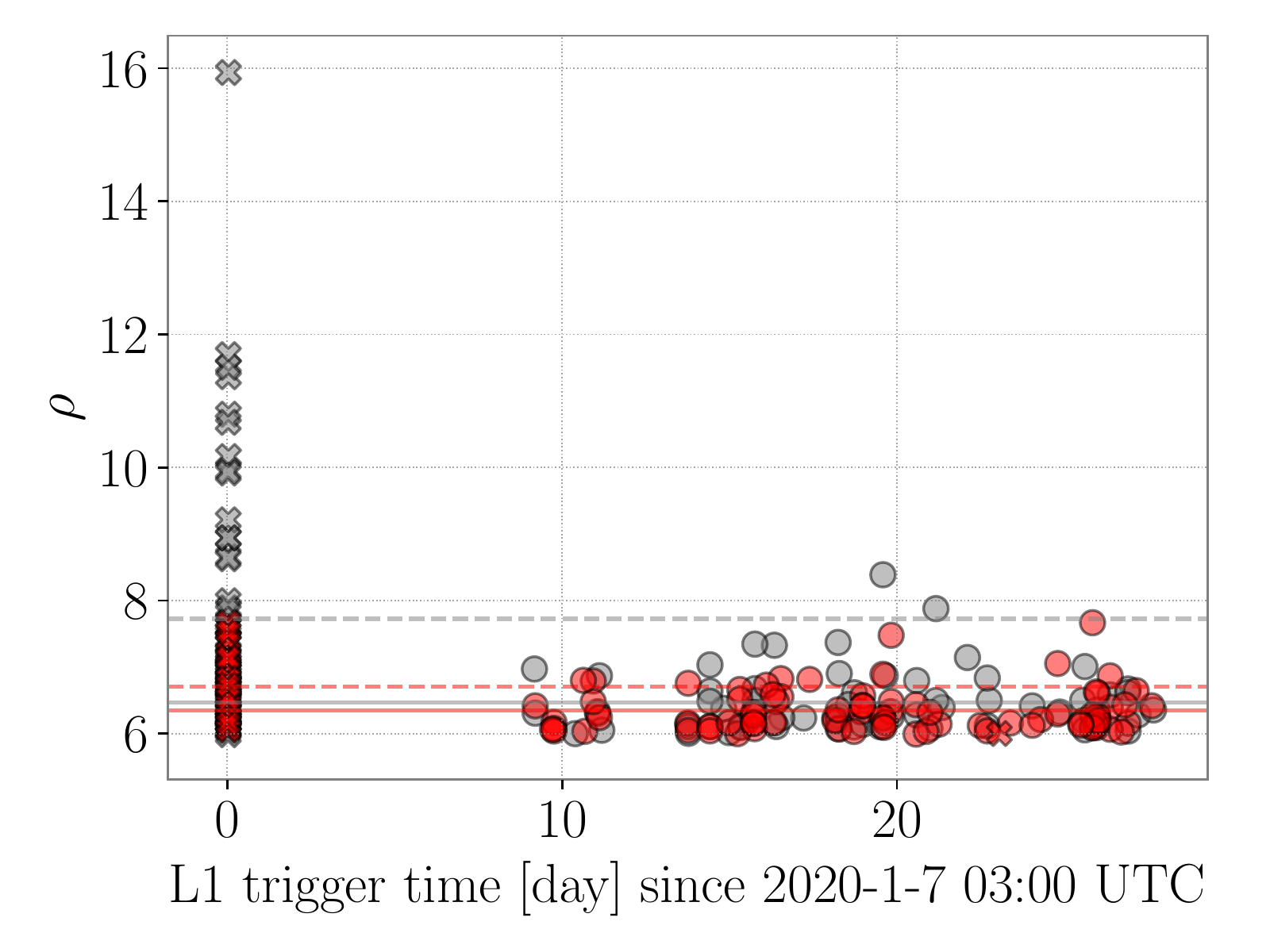}
    \caption[{\it Background} triggers within the interval of the subtracted data portions.]%
    {{\it Background} triggers within the intervals of the subtracted data portions for the data before (gray) and after (red) glitch subtraction for {\it Scattered light} (cross) and {\it Extremely loud} (circle) glitches, respectively. The dashed and solid lines denote the average values of $\rho$ for these triggers for {\it Scattered light} and {\it Extremely loud} glitches, respectively.}
    \label{fig:trig} 
\end{figure}

Figure \ref{fig:bk_far_comparison} shows the \ac{FAR} of background triggers before and after glitch subtraction. We find that the \ac{FAR} is typically reduced in the interval from $\rho\sim7$ to $\rho\sim12$. The reduced \ac{FAR} in this interval can be explained by the reduction of $\rho$ in the subtracted part of the data. Figure \ref{fig:trig} shows background triggers within the interval of the subtracted data portions. The distribution of these triggers is due to the quality of the \ac{L1} data. The average values of $\rho$ are reduced by 13.2\% and 1.9\% for {\it Scattered light} and {\it Extremely loud} glitches. Because {\it Scattered light} glitches are typically subtracted with our technique more effectively than {\it Extremely loud} glitches, the former has higher percentages than the latter. The \ac{FAR} is increased in the interval from  $\rho\sim12$ to $\rho\sim23$ in the glitch subtracted data because of two triggers with $\rho=18.2$ and $\rho=17.5$. However, the \ac{L1} times of these two triggers are not within the subtracted data portions so that it seems to be due to a realization of \ac{cWB} trigger-generation process.    

Using these two background sets, we first evaluate injections that are not nearby and overlap with glitches. Because we have created the no-glitch data sets in the previous section with the colored Gaussian noise using the \ac{PSD} of the real \ac{L1} data at the time of injections, we can consider injections in the no-glitch data to be those not nearby and overlap with glitches. As a lower limit, we set \acp{iFAR} of injections with $\rho$ greater than 53.8 to be ${\rm \ac{iFAR}}=21.2$ years, which is the maximum length of our background.

Across high- and low frequency, and \ac{BBH} injections, we find that $\sim90$\% of injections in the no-glitch data have non-reduced \acp{iFAR} after glitch subtraction. The \acp{iFAR} after glitch subtraction is greater than the \acp{iFAR} before glitch subtraction by a factor of $\sim1.02$ on average over injections because of the reduction of $\rho$ in the background. 

\begin{figure}[!ht]
    \begin{minipage}{1\hsize}
  \begin{center}
   \includegraphics[width=1\textwidth]{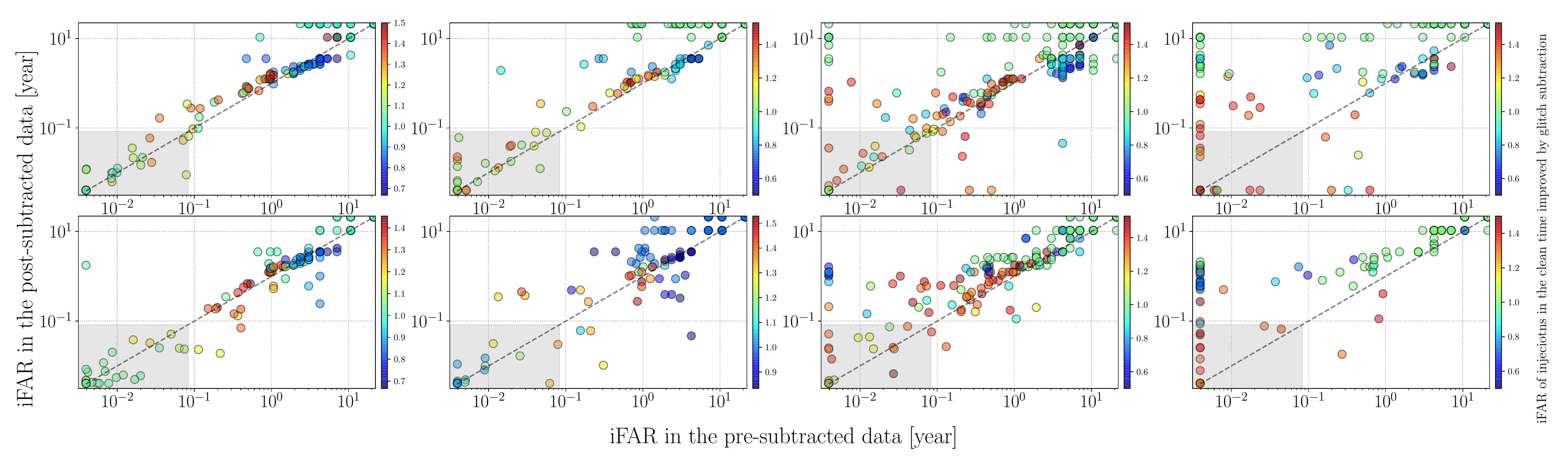}
  \end{center}
    \end{minipage}
    \begin{minipage}{1\hsize}
  \begin{center}
   \includegraphics[width=1\textwidth]{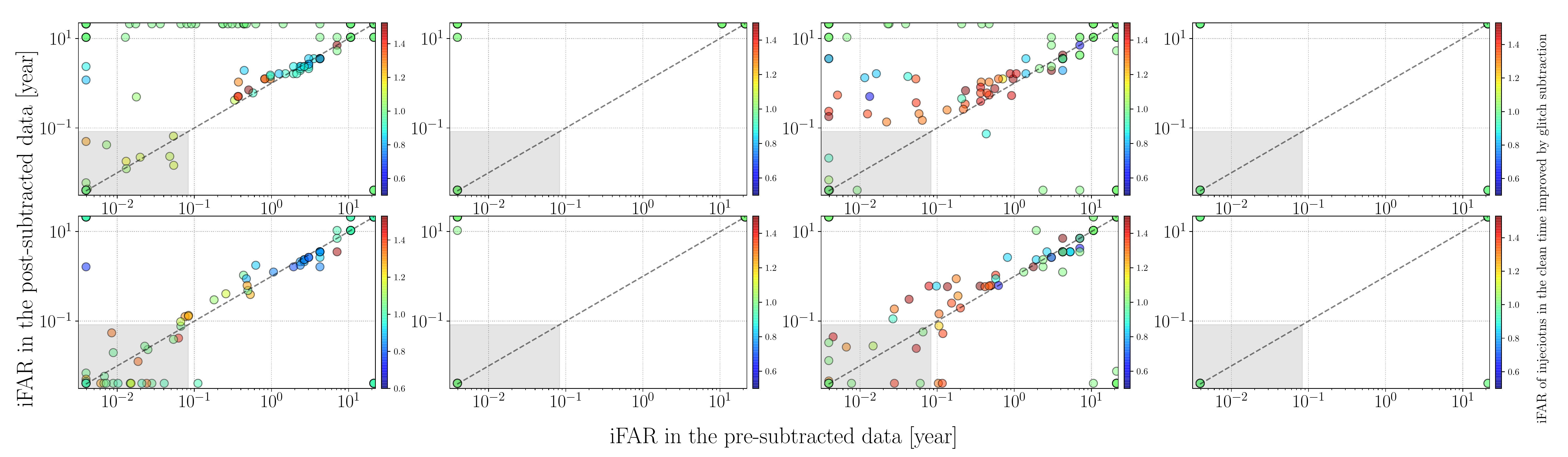}
  \end{center}
    \end{minipage}
\caption[Distributions of the \ac{iFAR} for Gaussian-modulated sinusoidal injections.]%
    {Distributions of the \ac{iFAR} for high-frequency (1-2$^{\rm th}$ columns) and low-frequency (3-4$^{\rm th}$ columns) Gaussian-modulated sinusoidal waveforms injected in the full (1,3$^{\rm th}$ columns) and partial (2,4$^{\rm th}$ columns) windows of the optimal (1,3$^{\rm th}$ rows) and median (2,4$^{\rm th}$ rows) testing samples of {\it Scattered light} (1-2$^{\rm th}$ rows) and {\it Extremely loud} (2-4$^{\rm th}$ rows) glitches, respectively. The color scale denotes the ratio of the \ac{iFAR} of injections in the no-glitch data evaluated with the background $\rho$ distribution obtained with the post-subtracted data to that obtained with the pre-subtracted data. The shaded area denotes the \ac{iFAR} less than 1 month.}
\label{fig:ifar_sin_gau}
\end{figure}

\begin{figure}[!ht]
    \centering
    \includegraphics[width=0.9\linewidth]{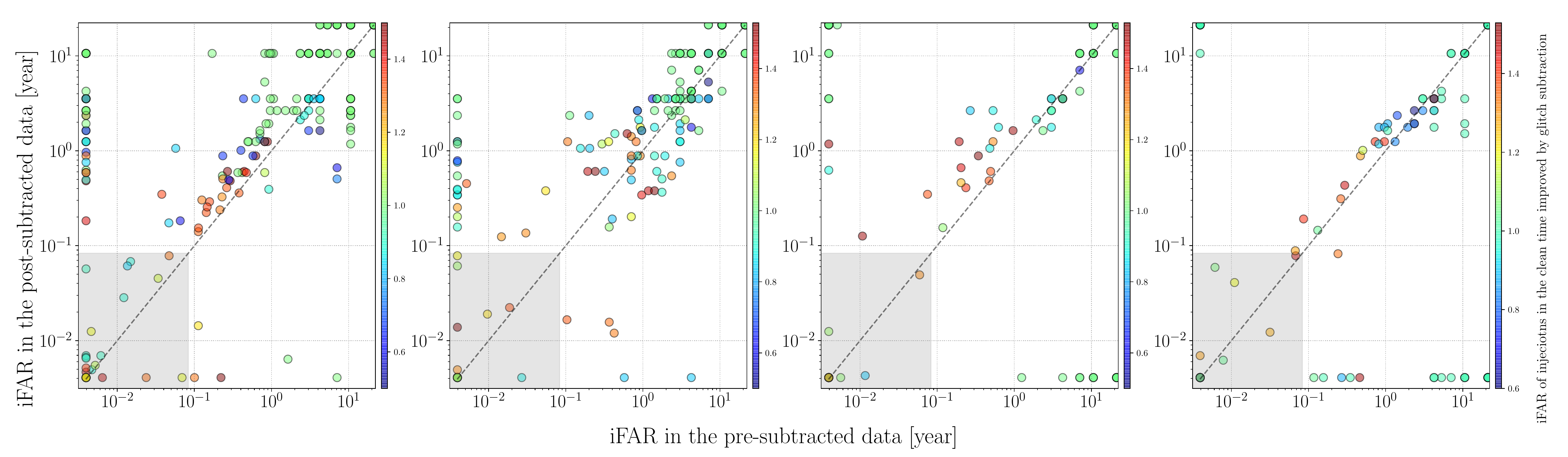}
    \caption[Distributions of \ac{iFAR} for \ac{BBH} injections.]%
    {Distributions of \ac{iFAR} of \ac{BBH} injections in the full window of the optimal (1/3$^{th}$ columns) and median (2/4$^{th}$ columns) testing sample of ({\it Scattered light}/{\it Extremely loud}) glitches, respectively. The color scale denotes the ratio of the \ac{iFAR} of injections in the no-glitch data evaluated with the background $\rho$ distribution obtained with the post-subtracted data to that obtained with the pre-subtracted data. The shaded area denotes the \ac{iFAR} less than 1 month.}
    \label{fig:ifar_bbh} 
\end{figure}

As mentioned above, the glitch subtraction technique may increase the $\rho$ of injections near to or overlapping with glitches, causing higher \acp{iFAR}. Figures \ref{fig:ifar_sin_gau} and \ref{fig:ifar_bbh} show \ac{iFAR} distributions before and after glitch subtraction using corresponding backgrounds for Gaussian modulated sinusoidal and \ac{BBH} injections, respectively. Percentages $P^{\rm i}_{\rm ab}$ of injections with non-reduced \ac{iFAR} after glitch subtraction range from 88\% (obtained from the set with the high-frequency injections in the full window of the median testing sample of {\it Scattered light} glitches) to 100\% (obtained from the set with high-frequency injections in the partial window of the optimal testing sample of {\it Extremely loud} glitches). The sets with the lowest and highest values of $P^{\rm i}_{\rm ab}$ respectively correspond to the sets with the lowest and highest values $P_\rho$ because the increases in $\rho$ for injections correspond to the increases in \ac{iFAR}. 

The ratio $\left<R_{\rm ab}\right>$ of \ac{iFAR} values after glitch subtraction to \ac{iFAR} values before glitch subtraction averaged over injections range from 1.03 (obtained from the set with high-frequency injections in the full window of the optimal testing sample of {\it Scattered light} glitches) to 1400 (obtained with high-frequency injections in the partial window of the optimal testing sample of {\it Extremely loud} glitches). Because high increases in $\rho$ of injections correspond to higher increases in $\ac{iFAR}$, values of $\left<R_{\rm ab}\right>$ for sets with the optimal testing sample typically are greater than values for sets with the median testing sample by a factor of $\sim0.7\sim150$ and $\sim2.1\sim4.3$ for {\it Scattered light} and {\it Extremely loud} glitches. The sets with high-frequency injections in the full window for {\it Scattered light} glitches corresponding to the lowest factor of $\sim0.7$ have comparable values of $\left<R_{\rm ab}\right>=1.3$ and $\left<R_{\rm ab}\right>=1.9$ for the optimal- and median-testing-sample sets. Subtracting glitches with their peak frequencies close to the characteristic frequencies of injections lets \ac{cWB} obtain higher values of $\rho$. Therefore, values of $\left<R_{\rm ab}\right>$ for sets with (low/high)-frequency injections are greater than values of $\left<R_{ab}\right>$ for sets with (high/low) frequency injections by a factor of ($\sim3.1\sim74$ /$\sim1.2\sim2.5$) for ({\it Scattered light}/{\it Extremely loud}) glitches.     

Weak signals (so-called sub-threshold triggers) near or overlapping with glitches that are missed by \ac{cWB} or are not confident enough to be classified as astrophysical signals can gain sufficient confidence after glitch subtraction. If we assume an \ac{iFAR} threshold for weak signals to be a month, percentages $P^{\rm i}_{\rm w}$ of injections with \ac{iFAR} above a month after glitch subtraction out of injections with \ac{iFAR} below a month before glitch subtraction range from 1\% (obtained from the set with high-frequency injections in the full window of the optimal testing sample of {\it Scattered light} glitches) to 57\% (obtained with the set with low-frequency injections in the partial window of the optimal testing sample of {\it Scattered light} glitches). For {\it Scattered light} glitches, sets with low-frequency injections have values of $P^{\rm i}_{\rm w}\sim40\sim57$\% and sets with high-frequency and \ac{BBH} injections have values of $P^{\rm i}_{\rm w}\sim1\sim4$\%, where values for the optimal and median-testing-sample sets are compatible. For {\it Extremely loud} glitches, sets with the optima-testing sample have values of $P^{\rm i}_{\rm w}\sim20-44$\% and sets with the median-testing-sample have values of $P^{\rm i}_{\rm w}\sim5-18$\%, where values for high-frequency sets are greater than values for the low-frequency sets by a factor of $\sim0.7\sim1.8$. Table \ref{table:ifar} shows values of ($P^{\rm i}_{\rm ab}$/$P^{\rm i}_{\rm n}$/$P^{\rm i}_{\rm w}$) and ($\left<R_{\rm ab}\right>$/$\,\left<R_{\rm n}\right>$).

\begin{table}[!ht]
\newcolumntype{g}{>{\columncolor{Gray}}c}
\renewcommand*{\arraystretch}{1.1}
\setlength{\tabcolsep}{2pt}
\caption[Percentages of injections with non-reduced \ac{iFAR} after glitch subtraction and ratios of \ac{iFAR} after and before subtraction.]%
{Percentages ($P^{\rm i}_{\rm ab}$/$P^{\rm i}_{\rm n}$) of injections (near to or overlapping with glitches/in the absence of glitches) with non-reduced \ac{iFAR} after glitch subtraction and percentages $P^{\rm i}_{\rm w}$ of injections with \ac{iFAR} above a month after glitch subtraction out of injections with \ac{iFAR} below a month before glitch subtraction, and ratios $(\left<R_{\rm ab}\right>/\left<R_{\rm n}\right>)$ of \ac{iFAR} after glitch subtraction to \ac{iFAR} before glitch subtraction averaged over injections (near to or overlapping with glitches/in the absence of glitches).}
\begin{center}
\begin{tabular}{c c c || c c c c c c} \toprule
\multirow{2}{*}{Glitch class} & \multirow{2}{*}{\makecell{Testing \\sample}} & \multirow{2}{*}{Injection} & \multicolumn{3}{ c }{Full window} & \multicolumn{3}{ c }{Partial window} \\ 
\cmidrule(rl){4-6} \cmidrule(rl){7-9}
 & &  & ($P^{\rm i}_{\rm ab}$/$P^{\rm i}_{\rm n}$/$P^{\rm i}_{\rm w}$) & $\left<R_{\rm ab}\right>$ & $\left<R_{\rm n}\right>$ & ($P^{\rm i}_{\rm ab}$/$P^{\rm i}_{\rm w}$/$P^{\rm i}_{\rm n}$) & $\left<R_{\rm ab}\right>$ & $\left<R_{\rm n}\right>$ \\ 
\midrule\midrule
\multirow{6}{*}{\makecell{{\it Scattered} \\ {\it light}}} & \multirow{3}{*}{Optimal} & High frequency & (90/1/91)\%  & 1.3 & 1.02 &   (90/3/89)\%& 3.5 & 1.02\\
& & \mycc Low frequency & \mycc (87/46/89) & \mycc 96  & \mycc 1.01 & \mycc (90/57/92)\% & \mycc 260 & \mycc 1.04\\
& & \ac{BBH} & (93/40/90)\% & 97 & 1.02 & -- & -- & --\\
\cmidrule(rl){2-2} \cmidrule(rl){3-3} \cmidrule(rl){4-4} \cmidrule(rl){5-5} \cmidrule(rl){6-6} \cmidrule(rl){7-7} \cmidrule(rl){8-8} \cmidrule(rl){9-9} 
&  \multirow{3}{*}{Median} & \mycc High frequency & \mycc (88/1/91)\% &\mycc 1.9  & \mycc 1.02 & \mycc (88/4/92)\% & \mycc 1.7 & \mycc 1.03\\
&  & Low frequency & (94/42/93)\% & 6.0  & 1.03 & (98/53/90)\% & 28 & 1.02\\
&  & \mycc \ac{BBH} & \mycc (90/45/90)\%  &\mycc 10& \mycc 1.01 & \mycc -- & \mycc -- & \mycc --  \\
\cmidrule(rl){1-1} \cmidrule(rl){2-2} \cmidrule(rl){3-3} \cmidrule(rl){4-4} \cmidrule(rl){5-5} \cmidrule(rl){6-6} \cmidrule(rl){7-7} \cmidrule(rl){8-8} \cmidrule(rl){9-9}
\multirow{6}{*}{\makecell{{\it Extremely} \\ {\it loud}}} & \multirow{3}{*}{Optimal} & High frequency & (94/31/92)\% & 800  & 1.01 & (100/30/91)\% & 1400 & 1.02 \\
& & \mycc Low frequency & \mycc (94/44/95)\% & \mycc 700  & \mycc 1.02 & \mycc (98/20/96)\% & \mycc 650 & \mycc 1.03   \\
& & \ac{BBH} & (94/41/96)\% & 760  & 1.02 & --  & -- & -- \\
\cmidrule(rl){2-2} \cmidrule(rl){3-3} \cmidrule(rl){4-4} \cmidrule(rl){5-5} \cmidrule(rl){6-6} \cmidrule(rl){7-7} \cmidrule(rl){8-8} \cmidrule(rl){9-9}
& \multirow{3}{*}{Median} & \mycc High frequency & \mycc (91/14/94)\% & \mycc 380  & \mycc 1.01 & \mycc (100/9/92)\% & \mycc 380 & \mycc 1.01 \\
&  & Low frequency & (91/13/93)\% & 210 & 1.01 & (99/5/97)\% & 150 & 1.02 \\
&  & \mycc \ac{BBH} & \mycc (91/18/95)\% & \mycc 320  & \mycc 1.02 & \mycc -- & \mycc -- & \mycc -- \\  
\bottomrule
\end{tabular}
\label{table:ifar}
\end{center}
\end{table}

\section{Conclusion} \label{ppr2_conclusion}
In this paper, we have presented a new machine learning-based algorithm to subtract glitches using a set of auxiliary channels. Glitches are the product of short-live linear and non-linear couplings due to interrelated sub-systems in the detector including the optic alignment systems and mitigation systems of ground motions. Because of the characteristic of glitches, modeling coupling mechanisms is typically challenging. Without prior knowledge of the physical coupling mechanisms, our algorithm takes the data from the sensors monitoring the instrumental and environmental noise transients and then estimates the glitch waveform in the detector's output, providing the glitch-subtracted data stream. Subtracting glitches improves the quality of the data and will enhance the detectability of astrophysical \ac{GW} signals. 

Using two classes of glitches with distinct noise couplings in the \ac{aLIGO} data, we find that our algorithm successfully reduces the \ac{SNR} of the data due to the presence of glitches by $10-70$\%. Subtracting glitches from the data enhances the \ac{cWB} ranking statistic by a factor of $\sim 1.03 \sim3.5$ and $\sim 1.2 \sim 2.7$ averaged over Gaussian modulated sinusoidal injections and \ac{BBH} injections, respectively. We find that the source-direction, central frequency and chirp mass estimated by \ac{cWB} after glitch subtraction are comparable or more accurate than that before glitch subtraction. The \ac{iFAR} of injections in the data portion in the absence of glitches is increased by $\sim1.02$ by subtracting glitches in $\sim0.1$\% of the 20.4-day data from the \ac{L1} detector. We find that injections near to or overlapping with glitches typically have significant enhancements with glitch subtraction. The \ac{iFAR} of those injections is increased by a factor $\sim1.3\sim1400$.   

In this paper, we focus on the two classes of glitches and apply the glitch subtraction technique to only $\sim0.1$\% of the \ac{L1} data so that we find no significant reduction of $\rho$ in the background. Creating the \ac{CNN} network models for other glitch classes and subtract a higher number of glitches both in the \ac{L1} and \ac{H1} could provide the statistically robust measure of the effect of the glitch subtraction technique on the data. 

Currently, the LIGO-Virgo collaboration vetoes glitch classes focused on this paper and other glitch classes with witness channels. For example, over the course of the 20.4-day data from the \ac{L1} data, $\sim 15000$ glitches with \ac{SNR} above 7.5 in the two classes are present and have a total period of $\sim1.8$\% (so-called {\it deadtime}), which would be vetoed. By accounting for the deadtime and injections removed by the veto, the comparison of the {\it volume-time} integrals \cite{Davis:2018yrz} between the vetoing method and the glitch subtraction technique allows us to find a better approach.    

We find that using the spectrograms of the data as the input for the network is more successful than using time series as the input. However, it might improve the glitch subtraction efficiency by using the \ac{FGL} transformation as well as the amplitude and phase corrections within the loss function to train the network. Improved glitch subtraction would allow us to detect astrophysical signals with higher confidence and brings us a better understanding of the physics in the universe.

%
%
%
%
%

\appendix
\section*{Appendix}
\section{Comparison between Scattered light glitches and quiet times} \label{apx:comp_sclight_quiet}

To show that the frequency region above 100 Hz in time periods containing \textit{Scattered light} glitches in the strain channel has no excess power and are compatible with the corresponding frequency region of the Gaussian noise, we compare 693 spectrograms containing \textit{Scattered light} glitches with 306 spectrograms when the strain channel is quiet, statistically evaluate them using the \ac{KS} test \cite{kolmogorov_1951}. 

\begin{figure}[!ht]
    \begin{minipage}{0.5\hsize}
  \begin{center}
   \includegraphics[width=1\linewidth]{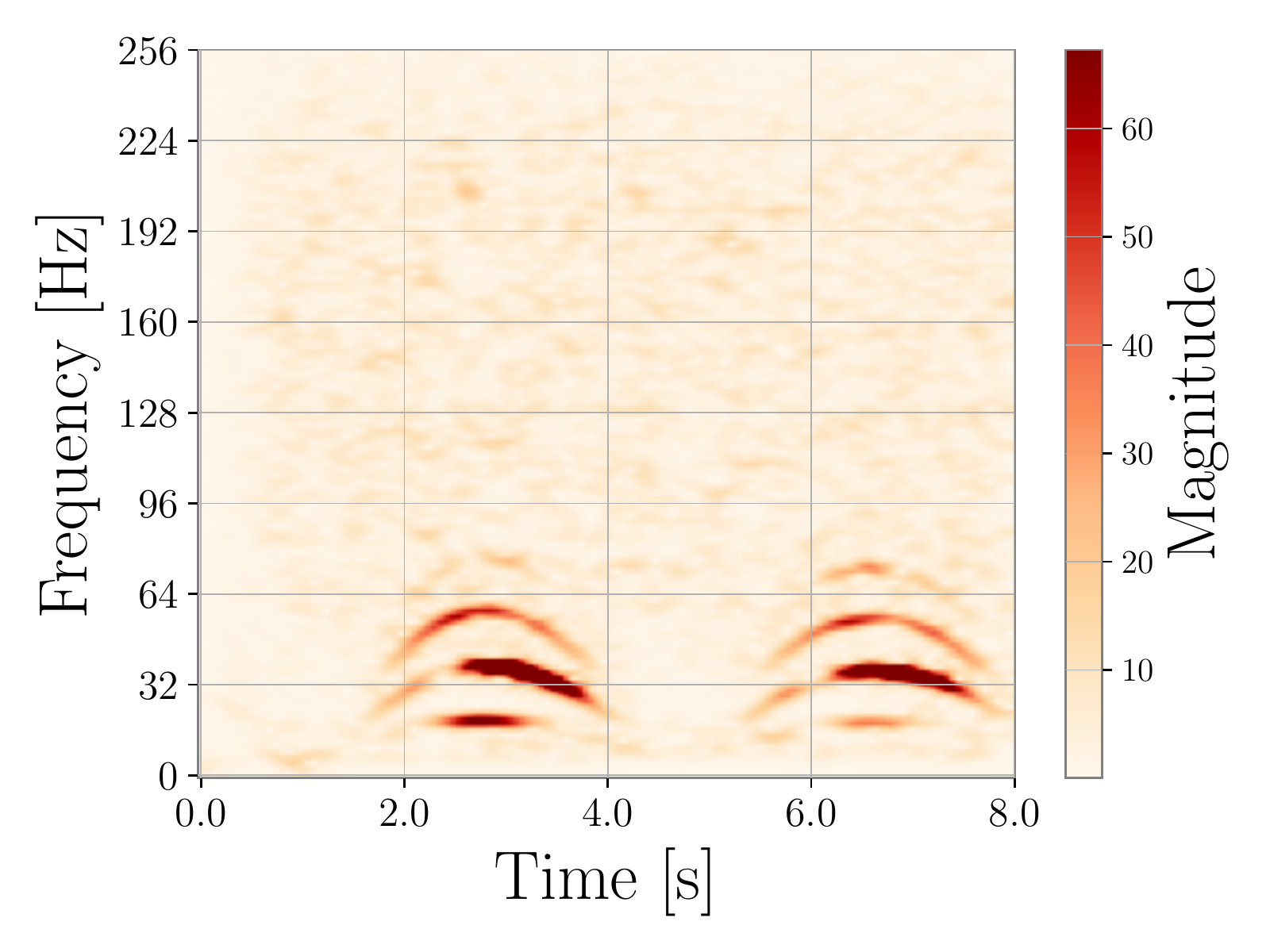}
  \end{center}
    \end{minipage}
    \begin{minipage}{0.5\hsize}
  \begin{center}
   \includegraphics[width=1\linewidth]{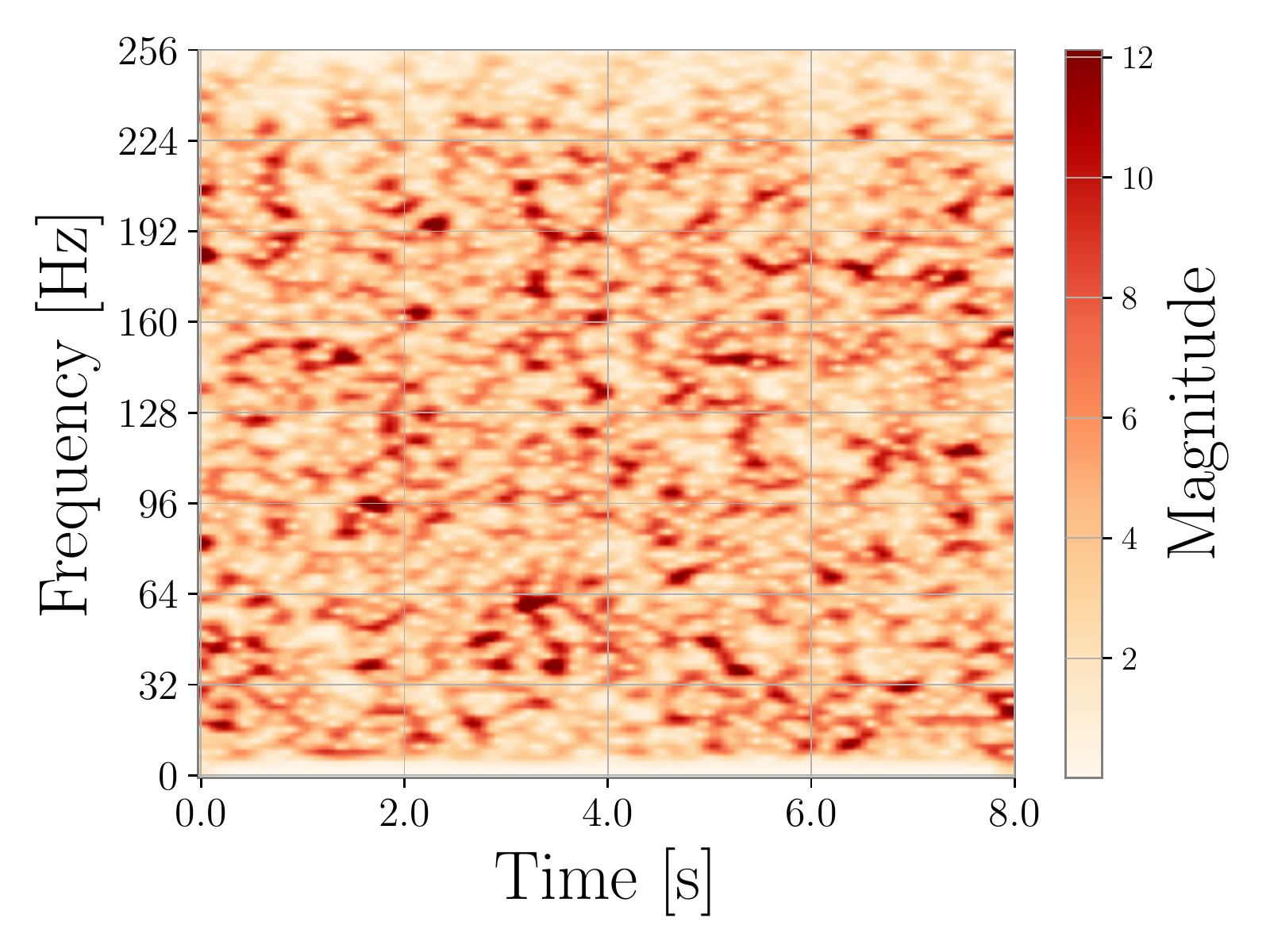}
  \end{center}
    \end{minipage}
    \caption[Magnitude of STFTs of a {\it Scattered light} glitch and a quiet time.]%
    {Magnitude of STFTs of {\it Scattered light} glitch (left) and a quiet time (right).}
\label{fig:spectrograms_sclight_quiet}
\end{figure}

To create the data set of \textit{Scattered light} glitches, we whiten the time series of the strain channel with a software called \textsc{GWpy} \cite{duncan_macleod_2020_4301851} and then apply a low-pass filter at 512 Hz as used in Sec. \ref{glitch_subtraction_pipeline}. We have the \textit{Scattered-light} set with a sample size of 693 by selecting time periods with a duration of 8 seconds that contains \textit{Scattered light} glitches. To have a set of quiet data, we use the observing-mode strain channel data with a duration of 4096 seconds beginning from April 2nd, 2019 at 5:04 UTC, without data quality issues such as the corrupting data, the presence of glitches, and hardware injections of simulated signals. We whiten and apply the high-pass filter to the quiet time series and then cut the edge of the whitened time series to remove artifacts of the Fourier transform. By diving the whitened time series into 8-second segments, we have the quiet-data set with a sample size of 301. We create \acp{mSTFT} of the \textit{Scattered-light} set and the \textit{quiet} set. Figure \ref{fig:spectrograms_sclight_quiet} shows the \ac{mSTFT} of a \textit{Scattered light} glitch and a quiet time.

\begin{figure}[ht!]
    \begin{minipage}{0.5\hsize}
  \begin{center}
   \includegraphics[width=1\linewidth]{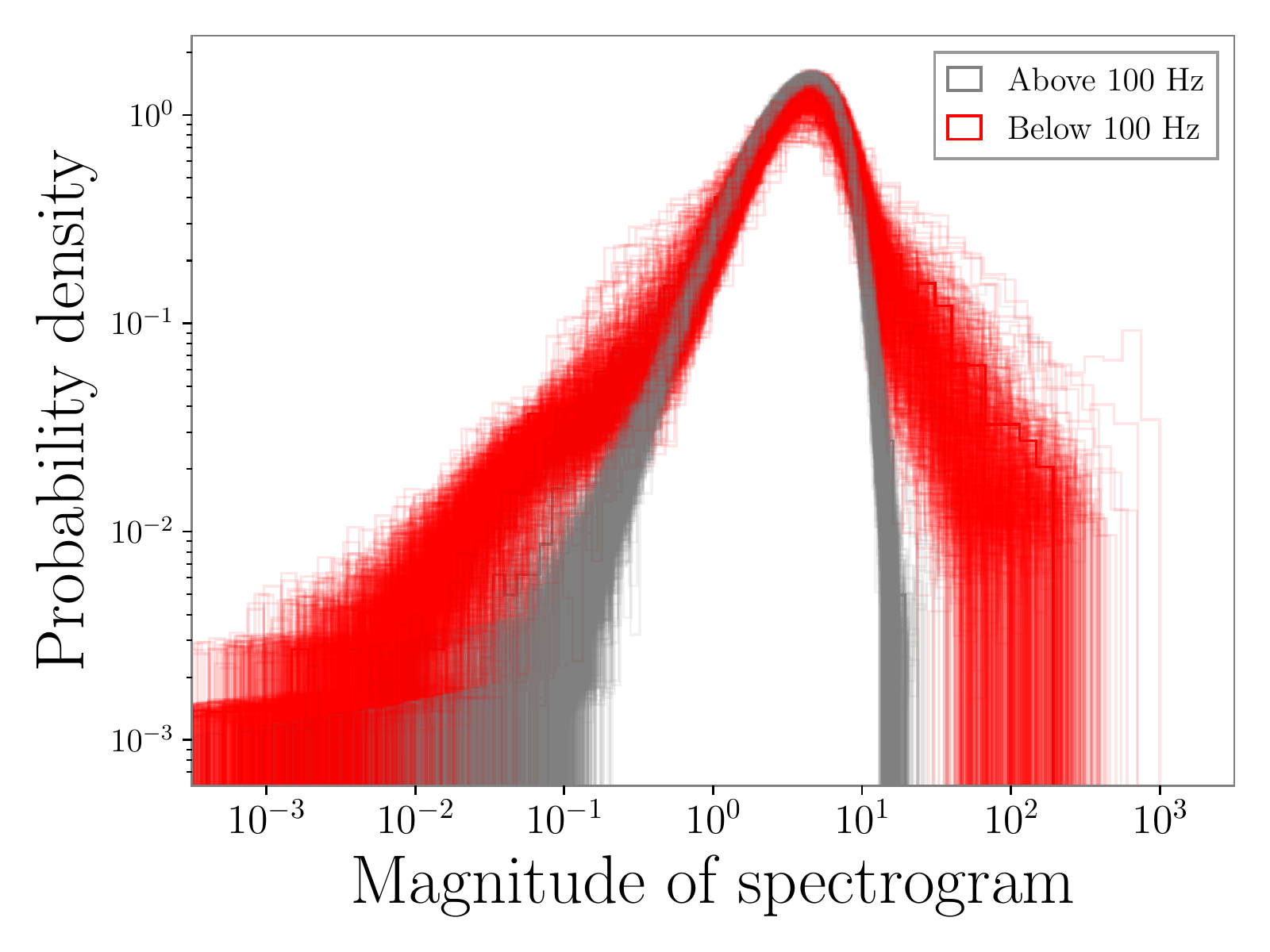}
  \end{center}
    \end{minipage}
    \begin{minipage}{0.5\hsize}
  \begin{center}
   \includegraphics[width=1\linewidth]{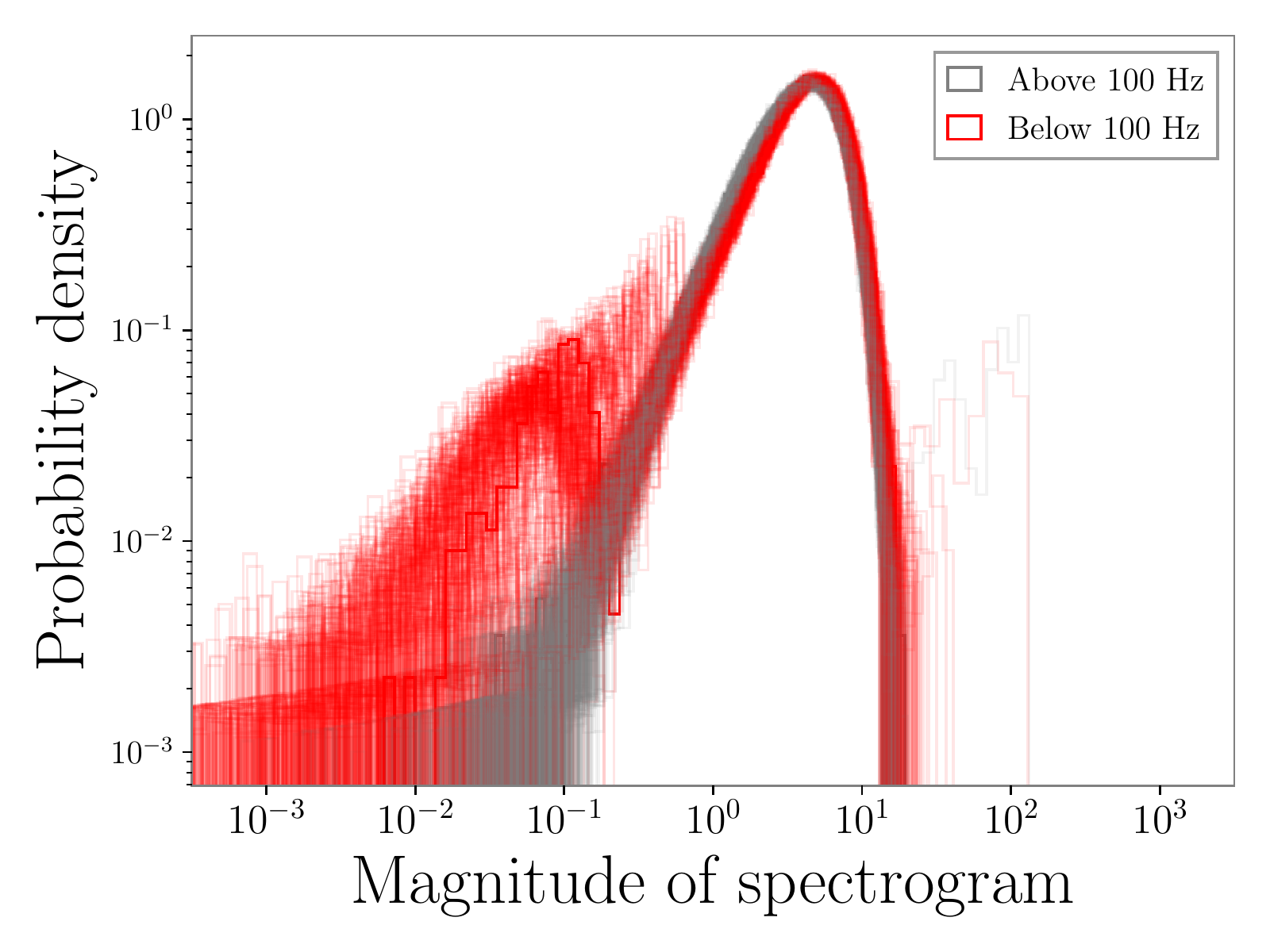}
  \end{center}
    \end{minipage}
    \caption[Distributions of \ac{mSTFT}-pixel values of {\rm Scattered light} glitches and quiet times.]%
    {Distributions of \ac{mSTFT}-pixel values of 693 {\rm Scattered light} glitches (left) and 306 quiet-time segments (right).}
\label{fig:distributinos_spectrograms}
\end{figure}

\begin{figure}[ht!]
    \begin{minipage}{0.5\hsize}
  \begin{center}
   \includegraphics[width=1\linewidth]{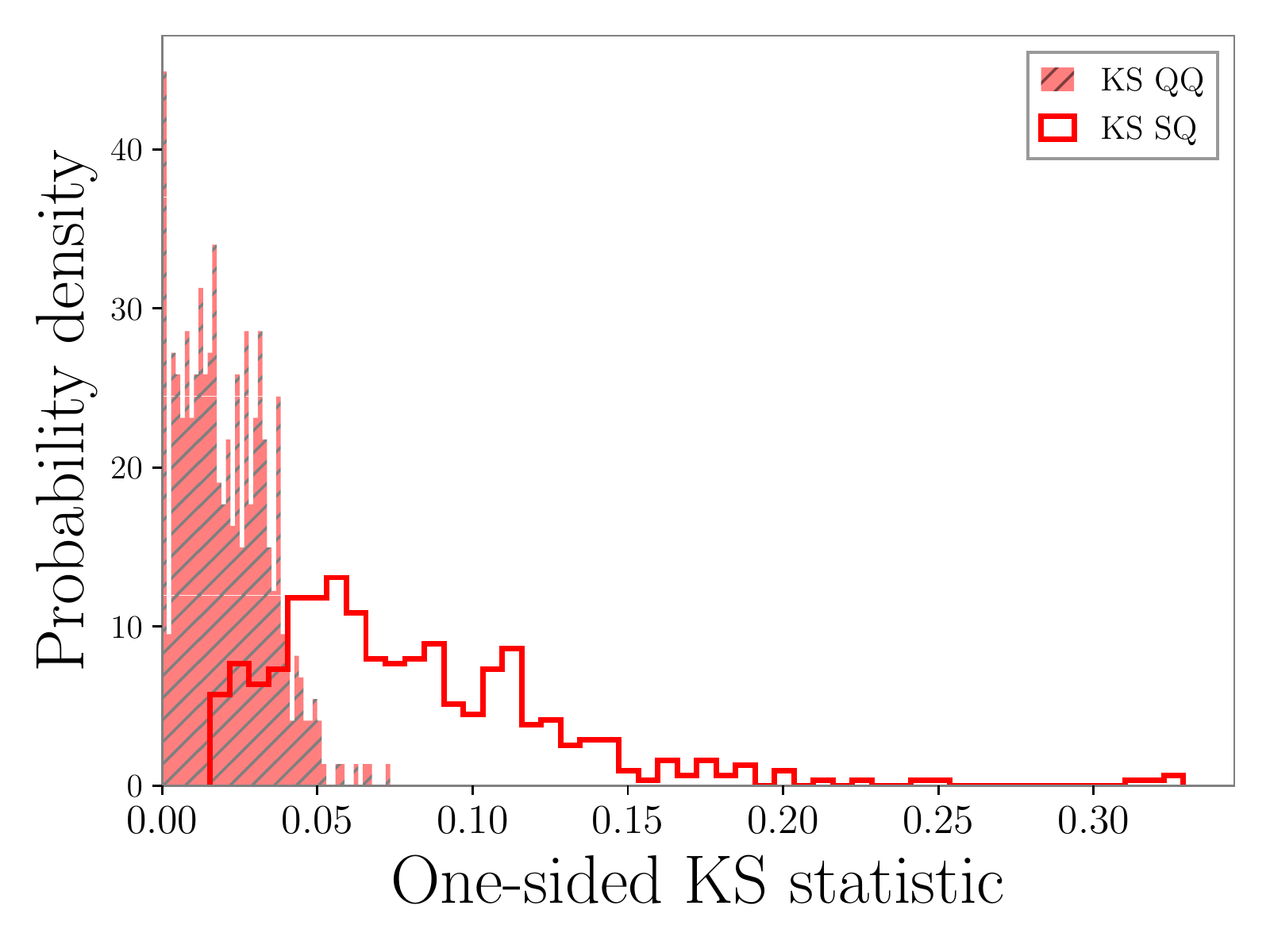}
  \end{center}
    \end{minipage}
    \begin{minipage}{0.5\hsize}
  \begin{center}
   \includegraphics[width=1\linewidth]{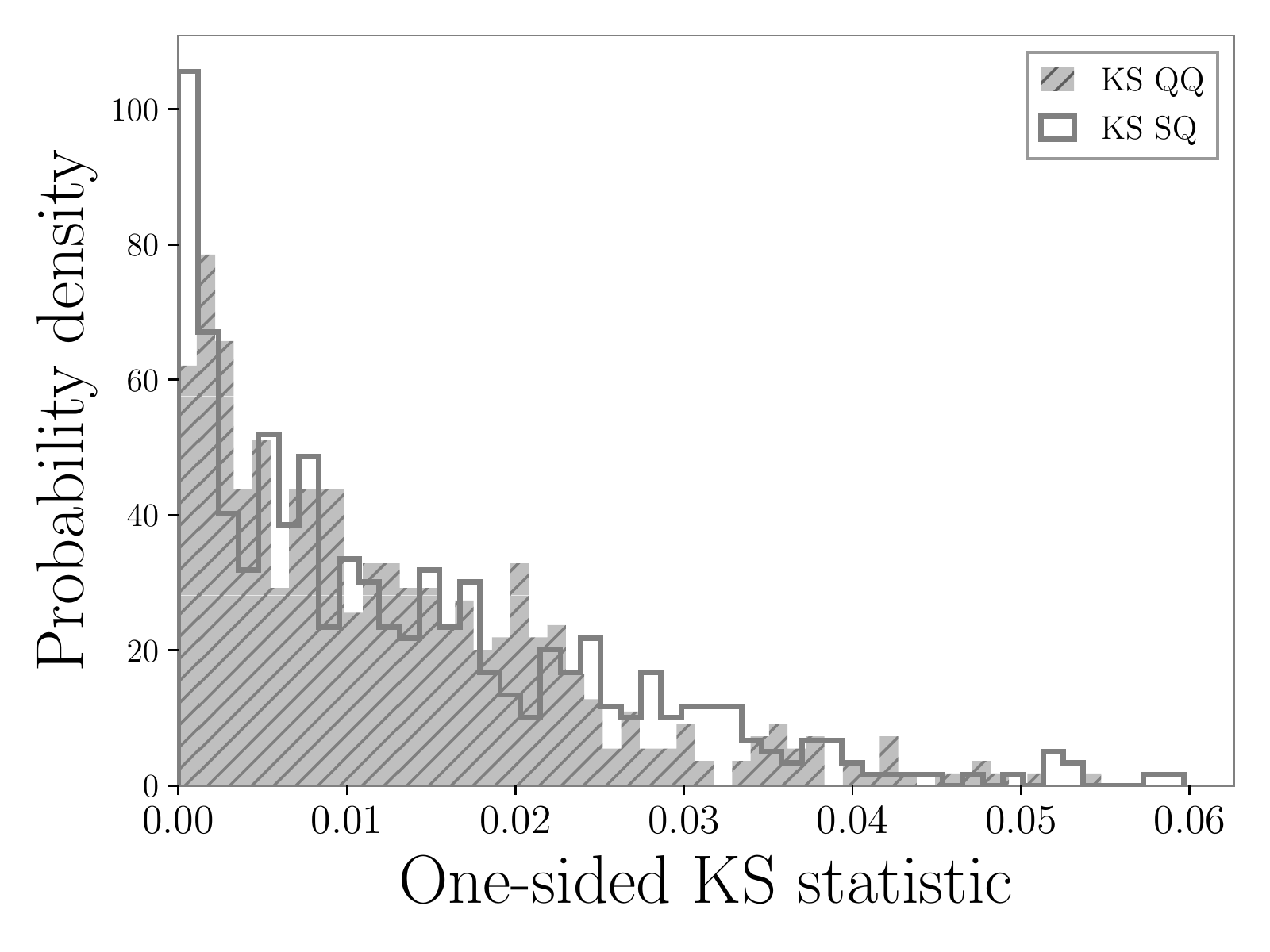}
  \end{center}
    \end{minipage}
    \caption[Distributions of \ac{KS} statistics of 500 random pairs of the data sample.]%
    {Distributions of \ac{KS} statistics calculated with 500 random pairs of \acp{mSTFT} in the frequency region below (left) and above (right) 100 Hz. QQ and SQ denote pairs of two \acp{mSTFT} from the \textit{quiet} set and pairs of a \ac{mSTFT} from the \textit{Scattered-light} set and a \ac{mSTFT} from the \textit{quiet} set, respectively.}
\label{fig:dist_ks}
\end{figure}

Figure \ref{fig:distributinos_spectrograms} shows distributions of the  \acp{mSTFT} in the frequency region above or below 100 Hz in the \textit{Scattered-light} set and the \textit{quiet} set. The \textit{Scattered-light} (\textit{quiet}) set has 1.6\% and 9.5\% (1.6\% and 2.1\%) of pixels with values above 10 for the frequency region above and below 100 Hz across the set, respectively. To verify the upper-frequency region of the \textit{Scattered light} set has no excess power above Gaussian fluctuations, we calculate one-sided \ac{KS}-test statistics for randomly selected 500 pairs of a \ac{mSTFT} from the \textit{Scattered-light} set and a \ac{mSTFT} from the \textit{quiet} set by taking the \ac{mSTFT} variations in both sets into account. As the null hypothesis in the one-sided \ac{KS} test, we consider \ac{mSTFT}-pixel values of a \textit{Scattered-light}-glitch data is lower than that of the \textit{quiet} data because we want to verify if the hypothesis that the upper-frequency region of the \textit{Scattered light} set has no excess power above Gaussian fluctuations can be rejected. Because the \ac{KS}-test statistics calculated above contain the variation of \ac{mSTFT} in both sets, we also calculate one-sided \ac{KS}-test statistics for randomly selected 500 pairs of two \acp{mSTFT} from the \textit{quiet} set. The right panel in Fig.\ \ref{fig:dist_ks} shows distributions of \ac{KS}-statistics of these pairs.

We then perform a one-sided \ac{KS} test for the above two distributions of \ac{KS} statistics. We find that the {\it p}-value of the test to be 0.099, which is not confident enough to reject the hypothesis that the upper-frequency region of the data containing \textit{Scattered light} glitches has no excess power. 

As a supplementary study, we perform the same procedure for the frequency region below 100 Hz. The left panel in Fig.\ \ref{fig:dist_ks} shows distributions of \ac{KS} statistics calculated with \ac{mSTFT} pairs in the frequency region below 100 Hz. We find the {\it p}-value of the test to be $2.4\times 10^{-141}$ so that \textit{Scattered light} glitches have excess power below 100 Hz.

To robustly verify that the null hypothesis that the \ac{mSTFT} of the data containing \textit{Scattered light} glitches has no excess power in the frequency region above 100 Hz, can not be rejected, we vary the number of pairs and random pair-selection realizations for $p$-values. 

\begin{figure}[!ht]
  \begin{center}
   \includegraphics[width=1\textwidth]{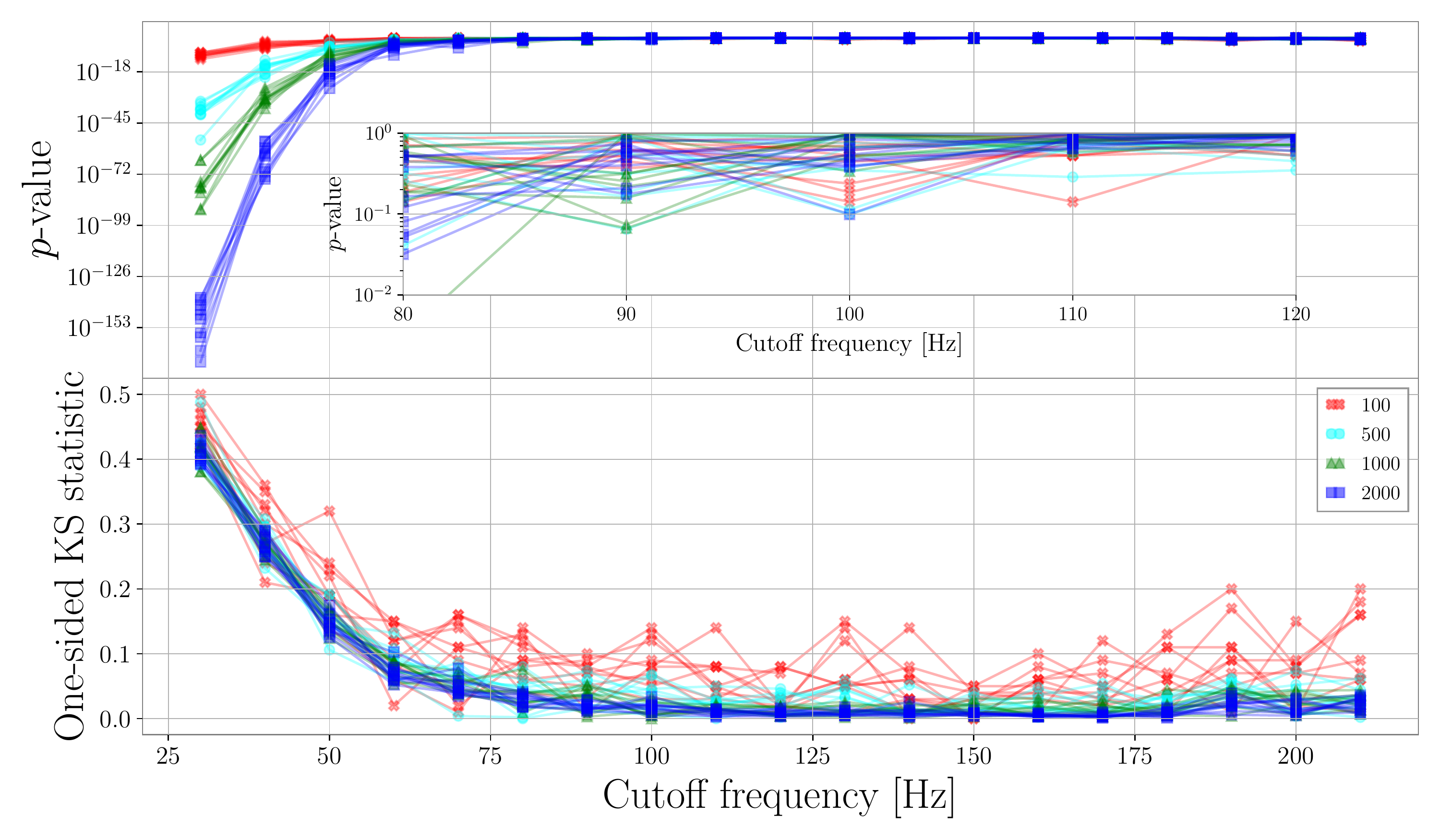}
  \end{center}
  \caption[Variations of $p$-values and one-sided \ac{KS} statistics for the \acp{mSTFT} above a given cutoff frequency.]%
  {The variations of $p$-values (top) and one-sided \ac{KS} statistics (bottom) for the \acp{mSTFT} above a given cutoff frequency. The one-sided \ac{KS}-statistics in this figure are the statistics calculated using the two \ac{KS}-statistic distributions from the {\it Scattered-light}-{\it quiet} pair and the {\it quiet}-{\it quiet} pair. Note the $y$-axis in the figure differs from the $x$-axis in Fig.\ \ref{fig:dist_ks}. Markers denote $N$ number of pairs to perform the \ac{KS}-test. Different curves with the same marker denote different realizations in the $N$-pair selection.}
\label{fig:variation_pvalue}
\end{figure}

Similar to the process mentioned above, we randomly select $N$ pairs of a \ac{mSTFT} from the {\it Scattered-light} set and a \ac{mSTFT} from the {\it quiet} set as well as $N$ pairs of two \acp{mSTFT} from the {\it quiet} set, subsequently calculate one-sided \ac{KS} statistics for paired \acp{mSTFT} in the frequency region above $x$ Hz for both {\it Scattered-light}-quiet and quiet-quiet pair sets. Then, we perform the one-sided \ac{KS} test for the two \ac{KS}-statistic distributions from two pair-sets. We vary values of $N$ between 100 and 2000 and $x$ between 30 Hz and 210 Hz to show the variation of $p$-values due to selected numbers of pairs $N$ and cutoff frequencies $x$. Also, to see the effect of the pair-selection realizations on $p$-values, we repeat 10 times each test for a given value of $N$ and $x$. Figure \ref{fig:variation_pvalue} shows the variations of $p$-values and the corresponding \ac{KS} statistics. The $p$-values with cutoff frequencies below 75 Hz significantly smaller, indicating high confidence in rejecting the null hypothesis. For cutoff frequencies below 75 Hz, larger values of $N$ corresponds to smaller $p$-values because the sample errors are smaller for a set with larger samples even if statistics with larger $N$ are slightly smaller than values with smaller $N$. The $p$-values with a cutoff frequency above 100 Hz are comparable, meaning that no excess power is observed in the frequency region above 100 Hz irrespective of values of $N$ and the pair-selection realizations.

\section{Peak time of glitches} \label{apx:peak_time}

To determine choices of splitting time-frequency regions in \acp{mSTFT} of the data containing glitches, one can use the peak time of glitches to identify if glitches are isolated in the time domain. 

\begin{figure}[!ht]
    \begin{minipage}{0.5\hsize}
  \begin{center}
   \includegraphics[width=1\linewidth]{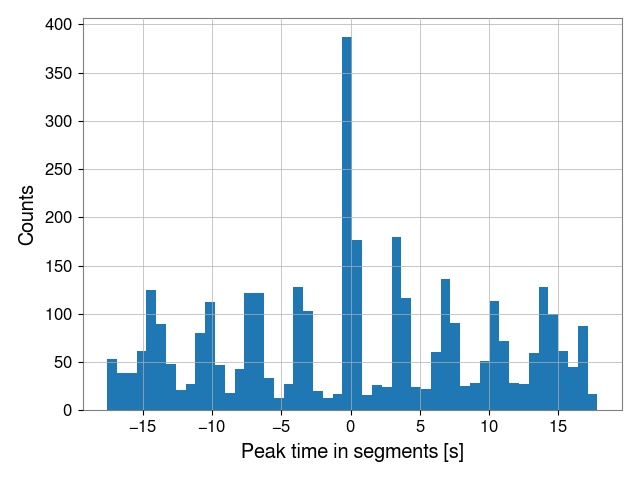}
  \end{center}
    \end{minipage}
    \begin{minipage}{0.5\hsize}
  \begin{center}
   \includegraphics[width=1\linewidth]{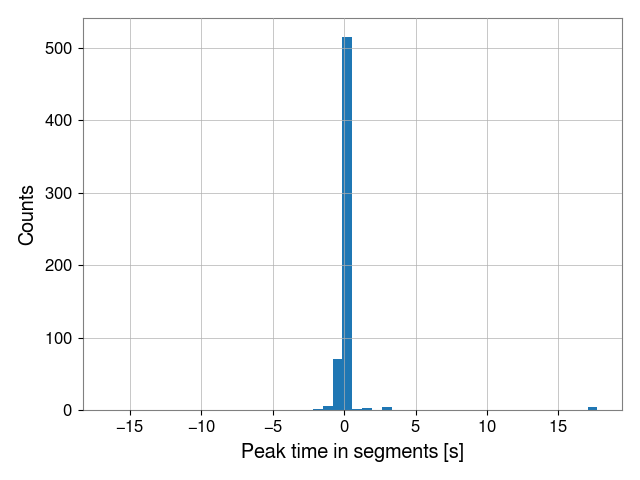}
  \end{center}
    \end{minipage}
    \caption[Peak time of glitches in 36-second time periods.]%
    {Peak time of {\it Scattered light} (left) and {\it Extremely loud} (right) glitches in 36-second time periods.}
\label{fig:peak_time}
\end{figure}

Figure \ref{fig:peak_time} shows peak times of {\it Scattered light} glitches and {\it Extremely loud} glitches in 36-seconds time periods. {\it Extremely loud} glitches are isolated in the time domain and their peak times are generally within $\pm 1$ around the \textsc{Omicron}-trigger \cite{ROBINET2020100620} times. {\it Scattered light} glitches are repeatedly present because this glitch class is generated due to swinging mirror motions caused by seismic activities. Therefore, to estimate the \ac{mSTFT} of the Gaussian noise, we use the frequency region above 100 Hz for {\it Scattered light} glitches and the time region outside of $\pm 2.5$ seconds around the trigger time for {\it Extremely loud} glitches.

\section{Pixel threshold to extract glitch waveforms} \label{apx:pixel_percentile_threhold}

\begin{figure}[!ht]
    \begin{minipage}{0.5\hsize}
  \begin{center}
   \includegraphics[width=1\linewidth]{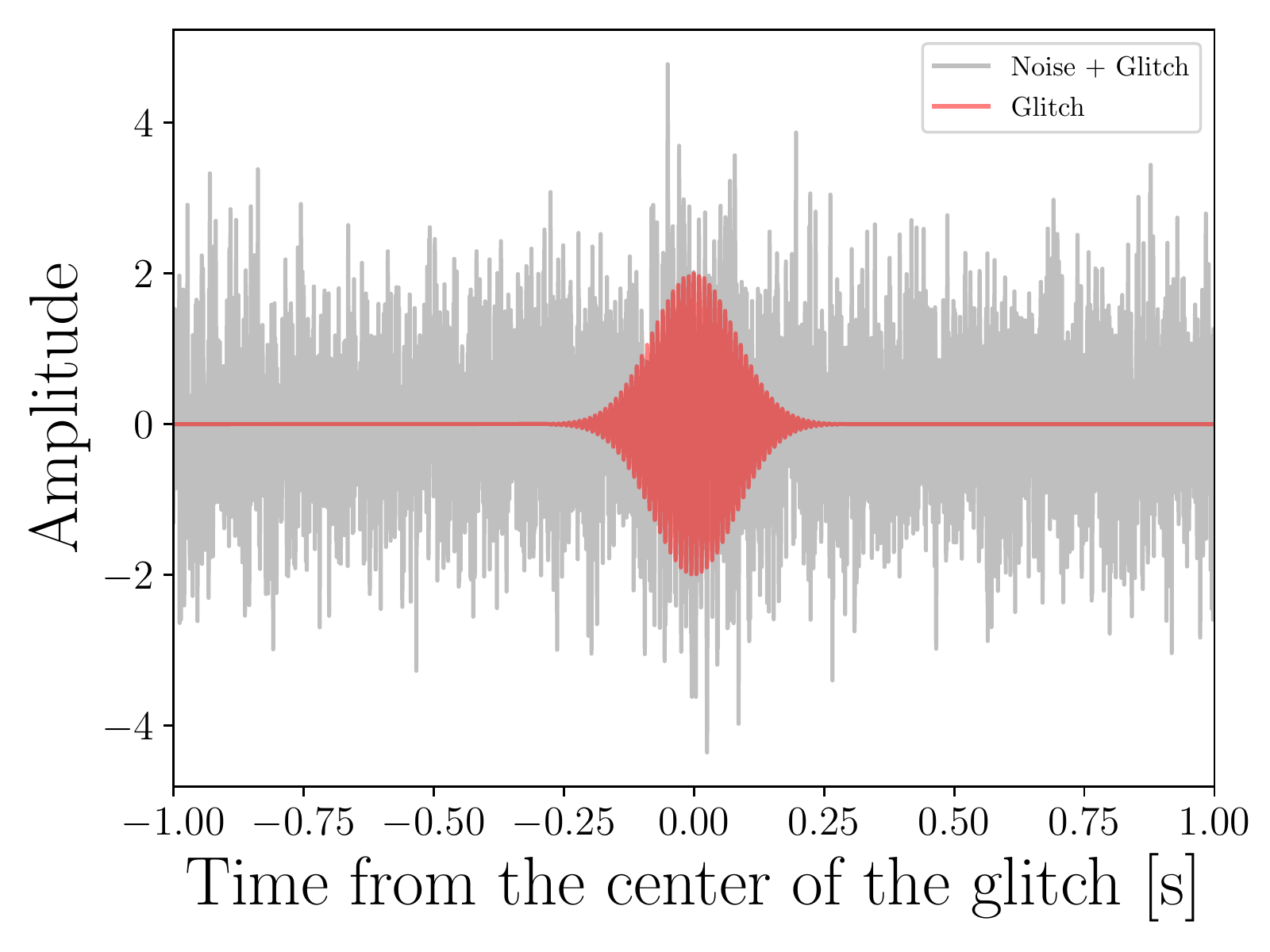}
  \end{center}
    \end{minipage}
    \begin{minipage}{0.5\hsize}
  \begin{center}
   \includegraphics[width=1\linewidth]{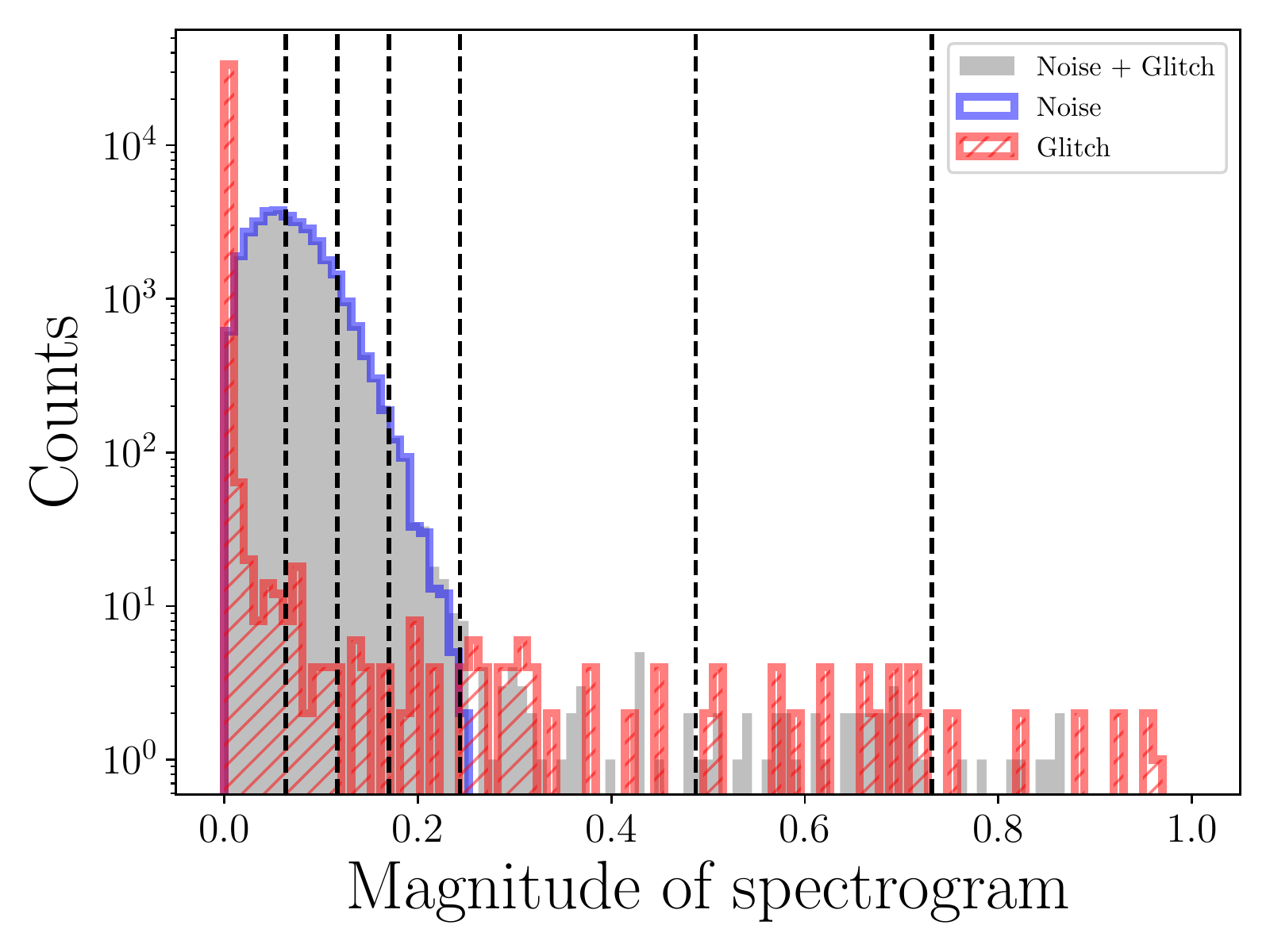}
  \end{center}
    \end{minipage}
    \caption[Hypothetical glitch injected into the simulated Gaussian noise.]%
    {Time series (left) of a hypothetical glitch injected into the simulated Gaussian noise and the histogram of \ac{mSTFT}-pixel values (right). The dashed vertical lines are 50, 90, 99, 100 percentiles of pixel values of the noise \ac{mSTFT}, and 100 percentile multiplied by 2 and 3, from left to right in the right panel.}
\label{fig:simulated_glitch}
\end{figure}

After determining the time-frequency region of the data to estimate the \ac{STFT} of the noise, we extract a glitch waveform from the \ac{STFT}. Because glitches generally can not be modeled, we choose a threshold for pixel values of the  \ac{STFT} to extract glitch waveforms. We keep pixels of the \ac{STFT} (hereafter called excess pixels) with their magnitude values above a threshold estimated from the \ac{STFT} representing the noise. The left panel of Fig.\ \ref{fig:simulated_glitch} shows a hypothetical glitch injected into the simulated Gaussian data. Its right panel shows the histograms of \acp{mSTFT} of the injected data, the noise only, and the glitch only. Smaller thresholds let excess pixels have a larger number of noise and glitch pixels while larger thresholds let excess pixels have a smaller number of the noise pixels but smaller glitch pixels.

To determine the best pixel threshold, we take \ac{mSTFT} pixels with their magnitude values below a threshold (hereafter called un-excess pixels) and then use the two-sided \ac{KS} test between the un-excess pixels and \ac{mSTFT} pixels of quiet times. If un-excess pixels are similar to quiet pixels, excess pixels tend to have the majority of glitch pixels and smaller numbers of noise pixels. To quantitatively determine the best threshold, we randomly select 200 pairs of un-excess-pixel \acp{mSTFT} with quiet \acp{mSTFT} and calculate two-sided \ac{KS} statistics for each pair. We use the two-sided \ac{KS} statistic because we want to see the similarity of two \ac{mSTFT} in a pair. Likewise, we randomly select 200 pairs of two quiet \acp{mSTFT} and calculate two-sided \ac{KS} statistics. We take the ratio of the \ac{KS} statistic averaged over the un-excess-quiet pairs to the \ac{KS} statistic averaged over the quiet-quiet pairs. 

\begin{figure}[!ht]
    \begin{minipage}{0.33\hsize}
  \begin{center}
   \includegraphics[width=\linewidth]{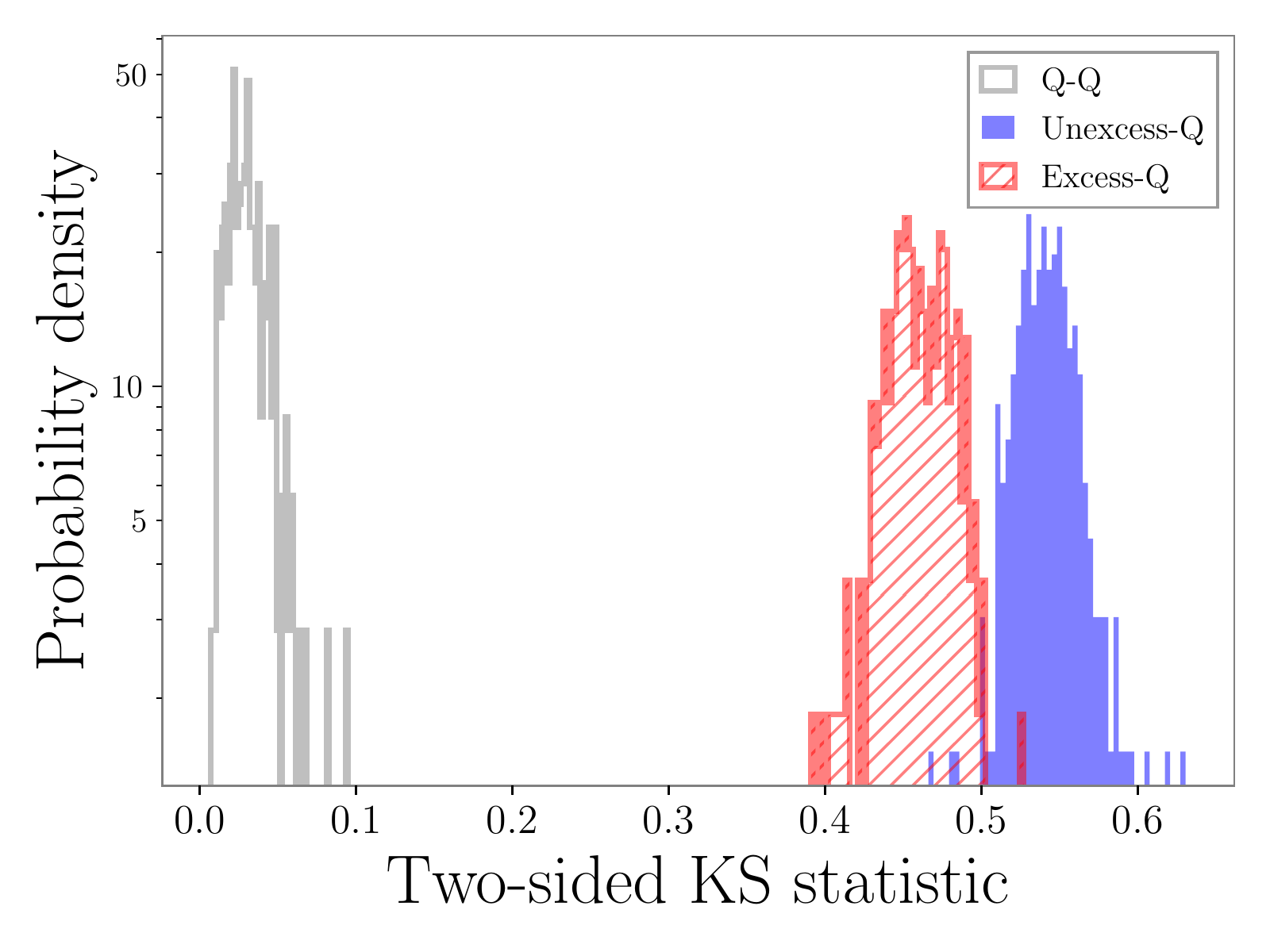}
  \end{center}
    \end{minipage}
    \begin{minipage}{0.33\hsize}
  \begin{center}
   \includegraphics[width=1\linewidth]{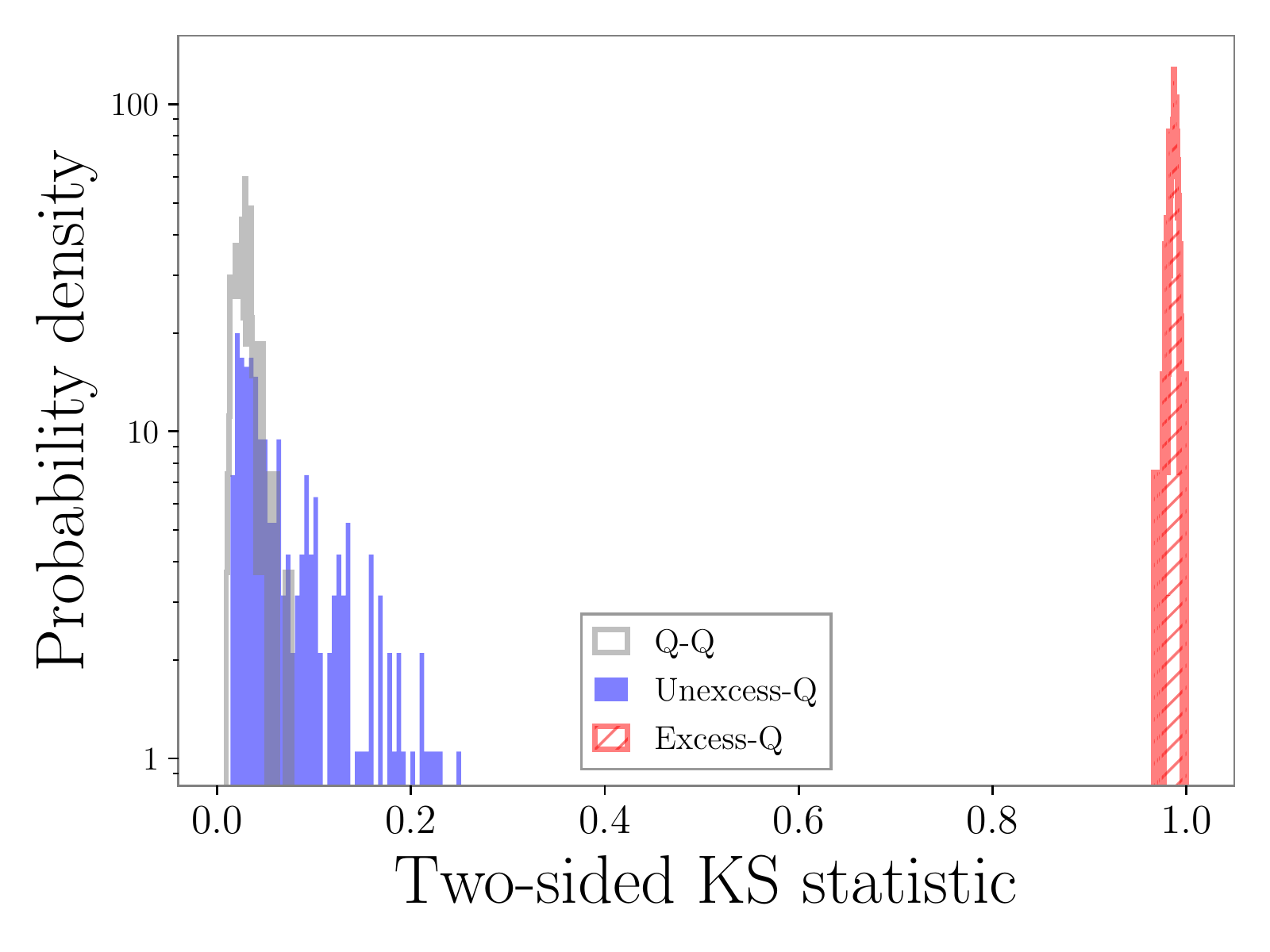}
  \end{center}
    \end{minipage}
    \begin{minipage}{0.33\hsize}
  \begin{center}
   \includegraphics[width=1\linewidth]{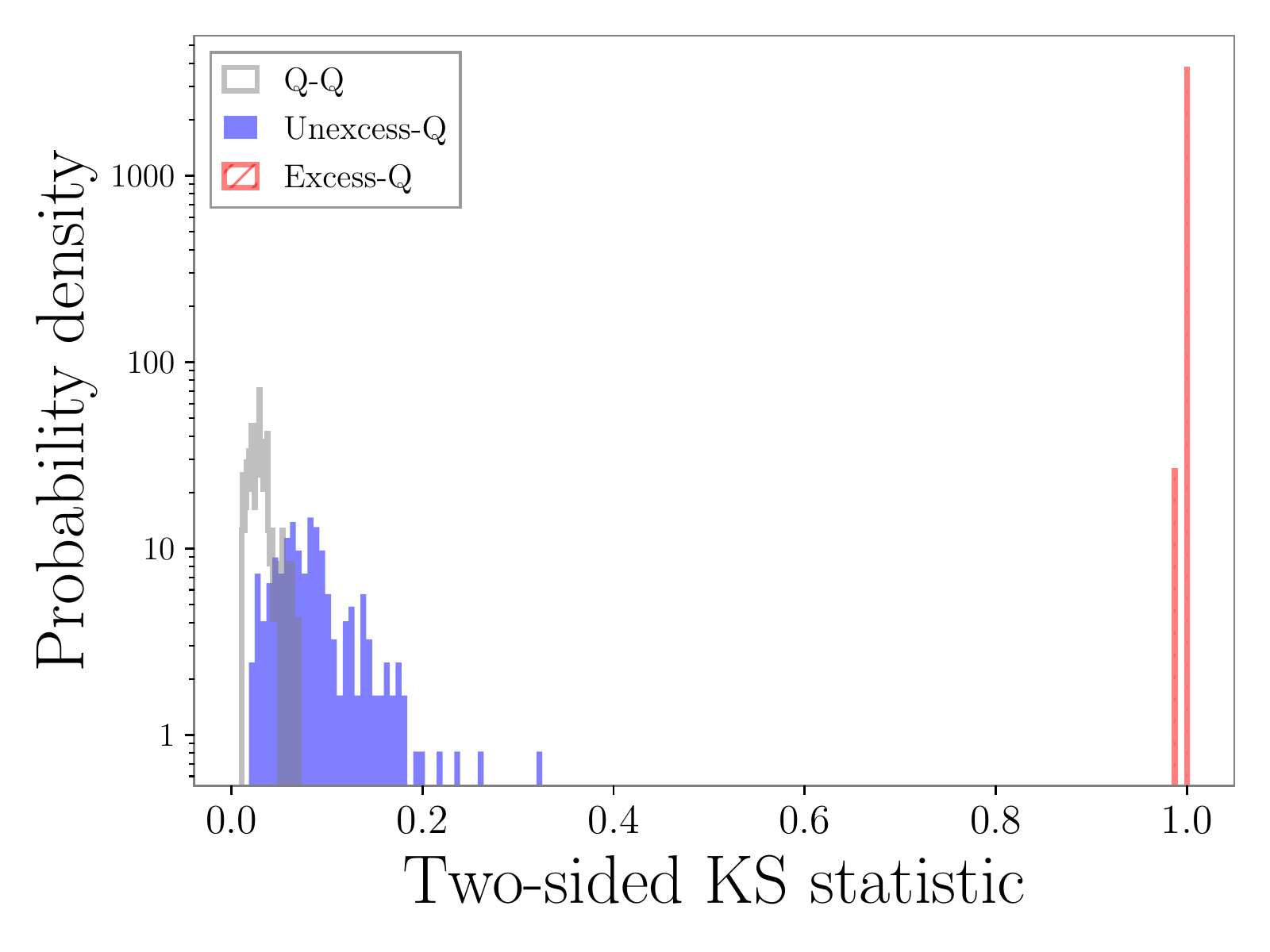}
  \end{center}
    \end{minipage}
      \begin{minipage}{0.33\hsize}
  \begin{center}
   \includegraphics[width=1\linewidth]{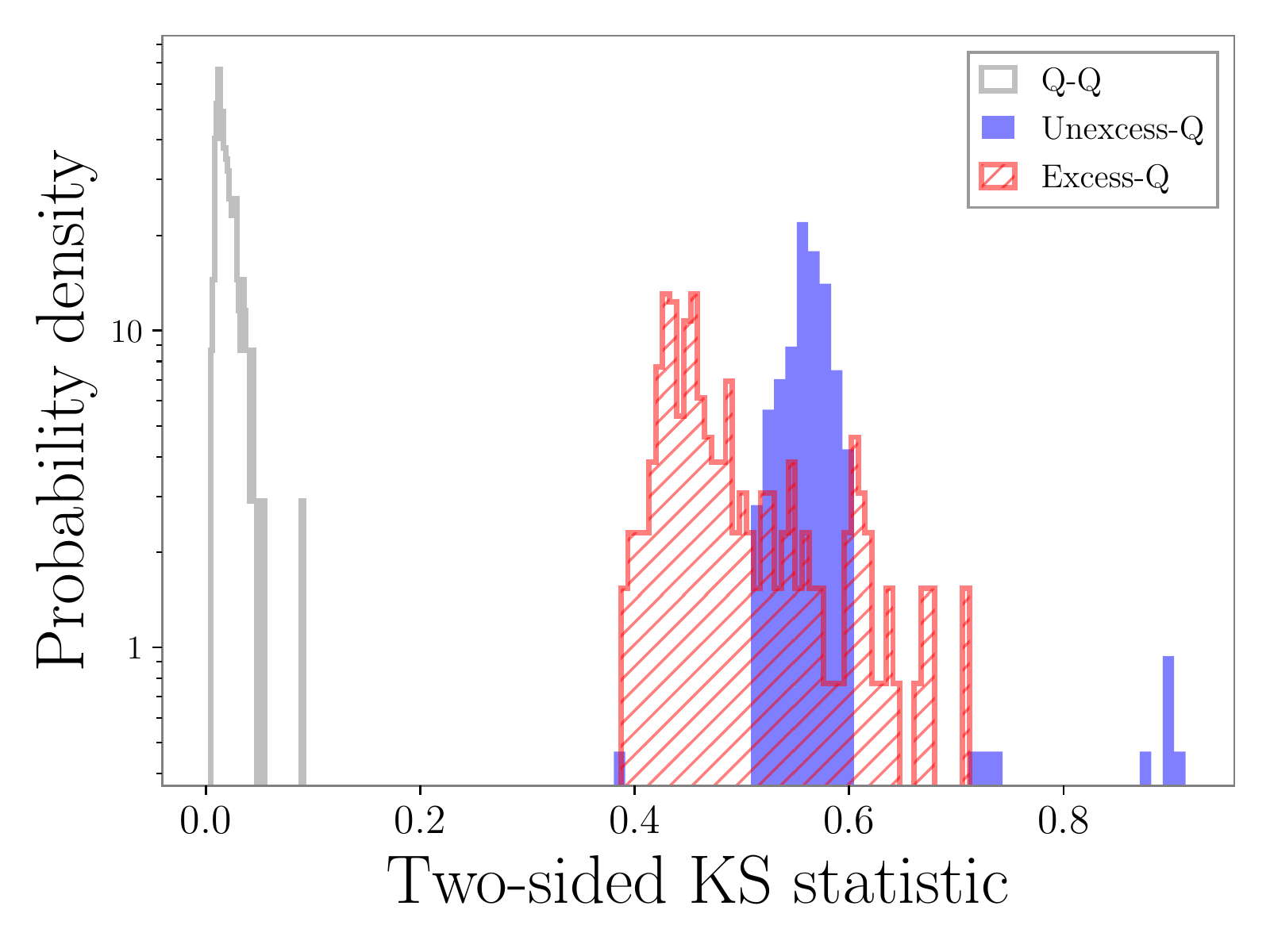}
  \end{center}
    \end{minipage}
      \begin{minipage}{0.33\hsize}
  \begin{center}
   \includegraphics[width=1\linewidth]{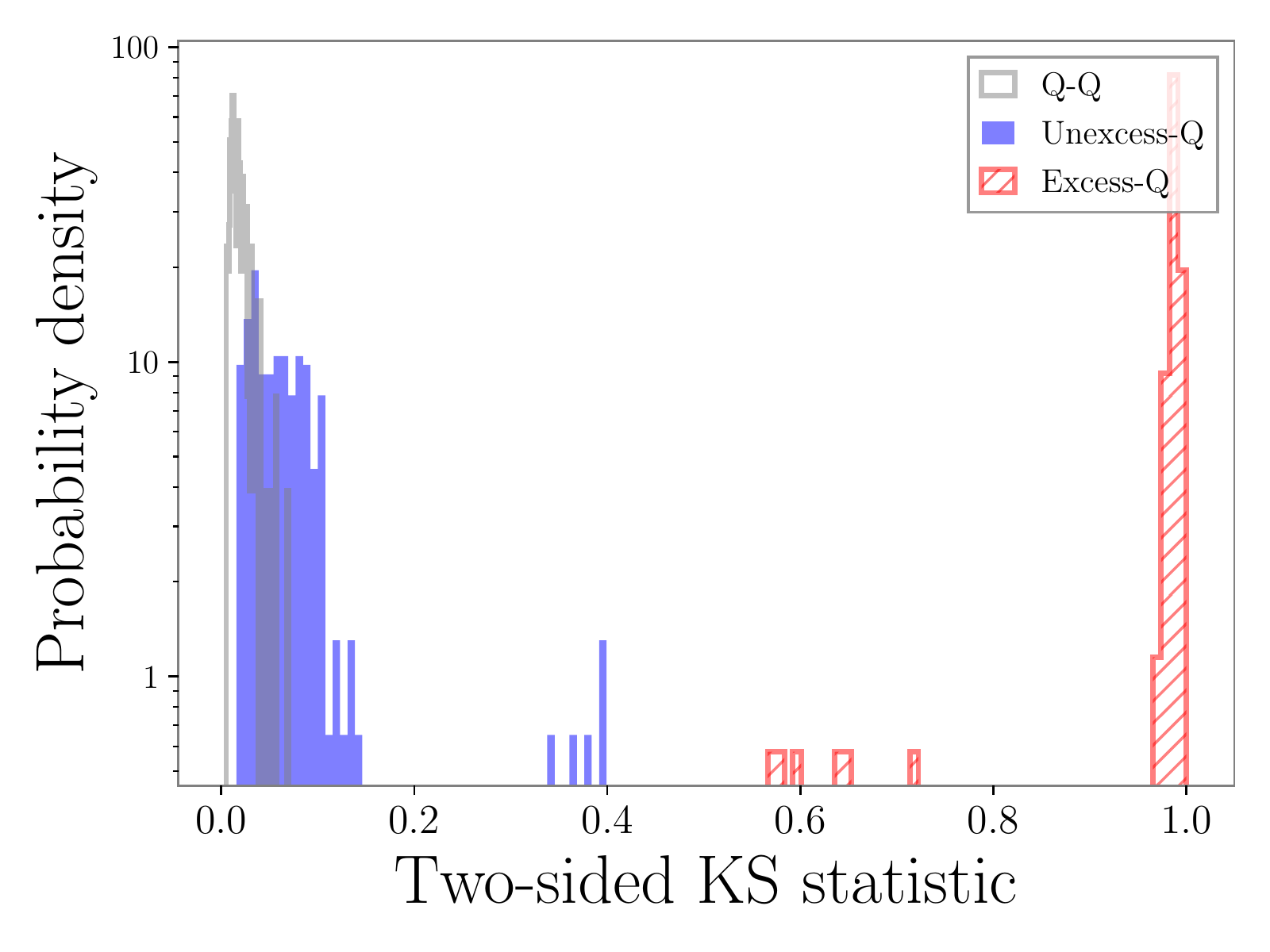}
  \end{center}
    \end{minipage}
      \begin{minipage}{0.33\hsize}
  \begin{center}
   \includegraphics[width=1\linewidth]{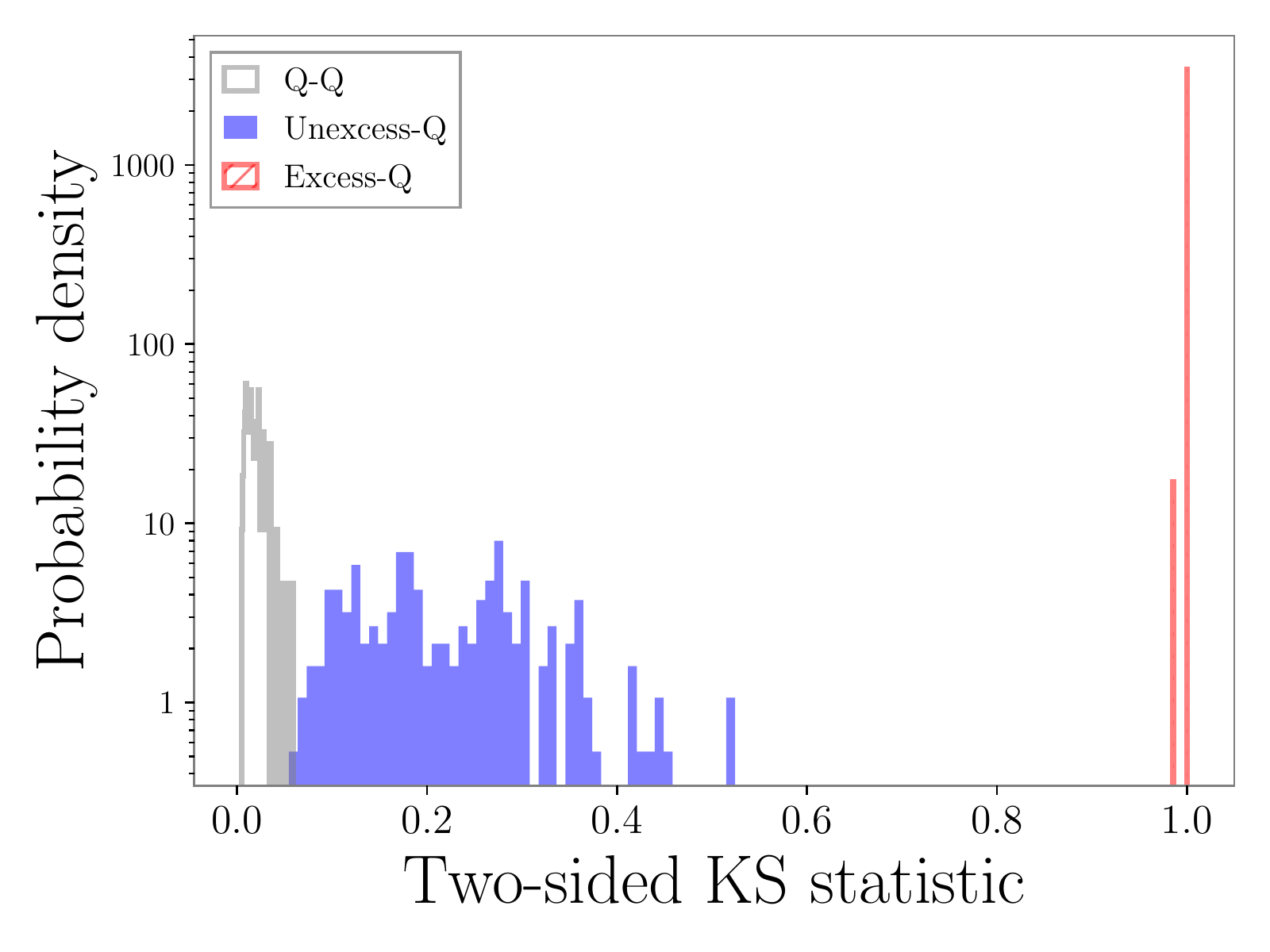}
  \end{center}
    \end{minipage}
    \caption[Histograms of two-sided \ac{KS} statistics of quiet-quiet, un-exess-quiet, excess-quiet pairs.]%
    {Histograms of two-sided \ac{KS} statistics of quiet-quiet, un-exess-quiet, excess-quiet pairs with 50 percentile (left), 99 percentile (middle), and 100 percentile multiplied by 5 (right) as pixel thresholds for {\it Scattered light} (top) and {\it Extremely loud} (bottom) glitches.}
\label{fig:hist_two_sided_ks}
\end{figure}

Figure \ref{fig:hist_two_sided_ks} shows histograms of \ac{KS} statistics of 200 quiet-quiet pairs, 200 un-excess-quiet pairs, and 200 excess-quiet pairs with different pixel thresholds. The ratio is closed to one when the un-excess pixels are similar to the quiet pixels, i.e., excess pixels contain the majority of glitch pixels and fewer noise-pixel. Whereas, larger values of the ratio imply that 1) only smaller amplitude noise pixels are contained in un-excess pixels, i.e., excess pixels contain a larger amount of noise pixels (corresponding smaller pixel thresholds), or 2) un-excess pixels have a higher number of glitch pixels, i.e., excess pixels have only a few glitch pixels (corresponding to larger pixel thresholds). We vary the pixel threshold from 50, 90, 99, 99.9, 100 percentiles of the \ac{mSTFT} in the time-frequency region that is expected to contain no glitches (see the above sections). Also, we consider values of 100-percentile multiplied by 2, 3 and, 5 as the threshold. 

Therefore, minimizing the ratio over pixel thresholds allows us to find the best pixel threshold. For {\it Scattered light} glitches, the ratio reaches the lowest values $\sim 2.1\sim2.2$ with thresholds of 99, 99.9, and 100 percentiles. For {\it Extremely loud} glitches, the ratio has the minimum value of $\sim3.9$ with 99 percentile. {\it Extremely loud} glitches have extremely high \ac{SNR} so that the ratio is more sensitive to the pixel threshold. According to the above study, we set the best threshold to be 99 percentile for both classes of glitches.               

\begin{figure}[!ht]
  \begin{center}
   \includegraphics[width=\textwidth]{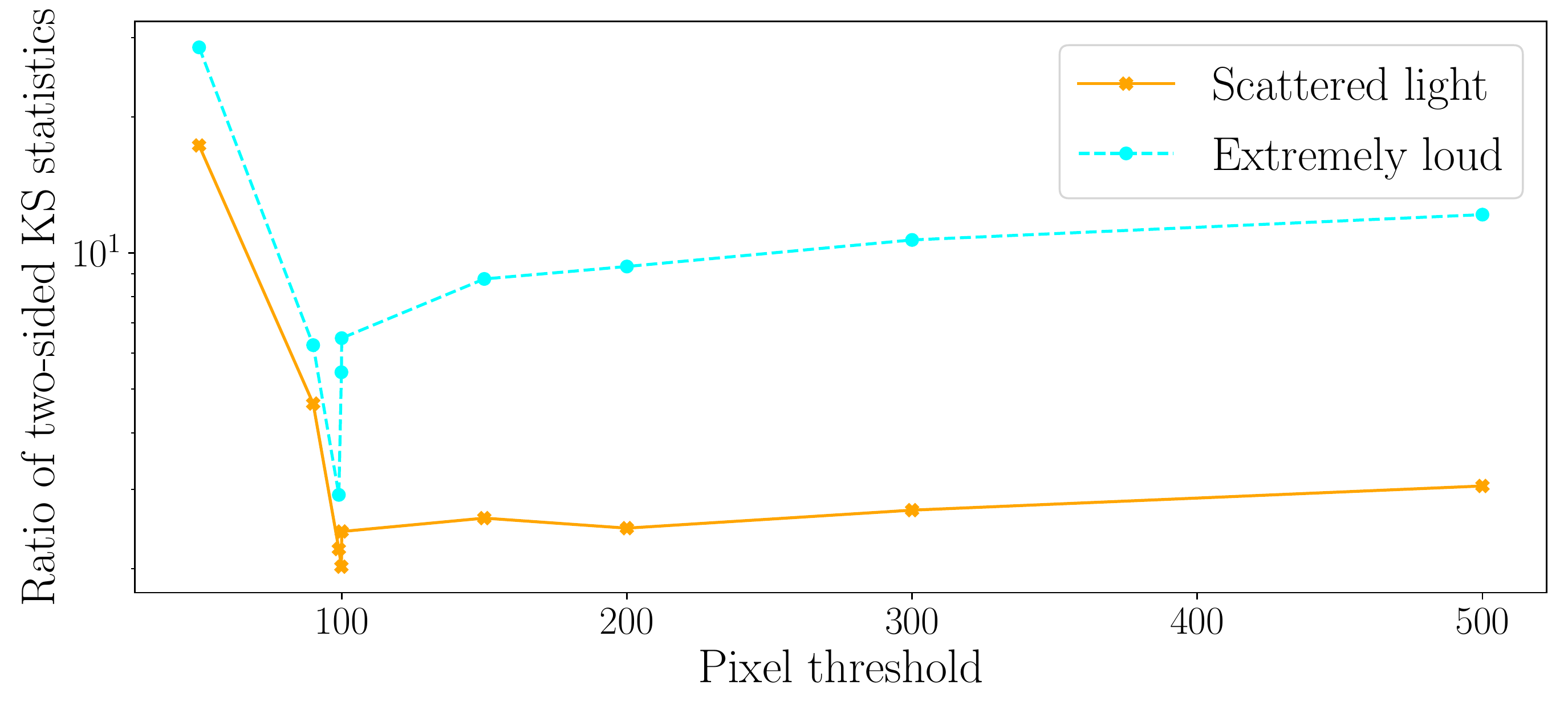}
  \end{center}
  \caption[The ratio of the unexcess-quiet pair \ac{KS} statistics to the quiet-quiet pair \ac{KS} statistic.]%
  {The ratio of the unexcess-quiet pair \ac{KS} statistic to the quiet-quiet pair \ac{KS} statistic. Pixels thresholds below 100 denote percentiles of \ac{mSTFT} in the time-frequency region that is expected to contain no glitches. Pixel thresholds above 100 denote 100 percentile multiplied by 2, 3, or 5.}
\label{fig:ratio_ks}
\end{figure} 

\section{Fast Griffin-Lim transform} \label{apx:FGL_transform}

\section{Iteration in Fast Griffin-Lim Transformation} \label{apx:iter_FGL_transform}

\begin{figure}[!ht]
    \begin{minipage}{0.5\hsize}
  \begin{center}
   \includegraphics[width=1\linewidth]{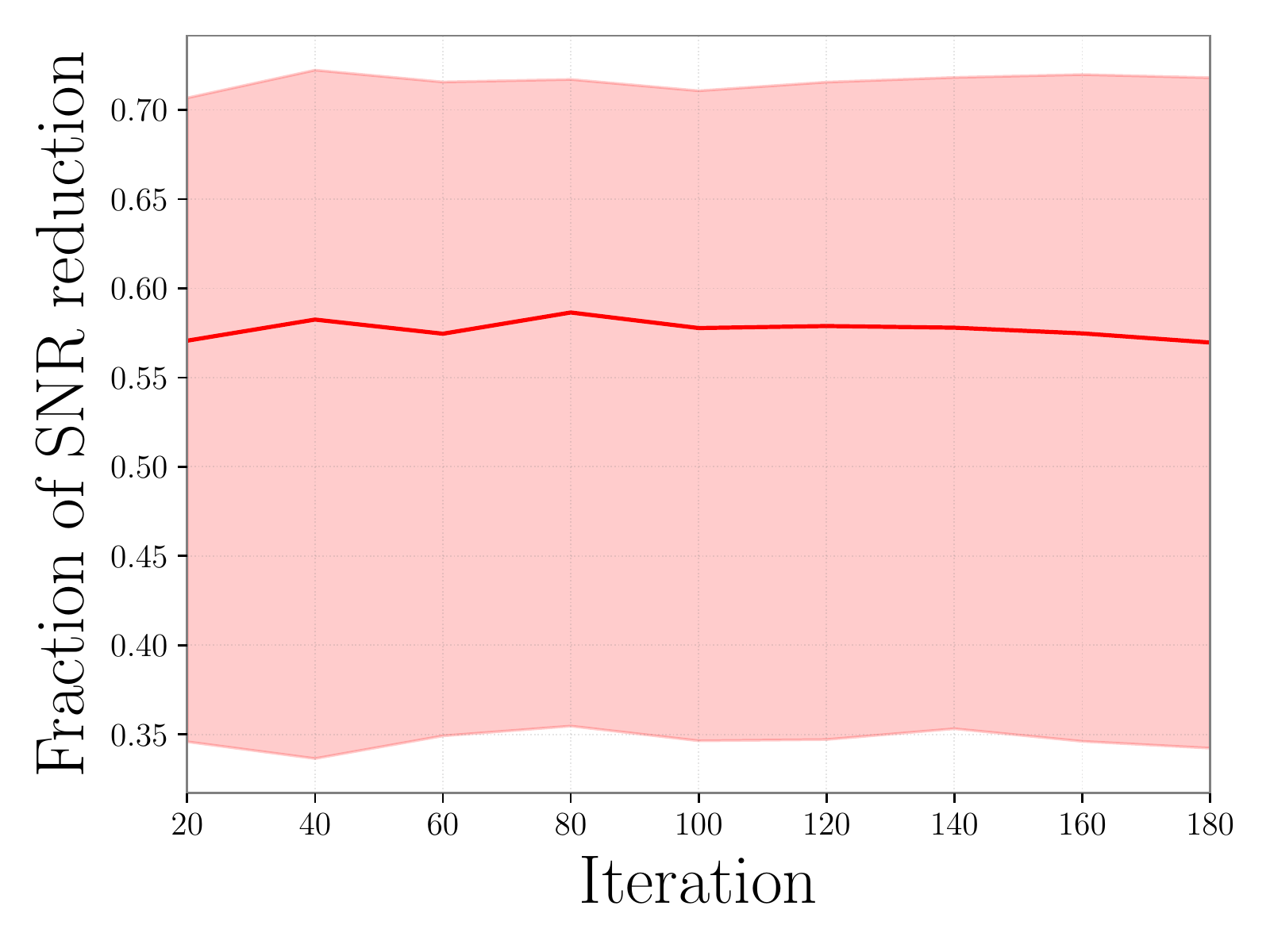}
  \end{center}
    \end{minipage}
    \begin{minipage}{0.5\hsize}
  \begin{center}
   \includegraphics[width=1\linewidth]{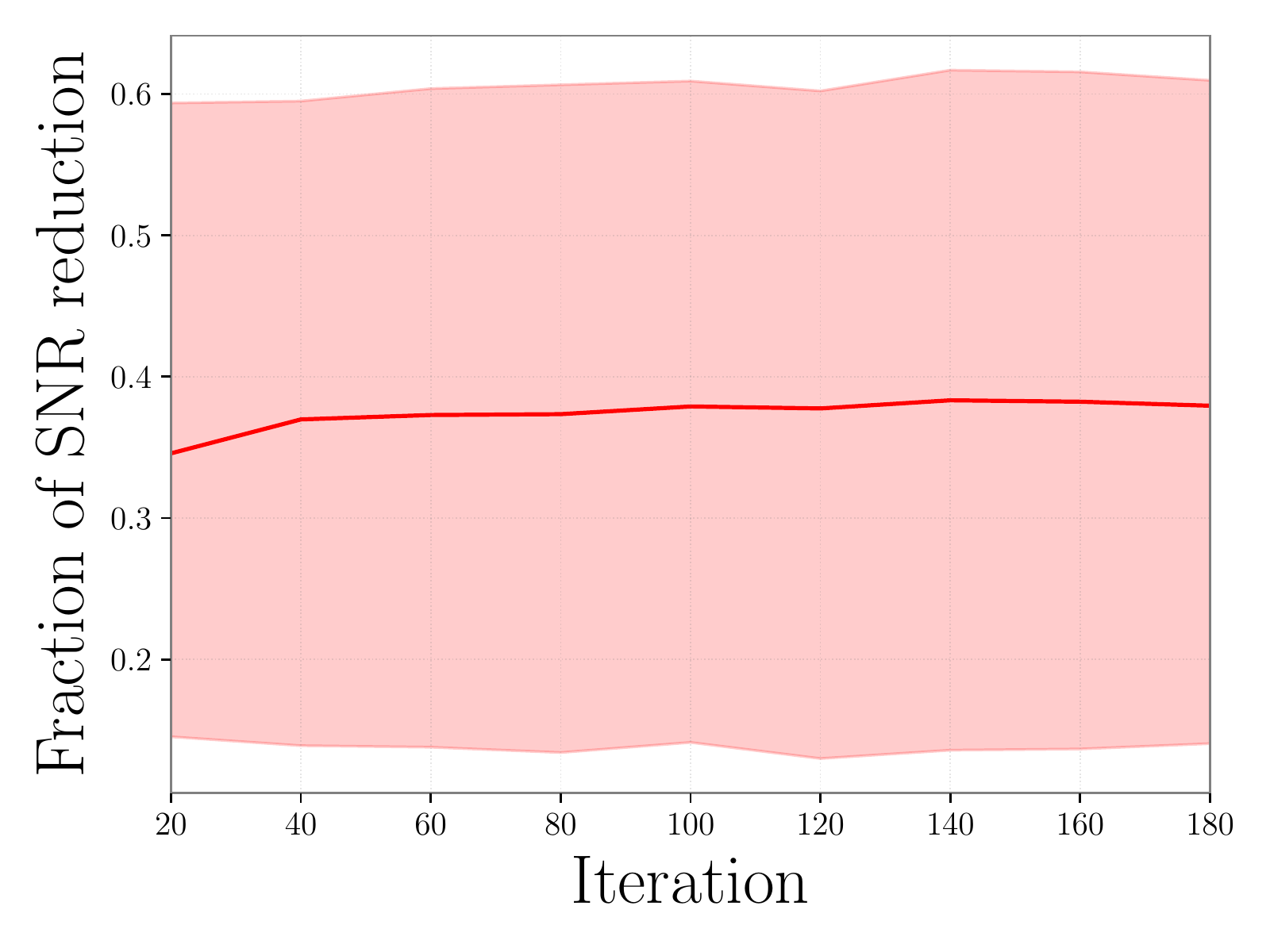}
  \end{center}
    \end{minipage}
    \caption[Fraction of \ac{SNR} reduction versus numbers of \ac{FGL} iteration.]%
    {Fraction of \ac{SNR} reduction versus the number of \ac{FGL} iterations for {\it Scattered light} (left) and {\it Extremely loud} glitches.  The solid curves denote the median values and the shaded areas denote 1-$\sigma$ percentiles.}
\label{fig:fnr_FGLiteration}
\end{figure}

Figure \ref{fig:fnr_FGLiteration} shows the \ac{FNR} as a function of \ac{FGL} iteration with $\alpha=0.99$ (recommended value in \cite{6701851}). Values \ac{FNR} are comparable after 20 iterations. We choose 32 iterations (default value in \cite{9174990}) and $\alpha=0.99$ in this paper.

The data used in this paper does not guarantee the theoretical convergence inherited from the algorithm of the \ac{FGL} transformation and the convergent point to be the global minimum inherited from the \ac{GL} transformation. We address the accuracy of the \ac{FGL} transformation on the data used in this paper in the next section.

\subsection{Accuracy of Fast Griffin-Lim transform}\label{apx:accurary_GL}

To quantify the accuracy of \ac{FGL} transform, we calculate {\it match} $M$ \cite{alex_nitz_2020_3904502} between extracted glitch waveforms $g(t)$ and the \ac{FGL} transformed time series $\hat{g}(t)$ from the \ac{mSTFT} of the same extracted glitch waveforms:
\begin{equation} \label{eq:match}
    M = \frac{( \tilde{g}_i| \tilde{\hat{g}})}{\sqrt{ ( \tilde{g}| \tilde{\hat{g}}) (\tilde{h}|\tilde{h})}}\,,
\end{equation}
where $(\tilde{g}_i| \tilde{\hat{g}})$ is the noise-weighted inner product \cite{PhysRevD.46.5236, Brown:2004vh} defined as 
\begin{equation} \label{eq:innerproduct}
    (\tilde{a} | \tilde{b}) = \int_{-\infty}^{\infty}df\frac{\tilde{a}^*(f)\tilde{b}(f) + \tilde{a}(f)\tilde{b}^*(f)}{S_n(|f|)}  \,,
\end{equation}
where $S_n(|f|)$ is the one-sided noise \ac{PSD}$, \tilde{a}$ and $\tilde{b}$ are the Fourier transform of the time series $a(t)$ and $b(t)$, respectively. 
The \ac{FGL} transformed time series has a phase error. Also, we apply the phase correction before subtracting glitches from the data before subtracting glitches. we also calculate the match maximized over phase and time, defined as 
\begin{equation}\label{eq:match_max}
    M_{\rm max} = {\rm argmax}_{t_0}\frac{|(\tilde{a}, \tilde{b}e^{2\pi if t_0})_{\rm complex}|}{\sqrt{(\tilde{a}, \tilde{a})(\tilde{b}, \tilde{b})}}\,,
\end{equation}
where the complex inner product $(\tilde{a}, \tilde{b}e^{2\pi if t_0})_{\rm complex}$ is defined as 
\begin{equation}\label{eq:complex_inner_product}
   (\tilde{a}, \tilde{b}e^{2\pi if t_0})_{\rm complex} =  4 \int_{0}^{\infty} df \frac{\tilde{a}^*(f) \tilde{b}(f) e^{2\pi i ft_0}}{S_n(f)}\,.
\end{equation}
Values of $M$ range from -1 (fully anti-correlated) to 1 (perfect match). 

\begin{figure}[!ht]
    \begin{minipage}{0.5\hsize}
  \begin{center}
   \includegraphics[width=1\linewidth]{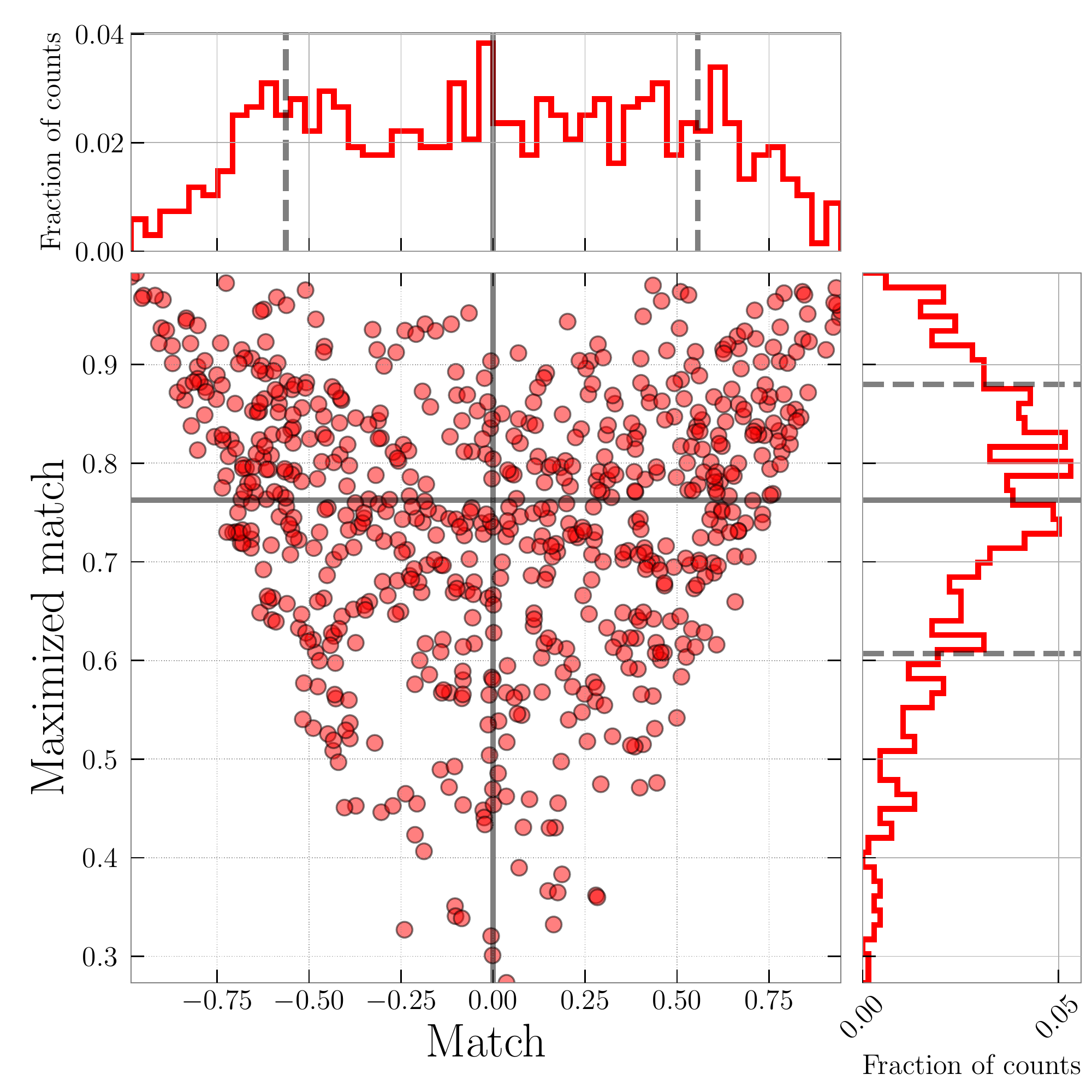}
  \end{center}
    \end{minipage}
    \begin{minipage}{0.5\hsize}
  \begin{center}
   \includegraphics[width=1\linewidth]{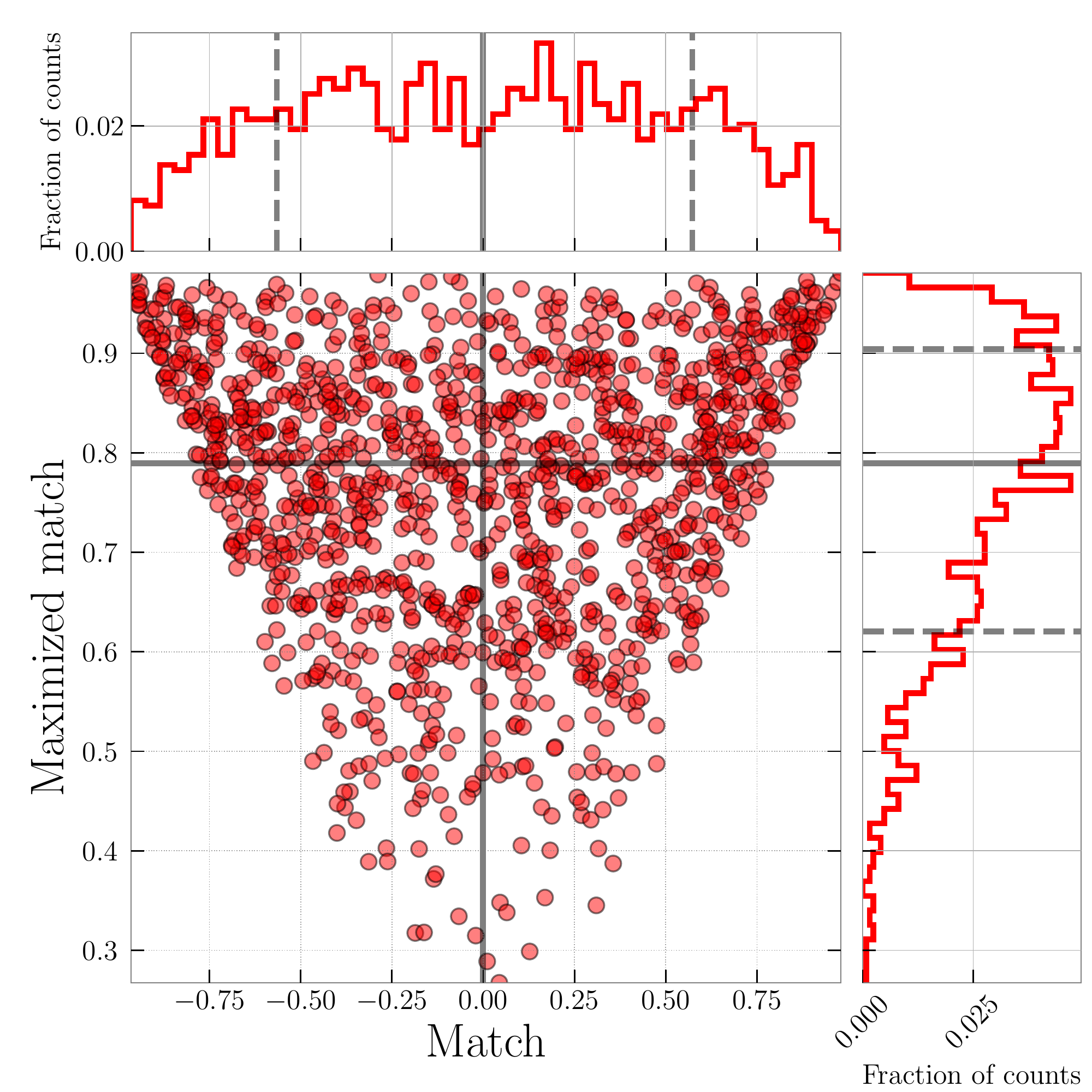}
  \end{center}
    \end{minipage}
    \caption[Distributions of the match and maximized match between extracted waveforms and \ac{FGL} transformed waveforms.]%
    {Distributions of the match $M$ and maximized match $M_{\rm max}$ between extracted waveforms and \ac{FGL} transformed waveforms for {\it Scattered light} (left) and {\it Extremely loud} (right) glitches. Black lines denote the median values and dashed lines denote 1-$\sigma$ percentiles.}
\label{fig:match_maxmatch_GL}
\end{figure}

We use extracted glitch waveforms in the testing set of {\it Scattered light} and {\it Extremely loud} glitches with sample sizes of 678 and 1233, respectively. Figure \ref{fig:match_maxmatch_GL} shows distributions of the match and maximized match. 

Samples with values of $M\sim \pm 1$ and of $M_{\rm max}\sim 1$ indicate that the \ac{FGL} transformed time series are similar to the original extracted-glitch waveforms with some degree of phase shifts. Samples with values $M\sim 0$ and $M_{\rm max}\sim 1$ indicate that the \ac{FGL} transformed time series have phase shifts such that they mismatch with the extracted glitch waveforms. Samples with values of $M\sim 0$ and $M_{\rm max} \sim 0$ indicate that the \ac{FGL} transformed time series mismatch with the original extracted-glitch waveforms with both phases and amplitudes. The median and 90 percentile of absolute values of the time shift $t_0$ which maximizes the match are $\sim0.02$ ($\sim 0.003$) and $\sim 0.06$ ($\sim 0.0034$) seconds for {\it Scattered light} ({\it Extremely loud}) glitches, respectively.    

\begin{figure}[!ht]
    \begin{minipage}{0.5\hsize}
  \begin{center}
   \includegraphics[width=1\linewidth]{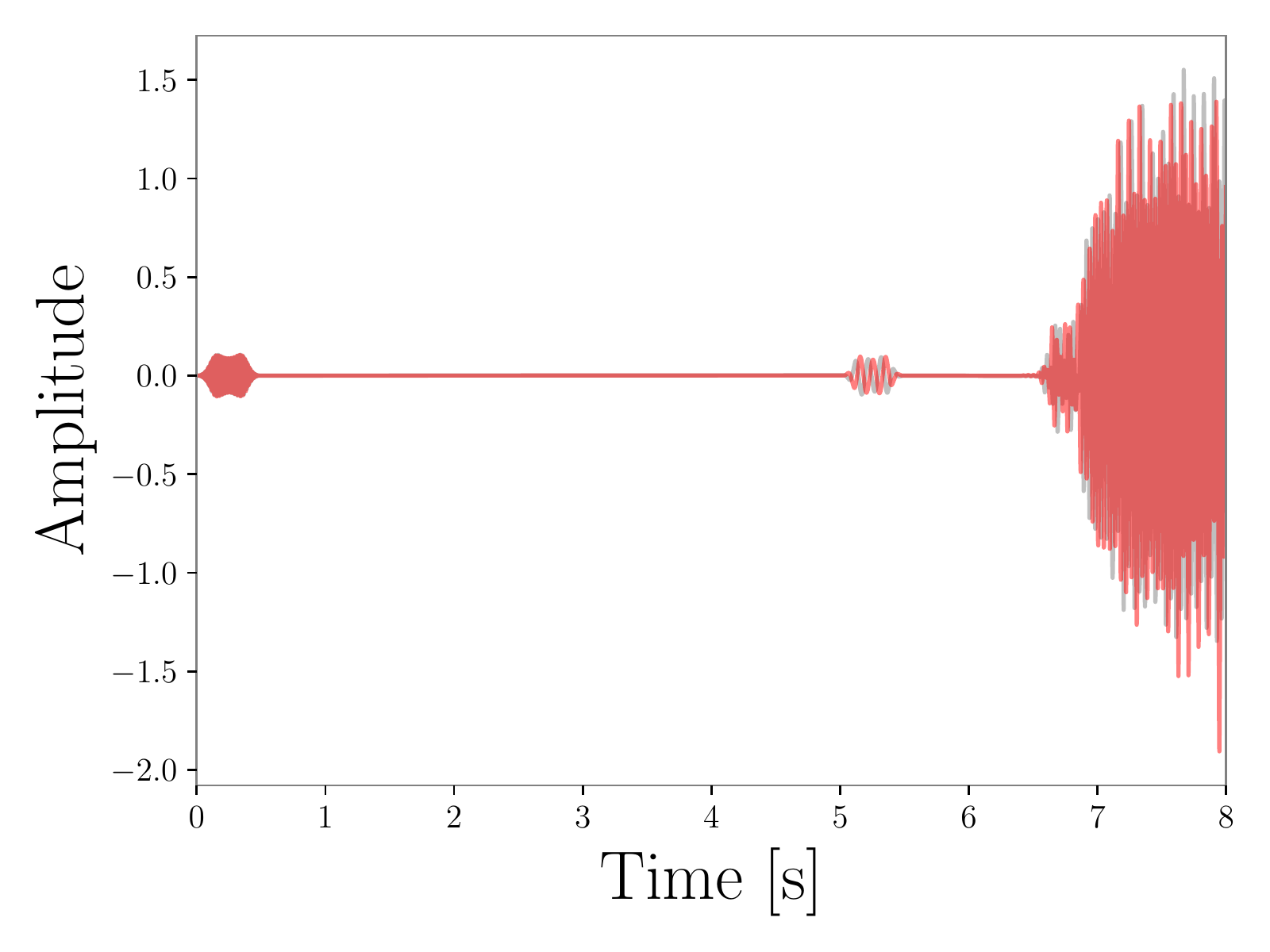}
  \end{center}
    \end{minipage}
    \begin{minipage}{0.5\hsize}
  \begin{center}
   \includegraphics[width=1\linewidth]{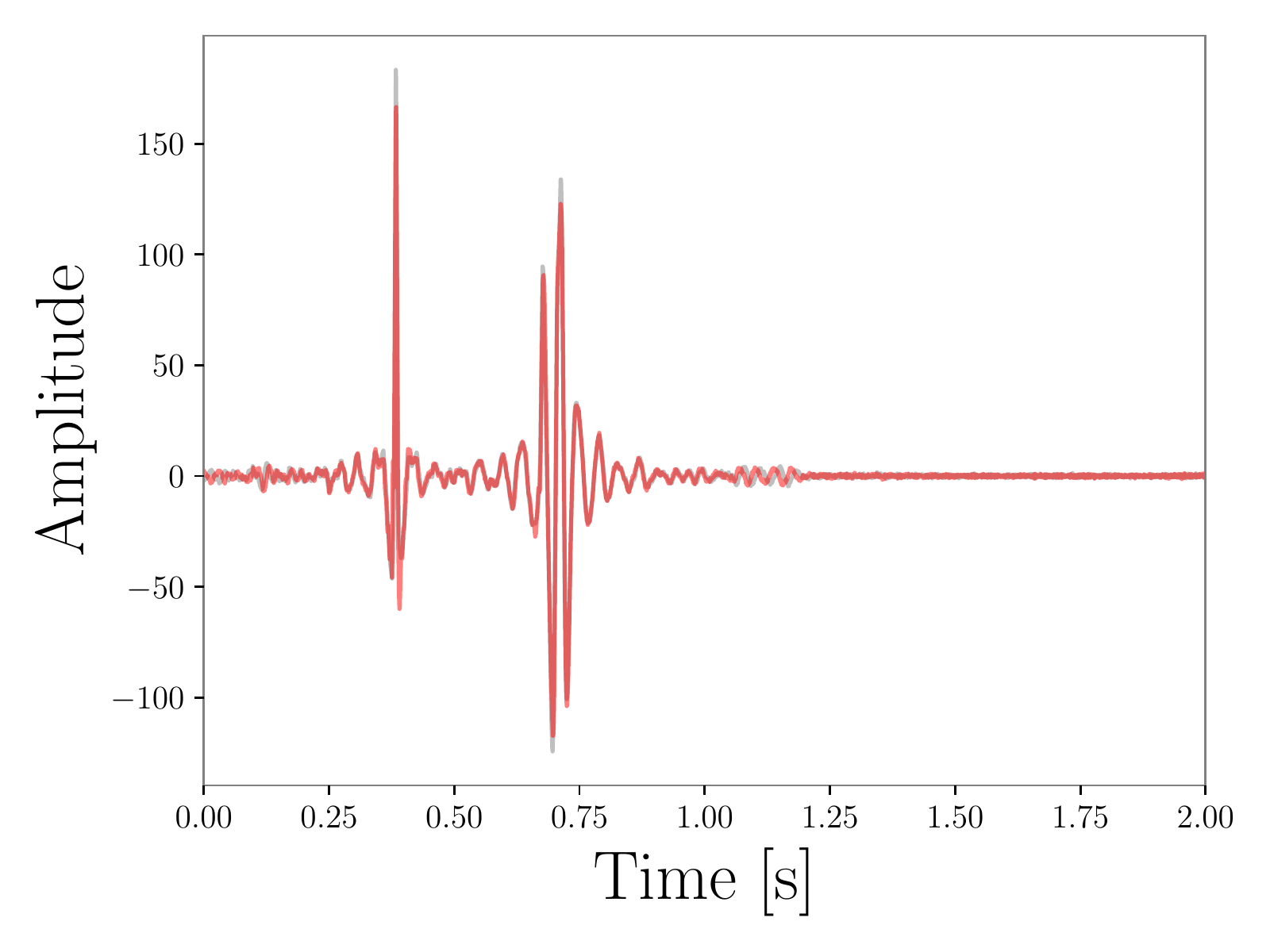}
  \end{center}
    \end{minipage}
    \begin{minipage}{0.5\hsize}
  \begin{center}
   \includegraphics[width=1\linewidth]{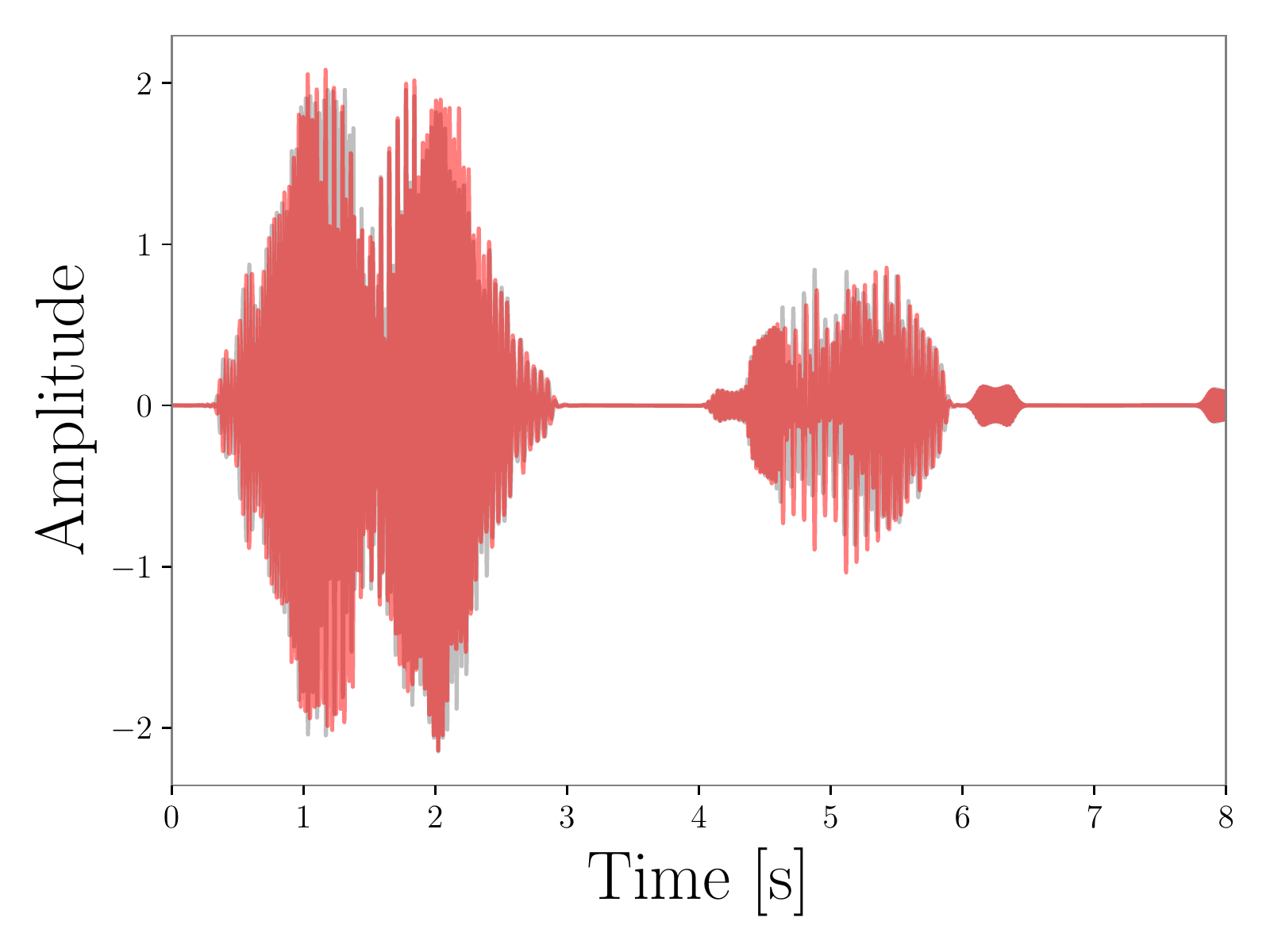}
  \end{center}
    \end{minipage}
    \begin{minipage}{0.5\hsize}
  \begin{center}
   \includegraphics[width=1\linewidth]{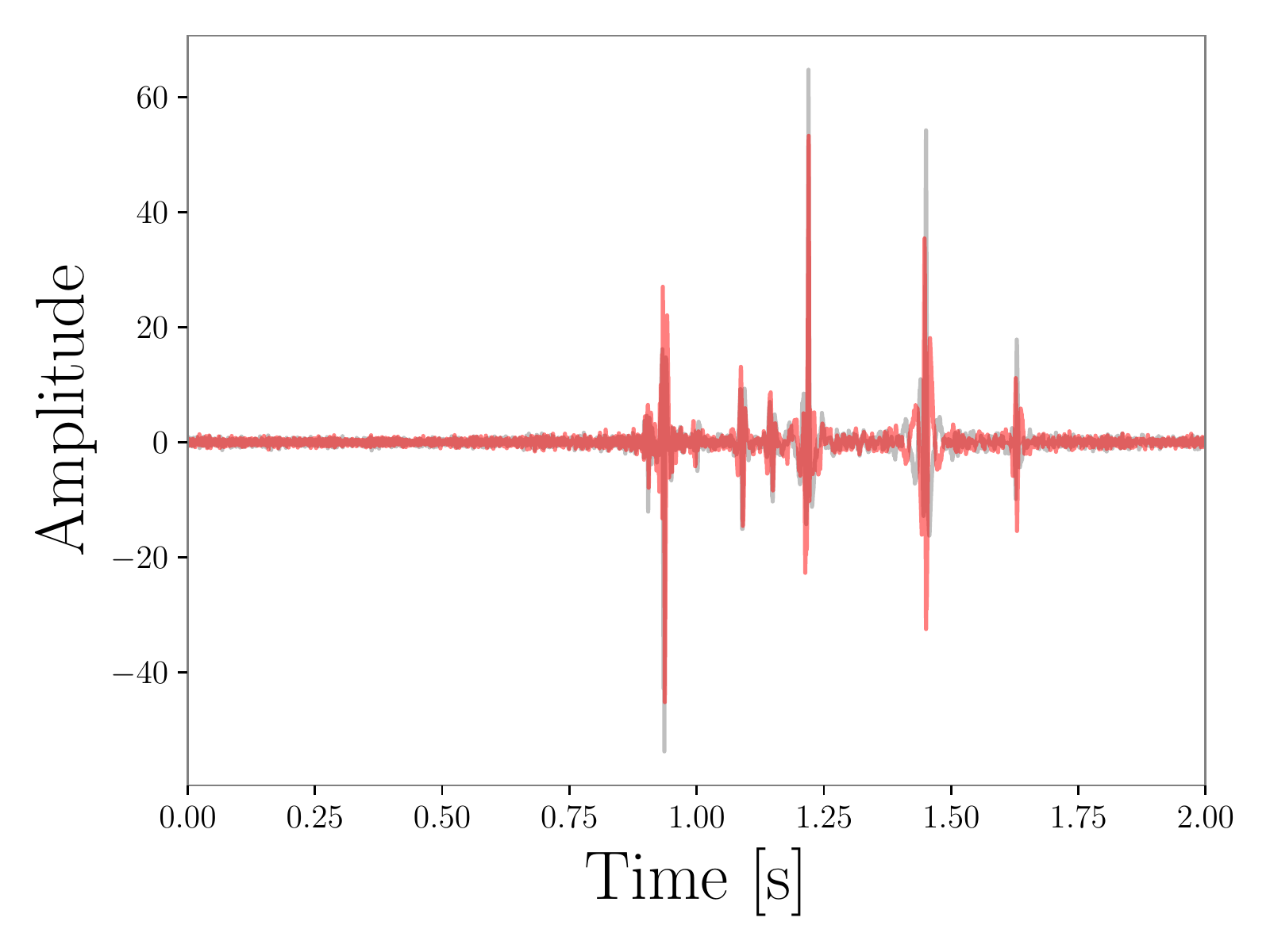}
  \end{center}
    \end{minipage}
    \caption[\ac{FGL} transformed time series and extracted glitch waveforms]%
    {Optimally (top) and least (bottom) accurate \ac{FGL} transformed time series (red) and extracted glitch waveforms (gray) of {\it Scattered light} (left) and {\it Extremely loud} (right) glitches. The optimal and least values of $M_{\rm max}$ are $\sim 0.27$ ($\sim 0.27$) and $\sim 0.99$ ($\sim 0.97$) for {\it Scattered light} ({\it Extremely loud}) glitches, respectively.}
\label{fig:best_worst_GL}
\end{figure}

Figure \ref{fig:best_worst_GL} shows the optimal and least accurate \ac{FGL} transformed time series. The optimal and least values of  $M_{\rm max}$ are $\sim 0.27$ ($\sim 0.27$) and $\sim 0.99$ ($\sim 0.97$) for {\it Scattered light} ({\it Extremely loud}) glitches, respectively. The \ac{FGL} transform seems to produce no significant deviations on the amplitude in the portion where the extracted glitch waveforms have amplitudes close to zero. Therefore, the amount of the mismatch seems to be due to the amplitude difference in the portion where extracted glitch waveforms have large amplitudes.

\begin{figure}[!ht]
    \begin{minipage}{0.5\hsize}
  \begin{center}
   \includegraphics[width=1\linewidth]{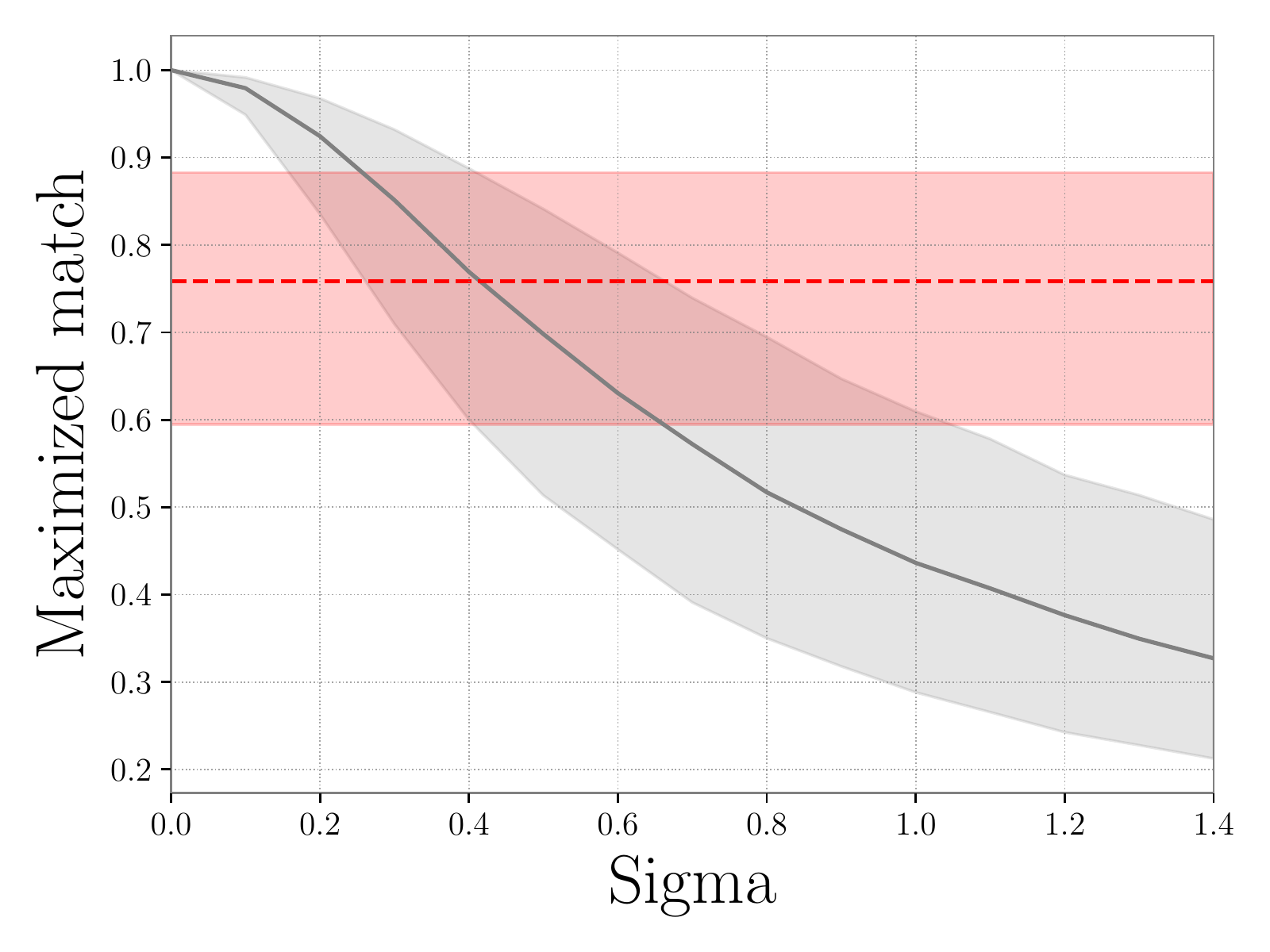}
  \end{center}
    \end{minipage}
    \begin{minipage}{0.5\hsize}
  \begin{center}
   \includegraphics[width=1\linewidth]{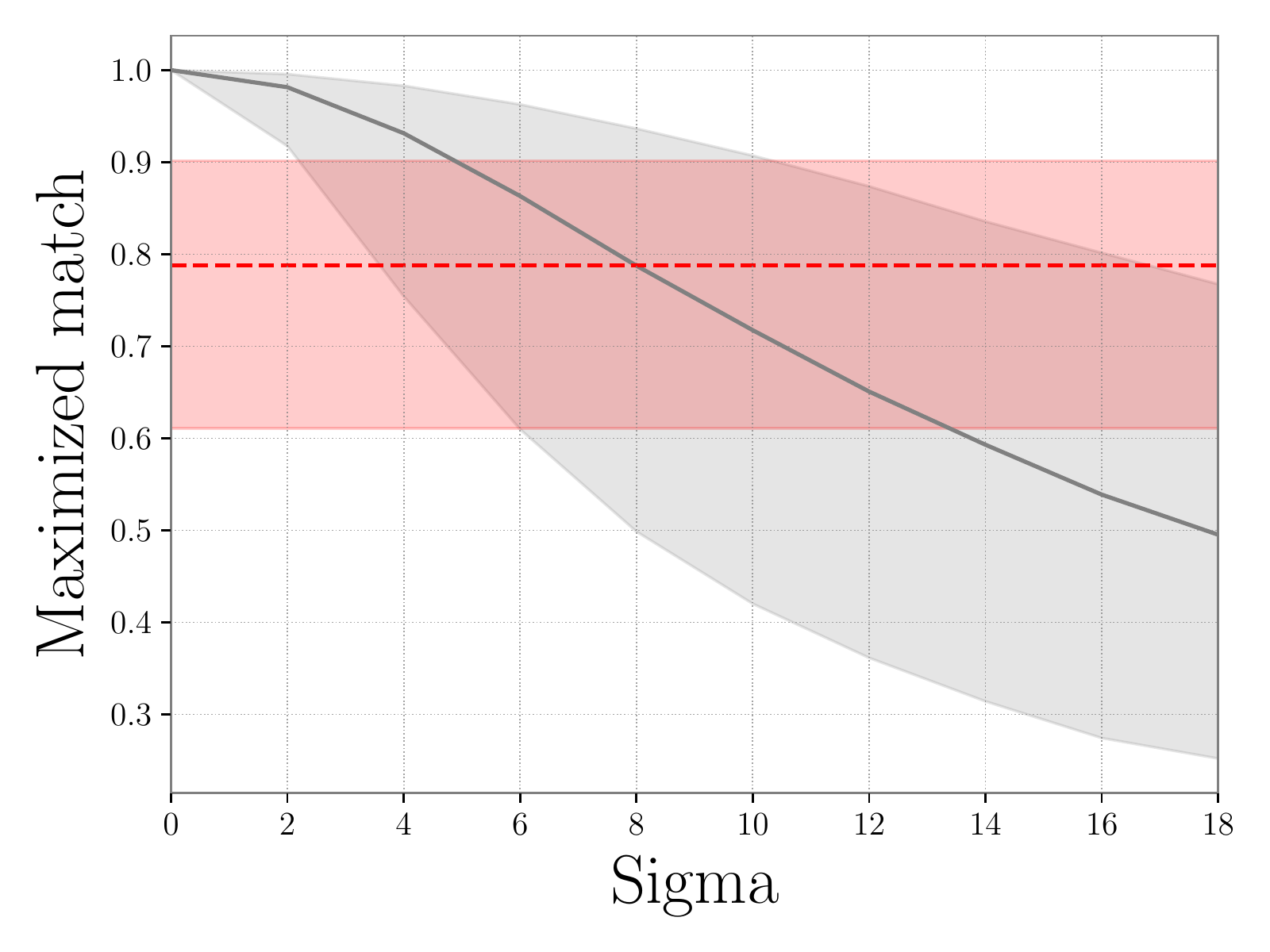}
  \end{center}
    \end{minipage}
    \caption[Maximized match between the extracted glitch waveforms and themselves injected into the Gaussian distributed noise.]%
    {Median of maximized matches (gray solid) between the extracted glitch waveforms and themselves injected into the Gaussian distributed noise with zero mean and a given standard deviation as well as the median of maximized matches (red dashed) between the \ac{FGL} transformed time series and the extracted glitch waveforms for {\it Scattered light} (left) {\it Extremely loud} (right) glitches. The shaded areas denote 1-$\sigma$ percentiles.}
\label{fig:amplitude_error}
\end{figure}

To understand the meaning of values of $M_{\rm max}$ between the \ac{FGL} transformed time series and the extracted glitch waveforms in terms of amplitude uncertainty of the \ac{FGL} transform, we calculate $M_{\rm max}$ between the extracted waveform and themselves injected into the Gaussian noise with zero mean and a given standard deviation. Assuming the \ac{FGL} transformation provides the amplitude error following a Gaussian distribution with zero mean throughout time series, we can find the standard deviation of the Gaussian noise corresponding to the amplitude uncertainty of the \ac{FGL} transform. We find that {\it Scattered light} and {\it Extremely loud} glitches have the amplitude uncertainties equivalent to the $\sim 0.4\sigma$ and $\sim 8 \sigma$ noise, respectively, as shown in Fig.\ \ref{fig:amplitude_error}. As discussed above, the \ac{FGL} transformation produces no significant amplitude errors in the portion where the extracted glitch waveforms have amplitudes close to zero. Therefore, the above estimates using the noise with a given $\sigma$ has a bias for waveforms with larger amplitudes in limited time portions such as {\it Extremely loud} glitches. One of the possible approaches to correct the phase and amplitudes of the \ac{FGL} transformed time series is to split the time series into small segments and change the phase and amplitude in each segment (see details in Sec. \ref{output-data_postprocessing}).

\subsection{Time window to subtract glitches}\label{apx:time_window_subtract_glitches}

\begin{figure}[!ht]
    \begin{minipage}{0.5\hsize}
  \begin{center}
   \includegraphics[width=1\linewidth]{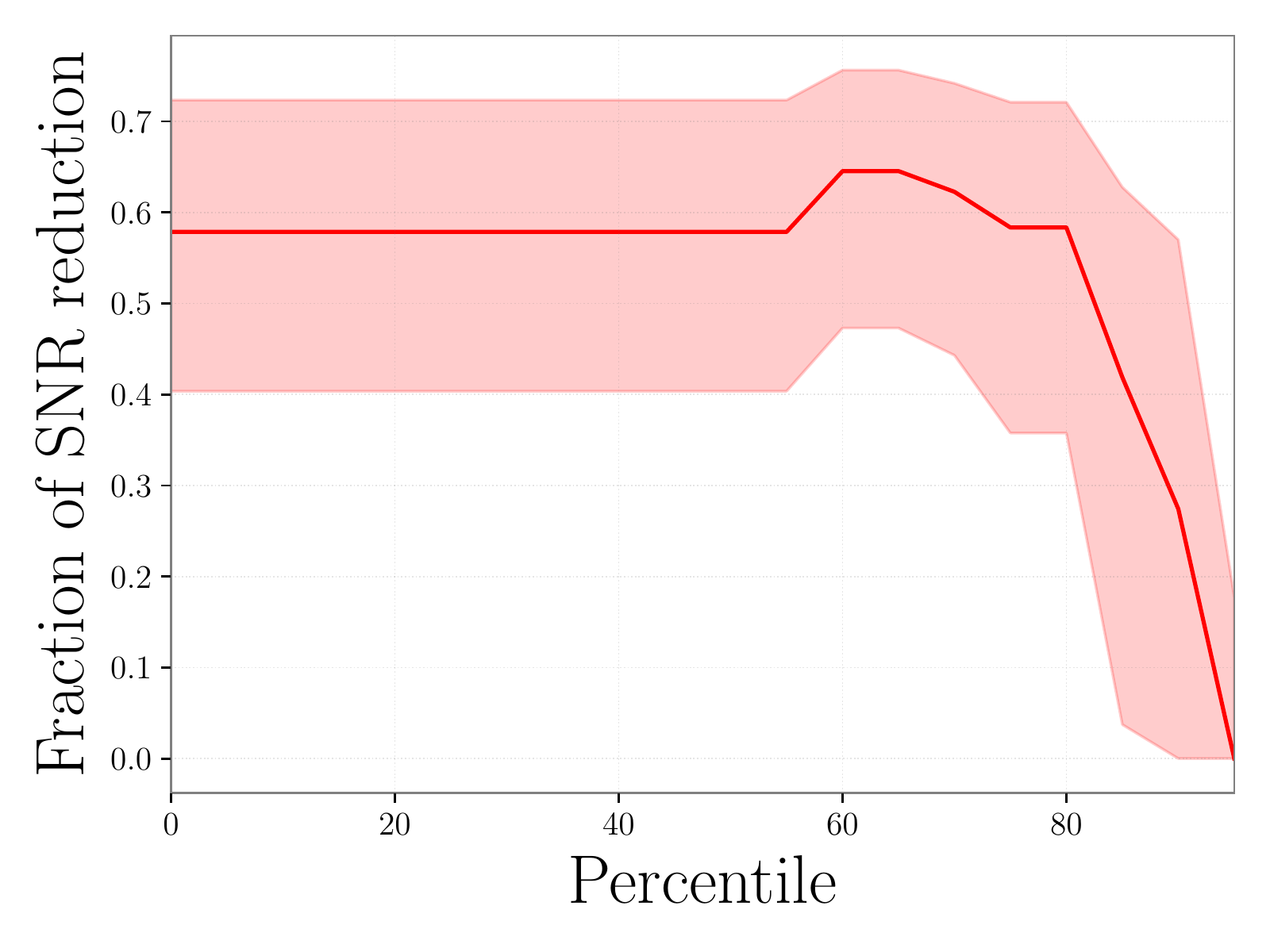}
  \end{center}
    \end{minipage}
    \begin{minipage}{0.5\hsize}
  \begin{center}
   \includegraphics[width=1\linewidth]{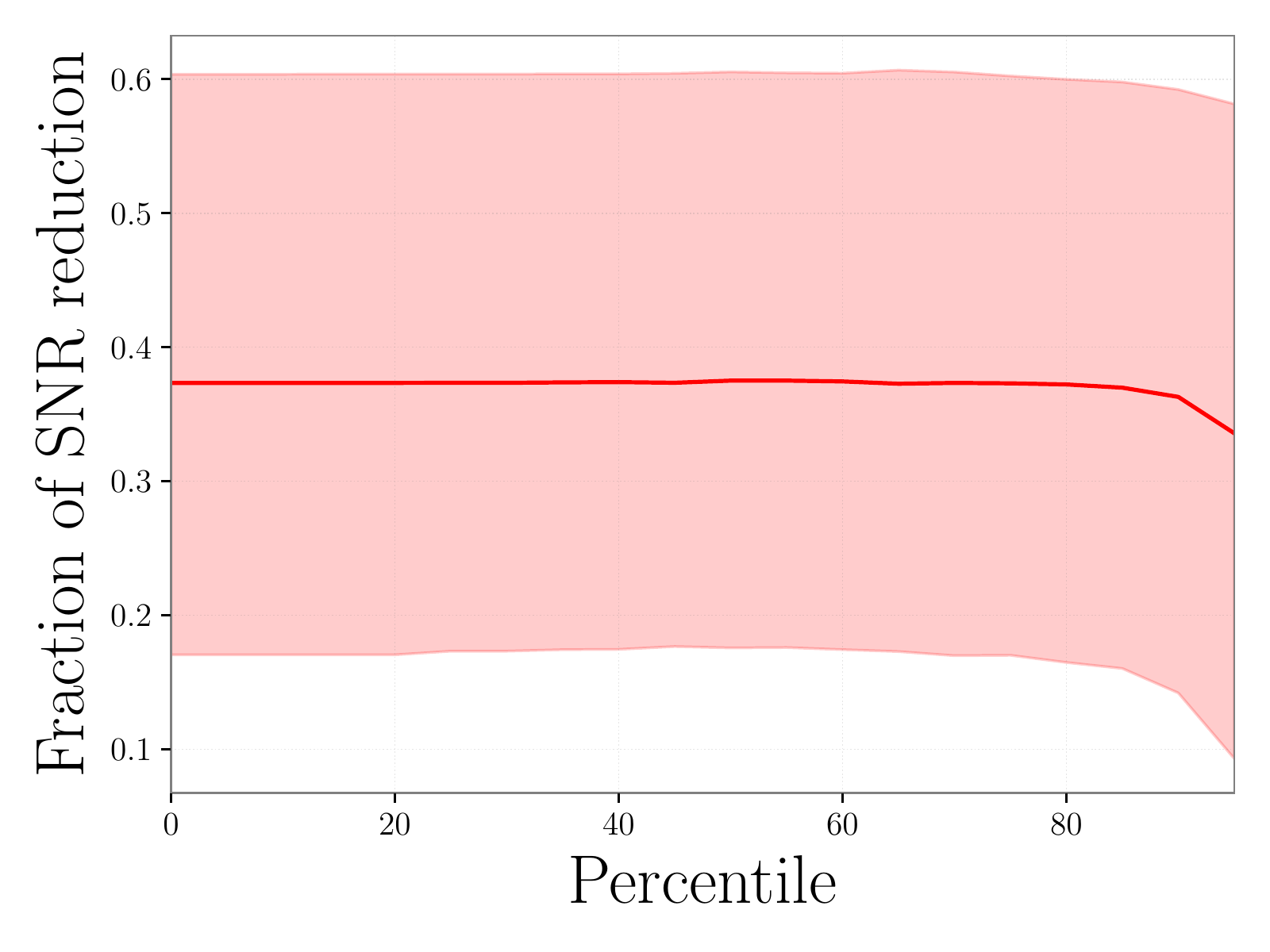}
  \end{center}
    \end{minipage}
    \caption[Fraction of \ac{SNR} reduction as a function of thresholds for the glitch present portions.]%
    {Fraction of \ac{SNR} reduction as a function of percentile as thresholds for the glitch present portions for {\it Scattered light} (left) and {\it Extremely loud} (right) glitches. The solid curves denote the median values and the shaded areas denote 1-$\sigma$ percentiles.}
\label{fig:fnr_percentile}
\end{figure}

Applying the \ac{FGL} transform to the estimated \ac{mSTFT} from the trained network, we obtain the estimated glitch waveforms. Using the estimated glitch waveforms, we subtract glitches from the data. In the glitch-subtraction process, we determine portions of the data containing g glitches based on the estimated glitch waveforms. We only subtract the data in portions containing glitches and use the original data in the rest portions without the glitch subtraction. To determine the data portions, we use the estimated glitch waveforms in the testing set with sample sizes of 678 and 1233 for {\it Scattered light} and {\it Extremely loud} glitches. We first calculate the absolute values of the estimated glitch waveforms and then smooth the curve. We consider data portions to be the regions where the smoothed curves are above thresholds. Smaller thresholds make the data portions to be larger so that larger data are used in the subtraction process, where larger fractions of the data portions have no glitches and no need to be subtracted. Larger thresholds make the data portions to be smaller so that only small fractions of glitches are subtracted. We consider various percentiles of the absolute value of the estimated glitch waveforms as thresholds. Figure \ref{fig:fnr_percentile} shows the variation of \acp{FNR} due to the choices of percentiles. We find that the peak of median \acp{FNR} are obtained with and 60 and 55 percentiles for {\it Scattered light} and {\it Extremely loud}, respectively. Below the peak values, values of \acp{FNR} are compatible within the 1-$\sigma$ uncertainty because the regions with the absence of glitches are not similar to the corresponding portions of the estimated glitch waveform other than the Gaussian fluctuations, causing no glitches to be subtracted subtraction in that regions. Larger percentiles corresponding to larger thresholds let only small fractions of glitches be subtracted, causing smaller values of $\ac{FNR}$. Threshold values used in Sec. \ref{scattered_light_result} and Sec. \ref{extremely_loud_results} are 70 percentile and 90 percentile for {\it Scattered light} and {\it Extremely loud}, respectively, whose values of \ac{FNR} are smaller than the peak median values by only $\sim 3$\%.    

\setcounter{section}{1}

\section*{Acknowledgements}

K.M. is supported by the U.S.\ National Science Foundation grant PHY-1921006. The author would like to thank their LIGO Scientific Collaboration and Virgo Collaboration colleagues for their help and useful comments, in particular Yanyan Zheng, Stuart B. Anderson, Duncan MacLeod and Marco Cavaglia. The objective was discussed in the group meeting at the Missouri University of Science and Technology. The author is grateful for computational resources provided by the LIGO Laboratory and supported by the U.S.\ National Science Foundation Grants PHY-0757058 and PHY-0823459, as well as a service of the LIGO Laboratory, the LIGO Scientific Collaboration and the Virgo Collaboration. LIGO was constructed and is operated by the California Institute of Technology and Massachusetts Institute of Technology with funding from the U.S.\ National Science Foundation under grant PHY-0757058. Virgo is funded by the French Centre National de la Recherche Scientifique (CNRS), the Italian Istituto Nazionale di Fisica Nucleare (INFN) and the Dutch Nikhef, with contributions by Polish and Hungarian institutes. Our method is made use of python packages including \textsc{PyTorch} \cite{NEURIPS2019_9015}, \textsc{Scipy} \cite{2020SciPy-NMeth}, \textsc{GWpy}  \cite{duncan_macleod_2020_4301851}, \textsc{Pandas} \cite{reback2020pandas}, \textsc{Matplotlib} \cite{Hunter:2007}, \textsc{nds2utils} \cite{nds2utils}, and \textsc{scikit-learn} \cite{scikit-learn}. This manuscript has been assigned LIGO Document Control Center number LIGO-P2100159.

\section*{References}
\bibliography{BibFile}

\providecommand{\newblock}{}
\begin{thebibliography}{10}
\expandafter\ifx\csname url\endcsname\relax
  \def\url#1{{\tt #1}}\fi
\expandafter\ifx\csname urlprefix\endcsname\relax\def\urlprefix{URL }\fi
\providecommand{\eprint}[2][]{\url{#2}}

\bibitem{Abbott:2016GW150914}
Abbott B~P {\em et~al.\/} (Virgo, LIGO Scientific) 2016 {\em Phys. Rev.
  Lett.\/} {\bf 116} 061102 (\textit{Preprint} \eprint{1602.03837})

\bibitem{GWTC-1}
Abbott B~P {\em et~al.\/} (LIGO Scientific Collaboration and Virgo
  Collaboration) 2019 {\em Phys. Rev. X\/} {\bf 9}(3) 031040
  \urlprefix\url{https://link.aps.org/doi/10.1103/PhysRevX.9.031040}

\bibitem{Abbott:2020niy}
Abbott R {\em et~al.\/} (LIGO Scientific, Virgo) 2021 {\em Phys. Rev. X\/} {\bf
  11} 021053 (\textit{Preprint} \eprint{2010.14527})

\bibitem{LIGOScientific:2020stg}
Abbott R {\em et~al.\/} (LIGO Scientific, Virgo) 2020 {\em Phys. Rev. D\/} {\bf
  102} 043015 (\textit{Preprint} \eprint{2004.08342})

\bibitem{PhysRevLett.125.101102}
Abbott R {\em et~al.\/} (LIGO Scientific Collaboration and Virgo Collaboration)
  2020 {\em Phys. Rev. Lett.\/} {\bf 125}(10) 101102
  \urlprefix\url{https://link.aps.org/doi/10.1103/PhysRevLett.125.101102}

\bibitem{Abbott_2020}
Abbott R {\em et~al.\/} 2020 {\em The Astrophysical Journal\/} {\bf 900} L13
  \urlprefix\url{https://doi.org/10.3847%2F2041-8213%2Faba493}

\bibitem{Abbott:2020khf}
Abbott R {\em et~al.\/} (LIGO Scientific, Virgo) 2020 {\em Astrophys. J.\/}
  {\bf 896} L44 (\textit{Preprint} \eprint{2006.12611})

\bibitem{advLigo2015}
Aasi J {\em et~al.\/} (LIGO Scientific) 2015 {\em Class. Quant. Grav.\/} {\bf
  32} 074001 (\textit{Preprint} \eprint{1411.4547})
  \urlprefix\url{http://stacks.iop.org/0264-9381/32/i=7/a=074001}

\bibitem{advVirgo2015}
Acernese F {\em et~al.\/} (VIRGO) 2015 {\em Class. Quant. Grav.\/} {\bf 32}
  024001 (\textit{Preprint} \eprint{1408.3978})
  \urlprefix\url{http://stacks.iop.org/0264-9381/32/i=2/a=024001}

\bibitem{Luck:2010rt}
Luck H {\em et~al.\/} 2010 {\em J. Phys. Conf. Ser.\/} {\bf 228} 012012
  (\textit{Preprint} \eprint{1004.0339})

\bibitem{Akutsu:2020his}
Akutsu T {\em et~al.\/} 2020 {\em Progress of Theoretical and Experimental
  Physics\/} {\bf 2021} ISSN 2050-3911 05A101 (\textit{Preprint}
  \eprint{https://academic.oup.com/ptep/article-pdf/2021/5/05A101/37974994/ptaa125.pdf})
  \urlprefix\url{https://doi.org/10.1093/ptep/ptaa125}

\bibitem{LIGO-India}
Iyer B {\em et~al.\/} 2011  (\textit{Preprint} \eprint{LIGO-M1100296})
  \urlprefix\url{https://dcc.ligo.org/LIGO-M1100296/public}

\bibitem{Riles:2012yw}
Riles K 2013 {\em Prog. Part. Nucl. Phys.\/} {\bf 68} 1--54 (\textit{Preprint}
  \eprint{1209.0667})

\bibitem{TheLIGOScientific:2014jea}
Aasi J {\em et~al.\/} (LIGO Scientific) 2015 {\em Class. Quant. Grav.\/} {\bf
  32} 074001 (\textit{Preprint} \eprint{1411.4547})

\bibitem{TheLIGOScientific:2016zmo}
Abbott B~P {\em et~al.\/} (LIGO Scientific, Virgo) 2016 {\em Class. Quant.
  Grav.\/} {\bf 33} 134001 (\textit{Preprint} \eprint{1602.03844})

\bibitem{Ormiston:2020ele}
Ormiston R, Nguyen T, Coughlin M, Adhikari R~X and Katsavounidis E 2020 {\em
  Phys. Rev. Res.\/} {\bf 2} 033066 (\textit{Preprint} \eprint{2005.06534})

\bibitem{Driggers:2018gii}
Driggers J {\em et~al.\/} (LIGO Scientific) 2019 {\em Phys. Rev. D\/} {\bf 99}
  042001 (\textit{Preprint} \eprint{1806.00532})

\bibitem{Kwee:12}
Kwee P, Bogan C, Danzmann K, Frede M, Kim H, King P, P\"{o}ld J, Puncken O,
  Savage R~L, Seifert F, Wessels P, Winkelmann L and Willke B 2012 {\em Opt.
  Express\/} {\bf 20} 10617--10634
  \urlprefix\url{http://www.opticsexpress.org/abstract.cfm?URI=oe-20-10-10617}

\bibitem{Robert:2016gii}
Schofield R 2016
  \urlprefix\url{https://alog.ligo-wa.caltech.edu/aLOG/index.php?callRep=30290}

\bibitem{Davis:2018yrz}
Davis D, Massinger T~J, Lundgren A~P, Driggers J~C, Urban A~L and Nuttall L~K
  2019 {\em Class. Quant. Grav.\/} {\bf 36} 055011 (\textit{Preprint}
  \eprint{1809.05348})

\bibitem{Tiwari:2015ofa}
Tiwari V {\em et~al.\/} 2015 {\em Class. Quant. Grav.\/} {\bf 32} 165014
  (\textit{Preprint} \eprint{1503.07476})

\bibitem{Meadors:2013lja}
Meadors G~D, Kawabe K and Riles K 2014 {\em Class. Quant. Grav.\/} {\bf 31}
  105014 (\textit{Preprint} \eprint{1311.6835})

\bibitem{Mukund:2020lby}
Mukund N {\em et~al.\/} 2020 {\em Phys. Rev. D\/} {\bf 101} 102006
  (\textit{Preprint} \eprint{2001.00242})

\bibitem{Vajente:2019ycy}
Vajente G, Huang Y, Isi M, Driggers J~C, Kissel J~S, Szczepanczyk M~J and
  Vitale S 2020 {\em Phys. Rev. D\/} {\bf 101} 042003 (\textit{Preprint}
  \eprint{1911.09083})

\bibitem{Davis:2021ecd}
Davis D {\em et~al.\/} (LIGO) 2021 {\em Class. Quant. Grav.\/} {\bf 38} 135014
  (\textit{Preprint} \eprint{2101.11673})

\bibitem{Godwin:2020weu}
Godwin P {\em et~al.\/} 2020  (\textit{Preprint} \eprint{2010.15282})

\bibitem{Isogai:2010zz(UPV)}
Isogai T (LIGO Scientific, Virgo) 2010 {\em J. Phys. Conf. Ser.\/} {\bf 243}
  012005

\bibitem{Smith:2011an(hVETO)}
Smith J~R, Abbott T, Hirose E, Leroy N, Macleod D, McIver J, Saulson P and
  Shawhan P 2011 {\em Class. Quant. Grav.\/} {\bf 28} 235005 (\textit{Preprint}
  \eprint{1107.2948})

\bibitem{Essick:2020cyv}
Essick R, Mo G and Katsavounidis E 2021 {\em Phys. Rev. D\/} {\bf 103} 042003
  (\textit{Preprint} \eprint{2011.13787})

\bibitem{mogushi2021application}
Mogushi K 2021 {\em Classical and Quantum Gravity\/}
  \urlprefix\url{http://iopscience.iop.org/article/10.1088/1361-6382/ac08a7}

\bibitem{Biswas:2013wfa}
Biswas R {\em et~al.\/} 2013 {\em Phys. Rev. D\/} {\bf 88} 062003
  (\textit{Preprint} \eprint{1303.6984})

\bibitem{Essick:2020qpo}
Essick R, Godwin P, Hanna C, Blackburn L and Katsavounidis E 2020 {\em Machine
  Learning: Science and Technology\/} {\bf 2} 015004
  \urlprefix\url{https://doi.org/10.1088/2632-2153/abab5f}

\bibitem{Was:2020ziy}
Was M, Gouaty R and Bonnand R 2021 {\em Class. Quant. Grav.\/} {\bf 38} 075020
  (\textit{Preprint} \eprint{2011.03539})

\bibitem{Cuoco:2020ogp}
Cuoco E {\em et~al.\/} 2021 {\em Mach. Learn. Sci. Tech.\/} {\bf 2} 011002
  (\textit{Preprint} \eprint{2005.03745})

\bibitem{Mukund:2016thr}
Mukund N, Abraham S, Kandhasamy S, Mitra S and Philip N~S 2017 {\em Phys. Rev.
  D\/} {\bf 95} 104059 (\textit{Preprint} \eprint{1609.07259})

\bibitem{Soni:2021cjy}
Soni S {\em et~al.\/} 2021  (\textit{Preprint} \eprint{2103.12104})

\bibitem{Zevin:2016qwy}
Zevin M {\em et~al.\/} 2017 {\em Class. Quant. Grav.\/} {\bf 34} 064003
  (\textit{Preprint} \eprint{1611.04596})

\bibitem{Mogushi:2021cpw}
Mogushi K, Quitzow-James R, Cavaglia M, Kulkarni S and Hayes F 2021 {\em
  Machine Learning: Science and Technology\/}
  \urlprefix\url{http://iopscience.iop.org/article/10.1088/2632-2153/abea69}

\bibitem{1164317}
{Griffin} D and {Jae Lim} 1984 {\em IEEE Transactions on Acoustics, Speech, and
  Signal Processing\/} {\bf 32} 236--243

\bibitem{6701851}
Perraudin N, Balazs P and Søndergaard P~L 2013 A fast griffin-lim algorithm
  {\em 2013 IEEE Workshop on Applications of Signal Processing to Audio and
  Acoustics\/} pp 1--4

\bibitem{ROBINET2020100620}
Robinet F, Arnaud N, Leroy N, Lundgren A, Macleod D and McIver J 2020 {\em
  SoftwareX\/}  100620 ISSN 2352-7110
  \urlprefix\url{http://www.sciencedirect.com/science/article/pii/S2352711020303332}

\bibitem{essenwanger1986elements}
Essenwanger O 1986 {\em Elements of Statistical Analysis\/} General climatology
  (Elsevier) ISBN 9780444424266
  \urlprefix\url{https://books.google.com/books?id=P6Y4vgAACAAJ}

\bibitem{duncan_macleod_2020_4301851}
Macleod D, Urban A~L, Coughlin S, Massinger T, Pitkin M, rngeorge, paulaltin,
  Areeda J, Singer L, Quintero E, Leinweber K and Badger T~G 2020 gwpy/gwpy:
  2.0.2 \urlprefix\url{https://doi.org/10.5281/zenodo.4301851}

\bibitem{9174990}
{Cheuk} K~W, {Anderson} H, {Agres} K and {Herremans} D 2020 {\em IEEE Access\/}
  {\bf 8} 161981--162003

\bibitem{bank2021autoencoders}
Bank D, Koenigstein N and Giryes R 2021 Autoencoders (\textit{Preprint}
  \eprint{2003.05991})

\bibitem{RUMELHART1988399}
Rumelhart D, HINTON G and WILLIAMS R 1988 Learning internal representations by
  error propagation {\em Readings in Cognitive Science\/} ed Collins A and
  Smith E~E (Morgan Kaufmann) pp 399--421 ISBN 978-1-4832-1446-7
  \urlprefix\url{https://www.sciencedirect.com/science/article/pii/B9781483214467500352}

\bibitem{pmlr-v27-baldi12a}
Baldi P 2012 Autoencoders, unsupervised learning, and deep architectures {\em
  Proceedings of ICML Workshop on Unsupervised and Transfer Learning\/} ({\em
  Proceedings of Machine Learning Research\/} vol~27) ed Guyon I, Dror G,
  Lemaire V, Taylor G and Silver D (Bellevue, Washington, USA: PMLR) pp 37--49
  \urlprefix\url{http://proceedings.mlr.press/v27/baldi12a.html}

\bibitem{10.1007/978-3-642-21735-7_7}
Masci J, Meier U, Cire{\c{s}}an D and Schmidhuber J 2011 Stacked convolutional
  auto-encoders for hierarchical feature extraction {\em Artificial Neural
  Networks and Machine Learning -- ICANN 2011\/} ed Honkela T, Duch W, Girolami
  M and Kaski S (Berlin, Heidelberg: Springer Berlin Heidelberg) pp 52--59 ISBN
  978-3-642-21735-7

\bibitem{JMLR:v11:vincent10a}
Vincent P, Larochelle H, Lajoie I, Bengio Y and Manzagol P~A 2010 {\em Journal
  of Machine Learning Research\/} {\bf 11} 3371--3408
  \urlprefix\url{http://jmlr.org/papers/v11/vincent10a.html}

\bibitem{ioffe2015batch}
Ioffe S and Szegedy C 2015 {\em ICML\/} {\bf 37} 448--456 (\textit{Preprint}
  \eprint{1502.03167})

\bibitem{Yamaguchi1990ANN}
Yamaguchi K, Sakamoto K, Akabane T and Fujimoto Y 1990 A neural network for
  speaker-independent isolated word recognition {\em ICSLP\/}

\bibitem{Hahnloser2000}
Hahnloser R~H~R, Sarpeshkar R, Mahowald M~A, Douglas R~J and Seung H~S 2000
  {\em Nature\/} {\bf 405} 947--51
  \urlprefix\url{https://doi.org/10.1038/35016072}

\bibitem{Nair2010RectifiedLU}
Nair V and Hinton G~E 2010 Rectified linear units improve restricted boltzmann
  machines {\em Proceedings of the 27th International Conference on
  International Conference on Machine Learning\/} ICML'10 (Madison, WI, USA:
  Omnipress) p 807–814 ISBN 9781605589077

\bibitem{adam}
Kingma D and Ba J 2014 {\em International Conference on Learning
  Representations\/}

\bibitem{Klimenko:2008fu}
Klimenko S, Yakushin I, Mercer A and Mitselmakher G 2008 {\em Class. Quant.
  Grav.\/} {\bf 25} 114029 (\textit{Preprint} \eprint{0802.3232})

\bibitem{Klimenko:2015ypf}
Klimenko S {\em et~al.\/} 2016 {\em Phys. Rev. D\/} {\bf 93} 042004
  (\textit{Preprint} \eprint{1511.05999})

\bibitem{Usman:2015kfa}
Usman S~A {\em et~al.\/} 2016 {\em Class. Quant. Grav.\/} {\bf 33} 215004
  (\textit{Preprint} \eprint{1508.02357})

\bibitem{Soni:2020rbu}
Soni S {\em et~al.\/} (LIGO) 2020 {\em Class. Quant. Grav.\/} {\bf 38} 025016
  (\textit{Preprint} \eprint{2007.14876})

\bibitem{data_aug1}
Shorten C and Khoshgoftaar T~M 2019 {\em Journal of Big Data\/} {\bf 6} 60

\bibitem{Perez2017TheEO}
Perez L and Wang J 2017 {\em ArXiv\/} {\bf abs/1712.04621}

\bibitem{Lemley2017SmartAL}
Lemley J, Bazrafkan S and Corcoran P 2017 {\em IEEE Access\/} {\bf 5}
  5858--5869

\bibitem{Was:2012zq}
Was M, Sutton P~J, Jones G and Leonor I 2012 {\em Phys. Rev. D\/} {\bf 86}
  022003 (\textit{Preprint} \eprint{1201.5599})

\bibitem{Abbott:2009kk}
Abbott B~P {\em et~al.\/} (VIRGO) 2010 {\em Astrophys. J.\/} {\bf 715}
  1438--1452 (\textit{Preprint} \eprint{0908.3824})

\bibitem{AdrianMartinez:2012tf}
Adrian-Martinez S {\em et~al.\/} (LIGO Scientific, VIRGO) 2013 {\em JCAP\/}
  {\bf 06} 008 (\textit{Preprint} \eprint{1205.3018})

\bibitem{Abadie:2012bz}
Abadie J {\em et~al.\/} (LIGO Scientific) 2012 {\em Astrophys. J.\/} {\bf 755}
  2 (\textit{Preprint} \eprint{1201.4413})

\bibitem{Briggs:2012ce}
Abadie J {\em et~al.\/} (LIGO Scientific) 2012 {\em Astrophys. J.\/} {\bf 760}
  12 (\textit{Preprint} \eprint{1205.2216})

\bibitem{Abbott:2008zzb}
Abbott B {\em et~al.\/} (LIGO Scientific) 2008 {\em Phys. Rev. D\/} {\bf 77}
  062004 (\textit{Preprint} \eprint{0709.0766})

\bibitem{Abbott:2020mjq}
Abbott R {\em et~al.\/} (LIGO Scientific, Virgo) 2020 {\em Astrophys. J.
  Lett.\/} {\bf 900} L13 (\textit{Preprint} \eprint{2009.01190})

\bibitem{ratnaparkhi-1996-maximum}
Ratnaparkhi A 1996 A maximum entropy model for part-of-speech tagging {\em
  Conference on Empirical Methods in Natural Language Processing\/}
  \urlprefix\url{https://www.aclweb.org/anthology/W96-0213}

\bibitem{reynar-ratnaparkhi-1997-maximum}
Reynar J~C and Ratnaparkhi A 1997 A maximum entropy approach to identifying
  sentence boundaries {\em Fifth Conference on Applied Natural Language
  Processing\/} (Washington, DC, USA: Association for Computational
  Linguistics) pp 16--19
  \urlprefix\url{https://www.aclweb.org/anthology/A97-1004}

\bibitem{kolmogorov_1951}
Massey F~J 1951 {\em Journal of the American Statistical Association\/} {\bf
  46} 68--78

\bibitem{IMRPhenomD}
Khan S, Husa S, Hannam M, Ohme F, P\"urrer M, Forteza X~J and Boh\'e A 2016
  {\em Phys. Rev. D\/} {\bf 93}(4) 044007
  \urlprefix\url{https://link.aps.org/doi/10.1103/PhysRevD.93.044007}

\bibitem{alex_nitz_2020_3904502}
Nitz A, Harry I, Brown D, Biwer C~M, Willis J, Canton T~D, Capano C, Pekowsky
  L, Dent T, Williamson A~R, Davies G~S, De S, Cabero M, Machenschalk B, Kumar
  P, Reyes S, Macleod D, Pannarale F, dfinstad, Massinger T, Tápai M, Singer
  L, Khan S, Fairhurst S, Kumar S, Nielsen A, shasvath, Dorrington I, Lenon A
  and Gabbard H 2020 gwastro/pycbc: Pycbc release 1.16.4
  \urlprefix\url{https://doi.org/10.5281/zenodo.3904502}

\bibitem{PhysRevD.46.5236}
Finn L~S 1992 {\em Phys. Rev. D\/} {\bf 46}(12) 5236--5249
  \urlprefix\url{https://link.aps.org/doi/10.1103/PhysRevD.46.5236}

\bibitem{Brown:2004vh}
Brown D~A 2004 {\em {Searching for gravitational radiation from binary black
  hole MACHOs in the galactic halo}\/} Other thesis (\textit{Preprint}
  \eprint{0705.1514})

\bibitem{NEURIPS2019_9015}
Paszke A, Gross S, Massa F, Lerer A, Bradbury J, Chanan G, Killeen T, Lin Z,
  Gimelshein N, Antiga L, Desmaison A, Kopf A, Yang E, DeVito Z, Raison M,
  Tejani A, Chilamkurthy S, Steiner B, Fang L, Bai J and Chintala S 2019
  Pytorch: An imperative style, high-performance deep learning library {\em
  Advances in Neural Information Processing Systems 32\/} ed Wallach H,
  Larochelle H, Beygelzimer A, d~Alch\'e-Buc F, Fox E and Garnett R (Curran
  Associates, Inc.) pp 8024--8035
  \urlprefix\url{http://papers.neurips.cc/paper/9015-pytorch-an-imperative-style-high-performance-deep-learning-library.pdf}

\bibitem{2020SciPy-NMeth}
Virtanen P, Gommers R, Oliphant T~E, Haberland M, Reddy T, Cournapeau D,
  Burovski E, Peterson P, Weckesser W, Bright J, {van der Walt} S~J, Brett M,
  Wilson J, Millman K~J, Mayorov N, Nelson A~R~J, Jones E, Kern R, Larson E,
  Carey C~J, Polat {\.I}, Feng Y, Moore E~W, {VanderPlas} J, Laxalde D,
  Perktold J, Cimrman R, Henriksen I, Quintero E~A, Harris C~R, Archibald A~M,
  Ribeiro A~H, Pedregosa F, {van Mulbregt} P and {SciPy 10 Contributors} 2020
  {\em Nature Methods\/} {\bf 17} 261--272

\bibitem{reback2020pandas}
Reback J, jbrockmendel, McKinney W, den Bossche J~V, Augspurger T, Cloud P,
  Hawkins S, gfyoung, Sinhrks, Roeschke M, Klein A, Petersen T, Tratner J, She
  C, Ayd W, Hoefler P, Naveh S, Garcia M, Schendel J, Hayden A, Saxton D,
  Gorelli M~E, Shadrach R, Jancauskas V, McMaster A, Battiston P, Li F, Seabold
  S, Dong K and chris b1 2021 pandas-dev/pandas: Pandas 1.3.0rc1
  \urlprefix\url{https://doi.org/10.5281/zenodo.4940217}

\bibitem{Hunter:2007}
Hunter J~D 2007 {\em Computing in Science \& Engineering\/} {\bf 9} 90--95

\bibitem{nds2utils}
Cahillane C 2020  Https://pypi.org/project/nds2utils/

\bibitem{scikit-learn}
Pedregosa F, Varoquaux G, Gramfort A, Michel V, Thirion B, Grisel O, Blondel M,
  Prettenhofer P, Weiss R, Dubourg V, Vanderplas J, Passos A, Cournapeau D,
  Brucher M, Perrot M and Duchesnay E 2011 {\em Journal of Machine Learning
  Research\/} {\bf 12} 2825--2830

\end{thebibliography}

\end{document}